%% file: version4.tex
\documentclass[12pt]{article}
\usepackage{latexsym}
\usepackage{amsmath}
\usepackage{epsfig,graphics}

\newcommand{\be}{\begin{equation}}
\newcommand{\ee}{\end{equation}}

\newlength{\figsize}
\begin{document}

\begin{titlepage}

\vspace*{0.60in}

\begin{center}
{\large\bf Strong to weak coupling transitions of SU(N) gauge theories
\\ in 2+1 dimensions\\ }
\vspace*{0.65in}
{Francis Bursa and Michael Teper\\
\vspace*{.25in}
Rudolf Peierls Centre for Theoretical Physics, University of Oxford,\\
1 Keble Road, Oxford OX1 3NP, U.K.
}
\end{center}

\vspace*{0.50in}

\begin{center}
{\bf Abstract}
\end{center}

We investigate strong-to-weak coupling transitions in 
$D=2+1$ $SU(N\to\infty)$ gauge theories,
by simulating the corresponding lattice theories
with a Wilson plaquette action. We find that there is a 
strong-to-weak coupling cross-over in the lattice theory
that appears to become a third-order phase transition at $N=\infty$,
in a manner that is essentially identical to the Gross-Witten 
transition in the $D=1+1$ $SU(\infty)$ lattice gauge theory. There 
is evidence of an additional second order transition developing 
at $N=\infty$ at approximately the same coupling, which is connected 
with $Z_N$ monopoles (instantons), thus making it an analogue of the 
first order bulk transition that occurs in $D=3+1$ lattice gauge 
theories for $N\geq 5$. 
We show that as the lattice spacing is reduced, 
the $N=\infty$ gauge theory on a finite 3-torus suffers a sequence of
(apparently) first-order $Z_N$ symmetry breaking transitions
associated with each of the tori (ordered by size). We discuss
how these transitions can be understood in terms of a sequence
of deconfining transitions on ever-more dimensionally reduced 
gauge theories.
We investigate whether the trace of the Wilson loop has a 
non-analyticity in the coupling at some critical area, but find 
no evidence for this. 
However we do find that, just as one can prove occurs in $D=1+1$, 
the eigenvalue density of a Wilson loop forms a gap at $N=\infty$ 
at a critical value of its trace. The physical
implications of this subtle non-analyticity are unclear. 
This gap formation is in fact a special case of a remarkable
similarity between the eigenvalue spectra of Wilson loops
in $D=1+1$ and $D=2+1$ (and indeed $D=3+1$): for the same value
of the trace, the eigenvalue spectra are nearly identical.
This holds for finite as well as infinite $N$; irrespective
of the Wilson loop size in lattice units;
and for Polyakov as well as Wilson loops.

\end{titlepage}

\setcounter{page}{1}
\newpage
\pagestyle{plain}

\section{Introduction}
\label{section_intro}

A phase transition is associated with a singularity in the partition 
function, and so requires an infinite number of degrees of freedom.
Usually that requires an infinite volume. One of the peculiarities
of large--$N$ field theories is that one can have phase transitions
on finite, or even infinitesimal, volumes at $N=\infty$ because in
this case we have an infinite number of degrees of freedom at each 
point in space. The classic example in the context of gauge field
theories is the Gross-Witten
transition 
\cite{GW}
that occurs in the $D=1+1$ $SU(\infty)$ lattice gauge theory (with the
standard Wilson action). In this case the theory is analytically 
soluble and one finds a third order phase transition at $N=\infty$
\cite{GW}
at a value of the bare coupling that separates the strong and weak 
coupling regions. 
The theoretical and practical interest of such phase 
transitions, particularly in $D=3+1$, has recently been reviewed in
\cite{NNlat05}.

In $D=3+1$ $SU(N)$ gauge theories numerical studies reveal the
existence for $N\geq 5$ of a first order `bulk' transition separating 
the weak and strong coupling regions 
\cite{Campo98,OxG01,OxT05}.
One also finds that the deconfinement transition, which is
first order for $N\geq 3$, becomes sharper on smaller volumes 
as $N$ increases suggesting
\cite{OxT03}
that here too one will have a phase transition on a finite volume 
at $N=\infty$. Indeed there appears to be a whole hierarchy of 
finite volume phase transitions at $N=\infty$
\cite{NN03,NNlat05}
which are, we shall argue below, related to the deconfinement
transition. 

These are all in some sense strong to weak coupling
transitions, and this has led to the conjecture
\cite{NNect04,NNlat05}
that Wilson loops in general will show such $N=\infty$ transitions
as the lattice spacing decreases, when the physical size of the
loop passes some critical value. 
Such a transition in $D=3+1$ could
have interesting implications for dual string approaches to
large-$N$ gauge theories, as well as providing a natural 
explanation for the rapid crossover between perturbative and 
non-perturbative physics that is  observed in the strong interactions
\cite{GW,MTect04,NNlat05}.
In fact it is known 
\cite{DurOle,BasGriVian}
that in the $N=\infty$ $D=1+1$ continuum theory
the eigenvalue spectrum of a Wilson loop suffers a
non-analyticity for a critical area that is very similar to 
that of the plaquette at the Gross-Witten transition.
However, in contrast to the Gross-Witten transition, there is 
no accompanying non-analyticity in the trace of the loop and it 
is unclear what, if any, are its physical implications.

In this paper we investigate the existence of such phase transitions
in $D=2+1$ $SU(N)$ gauge theories, as a step towards 
a unified understanding of these phenomena in all dimensions.

In the next Section we briefly describe the $SU(N)$ lattice gauge 
theory and how we simulate it. There follows a longer section in 
which we review in more detail what is known about the large-$N$
transitions and, in some cases, we extend the analysis. (We are 
interested in transitions that may be cross-overs or actual 
phase transitions, and when we refer to `transitions' in this
paper it may be either one of these.) Having established the
background, we move on to our detailed numerical results.
Our conclusions contain a summary of our main results.

\section{SU($N$) gauge theory on the lattice}
\label{section_lattice}

We discretise Euclidean space-time to a periodic cubic lattice
with lattice spacing $a$ and size $L_0\times L_1\times L_2$ in 
lattice units. We assign $SU(N)$ matrices, $U_l$, to the links 
$l$ of the lattice. (We sometimes write  $U_l$ as $U_\mu(n)$ where
the link $l$ emanates in the positive $\mu$ direction from
the site $n$.)  We use the standard Wilson plaquette action
\begin{equation}
S=\beta \sum_{p} \{1-\frac{1}{N}\mathrm{Re}\mathrm{Tr}U_p\}
\label{eqn_action}
\end{equation}
where $U_p$ is the ordered product of the $SU(N)$ link matrices 
around the boundary of the plaquette $p$. The partition function is 
\begin{equation}
Z= \int \prod_l dU_l \exp(- S) \qquad ;\qquad 
\lim_{a\to 0} \beta=\frac{2N}{ag^2}.
\label{eqn_Z}
\end{equation}
Exactly the same expression defines the lattice gauge theory
in $D=1+1$ and $D=3+1$ except that $\beta=2N/a^2g^2$  and
$\beta= 2N/g^2$ respectively. Eqn(\ref{eqn_Z}) also defines
the finite temperature partition function, if we choose
\begin{equation}
T=\frac{1}{aL_0} \qquad ;\qquad  L_1,L_2 \gg L_0.
\label{eqn_T}
\end{equation}
We simulate the above lattice theory using a conventional 
mixture of heat bath and over-relaxation steps applied to the 
$SU(2)$ subgroups of the $SU(N)$ link matrices.

It will sometimes be convenient to distinguish couplings,
inverse bare couplings and (critical) temperatures in 
different space-time dimensions, $D$, and we do so using 
subscripts or superscripts, e.g. $g^2_D$, $\beta_D$, $T^D_c$.
Where there is no ambiguity we will often omit such subscripts.

To obtain a smooth large $N$ limit we keep $g^2N$ fixed
\cite{thooft}.
It is therefore useful to define the bare 't Hooft coupling, 
$\lambda$, and the inverse bare 't Hooft coupling, $\gamma$,
\begin{equation}
\lambda = a g^2 N
 \qquad ,\qquad
\gamma = \frac{1}{\lambda} = \frac{\beta}{2N^2}.
\label{eqn_ggN}
\end{equation}
Various numerical calculations 
have confirmed that a smooth  $N\to\infty$ limit is indeed
obtained by keeping $\lambda$ fixed, both in $D=2+1$
\cite{Nd3}
and in $D=3+1$
\cite{OxG01,OxT03,OxG04}
and that to keep the cut-off $a$ fixed as  $N\to\infty$ one 
should keep $\gamma$ fixed.

A useful order parameter for finite volume phase transitions
is provided by taking the Polyakov loop, $l_\mu$, which is the 
ordered product of link matrices around the $\mu$-torus, and
averaging it over the space-time volume: 
\begin{equation}
{\bar l}_\mu
=
c_\mu \sum_{n_{\nu\neq\mu}}
\frac{1}{N} \mathrm{Tr} 
\left\{
\prod_{n_\mu=1}^{n_\mu=L_\mu}
U_\mu(n_0,n_1,n_2)
\right\}
\label{eqn_poly}
\end{equation}
where the normalisation is $c^{-1}_\mu = \prod_{\nu\neq\mu}L_\nu$.
When the system develops a non-zero value for 
$\langle{\bar l}_\mu\rangle$ this indicates
the spontaneous breaking of a global $Z_N$ symmetry 
associated with the $\mu$-torus. In particular such a symmetry
breaking occurs at the deconfining temperature, if the
$\mu$-torus defines the temperature $T$.

\section{Background}
\label{section_back}

\subsection{The `Gross-Witten' transition}
\label{subsection_GW}

By fixing gauge and making a change of variables, one can show
\cite{GW}
that the partition function of the $D=1+1$ $SU(N)$ lattice
gauge theory (with the Wilson plaquette action) factorises
into a product of integrals over $SU(N)$ matrices on the links
and the theory can be explicitly solved. One then finds a 
cross-over between weak and strong coupling that sharpens
with increasing $N$ into a third order phase transition at
$N=\infty$. In terms of the plaquette, $u_p = \mathrm{ReTr}U_p/N$,
this shows up in a change of functional behaviour
\begin{equation}
\langle u_p\rangle \stackrel{N\to\infty}{=} \begin{cases}
\frac{1}{\lambda}  & \lambda\geq 2, \\
1 - \frac{\lambda}{4}  &  \lambda\leq 2.
\end{cases}
\label{eqn_upGW}
\end{equation}
More detailed information about the behaviour of plaquettes and 
Wilson loops can be gained by considering not just their
traces but their eigenvalues. The eigenvalues of an $SU(N)$ matrix 
are just phases, $\lambda = \exp\{i\alpha\}$, and are 
gauge--invariant. (We also use $\lambda$ for the `t Hooft coupling:
which is intended should be clear from the context.)
As $\beta\to 0$ the eigenvalue distribution $\rho (\alpha )$ of 
a Wilson loop becomes uniform while as  $\beta\to\infty$ it 
becomes increasingly peaked around $\alpha=0$. As shown in
\cite{GW},
at the Gross--Witten transition a gap opens in the density
of plaquette eigenvalues: in the strongly--coupled phase
the eigenvalue density is non--zero for all angles
$-\pi \leq \alpha \leq \pi$, but in the weakly-coupled phase 
it is only non--zero in the range 
$-\alpha_c \leq \alpha \leq \alpha_c$, where $\alpha_c < \pi$
\cite{GW}. 

In $D=3+1$ it is known that at $N=\infty$
\cite{Campo98},
and indeed for $N\geq 5$
\cite{OxG01,OxT05},
there is a strong first order transition as $\beta$ is varied 
from strong to weak coupling. Calculations in progress
\cite{BurTepVai}
suggest that the plaquette eigenvalue distribution does indeed
show a gap formation at $N=\infty$ that is similar to the
$D=1+1$ Gross-Witten transition. However the first order
transition itself is usually believed to be a manifestation
of the phase structure one finds with a mixed adjoint-fundamental 
action as discussed below. This finite-$N$ phase transition `conceals'
any underlying $N=\infty$ Gross-Witten transition and makes the
latter hard to identify unambiguously.  

In $D=2+1$ there has been, as far as we are aware, no systematic
search for a Gross-Witten or `bulk' transition, and this is one
of the gaps that the present work intends to fill.

\subsection{Wilson loop transitions}
\label{subsection_WL}

The Gross-Witten transition involves the smallest possible
Wilson loop, the plaquette. On the weak coupling side the plaquette
can be calculated in terms of usual weak-coupling perturbation
theory; but this breaks down abruptly at the Gross-Witten transition, 
beyond which a strong coupling expansion becomes appropriate
\cite{GW}.
The coupling is the bare coupling and hence a coupling on the
length scale of the plaquette. Thus one might  interpret the transition
as saying that as one increases the length scale, there is a 
critical scale at which perturbation theory in the running coupling
will suddenly break down. 

One might imagine that this generalises to other Wilson loops: i.e.
when we scale up a Wilson loop, at some critical size, in 
`physical units', there is a non-analyticity. In fact precisely
such a scenario has been conjectured for $SU(N\to\infty)$ gauge
theories in $D=3+1$
\cite{NNect04,NNlat05}.
Unlike the lattice Gross-Witten transition, this would be a
property of the continuum theory.

Such a  non-analyticity does in fact occur for the $SU(N\to\infty)$
continuum theory in $D=1+1$
\cite{DurOle,BasGriVian}. 
The transition occurs at a fixed physical area 
\begin{equation}
A_{crit}=\frac{8}{g^2N}.
\label{eqn_d2Acrit}
\end{equation}
Very much larger Wilson loops have a flat 
eigenvalue spectrum $\rho(\alpha)$ which becomes peaked  
as $A \to A^{+}_{crit}$. As   $A$  decreases through $A_{crit}$
a gap appears in the spectrum near the extreme phases 
$\alpha = \pm \pi$. 
So for loops with $A < A^{+}_{crit}$ the eigenvalue density 
is only non-zero for 
$-\alpha_c \leq \alpha \leq \alpha_c$, where $\alpha_c < \pi$,
and $\alpha_c \to 0$ as $A\to 0$.
The non-analyticity at $A=A_{crit}$ is in fact more singular
than for the Gross-Witten transition, in that the derivative
$\partial{\rho}/\partial{\alpha}$ diverges at $\alpha_c=\pm\pi$.
However, unlike the Gross-Witten transition
this is not a phase transition: the partition function
is analytic. Moreover the trace of the Wilson loop,
and the traces of all powers of the Wilson loop,
remain analytic in the coupling. Thus it is unclear what if any
is the physical significance of this non-analyticity.

In this paper we shall investigate whether such a non-analyticity 
develops in  $D=2+1$ $SU(N)$ gauge theories and whether it is
accompanied by any non-analyticity of the trace. The implications
could be very interesting
\cite{MTect04}
and this makes a search in $D=2+1$ (and even more so in $D=3+1$
\cite{BurTepVai})
well worth while.

\subsection{Mixed actions}
\label{subsection_mixS}

In $D=3+1$ the strong-to-weak coupling transition occurs already at 
finite $N$. It is a cross-over for $N\leq 4$ and is first 
order for $N\geq 5$
\cite{OxG01,OxT05}. 
The conventional interpretation of this `bulk' transition 
proceeds by considering a generalised lattice action containing pieces 
in both fundamental and adjoint representations
\cite{Creutz}:
\begin{eqnarray}
S 
& = & 
\beta_f \sum_{p} \left\{1-\frac{1}{N}\mathrm{Re}\mathrm{Tr}_fU_p\right\}
+\beta^\prime_a\sum_{p} 
 \left\{1-\frac{1}{N^2-1}\mathrm{Tr}_a U_p\right\}
\nonumber\\
& = & 
\beta_f \sum_{p}  \left\{1-\frac{1}{N}\mathrm{Re}\mathrm{Tr}_f U_p\right\}
+\beta_a\sum_{p}  \left\{1-\frac{1}{N^2}
\mathrm{Tr}_f U^\dagger_p \mathrm{Tr}_f U_p\right\} 
\label{eqn_mixedS}
\end{eqnarray}
where we have used  $\mathrm{Tr}_a U_p = |\mathrm{Tr}_f U_p|^2 -1$.
By considering smooth fields one finds that at weak coupling, and
to leading order in $g^2$, one obtains constant physics by keeping 
constant the linear combination $\beta_f+2\beta_a$.  

Consider the limit $\beta_a\to\infty$ while keeping $\beta_f$ fixed. 
This requires $|\mathrm{Tr}_f U_p|^2/N^2 = 1$ which implies that
the link matrices are elements of the centre. Fluctuations
between different elements of the centre are controlled by the linear
plaquette term multiplied by $\beta_f$ which means that what we
have is a $Z_N$ gauge theory with coupling $\beta=\beta_f$.
When $N\to\infty$ at fixed $\beta_f$, this becomes a $U(1)$ gauge
theory. In $D=3+1$ this theory has a strong coupling confining phase 
that is separated from a weak coupling Coulomb phase by a phase
transition (probably first order) at $\beta_c = O(1)$ and a 
`freeze-out' transition at $\beta^\prime_c = O(N^2)$ where
fluctuations between neighbouring elements of the centre become
improbably small. These phase transitions will extend from
$\beta_a = \infty$ to some finite $\beta_a$. In addition numerical 
calculations suggest that there is a first order line that
crosses the $\beta_f$-axis (for $N\ge 5$, at the bulk transition) 
and also the $\beta_a$-axis (apparently for all $N$) and to 
which the first order line from $\beta_c$ is attached. This phase 
diagram has been explored in some detail for $SU(2)$
\cite{SmixedD4N2}
and $SU(3)$
\cite{SmixedD4N3}
One can interpret this phase structure in terms of the 
condensation of $Z_N$ monopoles and vortices
\cite{SmixedMono}.
These involve large plaquette values and so the first
order bulk transition involves a large jump in the
average plaquette. It appears, not surprisingly, that the 
would-be third-order Gross-Witten transition is
subsumed in this jump
\cite{BurTepVai}.

By contrast, in the analytically tractable case of $D=1+1$, the 
$\lim_{N\to\infty}Z_N \sim U(1)$
theory at $\beta_a = \infty$ will have no finite-$\beta$ phase
transition by the same arguments used in
\cite{GW}
for $SU(N<\infty)$ gauge theories. One can also readily show that
the finite-$N$ cross-over along the $\beta_f$ axis, which becomes 
the Gross-Witten transition at $N=\infty$, is matched by a similar 
cross-over and $N=\infty$ transition along the $\beta_a$ axis.
Indeed using the same change of variables and notation as in
\cite{GW}
one sees that $Z(\beta_f=0,\beta_a) = z_a^V$ where $V=L_0L_1$ and
\begin{eqnarray}
z_a
& = & 
\int dW \exp\{\frac{\beta_a}{N^2} 
\mathrm{Tr}_f W^\dagger \mathrm{Tr}_f W \}
\nonumber\\
& = & 
\frac{1}{1-\frac{\beta_a}{N^2}}.
\label{eqn_baD2}
\end{eqnarray}
Here we have expanded the exponential and then used eqn(41) of
\cite{GW}
which is valid for $N\to\infty$. (This argument is casual with 
limits and is at most valid coming from the strong coupling side
\cite{GW}.)
Now
\begin{eqnarray}
\langle \mathrm{Tr}_a U_p +1 \rangle  
& = & 
\langle |\mathrm{Tr}_f U_p|^2 \rangle
= \langle |\mathrm{Tr}_f W|^2\rangle 
\nonumber\\
& = &
\frac{\partial}{\partial\frac{\beta_a}{N^2}}\log z_a 
\nonumber\\
& = & 
\frac{1}{1-\frac{\beta_a}{N^2}} .
\label{eqn_UaD2}
\end{eqnarray}
Clearly as we increase $\beta_a$ from strong coupling there must 
be some non-analyticity in $\langle \mathrm{Tr}_a U_p \rangle$ 
at or before the value $\beta_a/N^2 =1$. The corresponding
non-analyticity in $Z=z^V_a$ represents a phase transition.
For a precise derivation we refer to 
\cite{d2beta_a}.

In $D=2+1$ the $\lim_{N\to\infty}Z_N \sim U(1)$ gauge theory 
at $\beta_a=\infty$ has no phase transition at finite $\beta$  and 
is linearly confining at all $\beta$ due to the the plasma of $U(1)$
monopole-instantons. In the $Z_N$ theory there is
freeze-out transition at some $\beta = O(N^2)$
\cite{Z_N2+1d}.
Numerical calculations in $SU(2)$ 
\cite{BaigCuervo}
suggest that the line of phase transitions descending from this
point into the finite $\beta_a$ plane ends before reaching 
the $\beta_a=0$ axis,
so that there is only a peak in the specific heat, but no phase 
transition, on the $\beta_f$ axis.
So all this suggests -- albeit on limited 
evidence -- that at finite $N$ the $D=2+1$
$(\beta_a,\beta_f)$ phase diagram contains only smooth cross-overs
(except for the freeze-out transitions), much like $D=1+1$ and 
in contrast to the finite-$N$ phase transitions in $D=3+1$.
In this paper we shall show that this is indeed the case along
the $\beta_f$ axis.

\subsection{Finite volume transitions}
\label{subsection_V}

Consider a $D=3+1$ $SU(N)$ gauge theory
on a $L_0L^3$ lattice where $L_0\ll L$. If we 
increase $\beta$ from  small values, then we will encounter 
a deconfining transition at $a(\beta)L_0=1/T_c$ (first
order for $N\geq 3$ and second order for $N=2$
\cite{OxT03,OxT05}). 
A convenient order parameter for this transition is the
Polyakov loop, $\langle {\bar l}_{\mu=0} \rangle $, 
which acquires a non-zero expectation value 
in the deconfined phase. The transition 
is a crossover for finite $L$ and sharpens to a 
phase transition as $L\to\infty$. As $N\uparrow$ the
transition becomes sharper on ever smaller volumes
\cite{OxT03,OxT05}. 
so that as $N\to\infty$ one will have a deconfining phase 
transition for $L=L_0+\epsilon$ where $\epsilon > 0$
is as small as we like. By continuity one would expect
an $N=\infty$ phase transition even as  $\epsilon \to 0$,
i.e. on an $L^4$ lattice. (Finessing any subtleties about 
orders of limits.)

This makes contact with calculations
\cite{NN03,NNect04} 
that have shown that if we are on a $L^4$ 
lattice and increase $\beta$, then there will be a
crossover, sharpening to a phase transition at $N=\infty$, 
which is characterised by one of the Polyakov loops, 
$\langle l_\mu\rangle$, with $\mu$ chosen at random,
acquiring a non-zero expectation value. In fact one finds
\cite{NN03,NNect04} 
that this is only the first of a sequence of phase transitions.
At a second higher value of $\beta$ there is a second phase
transition where another  Polyakov loop, $\langle l_\nu\rangle$ 
with $\nu\neq\mu$ again chosen at random, acquires a non zero 
expectation value. And there is some evidence that as $\beta$ is 
increased further there are similar transitions along a third 
and fourth direction
\cite{NN03,NNect04}.
Moreover one finds
\cite{NN03,NNect04}
that these transitions are associated with a gap forming in the 
eigenvalue spectrum of the appropriate Polyakov loop, just as
one finds for the plaquette at the Gross-Witten transition. It is 
therefore natural to think of these finite-volume transitions 
as being in some sense strong-to-weak coupling transitions.

As we argued above, the first of these $N=\infty$ transitions 
appears to be nothing but the $N=\infty$ deconfining transition
and should occur at $\beta=\beta_{c_0}$ where  
\begin{equation}
a(\beta_{c_0})L_0=1/T^{D=4}_c.
\label{eqn_abeta1c}
\end{equation}
One can make analogous arguments for the existence of a sequence 
of transitions on an $L_0L_1L_2L_3$ lattice with 
$L_0 \ll L_1 \ll L_2 \ll L_3$. To see this, consider the
following steps. \\
{\noindent}(1) Increase $\beta_4\equiv\beta$ to very high 
temperatures, $T=1/a(\beta)L_0 \gg T_c$.
In this regime we will have the familiar dimensional reduction 
to an effective $D=2+1$ $SU(N)$ gauge theory coupled to adjoint
scalars $\phi$ that are the time-translationally invariant remnant 
of the $A_0$ gauge field
\cite{dimred4}.
To leading order the gauge coupling and mass of the scalar of the 
effective $D=2+1$ gauge-scalar theory are
\cite{dimred4}
\begin{equation}
g^2_3 = g^2_4(T)T \qquad ; \qquad m^2_a \propto g^2_4(T) T^2.
\label{eqn_gmD3}
\end{equation}
So $m_a/g^2_3 = O(1/g_4(T))$ and at high enough $T$ the $D=3+1$ 
gauge theory reduces to the $SU(N)$
gauge theory in $D=2+1$ on a $L_1 \ll L_2, L_3$ lattice.
As we increase $\beta_4=\beta$ we simultaneously increase
$\beta_3 \equiv 2N/ag^2_3 \simeq \beta_4 L_0$ (neglecting the
difference between $g^2_4(a^{-1})$ and $g^2_4(T)$). 
This $D=2+1$ gauge theory will deconfine at
\begin{equation}
a(\beta_{c_1})L_1=1/T^{D=3}_c
\label{eqn_abeta2c}
\end{equation}
at which point $\langle l_{\mu=1}\rangle$ acquires a non-zero
vacuum expectation value. We can estimate the corresponding
critical value of $\beta (\equiv\beta_4)$ to be
\begin{equation}
\beta_{c_1} \sim 0.36 N^2 \frac{L_1}{L_0}.
\label{eqn_beta2c}
\end{equation}
To arrive at this estimate we use 
$(T_c/\surd\sigma )_{D=3} \sim 0.9$ 
\cite{TcD3} 
and $\surd\sigma/g^2_3N \simeq 0.198$
\cite{Nd3}, 
together with eqn(\ref{eqn_gmD3}). For finite $N$
this will be a cross-over, but we expect (for the same reasons
as in one higher dimension) that as $N\to\infty$ one will
have a phase transition on any volume where 
$L_2,L_3 = L_1 + \epsilon$, for any fixed $\epsilon$ however small.
In the limit we thus expect the transition on a symmetric $L_0L_1^3$ 
lattice with $L_1\gg L_0$. If we now reduce $L_1/L_0$ then we see from
eqn(\ref{eqn_beta2c}) that $\beta_{c_1}$ will begin to decrease. 
However long before its value reaches $\beta_{c_0}$
the value of $T$ will have become small enough that we cannot
neglect the adjoint scalar (and its self-interactions) and our
estimate in eqn(\ref{eqn_beta2c}) ceases to be useful. Nonetheless
it appears plausible that the second $N=\infty$ transition that has been
observed on $L^4$ lattices
\cite{NN03,NNect04}
is the continuation as $L_1\to L_0$ of this
three-dimensional deconfinement transition.\\
{\noindent}(2) As we increase $\beta$ beyond $\beta_{c_1}$
on our $L_0\ll L_1\ll L_2\ll L_3$ lattice, $T^{D=3} = 1/aL_1$ 
will become ever larger, and eventually the system will undergo 
a further dimensional reduction to a $D=1+1$ $SU(N)$ gauge theory
with adjoint scalars
\cite{dimred3}.
To leading order the gauge coupling and scalar mass of this
effective $D=1+1$ gauge-scalar theory are
\cite{dimred3}
\begin{equation}
g^2_2 = g^2_3 T^{D=3} \qquad ; \qquad m^2_a \propto g^2_2 \log (aT^{D=3}).
\label{eqn_gmD2}
\end{equation}
While the pure gauge $D=1+1$ theory has no propagating degrees of
freedom and is too trivial to deconfine, the presence of the
adjoint scalars renders
it a non-trivial confining field theory which we
would naively expect to deconfine at some 
$T^{D=2}_c = 1/a(\beta_{c_2})L_2$.  
We estimate the corresponding critical value of the coupling
$\beta (\equiv\beta_4)$ to be   
\begin{equation}
\beta_{c_2} \sim 0.43 r^2 N^2 \frac{L^2_2}{L_0L_1}
\qquad ; \qquad r = \frac{T^{D=2}_c}{\surd\sigma}
\label{eqn_beta3c}
\end{equation}
using the value $\sigma \sim (0.8)^2 g^2_3 T N/3$ extracted from
\cite{dimred3}.
Because we are in $D=1+1$ the high-$T$ phase cannot have
a true non-zero expectation value for $\langle l_{\mu=2}\rangle$,
as will be discussed more explictly when we come to our numerical
calculations below. Again we would expect that at $N=\infty$
this deconfining transition will appear on lattices with
finite $L_{3}$ and even as $L_3\to L_2$. It is again plausible
that this $N=\infty$ transition (although not our estimate in 
eqn(\ref{eqn_beta3c})) will survive as $L_3\to L_2 \to L_1 \to L_0$,
thus making the connection with the third transition observed on
$L^4$ lattices  
\cite{NN03,NNect04}. \\
{\noindent}(3) If we increase $\beta$ further we come to
consider a field theory with a finite Euclidean time extent given
by $aL_3$ living in an infinitesimal spatial volume $a^3L_0L_1L_2$.
Such systems can in principle have deconfining phase transitions
\cite{aharony}
although whether this one does or not we do not attempt to
make plausible by a simple argument. If it does exist
then it would provide the final step in our cascade of $N=\infty$
phase transitions $\beta_{c_0}\ll\beta_{c_1}\ll\beta_{c_2}\ll\beta_{c_3}$
on our  $L_0\ll L_1\ll L_2\ll L_3$ lattices.

The above discussion has taken the $D=3+1$ $SU(N)$ gauge theory
as its starting point. It is obvious that we could equally well have
started  with a $L_0\ll L_1 \ll L_2$ $D=2+1$ $SU(N)$ gauge theory and
followed that through a cascade of deconfining $N=\infty$ 
transitions.

\section{Results}
\label{section_results}

\subsection{Preliminaries}
\label{subsection_Rprelim}

\subsubsection{Phase transitions}
\label{subsubsection_phase}

At a phase transition appropriate derivatives of the partition 
function $Z$ diverge or are discontinuous. (Strictly speaking of
$\frac{1}{V}\log Z$ where $V$ is the volume.) The lowest order of such 
a singular derivative determines the order of the phase transition.
For  $Z$ or its derivatives to be singular, we require an
infinite number of degrees of freedom, and this usually demands
an infinite volume, with a cross-over at finite $V$ sharpening
to the appropriate singularity as $V\to\infty$. As $N\to\infty$ we 
have the possibility of a new kind of phase transition that takes 
place in a finite volume with the infinite number of degrees of
freedom being provided by $N$. 

With the standard plaquette action, a conventional first order 
transition has a discontinuity at $V = \infty$ in the average 
plaquette,
\begin{equation}
\langle u_p\rangle = N^{-1}_p \partial \log Z/\partial\beta
\label{eqn_avup}
\end{equation}
where $N_p$ is the number of plaquettes. (We denote the space-time
volume by $V$ and note that  $V = N_p/3$ in $D=2+1$.)
At finite $V$ this discontinuity is a rapid crossover so that
\begin{equation}
\partial \langle u_p\rangle / \partial\beta = 
N^{-1}_p \partial^2 \log Z/\partial\beta^2 
\equiv C
\label{eqn_avdup}
\end{equation}
diverges at the critical coupling $\beta = \beta_c$ as
$N_p\to\infty$. (Here $C$ is the specific heat.) This 
divergence is linear in $V$ since the cross-over between
the two distinct values of the plaquette occurs in the 
range of $\beta-\beta_c$ where there is back-and-forth tunnelling
and this range is $O(1/V)$  as we see from the linear
approximation 
\begin{equation}
\Delta F(\beta) = V \Delta f(\beta) 
\propto (\beta-\beta_c) V \sim O(1) 
\label{eqn_av1CV}
\end{equation}
to the free energy  (density) difference,  $\Delta F, f$,
between the two phases.

A conventional second order transition has a smooth first 
derivative of $Z$  but a diverging second derivative and specific heat
$C \to \infty$ as $V\to\infty$. Defining $\overline{u_p}$ to be the average  
value of $u_p$ over the space-time volume for a single lattice
field, we easily see that the specific heat can be written as
a correlation function:
\begin{eqnarray}
C 
& = &
N_p\langle(\overline{u_p}-\langle\overline{u_p}\rangle)^2\rangle
= 
N_p(\langle \overline{u_p}^2\rangle-\langle \overline{u_p} \rangle^2)
\nonumber\\
& = &
\sum_p \langle ( u_p-\langle u_p \rangle)
( u_{p_0} -\langle u_p \rangle)
\rangle
\label{eqn_C}
\end{eqnarray}
where $p_0$ is some arbitrary reference plaquette. It is clear
from eqn(\ref{eqn_C}) that the divergence of $C$ as $N_p\to\infty$ 
implies that there is a diverging correlation length  -- the
standard signal of a second order phase transition.

A conventional third order transition has smooth first and second 
order derivatives but a singular third-order derivative,
$C^{\prime}\equiv N^{-1}_p \partial^3 \log Z/\partial\beta^3$,
at $V =\infty$. This may be written as
\begin{eqnarray}
C^\prime 
& = &
\frac{\partial C}{\partial \beta}
=
N_p^2\langle(\overline{u_p}-\langle\overline{u_p}\rangle)^3\rangle
\nonumber\\
& = &
N_p^2 (\langle\overline{u_p}^3\rangle
-3\langle\overline{u_p}\rangle\langle\overline{u_p}^2\rangle
+2\langle\overline{u_p}\rangle^3).
\label{eqn_Cprime}
\end{eqnarray}
Note that if the fluctuations of  $\overline{u_p}$ were symmetric 
around $\langle\overline{u_p}\rangle$, as they are for $\beta = 0$, 
then $C^\prime$ would be zero, so
\begin{equation}
\lim_{\beta\to 0} C^\prime(\beta) = 0.
\label{eqn_C0b0}
\end{equation}

It should be clear that the higher the order of the transition,
the greater is the statistics needed to determine its properties
to a given precision. In particular, identifying third-order 
transitions is already a formidable numerical challenge,
and we do not attempt to look for transitions that are of
yet higher order. 

Since we are particularly interested in transitions that develop
as $N\to\infty$ and since we know that, in general, fluctuations 
in the pure gauge theory decrease by powers of $N$ in the 
large--$N$ limit
\cite{thooft,largeN}
it is convenient to define the rescaled quantities
\begin{equation}
C_2=N^2\times C \qquad ; \qquad C_3=N^4 \times C^\prime
\label{eqn_C23}
\end{equation}
which one expects generically to have finite non-zero limits
when $N\to\infty$. (Note that the increasing power of $N$ 
simply matches the increasing power of 
$\partial/\partial\beta = 2N^2 \partial/\partial\gamma$, 
where $\gamma$ is the inverse
't Hooft coupling defined in  eqn(\ref{eqn_ggN}).)
The signature of a phase transition which is only present for 
$N=\infty$ will be a crossover for finite $N$ at which
fluctuations decrease more slowly than the naive power of $1/N^2$.
If, therefore, we find a crossover in $C_2$ or $C_3$ which 
does not sharpen with increasing volume at fixed $N$, but 
rather becomes a divergence or a discontinuity 
only in the large--$N$ limit, then this will indicate a 
second-- or third--order $N=\infty$ phase transition 
respectively. (Provided of course that $\langle u_p \rangle$
is continuous so that there is no large-$N$ first order
transition.)

Large-$N$ phase transitions can have an unconventional behaviour.
Consider for example a second order transition characterised
by a value of $C_2$ that diverges at some $\lambda = \lambda_c$ as
$N\to\infty$. This may indeed be due to a correlation length 
$\xi$ that diverges (in lattice units) as $N\to\infty$:
$\xi(\lambda_c) \propto N^{\alpha} \ ; \ \alpha >0$. However there
is another, less conventional, possibility: the correlation length
may be finite and it may be that local plaquette fluctuations
have an anomalous $N$-dependence at the critical point:
$\langle u^2_p \rangle/\langle u_p \rangle^2 - 1 \propto
N^{\alpha-2} \ ; \ \alpha >0$.

Since large-$N$ phase transition can arise from fluctuations that
are completely local -- as in $D=1+1$ where the lattice partition
function factorises -- it is also useful to consider local versions 
of the quantities $C_2$ and $C_3$:  
\begin{equation}
P_2=N^2\times (\langle{u_p}^2\rangle-\langle{u_p}\rangle^2)
\label{eqn_P_2}
\end{equation}
and
\begin{equation}
P_3=N^4\times (\langle{u_p}^3\rangle-
3\langle{u_p}\rangle\langle{u_p}^2\rangle+2\langle{u_p}\rangle^3).
\label{eqn_P_3}
\end{equation}
These are the contributions to $C_2$ and $C_3$ from fluctuations
of individual plaquettes, i.e. neglecting correlations between plaquettes.
So calculations of $P_2$ and $P_3$ require much lower statistics than
$C_2$ and $C_3$ to achieve the same accuracy, and this will
be particularly useful at the largest values of $N$. Of course, 
divergences or discontinuities in $P_2$ or $P_3$ will normally
imply divergences or discontinuities in $C_2$ or $C_3$,
even if the latter are not visible in the statistical noise 
of the numerical calculation. 
Note that in 1+1 dimensions, where the theory factorises
and there are no propagating degrees of freedom, we 
have $P_2=C_2$ and $P_3=C_3$ at any $N$.
 
The eigenvalues of an $SU(N)$ matrix such as the plaquette, 
are gauge--invariant, and so we can use them to gain 
additional information to that encoded in the above 
correlators of low powers of traces of plaquettes. Indeed, 
at the $D=1+1$ $N=\infty$ Gross--Witten transition a gap 
opens in the eigenvalue density of the plaquette
\cite{GW}.
That is to say, while in the 
strongly--coupled phase the eigenvalue density is non--zero 
for all angles $-\pi \leq \alpha \leq \pi$,
in the weakly--coupled phase it is only non--zero in the range
$-\alpha_c \leq \alpha \leq \alpha_c$, where $\alpha_c < \pi$. 
To search for a similar transition in 2+1 dimensions
we will measure the total plaquette eigenvalue density.
Using the eigenvalue density directly
to search for a gap is difficult since for any finite $N$
there is not a true gap but instead a (near-)exponential 
drop in the eigenvalue density as $\alpha \to \pm \pi$,
so very good statistics are required to observe changes
in the exponentially suppressed tails of the density.
To avoid relying solely on the accurate calculation of these tails,
we also calculate the fluctuations of the eigenvalues around 
their average values,
$\langle \lambda_i^2\rangle-\langle \lambda_i\rangle^2$,
where $\lambda_i$ is the $i$th eigenvalue when ordered by its phase. 
(Recall that the eigenvalues of $SU(N)$ matrices are pure phases
$\lambda_j = \exp\{i\alpha_j\}$.)
In particular, we shall calculate the normalised ratio for
the extreme (smallest) eigenvalue:
\begin{equation}
R_p=\frac{\langle \lambda_1^2\rangle -\langle \lambda_1\rangle ^2}
{\langle \lambda_\frac{N}{2}^2\rangle -
\langle \lambda_\frac{N}{2}\rangle ^2}
\label{eqn_ratio}
\end{equation}
(for $N$ even). This is motivated by the situation in
the $N\to\infty$ limit in $D=1+1$ where the eigenvalue 
density at the Gross-Witten transition,
$\gamma=1/2$, is
\cite{GW} 
$\frac{1}{2\pi}(1+\mathrm{cos}\alpha)$. This density approaches zero
as $\alpha \to \pm \pi$ but is finite at $\alpha=0$, so we expect
the fluctuations to be $O(1/N)$ in $\lambda_\frac{N}{2}$ while
they can be up to $O(1)$ for $\lambda_1$. Thus we expect 
$R_p$ to diverge at the Gross--Witten transition, and it may 
provide a useful observable in our search for a similar transition 
in $D=2+1$.

\subsubsection{Wilson loop non-analyticities}
\label{subsubsection_WLtrans}

To investigate the possibility that Wilson loops undergo
some analogous non-analyticity as their area passes through
some critical value, $A_{crit}$, we calculate Wilson loops 
of a fixed size, $n_1\times n_2$, in lattice units and increase 
$\beta$ so as to decrease the lattice spacing $a$ and hence
the area, $A=an_1\times an_2$, in physical units. If there is 
a non-analyticity at $A(\beta_c(n_1,n_2))$ we can then vary 
$n_1,n_2$ so as to check whether the transition occurs at a 
fixed area in the continuum limit, when expressed in units of 
say $g^2N$, i.e. whether
\begin{equation}
\frac{A_{crit}}{(g^2N)^2}
=
\lim_{a\to 0} 
\left(\frac{\beta_c}{2N^2}\right)^2 A(\beta_c)      
\label{eqn_WLP1}
\end{equation}
is finite and non-zero. Since all the evidence is that
the $D=2+1$ SU($N$) lattice gauge theory has no phase
transition, at zero temperature, once $\lambda$ is
on the weak coupling side of the bulk transition, we
expect any Wilson loop non-analyticity not to correspond
to a phase transition of the whole system. This will
be an important constraint on what are the important
observables to calculate. We also expect that any such
transitions will be cross-overs at finite $N$, becoming
real non-analyticities only at $N=\infty$. This is
because we can imagine that they are driven by the degrees
of freedom close to the critical length scale, and that
we need these to be infinite in number for a real
non-analyticity.

\begin	{figure}[p]
\begin	{center}
\leavevmode
\input	{wilson48graph}

\end	{center}
\vskip 0.15in
\caption{The spectrum of eigenvalues, $e^{i\alpha}$, of a
$3\times 3$ Wilson loop for $SU(48)$ in $D=1+1$
at the critical coupling $\gamma = 1/\lambda = 1.255$,
together with the continuum spectrum ($- - -$).} 
\label{fig_wilson48graph}
\end 	{figure}
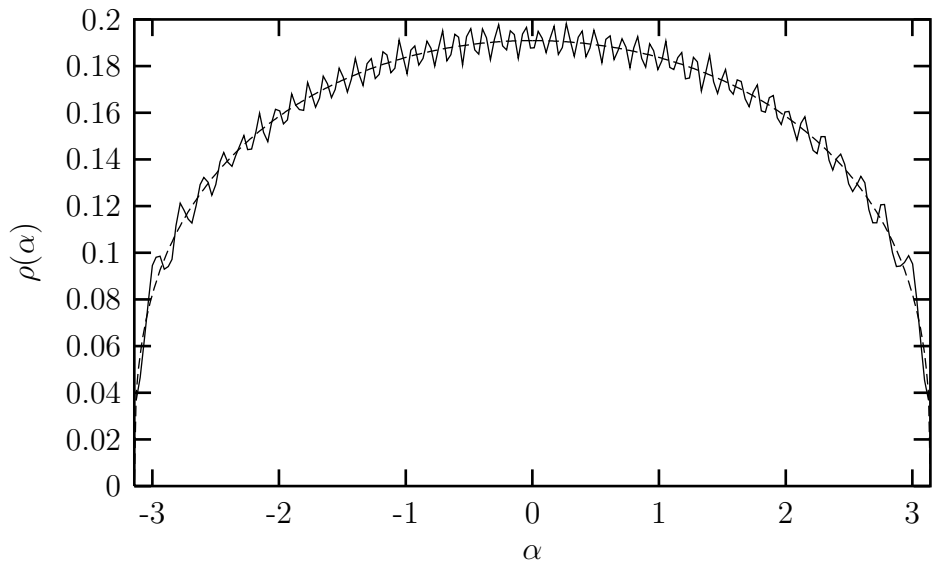

We remarked in Section~\ref{subsection_WL} that a non-analyticity
in the eigenvalue spectrum of the Wilson loop is known to occur 
\cite{DurOle,BasGriVian}
in the $D=1+1$ $N=\infty$ gauge theory at the critical area
given in eqn(\ref{eqn_d2Acrit}). We have performed 
numerical lattice calculations in this theory for large
$N$ and find that the lattice critical area is very
close to the continuum one for Wilson loops that
are $2\times 2$ or larger. To be more precise 
let us denote the product of link matrices around a square
$n \times n$ Wilson loop by $U_w^{n\times n}$ and its trace by  
$u_w^{n\times n}=\frac{1}{N}\mathrm{Re}\mathrm{Tr}\{U_w^{n\times n}\}$
which we generically write as $u_w$. Then we find that the
non-analyticity occurs when $u_w$ reaches a particular value 
\begin{equation}
\langle u_w \rangle \simeq e^{-2}.
\label{eqn_WLP3}
\end{equation}
As an example we show 
in Fig.~\ref{fig_wilson48graph} the eigenvalue spectrum of a $3\times 3$
Wilson loop in $D=1+1$ for $N=48$ at $\lambda = 0.7971$ where the 
trace satisfies eqn(\ref{eqn_WLP3}) and we compare it to the
continuum expression obtained from
\cite{DurOle,BasGriVian}
We clearly have a very good match (apart from the $N=48$ bumps that
arise  from the eigenvalue repulsion in the Haar measure).
Now we know that in $D=1+1$ the Wilson loop factorises into 
a product of plaquettes
\begin{equation}
\langle u_w\rangle = \langle u_p\rangle^{\frac{A}{a^2}}
\label{eqn_WLP5}
\end{equation}
and that  $\langle u_p\rangle = 1-\lambda/4$ at $N=\infty$
\cite{GW}.
Putting all this together, we have
\begin{equation}
\left(1-\frac{\lambda}{4}\right)^{\frac{A}{a^2}} 
=
e^{{\frac{A}{a^2}}\ln (1-\frac{\lambda}{4})}
\stackrel{a\to 0}{\simeq}  
e^{-\frac{A\lambda}{4a^2}}
\simeq  
e^{-2}
\label{eqn_WLP7}
\end{equation}
which we observe is nothing but the continuum relation in
eqn(\ref{eqn_d2Acrit}). These numerical calculations
show that lattice corrections are small except for
loops smaller than $2\times 2$, such as the plaquette that
has its non-analyticity at the Gross-Witten transition
where $\lambda=2$. Because of the factorisation in
eqn(\ref{eqn_WLP5}) the trace of $u_w$ will be analytic
in the (bare) coupling when this gap in the eigenvalue
spectrum forms (except for the very smallest loops 
where it occurs at the Gross-Witten transition) and so
it is not immediately obvious what is the significance of
this gap formation. What this tells us, nonetheless, is 
that we should not only search in $D=2+1$ for 
non-analyticities of traces of Wilson loops, but  
also for such eigenvalue gap formation. 

Our initial question will be whether $\langle u_w \rangle$
has a  non-analyticity in $\lambda$ for some value
of the area, $A/a^2$. To investigate this we can
look at $\langle u_w \rangle$ and its derivatives 
as a function of the bare coupling $\lambda$. 
The derivatives can be expressed as correlation functions
in the usual way e.g.
\begin{eqnarray}
\partial \langle u_w\rangle / \partial\beta 
& = &
N_p\left(\langle u_w \overline{u_p}\rangle
-\langle\overline{u_p}\rangle\langle u_w\rangle\right)
\nonumber\\
& = &
\sum_p \langle ( u_w-\langle u_w \rangle)
( u_p -\langle u_p \rangle) \rangle.
\label{eqn_WLP9}
\end{eqnarray}
Because the whole system has no phase transition we 
expect that the non-analyticity will be visible in 
the `local' correlators analogous to those in
eqns(\ref{eqn_P_2},\ref{eqn_P_3}) e.g. 
\begin{equation}
P^w_2=N^2\times (\langle{u_w} \hat{u}_p \rangle
-\langle{u_w}\rangle \langle \hat{u}_p \rangle )
\label{eqn_WLP11}
\end{equation}
where $ \hat{u}_p $ is the average of the plaquettes
that tile the minimal Wilson loop surface. We 
define $P^w_3$ in analogy to $P_3$ in the same way. 

It is also possible that some non-analyticity
might be present in just the fluctuations of Wilson
loops rather than in the derivatives with respect to
the coupling. Thus we also consider the correlators
\begin{equation}
W_2^{n\times n}=
N^2\times (\langle{u_w^{n\times n}}^2\rangle
-\langle{u_w^{n\times n}}\rangle^2)
\label{eqn_WLP13}
\end{equation}
and
\begin{equation}
W_3^{n\times n}=N^4\times (2\langle{u_w^{n\times n}}\rangle^3
-3\langle{u_w^{n\times n}}\rangle\langle{u_w^{n\times n}}^2\rangle
+\langle{u_w^{n\times n}}^3\rangle).
\label{eqn_WLP15}
\end{equation}
that represent an alternative generalisation to Wilson
loops of the quantities $P_2$ and $P_3$ defined in
eqns(\ref{eqn_P_2},\ref{eqn_P_3}).
 
In searching for a gap formation in the eigenvalue spectrum
of an  $n \times n$ Wilson loop, we define the quantity 
$R^{n \times n}$ in direct analogy to the quantity $R_p$
defined in eqn(\ref{eqn_ratio}). However this quantity
is only useful for the bulk transition because the eigenvalue
spectrum approaches the gap with a finite slope. We know
that for larger Wilson loops in $D=1+1$
\cite{DurOle,BasGriVian}
the approach is with a diverging slope (at $N=\infty$),
and that $R^{n \times n}$ is not a useful observable
in that case.
We shall in fact find it much more useful to match the
eigenvalue spectra in $D=2+1$ and $D=1+1$. That this is
in fact possible is one of our most interesting results.

\subsection{Bulk transition}
\label{subsection_Rbulk}

In 3+1 dimensions the bulk transition is easily visible as a large 
discontinuity in the action for $N\geq 5$ (where the transition
is first order) and as a (finite) peak in the specific heat for
$N\leq 4$ (where the transition is a crossover). We have searched 
for an analogous jump or rapid crossover in 2+1 dimensions, in
particular around $\gamma \equiv \beta/2N^2 \sim 1/2$.

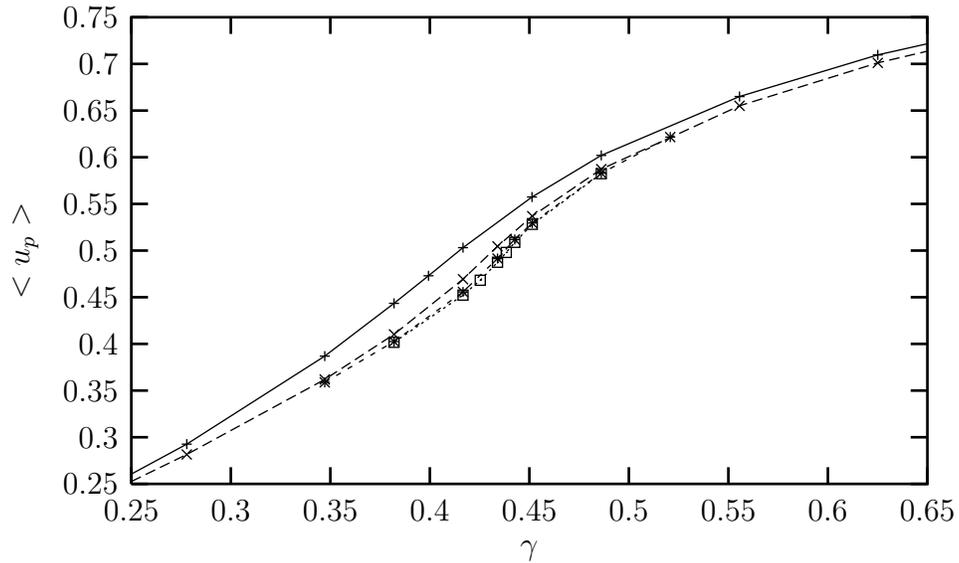
\begin	{figure}[p]
\begin	{center}
\leavevmode
\input	{actiongraph}

\end	{center}
\vskip 0.15in
\caption{The average plaquette as a function of 
$\gamma=\frac{1}{ag^2N}=\frac{\beta}{2N^2}$
for SU(6) ($+$), SU(12) ($\times$), 
SU(24) ($\ast$) and SU(48) ($\Box$).}  
\label{fig_action}
\end 	{figure}

In Fig.\ref{fig_action} we display the values of the average
plaquette for $SU(6)$, $SU(12)$, $SU(24)$ and $SU(48)$ as obtained on  
$6^3$ lattices. At $\gamma \sim 1/2$ an $L=6$ lattice
has a size $La\surd\sigma \sim 3$ and so is large enough
that it should display a very sharp cross-over for a conventional
first-order transition. This should be more so as $N\uparrow$
and (most) finite volume effects disappear. As a check we have
repeated our calculations on $12^3$ lattices for $SU(6)$ 
and have found no volume dependence. What we see in 
Fig.~\ref{fig_action} is that the action appears to be approaching 
a smooth crossover in the large--$N$ limit, with no evidence for 
a first order phase transition either at finite  $N$ or at
$N=\infty$.

Our results for the specific heat $C_2$ for $SU(6)$ and $SU(12)$
are shown in Fig.~\ref{fig_C_2}. (For $SU(24)$ and $SU(48)$ our 
accuracy is insufficient to get useful results for $C_2$.)
There is a clear peak around $\gamma\simeq 0.42$ which 
appears to be growing stronger with increasing $N$. For $SU(6)$ we 
repeated our calculations on $12^3$ lattices and found no
volume dependence. This tells us that we are not seeing a
conventional second--order phase transition at fixed $N$ for which the
specific heat peak grows as the volume increases (since a larger
volume can better accommodate the diverging correlation length).
So if there is a second--order phase transition here it would appear 
to be not at finite $N$, but only at $N=\infty$.

To search for a possible third order transition we calculate
$C_3$, but our calculations are not accurate enough to produce
anything significant, even for $SU(6)$.

\begin	{figure}[p]
\begin	{center}
\leavevmode
\input	{P_2graph}

\end	{center}
\vskip 0.15in
\caption{The `local' specific heat, $P_2$, as a function of 
$\gamma=\frac{\beta}{2N^2}$
for SU(6) ($+$), SU(12) ($\times$), 
SU(24) ($\ast$) and SU(48) ($\Box$).}  
\label{fig_P_2}
\end 	{figure}
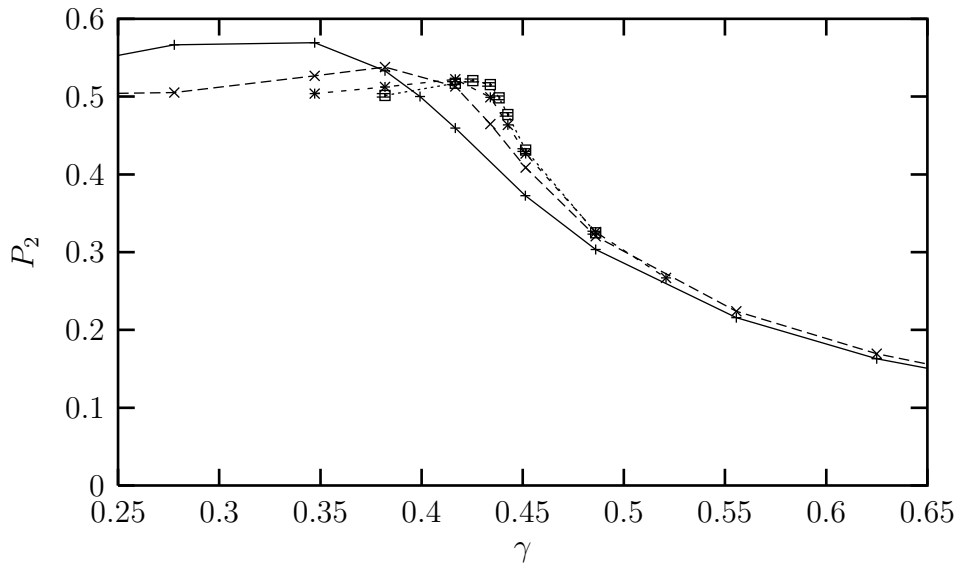

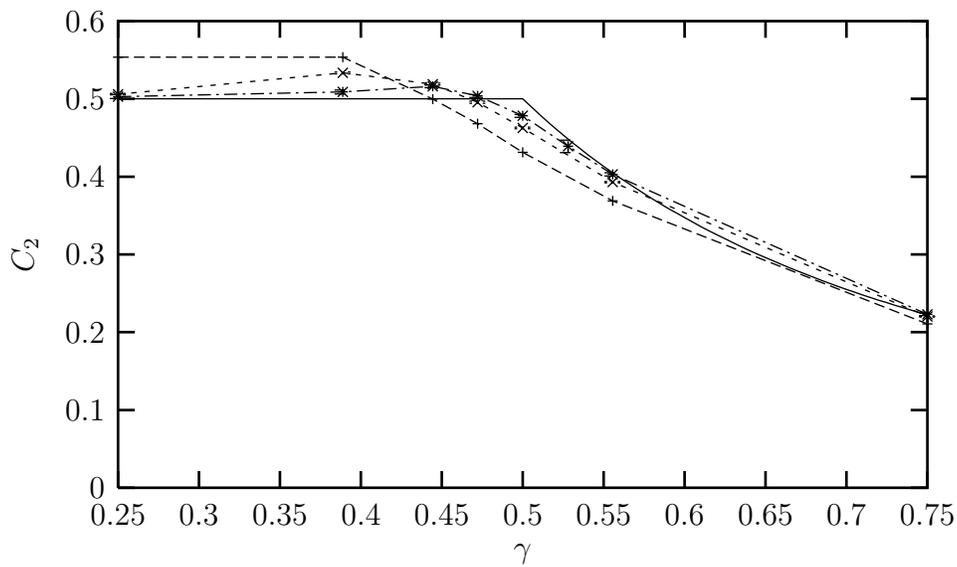
\begin	{figure}[p]
\begin	{center}
\leavevmode
\input	{1+1d_C_2graph}

\end	{center}
\vskip 0.15in
\caption{The specific heat, $C_2$  (equal to $P_2$ here) as 
a function of $\gamma=\frac{\beta}{2N^2}$ in 1+1 dimensions
for SU(6) ($+$), SU(12) ($\times$), SU(24) ($\ast$)
and the analytic result for SU($\infty$) (solid line).} 
\label{fig_1+1d_C_2}
\end 	{figure}

To improve our reach in $N$ we calculate the quantities $P_2$ 
and $P_3$ defined in eqns(\ref{eqn_P_2},\ref{eqn_P_3}). These
represent the contributions to $C_2$ and $C_3$ made by the 
fluctuations of individual plaquettes and are the quantities
that reveal the Gross-Witten transition in $D=1+1$. 
In Fig.~\ref{fig_P_2}
we show the values of $P_2$ obtained for $SU(6)$, $SU(12)$, $SU(24)$ 
and $SU(48)$. We observe, as expected, a dramatic reduction in the 
statistical errors as compared to $C_2$ in Fig.~\ref{fig_C_2},
enabling us to look for fine structure in the $\beta$-dependence. 
There is no significant evidence for a peak in $P_2$
which indicates that if there is a second order transition 
at $N=\infty$, as suggested by the peak in $C_2$, it will primarily
involve correlations between different plaquettes rather than
arising from the fluctuations of individual plaquettes.
What we do see in $P_2$ however 
is definite evidence for a cusp developing with increasing
$N$, at $\gamma \simeq 0.43$, where the derivative of $P_2$ will
suffer a discontinuity. This corresponds to a third-order transition
at $N=\infty$, just like the $D=1+1$ Gross-Witten transition
\cite{GW}.
It is therefore useful to compare the $D=1+1$
and $D=2+1$ cases in more detail. For this purpose we show
in Fig.~\ref{fig_1+1d_C_2} some numerically calculated values
of $P_2$ in $D=1+1$ $SU(6)$, $SU(12)$ and $SU(24)$ gauge theories.
(Recall that in $D=1+1$ the factorisation of the partition function
implies that $C_2=P_2$.) We also plot the analytic results for 
$SU(\infty)$
\cite{GW}:
\begin{equation}
P_2=C_2=\left\{ \begin{array}{ll}
\frac{1}{2}, & \gamma\leq 0.5 \\
\frac{1}{8\gamma^2}, & \gamma\geq 0.5.
\end{array}
\right.
\label{eqn_C_2_1+1d}
\end{equation}
Apart from a small relative shift in $\gamma$ the results for
$D=2+1$ and $D=1+1$ are remarkably similar, strengthening
the evidence for a third-order $N=\infty$ transition.

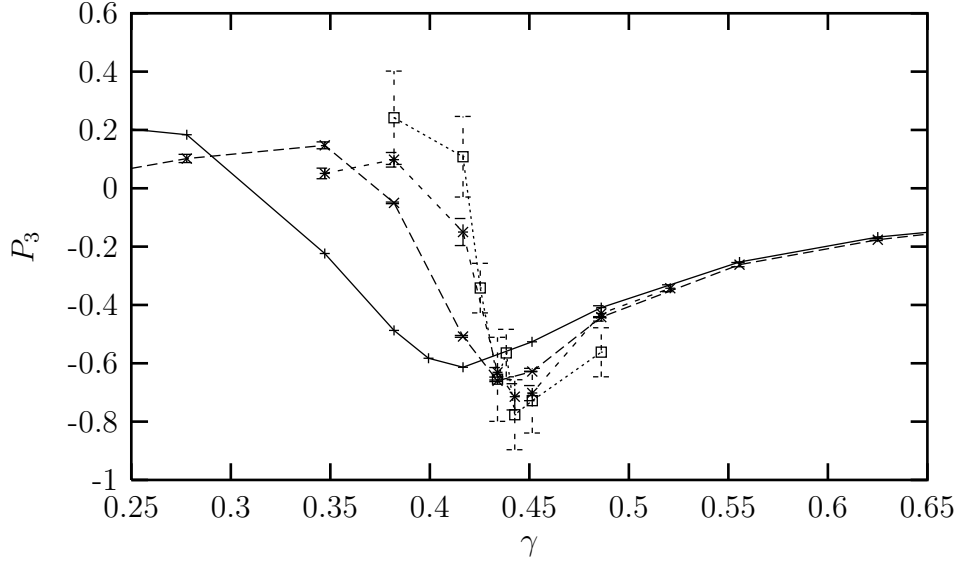
\begin	{figure}[p]
\begin	{center}
\leavevmode
\input	{P_3graph}

\end	{center}
\vskip 0.15in
\caption{The cubic local plaquette correlator, $P_3$, as a 
function of $\gamma=\frac{\beta}{2N^2}$
for SU(6) ($+$), SU(12) ($\times$), 
SU(24) ($\ast$) and SU(48) ($\Box$).}  
\label{fig_P_3}
\end 	{figure}

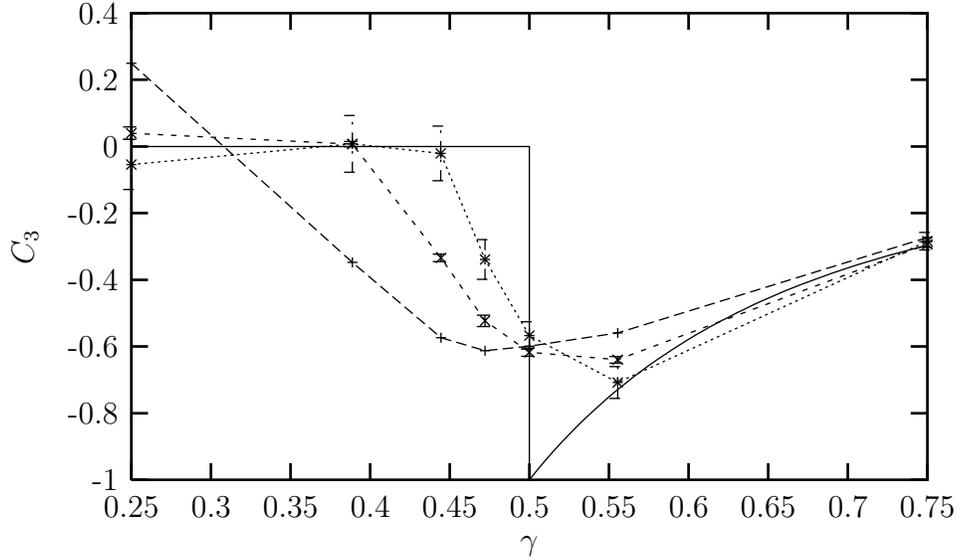
\begin	{figure}[p]
\begin	{center}
\leavevmode
\input	{1+1d_C_3graph}

\end	{center}
\vskip 0.15in
\caption{The cubic plaquette correlator, $C_3$  (equal to $P_3$ here) 
as a function of $\gamma=\frac{\beta}{2N^2}$ in 1+1 dimensions
for SU(6) (long dashes), SU(12) (short dashes),
for SU(6) ($+$), SU(12) ($\times$), SU(24) ($\ast$)
and the analytic result for SU($\infty$) (solid line).} 
\label{fig_1+1d_C_3}
\end 	{figure}

To investigate this further, we show in Fig.~\ref{fig_P_3} our 
results for $P_3$ for $SU(6)$, $SU(12)$, $SU(24)$ and $SU(48)$.
There is clearly an increasingly
sharp transition as $N$ increases around $\gamma\simeq 0.43$.
For comparison we show in  Fig.~\ref{fig_1+1d_C_3} corresponding
numerical results for $D=1+1$ (where $C_3=P_3$) together with 
the  analytic result for $SU(\infty)$
\cite{GW}:
\begin{equation}
C_3=\left\{ \begin{array}{ll}
0, & \gamma\leq 0.5 \\
- \frac{1}{8\gamma^3}, & \gamma\geq 0.5.
\end{array}
\right.
\label{eqn_C_3_1+1d}
\end{equation}
which has a discontinuity at  the Gross--Witten transition at
$\gamma=1/2$. It is clear that once again the
the behaviour in $D=2+1$ is remarkably similar to that in 1+1 
dimensions.

\begin	{figure}[p]
\begin	{center}
\leavevmode
\input	{ratiograph}

\end	{center}
\vskip 0.15in
\caption{Fluctuations of extreme plaquette eigenvalues, $R_p$, 
as a function of $\gamma=\frac{\beta}{2N^2}$
for SU(6) ($+$), SU(12) ($\times$), 
SU(24) ($\ast$) and SU(48) ($\Box$).}  
\label{fig_ratio}
\end 	{figure}
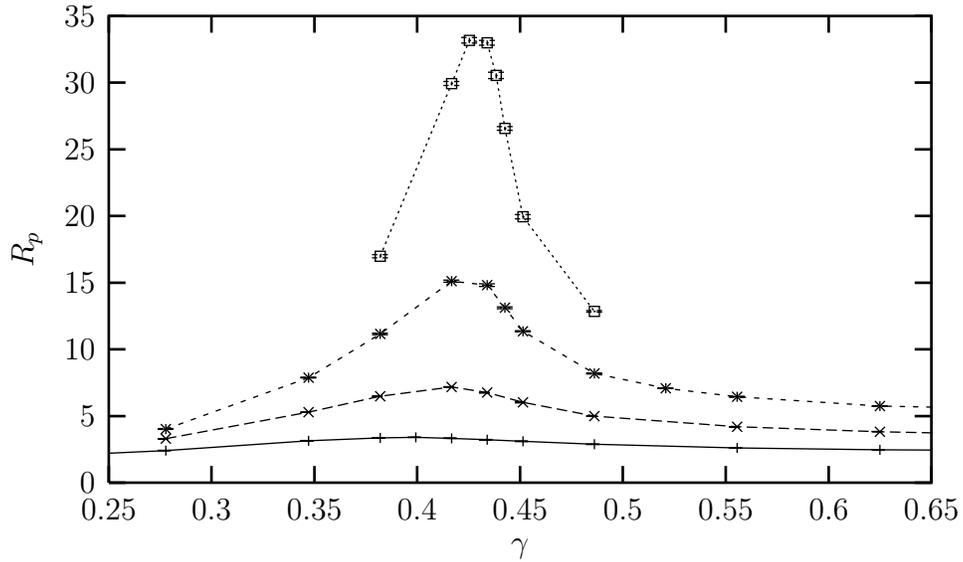

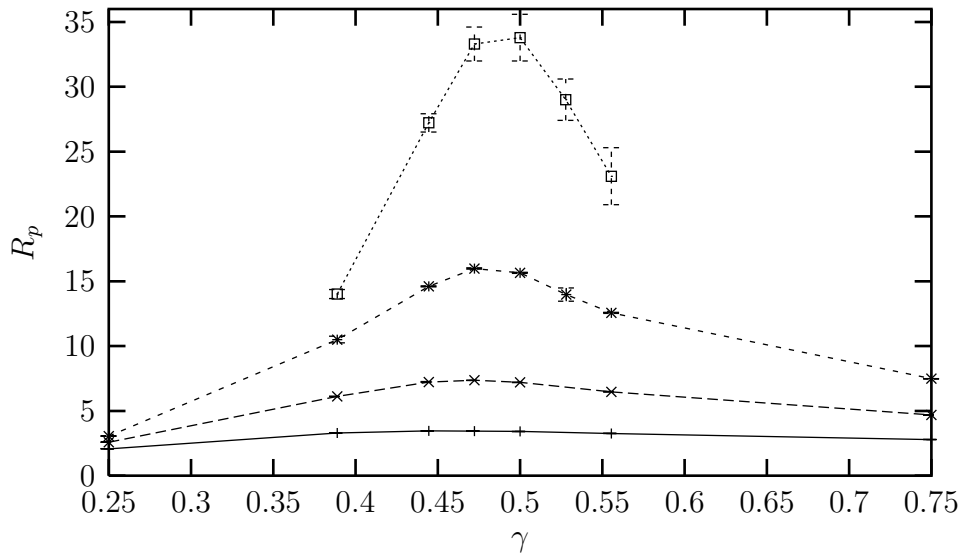
\begin	{figure}[p]
\begin	{center}
\leavevmode
\input	{1+1d_ratiograph}

\end	{center}
\vskip 0.15in
\caption{As in Fig.~\ref{fig_ratio}, but in  1+1 dimensions.}
\label{fig_1+1d_ratio}
\end 	{figure}

We see further evidence for a Gross--Witten--like transition in 
our results for the ratio $R_p$ defined in eq.~\ref{eqn_ratio}. 
Our results for $SU(6)$, $SU(12)$, 
$SU(24)$ and $SU(48)$ are plotted in Fig.~\ref{fig_ratio}.
There is a clear peak around $\gamma\simeq 0.43$ whose height
increases rapidly with $N$, indicating that the fluctations of the 
extreme eigenvalues are becoming much larger than fluctuations
of the `middle' eigenvalue (the one nearest $\alpha=0$).
Very similar behaviour occurs in 1+1 dimensions, for which our results
are shown in Fig.~\ref{fig_1+1d_ratio}. Indeed, in the $N\to\infty$ limit
in 1+1 dimensions we expect that $R_p$ will diverge, as discussed
below eq.~\ref{eqn_ratio}. This appears to be exactly
what we see in Fig.~\ref{fig_1+1d_ratio}.

Finally we compare the $D=2+1$ and $D=1+1$ transitions directly 
by comparing the eigenvalue densities across the transition. 
We do this for SU(12) in
Fig.~\ref{fig_eigenvaluedensity}. The eigenvalue densities
both below and above the transition are clearly very similar
in 1+1 and 2+1 dimensions.

All the above confirms that $D=2+1$ SU($N$) gauge theories
possess an $N=\infty$ third-order strong-to-weak coupling
transition that is remarkably similar, even in its details, 
to the $D=1+1$ Gross-Witten transition.

\begin	{figure}[p]
\begin	{center}
\leavevmode
\input	{outplanegraph}

\end	{center}
\vskip 0.15in
\caption{The plaquette correlator, $C_o$, as a function of 
$\gamma=\frac{\beta}{2N^2}$
for SU(6) ($+$), SU(12) ($\times$), 
SU(24) ($\ast$) and SU(48) ($\Box$).}  
\label{fig_outplane}
\end 	{figure}
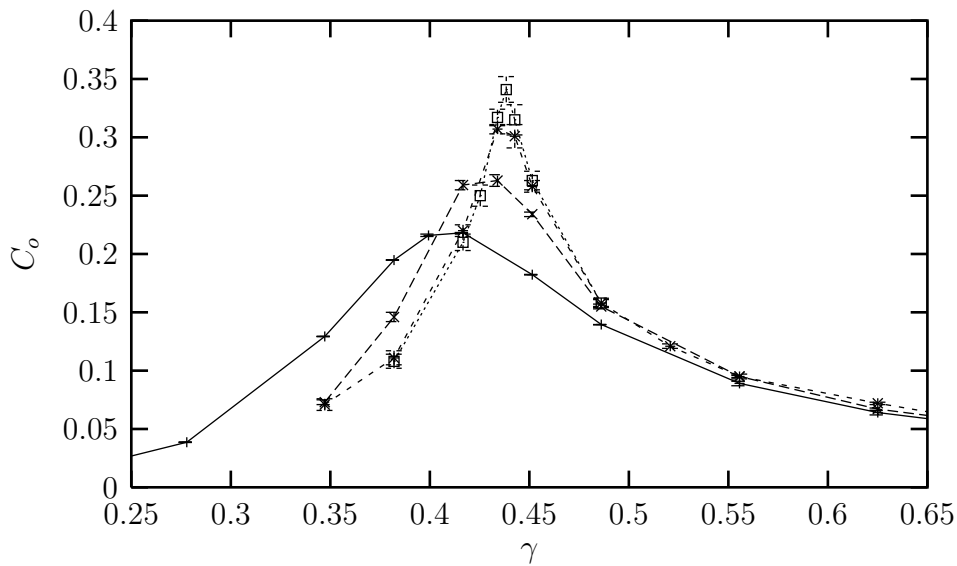

Despite this striking similarity, when we look in more
detail we also observe significant differences 
between the bulk transition in 2+1 dimensions and the 
Gross--Witten transition. Comparing
Figs.~\ref{fig_C_2} and~\ref{fig_1+1d_C_2}, we see that there
is a peak in the specific heat in $D=2+1$ which is simply
not present in $D=1+1$. From Fig.~\ref{fig_P_2} it is clear
that this peak does not come from fluctuations of individual 
plaquettes, but must come from correlations between different 
plaquettes. To investigate this we consider the following
particular contributions to the specific heat $C_2$:
the contribution from correlations between a plaquette and 
its neighbours in the same plane, which we label $C_i$; the 
contribution from correlations between a plaquette and its 
neighbours which share an edge but are not in the same plane,  
$C_o$; and finally $C_f$, the contribution from correlations 
between a plaquette and the plaquettes facing it across an 
elementary cube. We include a factor $N^2$, as for $C_2$.
We find a clear 
peak, growing with $N$, in our results for $C_o$, plotted 
in Fig.~\ref{fig_outplane}. The peak accounts for about half 
of the difference between $C_2$ and $P_2$. There is also 
a much weaker peak in $C_i$, approximately a factor of 15
times smaller, which also clearly grows with $N$, at least 
up to $N=24$. (The weakness of the signal means that
we lose statistical significance for larger $N$.)
For $C_f$, where we happen to have results only for SU(6) 
and SU(12), we see in both cases a clear peak. This is
almost exactly a factor of four lower than the corresponding 
peak for $C_o$. Since each plaquette has four times as many 
out--of--plane neighbours as it has neighbours facing it 
across an elementary cube, this shows the correlation of 
a plaquette with its individual out--of--plane neighbours 
is in fact the same 
as with a facing plaquette. By contrast the correlation
with the `nearer' neighbouring plaquettes that are in the
same plane (as measured by $C_i$) is very much weaker.
This pattern is precisely what one would expect
if the correlations were due to a flux emerging from the cube
symmetrically through every face, i.e. due to the presence
of monopole--instantons. 

If such monopoles are present, we would expect the correlation
of the plaquette with itself, $P_2$, to be also affected.
These correlations of the plaquette with itself should be as 
large as with each of its eight out--of--plane neighbours,
so this contribution to $P_2$ should about one eighth of $C_o$.
Even for $SU(48)$ this is only $\sim 0.04$, which is easily consistent
with our results in Fig.~\ref{fig_P_2}. Of course, if there really
is a second--order phase transition at $N=\infty$, then  eventually
we would expect to see a growing peak in $P_2$. It is of course 
possible that there is no second--order phase transition at
$N=\infty$, but only a rapid cross-over, so that $C_o$ asymptotes 
to a finite value, and in that case there would not need be a 
pronounced peak in $P_2$. This however seems a rather articial
scenario. 

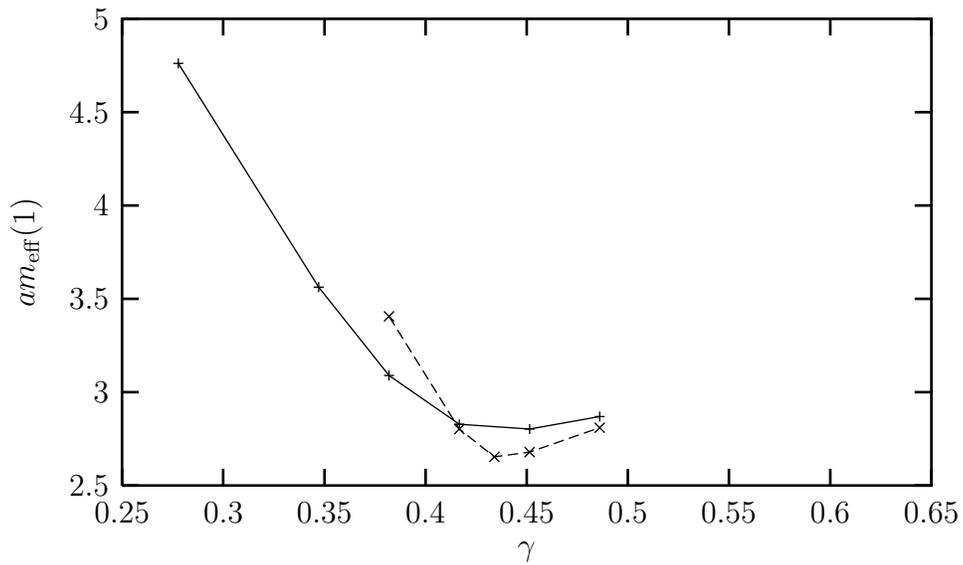
\begin	{figure}[p]
\begin	{center}
\leavevmode
\input	{massgraph}

\end	{center}
\vskip 0.15in
\caption{The lightest effective mass that couples to the plaquette
as a function of $\gamma=\frac{\beta}{2N^2}$
for SU(6) (solid line)and SU(12) (dashed line).} 
\label{fig_mass}
\end 	{figure}

Since a second-order transition is usually associated with
a diverging correlation length, we also measured the mass of the 
lightest particle that couples to the plaquette, in both $SU(6)$  
and $SU(12)$. We calculate an effective mass at a separation of 
$n$ lattice units:
\begin{equation}
a m_\mathrm{eff}(n)
=
-\mathrm{ln}\frac{\langle \phi_0 \phi_n \rangle - 
\langle \phi_0 \rangle^2}{\langle \phi_0 \phi_{n-1} \rangle - 
\langle \phi_0 \rangle^2}
\label{eqn_m_eff}
\end{equation}
where $\phi_0$ is the trace of a plaquette and 
$\phi_n$ is the trace of a facing plaquette lying in the same
plane $n$ lattice spacings away. This is not a zero--momentum 
correlator, so it will overestimate the masses. (We do not have
a statistically significant signal from zero-momentum
correlators.)  Our results for $a m_\mathrm{eff}(1)$ are plotted
in Fig.~\ref{fig_mass}. (Our results for  $n\geq 2$ do not have 
a useful statistical accuracy.) We observe a dip in the effective 
masses 
near the transition, which becomes more significant for $SU(12)$,
particularly when we take into account the expected weak-coupling
scaling behaviour, $am \propto 1/\beta$.
While this is certainly consistent with a second-order cross-over,
the masses are large, and if the correlation length
is going to show any sign of diverging it is clear that it 
will be at a much larger value of $N$ than are accessible
to our calculations.

\subsection{Wilson loops}
\label{subsection_RWL}

\subsubsection{Traces and correlators}
\label{subsubsection_RWLcor}

\begin	{figure}[p]
\begin	{center}
\leavevmode
\input	{wilsongraph}

\end	{center}
\vskip 0.15in
\caption{Trace of $2\times 2$ Wilson loop in
SU(6) ($+$) and SU(12) ($\times$) and of the
$3\times 3$ loop in SU(6) ($\star$) and SU(12) ($\Box$).}
\label{fig_wilsongraph}
\end 	{figure}
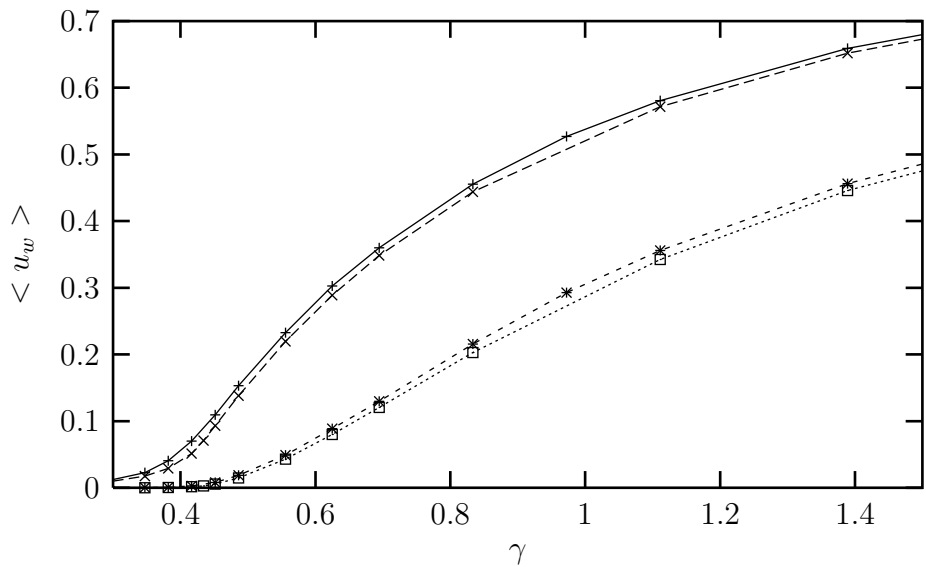

\begin	{figure}[p]
\begin	{center}
\leavevmode
\input	{P_2wgraph}

\end	{center}
\vskip 0.15in
\caption{Local `specific heat' of  $2\times 2$ and
$3\times 3$  Wilson loops in  SU(6) and SU(12).
Labels as in Fig.~\ref{fig_wilsongraph}.}
\label{fig_P_2wgraph}
\end 	{figure}
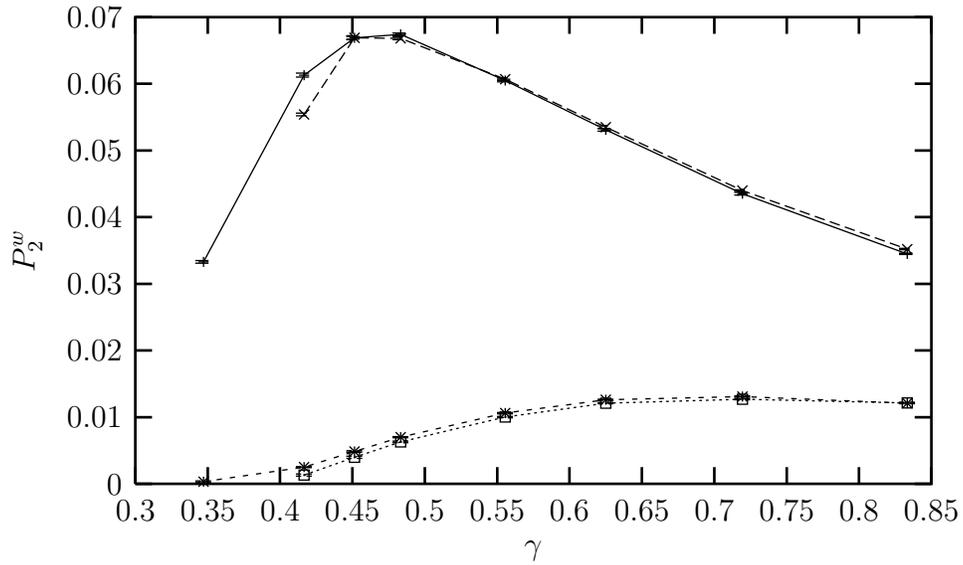

In Fig~\ref{fig_wilsongraph} we show how $\langle u_w \rangle$  
varies with $\lambda$ in some sample calculations. We see no 
sign of any singularity developing in this quantity, or in
our simulataneous calculations of 
$\partial \langle u_w \rangle / \partial\lambda$, in contrast
to the growing peak we saw for 
$C_2\propto \partial \langle u_w \rangle / \partial\lambda$ 
in Fig.~\ref{fig_C_2}. The local version  of the correlator
corresponding to the first derivative defined in eqn(\ref{eqn_WLP11}),
$P^w_2$, is more accurately calculated and its variation with 
$\lambda$ is shown in  Fig~\ref{fig_P_2wgraph}  and shows no evidence 
of a developing cusp that would suggest an $N=\infty$ singularity
in the second derivative. Thus, at this level of accuracy, we
see no evidence for any $N=\infty$ non-analyticity in the
variation of  $\langle u_w \rangle$ as a function of
the coupling $\lambda$.  

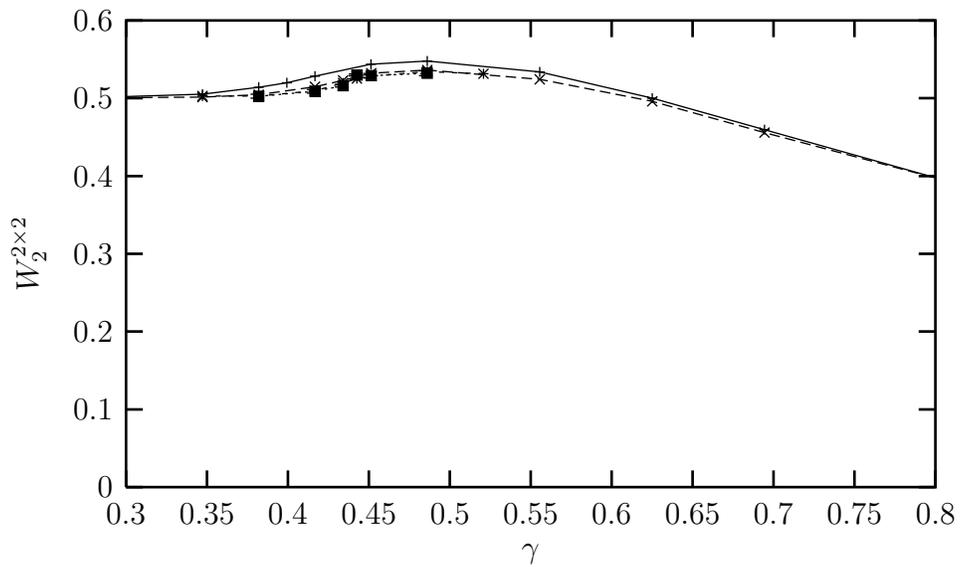
\begin	{figure}[p]
\begin	{center}
\leavevmode
\input	{W_2graph}

\end	{center}
\vskip 0.15in
\caption{Quadratic correlator of $2\times 2$ Wilson loops, 
$W_2^{2\times 2}$, as a function of  $\gamma=\frac{\beta}{2N^2}$
for SU(6) ($+$), SU(12) ($\times$), 
SU(24) ($\ast$) and SU(48) ($\Box$).}  
\label{fig_W_2}
\end 	{figure}

Given our uncertainty in the type of analyticity that might
occur we have also considered it worthwhile to look at the
quantities $W_2^{n\times n}$ and $W_3^{n\times n}$ which are defined 
in eqs.~\ref{eqn_WLP13} and ~\ref{eqn_WLP15} and which are 
alternative analogues of the quantities $P_2$ and $P_3$ for the plaquette.
Some results for  $SU(6)$, $SU(12)$, $SU(24)$ and $SU(48)$
are plotted in Fig.~\ref{fig_W_2}. While there is a range of 
$\gamma=1/\lambda$ over which $W_2^{2\times 2}$
stops being constant and starts to decline, this `transition'
does not become sharper with $N$, in contrast to the behaviour of
the plaquette equivalent, $P_2$, in Fig.~\ref{fig_P_2}.
We see the same behaviour for $3\times 3$ and $4\times 4$ Wilson loops,
only shifted to higher $\gamma$, with no sign of the
transitions becoming sharper as $N$ increases.
Our results for $W_3^{2\times 2}$ for SU(6) and SU(12)
are plotted in Fig.~\ref{fig_W_3}. While there is a transition region
in which $W_3^{2\times 2}$ becomes negative, just as one sees
for $P_3$ in Fig.~\ref{fig_P_3}, the transition does not become
sharper as $N$ increases, unlike $P_3$.
Again we see the same behaviour for $W_3^{3\times 3}$
and $W_3^{4\times 4}$, with no sign of the transitions becoming
sharper as $N$ increases. 

All these results are in fact essentially identical to those we obtain
in similar calculations in $D=1+1$, where we know that the 
$\langle u_w \rangle$ is analytic in $\lambda$ except at the Gross-Witten
transition. 

Finally we recall that for the plaquette the Gross-Witten 
transition is characterised by a divergence in the relative
fluctuation of extremal eigenvalues, as shown in
Fig.~\ref{fig_ratio}. For $n\times n$ Wilson loops the analogous 
quantity, $R^{n\times n}$, shows no such behaviour,
as we see, for the example of  $R^{2\times 2}$ for $SU(6)$, $SU(12)$ 
and $SU(24)$ plotted in Fig.~\ref{fig_wilsonratio}. This is perhaps
no surprise, given that in $D=1+1$ the eigenvalue gap formation
for Wilson loops larger than the plaquette does not involve
growing fluctuations of the extremal eigenvalues.

\subsubsection{Matching eigenvalue spectra}
\label{subsubsection_RWLeig}

Although we have found no evidence that the trace of a Wilson loop
is non-analytic in $\lambda$ at some critical area, it is possible
that there are more subtle non-analyticities of the kind that
exist in $D=1+1$ and which are associated with a gap forming in the
eigenvalue spectrum.

To determine numerically whether at some given $\lambda$ the spectrum 
$\rho(\alpha)$ in some region close to $\alpha =\pm \pi$ will
extrapolate exactly to zero when $N\to\infty$ is clearly a delicate 
matter, given that the values at finite $N$ from which we extrapolate
are already extremely small.

So to search for such non--analytic behaviour we explore the 
strategy of directly comparing Wilson loop eigenvalue spectra
in 1+1 and 2+1 dimensions. We first evaluate the 
spectrum in 1+1 dimensions at the critical coupling at which 
the gap forms. A true gap only forms at $N=\infty$; for finite
$N$ we use the same value of the critical 't~Hooft coupling,
\cite{DurOle,BasGriVian}
\begin{equation}
\lambda_c = \frac{1}{\gamma_c} = 4(1-e^{\frac{-2a^2}{A}}),
\label{eqn_WLr3}
\end{equation}
where $A$ is the area of the Wilson loop in physical units.
At this coupling the expectation value of the trace of the Wilson 
loop is, using eqn(\ref{eqn_upGW})
\cite{DurOle,BasGriVian},
\begin{equation}
\langle u_w \rangle = 
\left\{ \langle u_p \rangle \right\}^{A/a^2}
\stackrel{N\to\infty}{=}
\left( 1-\frac{\lambda}{4} \right)^{A/a^2}
=
e^{-2},
\label{eqn_WLr5}
\end{equation}
which is the same value as
at the critical coupling in the continuum limit.
Note also that as $a\to 0$ and $A/a^2 \to \infty$, 
eqn(\ref{eqn_WLr3}) reduces to eqn(\ref{eqn_d2Acrit}) as
it should. Having obtained the spectrum (numerically) in 
$D=1+1$ for a given size Wilson loop (in lattice units) 
and for a given value of $N$, we then calculate the eigenvalue 
spectrum in $D=2+1$ for the same size loop and for the same $N$,
varying the coupling to a value where the two eigenvalue spectra
match. 

We find that it is always possible to achieve such a match,
for any $N$ and for any size of Wilson loop.
We show an example in Fig.~\ref{fig_wilsoneigenvaluedensity},
where we compare the eigenvalue density of the $3\times 3$ 
Wilson loop in SU(12) in 1+1 dimensions to the density in 2+1 
dimensions, at a coupling chosen to give the best match. 
In Fig.~\ref{fig_wilsoneigenvaluedensity} the coupling in $D=1+1$ 
is $\lambda_c$, the coupling at which the gap forms.
The spectra are clearly very similar and indeed indistinguishable
on this plot. We also find that the spectra can be matched 
when they are away from the critical coupling:
we show this in Fig.~\ref{fig_wilsoneigenvaluedensity2},
where we choose a higher value of $\gamma$ to
illustrate that the matching continues to work after the 
gap forms. In Figs.~\ref{fig_wilsoneigenvaluedensity} and 
~\ref{fig_wilsoneigenvaluedensity2} we also plot the analytically
known spectra
\cite{DurOle,BasGriVian}.
in the continuum limit of the $N=\infty$ theory in  $D=1+1$
at the corresponding couplings. These clearly match
the corresponding finite-$N$ spectra very well, except
in two respects: the latter have  $N$ bumps which arise from
the eigenvalue repulsion that is a well-known characteristic 
of the Haar measure, and the finite-$N$ spectrum is not
precisely zero in the region of the `gap'.

The fact that at finite but large $N$ we can match so precisely 
the $D=1+1$ and $D=2+1$ eigenvalue spectra for couplings at and
above the $D=1+1$ transition, provides convincing evidence that 
the Wilson loops in the $D=2+1$ $N=\infty$ theory also undergo
a transition involving the formation of a gap in the eigenvalue
spectrum. 

In Fig.~\ref{fig_wilsonsizematch} we plot the eigenvalue spectra
of $2\times 2$, $3\times 3$, and $4\times 4$ loops in SU(6) in 
2+1 dimensions. The three couplings have been chosen so as to give 
the best match to the eigenvalue spectra of Wilson loops of the 
same size in $D=1+1$ at $\lambda_c$. We see that the three spectra
are essentially identical. Moreover the critical value of
the  $D=2+1$  coupling $\gamma_c = 1/\lambda_c$ 
appears to grow linearly with the size of the $L\times L$ loop, 
suggesting that there is a finite critical area 
for gap formation in the contimuum limit:
\cite{DurOle,BasGriVian}
\begin{equation}
\lambda_c^2 L^2 
\stackrel{a\to 0}{=} 
(ag^2 N)^2 L^2 
=  
(g^2 N)^2 A_{crit}. 
\label{eqn_WLr7}
\end{equation}
As we shall see below, in Section~\ref{subsubsection_RWLtheory},
this is nearly but not quite the case.
 
It turns out that all the above is an immediate corollary
of a much stronger 
and rather surprising result concerning the matching of Wilson loop
eigenvalue spectra in 1+1 and 2+1 (and indeed 3+1) dimensions.

The general statement is that if take an $n\times n$ Wilson
loop  $U_w^{n\times n}$ in the SU($N$) gauge theory and calculate
the eigenvalue spectra in $D$ and $D^\prime$ dimensions, we
find that the eigenvalue spectra match at the couplings $\lambda_D$ 
and $\lambda_{D^\prime}$ at which the averages of the traces 
$u_w^{n\times n}=\frac{1}{N}\mathrm{Re}\mathrm{Tr}\{U_w^{n\times n}\}$
are equal:
\begin{equation}
\langle u_w^{n\times n}(\lambda_D) \rangle_D
=
\langle u_w^{n\times n}(\lambda_{D^\prime})\rangle_{D^\prime}.
\label{eqn_WLr9}
\end{equation}
We have tested this matching for $D=1+1$ and $D=2+1$ over groups 
in the range $N=2$ to $N=48$ and for Wilson loops ranging in size 
from $1\times 1$ (the plaquette) to $8\times 8$ and, in
2+1 dimensional, for couplings from $\lambda = 4.0$ to 
$\lambda = 0.40$. We have in addition tested it in the deconfined
as well as in the confined phase. Some  sample calculations in
$D=3+1$ have also been performed
\cite{BurTepVai}
strongly suggesting that the same is true there.

The fact that such a precise matching is possible implies that the 
eigenvalue spectrum is completely determined by $N$, the size of 
the loop, and its trace. Hence the eigenvalues are not really 
independent degrees of freedom, which is unexpected. Moreover
we have seen in Fig.~\ref{fig_wilsonsizematch} a demonstration
of the fact that the spectra of Wilson loops that are $2\time 2$ 
and larger can also be matched with each other. The matching
occurs at values of the traces that are the same as those
in $D=1+1$ where they are calculable. In this sense, the
size of the Wilson loop is not really an extra variable here.
Finally, the $N$ dependence is weak, and consists mainly
of the two differences noted earlier.

Finally we remark that our results at this stage rely 
on a comparison that
is visual and impressionistic. Ideally one would like
to match the spectra by varying $\lambda$ continuously
and this can be done, from nearby calculated values of
the coupling,
by standard reweighting techniques. In addition it would
be useful to quantify any differences (which must be very 
small) with a standard error analysis. We intend to
provide such analyses elsewhere
\cite{BurTepVai}.

\subsubsection{Polyakov loops}
\label{subsubsection_RWLpoly}

We have also investigated the eigenvalue spectra of Polyakov loops
as defined in Section~\ref{section_lattice}.
These are products of link matrices that wrap around one 
of the space-time tori (and are of minimal length unless
specified otherwise) i.e. they
can be thought of as  non-contractible Wilson loops.
They provide the conventional order parameter for the deconfinement
phase transition. As one crosses this transition the Polyakov loop 
that winds around the time (temperature) torus  acquires a
non-zero expectation value. This corresponds to the spontaneous
breaking of a global centre symmetry in the Euclidean system.   
To simulate the system at temperature $T$ we use a $L_s^2 L_0$ 
lattice with $L_s \gg L_t$ so that  $T=1/aL_0$. As  $N$ grows
one can weaken the inequality, so that one can take $L_s \to L_t$ 
as $N\to \infty$ while still maintaining the thermodynamic 
interpretation and the sharp phase transition. 

We calculated the eigenvalue spectra of timelike Polyakov loops  
in $SU(12)$ on $L_s^2 L_0$ lattices. 
We found that it is always possible to match the Polyakov loop
eigenvalue spectra to those of Wilson loops in 1+1 dimensions
(and hence also to Wilson loops in 2+1 dimensions) by choosing
couplings at which the trace of the Polyakov loop equals that
of the Wilson loop   
\begin{equation}
|\langle {\bar l}_{\mu=0} \rangle|
=
\langle u_w \rangle
\label{eqn_WLr11}
\end{equation}
where, as we have seen, the size of the Wilson loop does not
matter to a very good approximation. (We take the modulus
because the Polyakov loop is proportional to some element of 
the centre in the deconfined phase and the modulus effectively 
rotates that element to unity. The eigenvalue spectrum also
needs to be rotated by the same centre element.) 
This matching has the corollary
that the Polyakov loop eigenvalue spectrum will develop a gap 
at $N=\infty$ when its trace crosses the critical value 
$ |\langle {\bar l}_{\mu=0} \rangle| = e^{-2}$. For $N > 4$
the deconfining transition at $T=T_c$ is strongly first order
and the value of $ |\langle {\bar l}_{\mu=0} \rangle| $ will jump 
from  $ |\langle {\bar l}_{\mu=0} \rangle| = 0$ at $T< T_c$ to
some non-zero value for $T = T_c^+$. The latter value will
typically be greater than $e^{-2}$ for small $L_0$, i.e. for coarse
lattice spacings, and will $\to 0$ as $a \to 0$ and hence 
$L_0 \to \infty$. Moreover for fixed $L_0$ the trace increases
with increasing $T$. (See Section~\ref{subsubsection_RWLtheory}
for why this is so.) Thus for coarse lattice spacings we expect
the gap formation to occur at the phase transition, $T=T_c$, 
while for larger $L_0$ it will not coincide with the deconfining
transition; instead it will occur at some $T > T_c$. The critical
value  turns out to be $L_0 = 7$. Thus in the continuum limit
the gap formation in timelike Polyakov loops does not occur 
at $T=T_c$ but rather at $T=\infty$.
 
As a numerical example of the eigenvalue matching 
we show in Fig~\ref{fig_polyakovwilsonmatch},
the eigenvalue spectra of the timelike Polyakov
loop just below and just above the deconfinement transition for $L_t=4$,
together with a $3\times 3$ Wilson loop spectrum in 1+1 dimensions
at a coupling chosen to match the spectrum of the deconfined 
Polyakov loop. The spectra clearly match closely.
Since the 1+1 dimensional $\gamma$
is above $\gamma_c$ for the $3\times 3$ loop, the Wilson loop
will develop a gap at this coupling in the large--$N$ limit. Hence
the Polyakov loop will presumably also develop  a gap.

Finally we recall that as $N\to\infty$ the deconfining transition
occurs on smaller spatial volumes $L_s \to L_0$ so that
at $N=\infty$ one can discuss the transition on a $L^3$ lattice.
Taking into account the fact that our preliminary results
\cite{BurTepVai}
indicate that all the above carries over to Wilson and Polyakov
loops in $D=3+1$, we can make direct contact with the observation in
\cite{NN03,NNect04}
that  on an $L^4$ lattice the Polyakov loop develops a gap
when it develops a non-zero expectation value.

\subsubsection{Theoretical interpretation}
\label{subsubsection_RWLtheory}

The fact that at $N=\infty$ there is a gap at weak coupling 
in the eigenvalue spectra of Wilson loops, has a simple
explanation in the theory of Random Matrices. (See e.g.
\cite{RMT}
for a recent review.) At $N=\infty$ the Gaussian Unitary 
Ensemble (GUE) of complex Hermitian  $N\times N$ matrices
generates an eigenvalue spectrum that is the 
well-known Wigner semicircle 
\begin{equation}
\rho(\lambda) 
\stackrel{N\to\infty}{\propto}
\left(1-\frac{\lambda^2}{4}\right)^{\frac{1}{2}}.
\label{eqn_WLr15}
\end{equation}
In weak coupling, when $\beta\to\infty$, the SU(N) link matrices
can be expanded in terms of the Hermitian gauge potentials
and it is very plausible that the averages involved in
the calculation of Wilson loops fall into the same
`universality class' as the GUE. That is to say, once the
eigenvalues of Wilson loops are clustered close to unity,
the fact that the phases are on a circle rather than on 
the line becomes irrelevant and the phases (suitably rescaled 
by the coupling) should be distributed according to the semi-circle 
in eqn(\ref{eqn_WLr15}). In fact this is precisely what we find. 
Thus the existence of a gap in the eigenvalue spectrum at
weak coupling has a rather general origin in terms of Random
Matrix Theory.

On the other hand we know that in a confining theory 
\begin{equation}
\langle u_w \rangle
\propto
e^{-\sigma A}
\stackrel{A\to\infty}{\longrightarrow}
0
\label{eqn_WLr17}
\end{equation}
which requires a nearly flat eigenvalue spectrum in $[-\pi,+\pi]$.
Thus as we decrease the lattice spacing, the eigenvalue spectrum
of a $L\times L$ Wilson loop must change from being
nearly uniform to eventually having a  Wigner semicircle gap.
Thus at some bare coupling it must pass through a transition
where the gap forms. 

For this gap to be physically significant, it must occur at
a fixed physical area in the continuum limit. However, as we 
shall now see, this is not the case for either $2+1$ or  
$3+1$ dimensions (in contrast to $D=1+1$). The reason is the
perturbative self-energy of the sources whose propagators
are the straight-line sections of the Wilson loop. (Often
referred to as the `perimeter term'.)
The leading correction is given by the Coulomb potential $V_c(r)$
at the `cutoff' $r=a$. For a Wilson loop whose size is 
$l\times l = aL\times aL$ in physical units, this correction is
\begin{equation}
\delta \log \langle u_w \rangle
\propto
l V_c(a)
\propto
\begin{cases}
\lambda L \log a  &  D=2+1 \\ 
\lambda L   &  D=3+1 
\end{cases}
\label{eqn_WLr19}
\end{equation}
using the fact that $V_c(r) \propto g^2N\log r, g^2N/r$ and
$\lambda = ag^2N, g^2N$ in $D=2+1, 3+1$ respectively.
Let us, for illustrative purposes, assume that the full potential
is given by this self-energy and the area piece, 
$\sigma A = a^2 \sigma L^2$, that comes from linear confinement.
Then we have
\begin{equation}
\langle u_w \rangle
\propto
\exp\left\{c \lambda L \log\lambda - c^\prime  \lambda^2 L^2\right\} 
 \qquad : \qquad D=2+1 
\label{eqn_WLr21}
\end{equation}
using the fact that $a^2\sigma \propto (ag^2N)^2 = \lambda^2$
and $\log a = \log\lambda + \cdots $ in  $D=2+1$, and
\begin{equation}
\langle u_w \rangle
\propto
\exp\left\{c \lambda L  - c^\prime  e^{-\frac{c_r}{\lambda} }
L^2\right\}
 \qquad : \qquad D=3+1 
\label{eqn_WLr23}
\end{equation}
using the fact that $a^2\sigma \propto \exp\{-c_r/g^2N\}$ in 
$D=3+1$, where $c_r$ is given by the coefficients of the 2-loop 
renormalisation group equation. 

Consider first the $D=2+1$ case in eqn(\ref{eqn_WLr21}). Since
$\lambda L = ag^2N L = g^2Nl$ is the length scale in physical units,
we see that if it were not for the weakly varying $\log\lambda$ 
term in eqn(\ref{eqn_WLr21}), the Wilson loop trace would 
be the same on the lattice and in the continuum (up to the usual
$O(a^2)$ lattice corrections). That is to say, we expect
that as we approach the continuum limit, $\lambda \to 0$, 
the critical area for gap formation will vanish
\begin{equation}
A_{crit} \propto \frac{1}{(\log\lambda)^2}
\stackrel{a\to 0}{\longrightarrow} 0
\label{eqn_WLr25}
\end{equation}
rather than tending to some finite limit. At coarse $a$ the
logarithmic correction will be weak and one might well be
tempted to perform an extrapolation to the continuum limit
that does not include it. We illustrate all this with a numerical 
calculation of the coupling, and hence lattice spacing, at which 
$L\times L$ loops develop a gap. We define the appearance of a `gap'
in our finite--$N$ calculations as the coupling at which the 
spectrum is closest to the spectrum
of the $L\times L$ loop in 1+1 dimensions at the coupling $\gamma_c$,
for the same $N$. For SU(2) we calculated this coupling for $L$ up to 8
on $16^3$ lattices. For SU(6) we calculated up to $L=4$ on $6^3$ lattices.
We show our results in Fig.~\ref{fig_wilsoncriticalbeta}, together 
with a best fit to the SU(2) data which has the asymptotic 
behaviour in eqn(\ref{eqn_WLr25}). The numerical data shows deviations
from linearity which could either be interpreted as low $L$
corrections to an asymptotic scaling behaviour 
$\gamma = \lambda^{-1} \propto L$, or as a logarithmic violation
of this asymptotic scaling. From our above analysis we know the
latter to be the correct interpretation.

In contrast to the anomalous behaviour we see when taking the
continuum limit of $\lambda_c(A)$, the large--$N$ limit
is achieved rapidly and smoothly. To illustrate this we
list in Table~\ref{table_Ngap} the coupling for which the 
$3\times 3$ loop develops a gap for $N\in [2,48]$. 
The critical coupling is essentially constant from $SU(6)$ onwards, 
showing that are in the large--$N$ limit. Indeed, even for SU(2) the
corrections are small.

In the case of $D=3+1$ the self-energy diverges linearly
and will normally dominate the trace for all $\lambda$
in the weak coupling region. Thus we expect 
$A_{crit} \propto a^2$ up to logarithmic corrections from
the running coupling, so that the gap formation occurs
in the deep ultraviolet as we approach the continuum limit.

In contrast, in $D=1+1$ where the Coulomb potential is linear
$V_c(r) \propto g^2 N r$, the self-energy term contributes
at most a mere lattice spacing correction that vanishes
in the continuum limit.

From the above discussion we see that the anomalous 
behaviour of $A_{crit}$ as $a\to 0$ arises from divergent
self-energy contributions. If the source had a finite
mass, so that the propagator was smeared over some range
$\delta r \sim 1/\mu$, we would evaluate the Coulomb
self-interaction at $r=1/\mu$ rather than at $r=a$
and hence would replace
$\lambda L \log\lambda \to \lambda L \log \mu$ 
in eqn(\ref{eqn_WLr21}), and 
$\lambda L \to \lambda/\mu$ in eqn(\ref{eqn_WLr23}).
Assuming the univerality of the gap formation persists
for such loops, we would then expect them to form a
gap at a value of $A_{crit}$ that is finite in the continuum
limit if we have chosen $\mu$ to be finite in physical units.
The value of  $A_{crit}$ will of course depend on the value
of $\mu$.

Similar considerations apply to Polyakov loops.

While the above considerations make plausible the universality
aspect of the gap formation in Wilson and Polyakov loops,
they do not explain our most striking result which is that
the complete eigenvalue spectra can be matched across space-time 
dimension and loop size by merely matching traces.

Finally, whether the gap formation, and the associated non-analyticity, 
has any significant physical implications is unclear. For that to be
so one would require that the gap should form at a fixed physical
area $A_{crit}$ in the continuum limit. As we have seen that is not
the case in $D=2+1$ or in $D=3+1$ and is only the case in
$D=1+1$, where there are no propagating degrees of
freedom and so no `physics' in the usual sense. One can imagine
regularising the divergent self-energies so that  $A_{crit}$
is finite and non-zero in the continuum limit, but then it would 
appear to depend on the regularisation mass scale $\mu$ used.

\subsection{Finite volume}
\label{subsection_Rvol}

In Section~\ref{subsection_V} we argued that $D=3+1$ $SU(N)$
gauge theories on $L_0 \ll L_1 \ll L_2 \ll L_3$ lattices
undergo a series of $N=\infty$ phase transitions at 
$\beta_{c_0} \ll \beta_{c_1} \ll\beta_{c_2} \ll\beta_{c_3}$.
These phase transitions are essentially deconfining
transitions, $a(\beta_{c_i})L_i=1/T^{D=4-i}_{c}$, in a dimensionally
reduced theory. We argued that continuity, and vanishing
finite size corrections at large $N$, link these transitions 
to the $N=\infty$ phase transitions on $L^4$ lattices
discussed in
\cite{NNlat05,NNect04,NN03}. 

The same argument clearly holds for $SU(N)$ gauge theories on 
$L_0 \ll L_1 \ll L_2$ lattices in $D=2+1$. Here we provide some
(very) exploratory numerical results in support of this scenario.
We do so on lattices with a less than asymptotic ordering,
$L_0 < L_1 < L_2$. 

\subsubsection{First transition}
\label{subsubsection_firstV}

The first transition is the usual deconfining phase transition
when $L_1,L_2\to\infty$. It is second order for SU(2) and SU(3),
either second or first order for SU(4),
and first order for $N\geq 5$
\cite{TcD3}.  
Because the latent heat for $N\geq 5$ is $\propto N^2$, the 
cross-over on a finite $L_0 < L_1, L_2$ lattice will become a 
first-order phase transition at $N=\infty$. All this is 
well-established and does not require further numerical 
confirmation in this paper.

\subsubsection{Second transition}
\label{subsubsection_secondV}

To search for the second transition we simulate $SU(12)$ gauge 
fields on a $L_0 L_1 L_2 = 2\times 4\times 40$ lattice over a large 
range of $\gamma = \beta/2N^2$.  We calculate the Polyakov 
loop around the $\mu=1$ torus, average it over the given lattice 
field, and take the modulus: $|{\bar l}_1|$. This provides the
conventional order parameter for a deconfining transition 
with the $L_1=4$ torus providing the (inverse) temeprature.
We plot results for the average of this, 
$\langle |{\bar l}_{\mu=1}|\rangle$, at each value of $\gamma$
in Fig.~\ref{fig_l1_40}. We also plot values of the plaquette 
difference, $\langle (u_{01}-u_{02})\rangle$, which should
also reflect such a transition. We see in  Fig.~\ref{fig_l1_40}
a very clear signal for a transition at $\gamma \sim 3.2$
in both quantities. This occurs at a temperature
$T_{c_1} \equiv T^{D=1+1}_{c} = 1/4a(\gamma \sim 3.2)$ in the 
reduced theory. 
In units of the usual deconfining temperature of the $D=2+1$ 
gauge theory, $T_{c_0} = T^{D=2+1}_{c}$, this amounts to
$T_{c_1} \sim 3 T_{c_0}$. 

The rapid, steep crossover suggests that the transition is 
first order. We find support for this when we plot a histogram
of the values of  $|{\bar l}_1|$ at some of the values of
$\gamma$ in the cross-over region. This is illustrated in
Fig.~\ref{fig_histl1b} for an ensemble of $SU(12)$ fields on a
$2\times 4\times 80$ lattice at $\gamma = 3.368$. We see
a clear two state signal, with the peak at small $|{\bar l}_1|$ 
being naturally interpreted as coming from fields in the confining
phase and the peak at large  $|{\bar l}_1|$ as coming from fields
in the deconfined phase. Such a two-state signal is typical
of a first order deconfining transition. 

The next question is whether this cross-over will sharpen into an 
actual  phase transition in the two interesting limits: when we 
increase the spatial volume (here just $aL_2$) at fixed $N$; 
or when we increase $N$ at fixed volume. 
In addressing the former question we need to remark upon 
some special features of first-order transitions in the
effective $D=1+1$ theory that we are discussing here. The 
high temperature deconfined phase is normally characterised
by a centre symmetry breaking so that 
${\bar l} \sim c(\beta) \exp\{i2\pi n/N\}$ where  $c(\beta)$
is a self-energy renormalisation factor. 
Two such phases, characterised
by $n$ and $n^\prime$ say, can coexist and will be separated 
by a domain wall whose tension we expect 
\cite{sigkref}
to be
\begin{equation}
\sigma_k \propto k(N-k)\frac{T^2}{\sqrt{g^2_2 N}} 
\qquad ; \qquad k=|n-n^\prime| \ , \ T = \frac{1}{aL_1}.
\label{eqn_sigk}
\end{equation}
In one spatial dimension the domain wall is just a `point'
and so the usual energy/entropy arguments tell us that
at $T = 1/aL_1 \geq T_{c_1}$ the field will break up
into domains of typical size 
\begin{equation}
\Delta r 
\propto 
\exp\left\{+\frac{\sigma_k}{T}\right\}
\label{eqn_Dsize}
\end{equation}
Thus at any $T$ if we take $L_2\to\infty$ the volume will
consist of a `gas' of domain `walls', and hence domains, and
on the average these will be equally distributed
amongst all the centre phases, so that
$\langle |{\bar l}_{\mu=1}|\rangle\to 0$. However on
volumes that satisfy $L_1 \ll L_2 \ll \Delta r$ we
will typically be in one domain and will thus have the usual 
deconfining signal of a non-zero value for $|{\bar l}_{\mu=1}|$.
In addition it is clear that the lightest mass, $m_p$, coupling
to the $\mu=1$ Polyakov loop will not vanish at $T > T_{c_1}$
but will approximately satisfy 
$m_p \propto \exp\{-cNT/\surd\lambda\}$. Note that this
mass decreases with increasing $L_1=1/aT_1$ in contrast
to the stringy behaviour, $m_p \propto L_1$, in the
confining phase. It is clear from the above that  our conventional 
signals for being in a deconfined phase become more complicated to 
interpret in $D=1+1$.

Similar considerations apply to the deconfining transition
itself. Suppose that there are confining and deconfining 
phases that differ by a free energy density $f$. At 
$T=T_{c_1}$ we have $f=0$ so that the typical field will
consist of a `gas' of domain walls of typical size
$\Delta r \propto \exp\{+\sigma_{cd}/T\}$ where
$\sigma_{cd}$ is the free energy of the confining-deconfining
interface. This is the essential difference with higher
dimensions. For large enough volume ($=L_2$) half the domains
will be confining and half will be deconfining. Let us now
increase the temperature $T$ a little above  $T_{c_1}$.
Then $f= \epsilon_0 (T-T_{c_1})$ near $T_{c_1}$,
where $\epsilon_0 = [m]^0$ in $D=1+1$. If $T-T_{c_1}$
is small enough, then $\Delta r f(T)/T \ll 1$ and
the fraction of the volume that is still in the confined
phase will be $\propto \exp\{- \Delta r f(T)/T\} \sim O(1)$.
That is to say, the transition will take place over a
range of temperatures $\Delta T$ that is no smaller than
\begin{equation}
\frac{\Delta T}{T_{c_1}}
\propto 
\epsilon^{-1}_0 \exp\{- \sigma_{cd}/T_{c_1}\}
\label{eqn_Tsized2}
\end{equation}
and this remains non-zero in the infinite volume limit.
This implies that in $D=1+1$ there cannot be an infinitely
sharp first order transition even in the large volume limit.
However, because both $\sigma_{cd}$ (probably) and
$f$ (certainly) grow $\propto N^2$, there can be a
phase transition at $N=\infty$, and this can occur 
at finite volume. 

\begin	{figure}[p]
\begin	{center}
\leavevmode
\input	{plot_a1V}
\end	{center}
\vskip 0.015in
\caption{The average plaquette difference 
$\langle \delta u \rangle = 10^3\times\langle(u_{01}-u_{02})\rangle$
in SU(12) on $2\times 4\times L_2$ lattices with 
$L_2=10$ ($+$),  $L_2=40$ ($\circ$), and $L_2=80$ ($\bullet$),
versus $\gamma=\beta/2N^2$.}
\label{fig_a1V}
\end 	{figure}
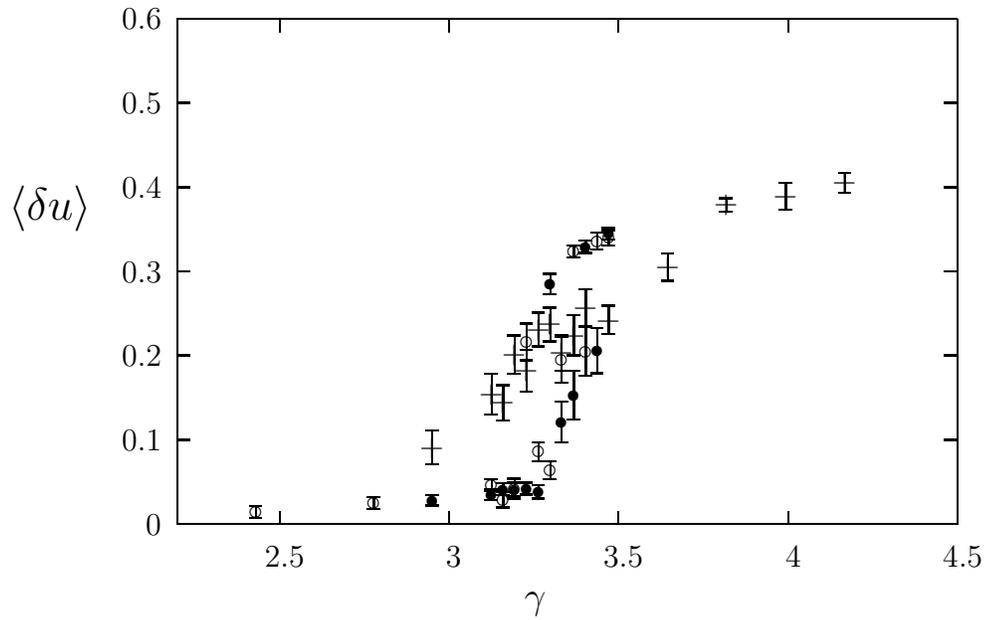

Returning to our numerical results, we begin with $SU(12)$
and show in Fig.~\ref{fig_a1V} how the average plaquette 
difference $\langle(u_{01}-u_{02})\rangle$ varies
across the transition when we vary the `spatial' volume, $L_2$.
(We expect the plaquette difference to be less sensitive
to domain formation than the Polyakov loop.)   
It is clear that the transition does become much
sharper when we pass from $L_2 =10$ to  $L_2 =40$ although
the nature of the change between $L_2 =40$ and  $L_2 =80$ 
is less clear. The evidence is for a would-be first-order
transition inhibited by the domain formation described in
the previous paragraph.  

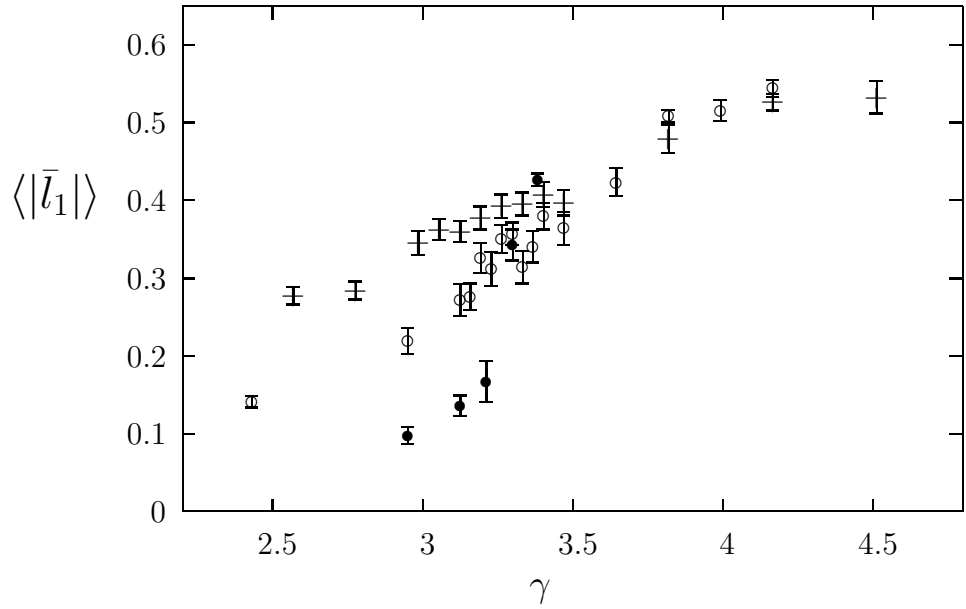
\begin	{figure}[p]
\begin	{center}
\leavevmode
\input	{plot_l1}
\end	{center}
\vskip 0.015in
\caption{The average $\mu=1$ Polyakov loop 
for $SU(6)$  ($+$), $SU(12)$ ($\circ$), and $SU(24)$ ($\bullet$)
versus the inverse bare 't Hooft coupling $\gamma=\beta/2N^2$,
all on  $2\times 4\times 10$ lattices.}
\label{fig_l1}
\end 	{figure}

Turning now to the $N$-dependence of the transition,
we show in Fig.~\ref{fig_l1} how 
$\langle |{\bar l}_{\mu=1}|\rangle$ 
varies with $\gamma$ on a $2\times 4\times 10$ lattice for
$SU(6)$, $SU(12)$ and $SU(24)$ gauge theories. We see a
rapid sharpening of the transition with increasing 
$N$ which leaves little doubt that there is a first-order
transition at $N=\infty$ at $L_2=10$, and presumably 
at other values of $L_2$ as well.

\subsubsection{Third transition}
\label{subsubsection_thirdV}

To search for a third transition, characterised by
a non-zero expectation value for $|{\bar l}_{\mu=2}|$,
we take our $SU(12)$ gauge theory on an $2\times 4\times 10$
lattice and increase $\gamma$ beyond the values associated
with the transitions discussed above. In Fig.~\ref{fig_l2}
we plot the resulting values of  $\langle|{\bar l}_{\mu=2}|\rangle$.
We see a transition of the kind that we are looking for,
but one which is very smooth. Increasing $N$ to $N=24$ we
see what appears to be a significant sharpening of the
transition, suggesting that it might become an actual
phase transition at $N=\infty$.
 
\begin	{figure}[p]
\begin	{center}
\leavevmode
\input	{plot_histl2b}
\end	{center}
\vskip 0.015in
\caption{Histogram of values of $|{\bar l}_{\mu=2}|$ in SU(24)
on a  $2\times 4\times 10$ lattice at $\gamma = 156.25$.}
\label{fig_histl2}
\end 	{figure}
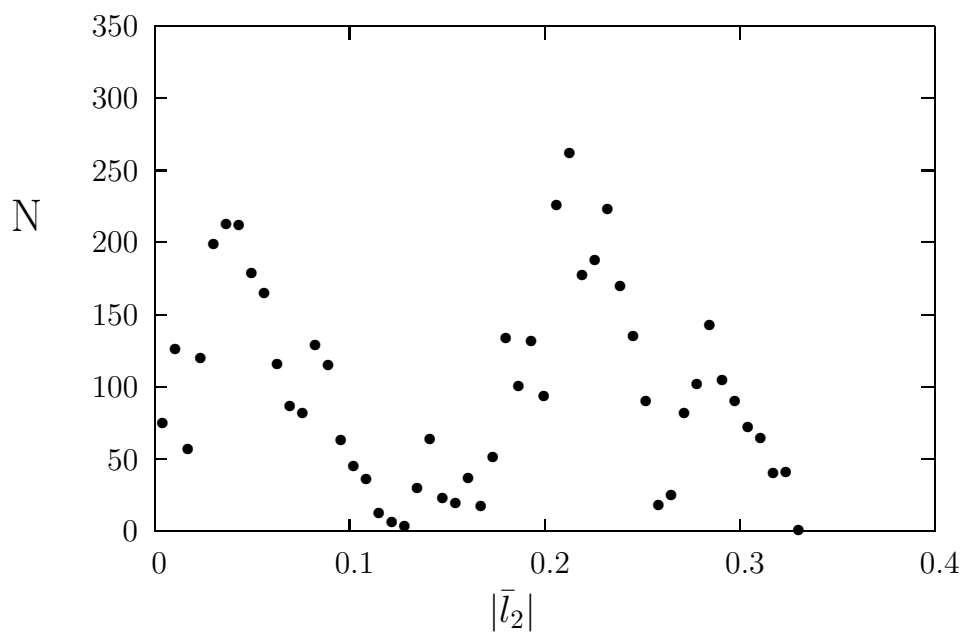

In  Fig.~\ref{fig_histl2} we show a histogram of the values
of $|{\bar l}_{\mu=2}|$ obtained in $SU(24)$ on a $2\times 4\times 10$
lattice at $\gamma = 156.25$. This shows a clear peak at low
values that one naturally interprets as belonging to the confined 
phase, and a further peak (or peaks) at larger values that
one naturally associates with the deconfined phase.
This suggests that if this is a phase transition 
at $N=\infty$ then it is first order.

\section{Conclusions}
\label{section_conclusions}

We have shown that there is a very close match in the behaviour
of several observables across the  bulk transition that separates
strong and weak coupling in 2+1 dimensions, and the 
Gross--Witten transition in 1+1 dimensions. In particular
the third derivative of the partition function, 
$C_3 \propto N^4 \partial^3 \log Z/\partial\beta^3$, appears
to develop a  discontinuity as $N\to\infty$, just as it does
across the Gross-Witten transition, providing strong
evidence for a third order transition in the large--$N$ limit
of $D=2+1$ $SU(N)$ gauge theories. 

When we expressed  $\partial^3 \log Z/\partial\beta^3$ as a
cubic correlator of plaquettes, we saw that the discontinuity
arose from fluctuations of plaquettes at the same position.
It is thus a genuine $N=\infty$ phase transition that arises 
from the $N^2\to\infty$ degrees of freedom on each 
plaquette rather than from the collective behaviour of a large
number of separated plaquettes. This motivated us to study 
the eigenvalue spectrum of the plaquette. We found that at 
the critical inverse 'tHooft coupling, $\gamma = \gamma_c$, 
the spectrum develops a gap at the boundary of its range 
$e^{i\alpha} = \pm 1$ and this gap grows as $\gamma$ increases.
While the gap formation does not, in itself, lead to a 
nonanalyticity in $Z$, it possesses a feature that does.
At  $\gamma = \gamma_c$ and for $N\to\infty$ the spectrum 
$\rho(\alpha)$ approaches its end-points with a vanishing
derivative. This means that the extreme eigenvalues possess
fluctuations that diverge compared to the $O(1/N)$ fluctuations
of the eigenvalues in the bulk of the spectrum, and this is
directly related to the singularity in the partition function.
All these features are exactly the same as in the $D=1+1$
Gross-Witten transition. In addition 
we find that there is a very close match in the behaviour
of the plaquette eigenvalue density and in the ratio
of plaquette eigenvalue fluctuations when we compare
the transitions in $D=2+1$ and in $D=1+1$. 
Thus it would appear that the bulk transition in 2+1 dimensions is 
very much like the Gross--Witten transition.

However, there is clearly more than this going on.
The Gross--Witten transition has no peak in the specific heat,
but we see in 2+1 dimensions a clear peak that coincides 
with (or is very close to) the third order transition.
The contribution from
neighbouring or nearly neighbouring plaquettes appears to
grow with $N$, indicating a possible second--order phase 
transition in the large--$N$ limit. Whether this is
due to a correlation length that diverges as $N\to\infty$
(we see a slight decrease in the lightest mass that couples
to the plaquette when we go from $SU(6)$ to $SU(12)$)
or to the plaquette fluctuations decreasing more
slowly than $1/N^2$ at the critical point,
is not clear at present. In any case, the fact that the correlations
between nearby plaquettes behave as if due to a flux
emerging from an elementary cube, suggests that the transition
may be due to centre monopole(-instanton) and vortex condensation. 
It is therefore 
plausible that this (possible) second--order phase transition 
is connected to the line of specific heat peaks in the 
fundamental--adjoint plane found in $SU(2)$
\cite{BaigCuervo}, 
which may also become a line of second--order phase transitions 
in the large--$N$ limit, and which, just as in $D=3+1$
\cite{SmixedMono}, 
can be understood in terms of condensation
of $Z_N$ monopoles and vortices. In $D=3+1$ this phase structure 
is believed to lead to the observed first order bulk transition.
Our $N=\infty$ second-order transition would appear to be
a manifestation of the same dynamics, but in one lower dimension.
Numerical calculations that are both more accurate and extend to
larger $N$ are clearly needed here.

We have also investigated the sequence of finite volume 
transitions on $L_0 L_1 L_2$ lattices. We argued
that when the tori are strongly ordered, $L_0\ll L_1 \ll L_2$, 
these can be understood in terms of deconfinement, 
followed by high-$T$ dimensional reduction as $\beta$ 
is increased, followed by deconfinement in the reduced system,
and so on. So the first transition, which is first-order
for large $N$, occurs at $\beta=\beta_{c_0}$
where $a(\beta_{c_0})L_0 = 1/T_{c_0}$. Then as we increase
$\beta$, and hence $T$, the system will eventually be dimensionally  
reduced, $L_0 L_1 L_2 \to L_1 L_2$ for $T\gg T_{c_0}$. This
$L_1\ll  L_2$ system will undergo a deconfining transition 
when $\beta=\beta_{c_1}\gg \beta_{c_0}$ where
$a(\beta_{c_1})L_1 = 1/T_{c_1}$. Due to the fragmentation
of the high-$T$ phase into domains (a feature of 1 spatial
dimension) this transition is a finite cross-over at
finite $N$. As $N\to\infty$ the domain `wall' tension
should diverge (probably as $N^2$), the domain structure
will be suppressed, and the cross-over appears to become a
genuine first order transition. At higher $\beta$ and
hence higher $T^{D=1+1}$, we can again expect dimensional
reduction to occur $L_1 L_2 \to L_2$ at some $T^{D=1+1}\gg T_{c_1}$.
and there is some evidence that the `infinitesimal' $L_0\times L_1$
system undergoes a transition at $a(\beta_{c_2})L_2 = 1/T_{c_2}$ 
with  $\beta_{c_2}\gg \beta_{c_1}$. 

The first of these finite volume transitions becomes a phase 
transition as $L_1,L_2 \to \infty$ at fixed $N$. It also becomes 
a phase transition at fixed $L_1,L_2$ as $N\to\infty$. The latter
will occur even as $L_1,L_2 \to L_0$. The second
transition, which is a deconfining transition in the $D=1+1$ 
effective theory, is a sharp crossover at finite $N$
and appears to be first-order. As we remarked above, an 
actual phase transition
is not possible at finite $N$ in 1 spatial dimension
because of domain formation.
However, as we pointed out, this does not preclude a
first-order  phase transition at $N=\infty$, as suggested
by our numerical computations. Such a  $N=\infty$ phase 
transition will continue to occur for $L_2 \to L_1$,
and it may be that it also 
survives the $L_1,L_2 \to L_0$ limit, although
we have not investigated this possibility. Finally
we saw some numerical evidence for a third  $N=\infty$ 
transition in the effective $D=0+1$ theory, although
here the calculations are no more than suggestive. These
arguments can trivially be lifted to $SU(N)$ gauge theories in
3+1 dimensions, where they clearly have some relation to the 
$N=\infty$ finite volume transitions on $L^4$ lattices 
transitions discussed in
\cite{NNlat05,NNect04,NN03}.

In view of conjectures 
\cite{NNlat05,NNect04,NN03}
that Wilson loops in $D=3+1$ may undergo $N=\infty$ 
non-analyticities, when their area, in physical units, 
reaches a critical value, we have analysed the behaviour
of Wilson loops in $D=2+1$ $SU(N)$ gauge theories. 
Our results show a remarkable match between the behaviour of
Wilson loops in $D=2+1$ and in $D=1+1$,
where a gap in the Wilson loop eigenvalue spectrum 
is known to open at a critical area at $N=\infty$, in
both the lattice and continuum theories
\cite{DurOle,BasGriVian}.
We find that
the eigenvalue spectra of Wilson loops (and indeed Polyakov 
loops) in $D=2+1$ match those of Wilson loops in $D=1+1$ when 
the traces are equal. Moreover the spectra of Wilson
loops of any size (in lattice units and when larger than about
$2\times 2$) also match if the couplings are tuned to values 
where their traces are equal. This is true for any fixed $N$.
As a corollary, it immediately follows that in $D=2+1$ at 
$N=\infty$ a gap will form in the eigenvalue spectrum of a 
Wilson loop at a critical coupling that depends on the
size of the loop. However because of a logarithmically
divergent self-energy piece, this non-analyticity in
the spectrum will not occur at a finite non-zero value
of the area in the continuum limit. This is in contrast
to the case in $D=1+1$. We have preliminary evidence 
\cite{BurTepVai}
for a similar matching between Wilson loops in $D=3+1$
and those with the same trace in lower dimensions.
Here the self-energy divergence is even more severe
and the gap forms deep in the ultraviolet. As in 
$D=2+1$ one can imagine regularising this self-energy
by using finite mass sources in constructing the
`Wilson/Polyakov loops', so as to obtain a gap
formation at a fixed physical area.

The appearance of the gap at $N=\infty$ follows 
quite generally if we make plausible connections with
Random Matrix Theory. The spectrum of an $l\times l = aL\times aL$ 
Wilson loop should be flat at large $a$, since linear confinement 
demands its trace to be $\propto \exp\{-\sigma l^2\} \sim 0$,
while at sufficiently small $a$ we expect to find the 
Wigner semi-circle of the $N=\infty$ Guassian Unitary Ensemble.
Somewhere in between a gap must form. Because the
derivative of the spectrum diverges at its end-point,
in contrast to the plaquette at the bulk transition, there
are no anomalous fluctations of the extreme  eigenvalues 
and no non-analytic behaviour in the correlators that
are related to derivatives of the Wilson loop with respect to
the coupling. And indeed we find the traces of Wilson loops 
to be analytic in the coupling just as they are in $D=1+1$. 
Thus the physical implications of this nonanalyticity in the
eigenvalue spectrum remain unclear.

The remarkable  similarity between the eigenvalue spectra
of Wilson loops in different dimensions does not
appear to have a simple explanation within Random Matrix
Theory and merits a more careful and quantitative investigation
than the one provided in this paper.

The  $D=2+1$ large-$N$ phase structure that we have investigated 
in this paper can be understood, as we 
have argued above, in terms that appear to allow a unified
understanding of these phase transitions in $D=1+1$,  
$D=2+1$ and $D=3+1$  $SU(N)$ gauge theories.

\vspace*{0.30in}

\leftline{Note added.}

This revised version arose from our discovery, immediately
after sending the original version to the archive, of the
papers
\cite{DurOle,BasGriVian}
which then motivated our revised and extended study of Wilson
loops in this paper. As this revision was in progress 
an interesting paper 
\cite{NN06}
on gap formation in smeared Wilson loops in $D=3+1$ has 
appeared.

\vspace*{0.30in}

\section*{Acknowledgements}

We are grateful to Barak Bringoltz and Helvio Vairinhos
for useful discussions throughout this work and to
Gernot Akemann for convincing us of the potential relevance
of random matrix theory.
We are also grateful to Rajamani Narayanan and Herbert Neuberger 
for discussions much earlier that sparked our original 
interest in these problems.
Our lattice calculations were carried out on PPARC and EPSRC 
funded computers in Oxford Theoretical Physics. FB acknowledges 
the support of a PPARC graduate studentship.

\vfill\eject

\begin{table}
\begin{center}
\begin{tabular}{|c|c|}\hline
$N$ & $\gamma_c$ \\ \hline
2 & 0.700(3) \\
6 & 0.719(3) \\
12 & 0.722(2) \\
24 & 0.722(1) \\
48 & 0.722(2) \\ \hline
\end{tabular}
\caption{\label{table_Ngap}
Inverse coupling at which gap forms for $3\times 3$ loops in $SU(N)$.}
\end{center}
\end{table}

\begin	{figure}[p]
\begin	{center}
\leavevmode
\input	{C_2graph}

\end	{center}
\vskip 0.15in
\caption{The specific heat, $C_2$, as a function of 
$\gamma=\frac{1}{ag^2N}=\frac{\beta}{2N^2}$
for SU(6) (solid line) and SU(12) (dashed line).} 
\label{fig_C_2}
\end 	{figure}
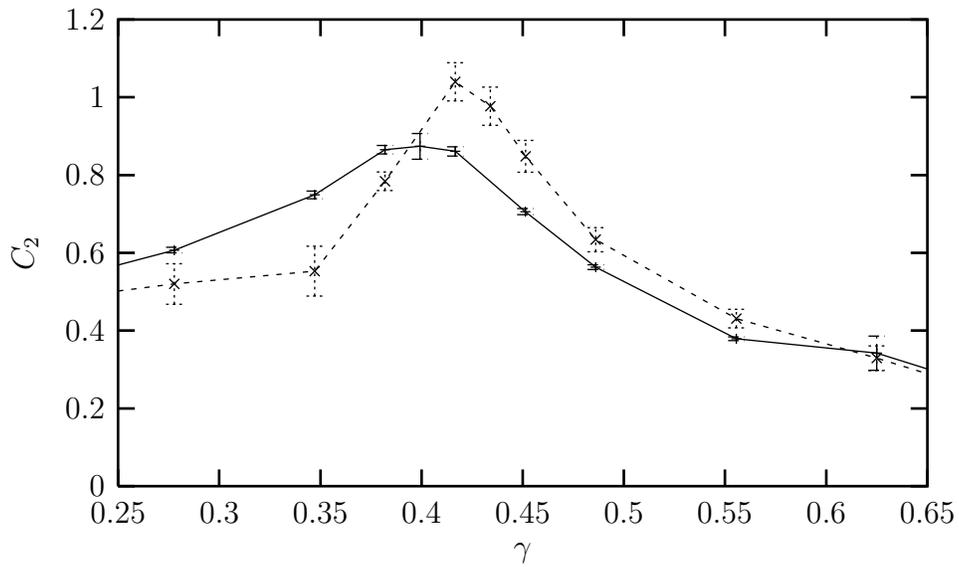

\begin	{figure}[p]
\begin	{center}
\leavevmode
\input	{eigenvaluedensitygraph}

\end	{center}
\vskip 0.15in
\caption{Density of plaquette eigenvalues, $e^{i\alpha}$, for SU(12)
in 1+1 dimensions at $\gamma=\frac{\beta}{2N^2}=0.462$ 
(long dashes) and $\gamma=0.542$ (dots)
and in 2+1 dimensions at $\gamma=0.417$ (solid line)
and $\gamma=0.451$ (short dashes).}
\label{fig_eigenvaluedensity}
\end 	{figure}
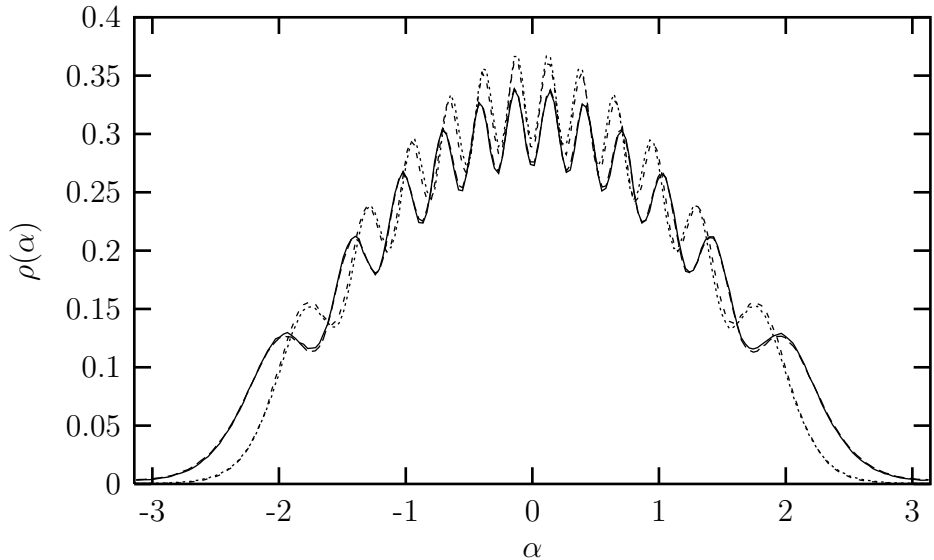

\clearpage

\clearpage

\begin	{figure}[p]
\begin	{center}
\leavevmode
\input	{W_3graph}

\end	{center}
\vskip 0.15in
\caption{Cubic correlator of $2\times 2$ Wilson loops, 
$W_3^{2\times 2}$, as a function of $\gamma=\frac{\beta}{2N^2}$
for SU(6) (solid line) and SU(12) (long dashes).} 
\label{fig_W_3}
\end 	{figure}
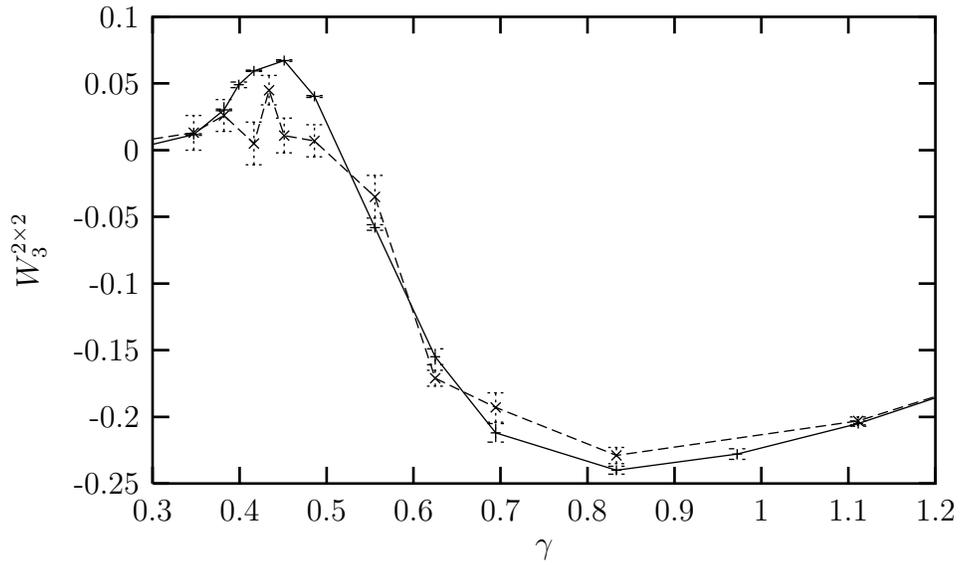

\begin	{figure}[p]
\begin	{center}
\leavevmode
\input	{wilsonratiograph}

\end	{center}
\vskip 0.15in
\caption{$R^{2\times 2}$ as a function of 
$\gamma=\frac{\beta}{2N^2}$
for SU(6) (solid line), SU(12) (long dashes) and 
SU(24) (short dashes).}  
\label{fig_wilsonratio}
\end 	{figure}
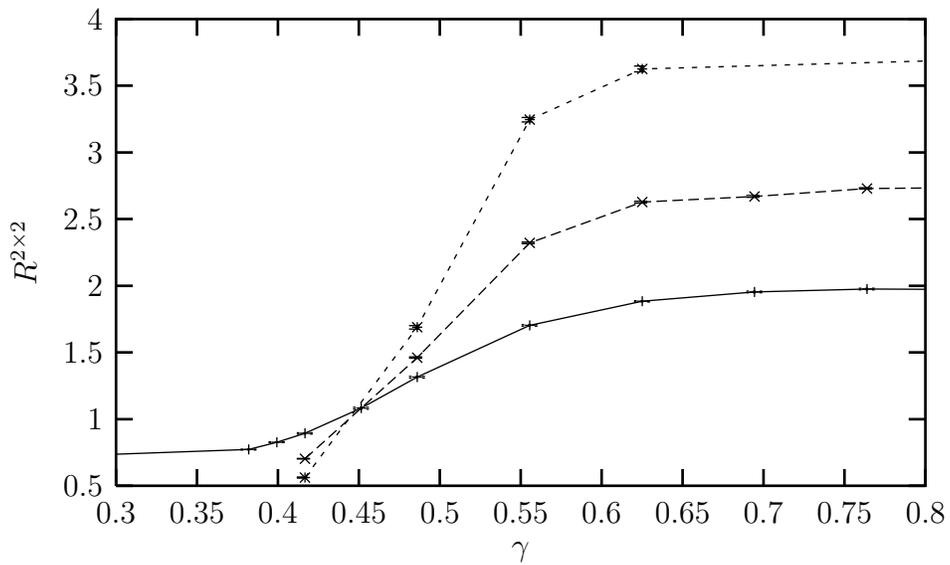

\begin	{figure}[p]
\begin	{center}
\leavevmode
\input	{wilsoneigenvaluedensitygraph}

\end	{center}
\vskip 0.15in
\caption{$3\times 3$ Wilson loop eigenvalue density, $e^{i\alpha}$, 
for SU(12) in 1+1 dimensions at $\gamma=\frac{\beta}{2N^2}=1.255$ 
(solid line) and in 2+1 dimensions at $\gamma=0.722$ (long dashes),
and the continuum large--N distribution in 1+1 dimensions at
$A=A_{crit}$ (short dashes).}
\label{fig_wilsoneigenvaluedensity}
\end 	{figure}
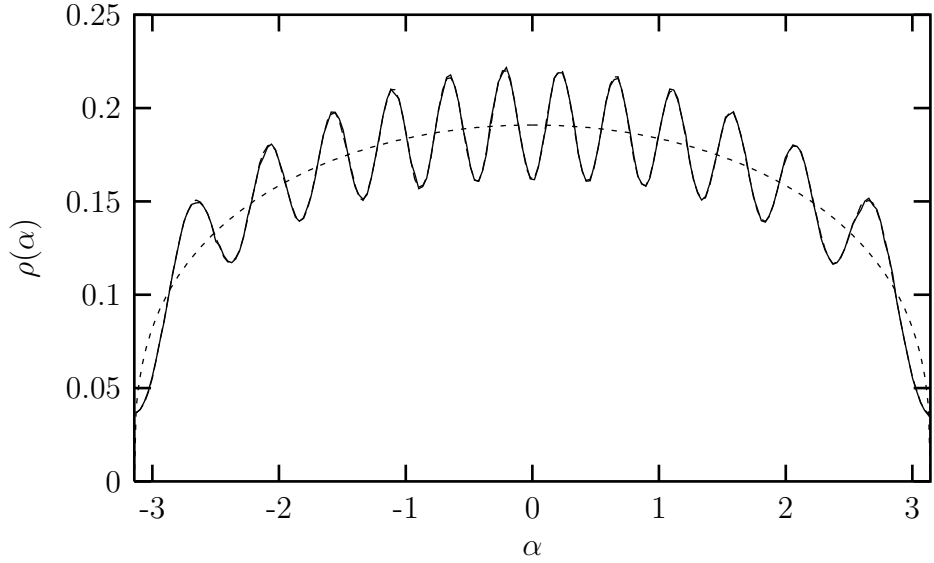

\begin	{figure}[p]
\begin	{center}
\leavevmode
\input	{wilsoneigenvaluedensitygraph2}

\end	{center}
\vskip 0.15in
\caption{$3\times 3$ Wilson loop eigenvalue density, $e^{i\alpha}$, 
for SU(12) in 1+1 dimensions at $\gamma=\frac{\beta}{2N^2}=2.215$ 
(solid line) and in 2+1 dimensions at $\gamma=1.111$ (long dashes),
and the continuum large--N distribution in 1+1 dimensions at
$A=0.539 A_{crit}$ (short dashes).}
\label{fig_wilsoneigenvaluedensity2}
\end 	{figure}

\begin	{figure}[p]
\begin	{center}
\leavevmode
\input	{wilsonsizematchgraph}

\end	{center}
\vskip 0.15in
\caption{Eigenvalue density in SU(6) in 2+1 dimensions
for the $2\times 2$ loop at $\gamma=0.483$ (solid line),
the $3\times 3$ loop at $\gamma=0.719$ (long dashes),
the $4\times 4$ loop at $\gamma=0.965$ (short dashes)
and the continuum large--N distribution in 1+1 dimensions at
$A=A_{crit}$ (dots).}
\label{fig_wilsonsizematch}
\end 	{figure}
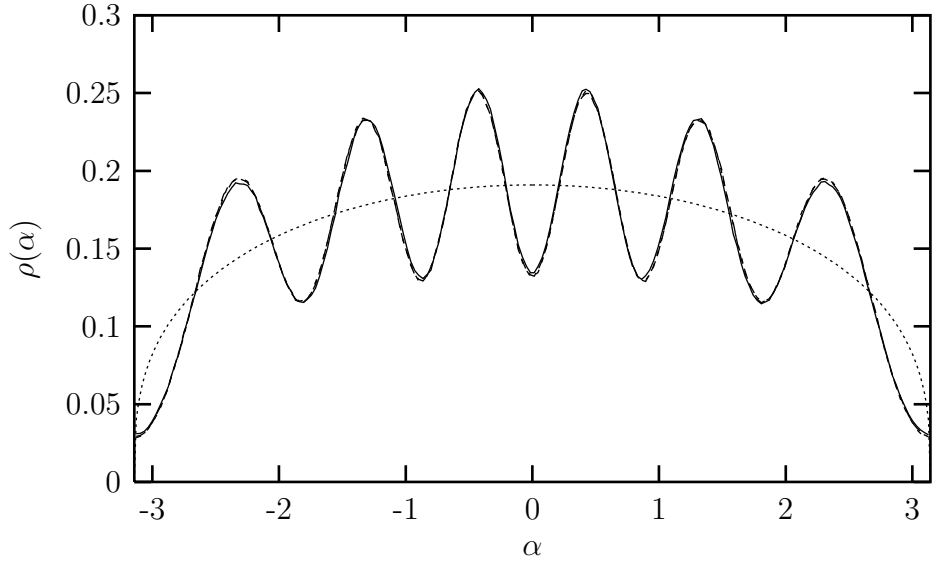

\begin	{figure}[p]
\begin	{center}
\leavevmode
\input	{wilsoncriticalbetagraph}

\end	{center}
\vskip 0.15in
\caption{Couplings for which the gap forms for $L\times L$ Wilson loops
in SU(2) ($+$) and SU(6) ($\ast$), and fit to SU(2) data (dashed line).}
\label{fig_wilsoncriticalbeta}
\end 	{figure}
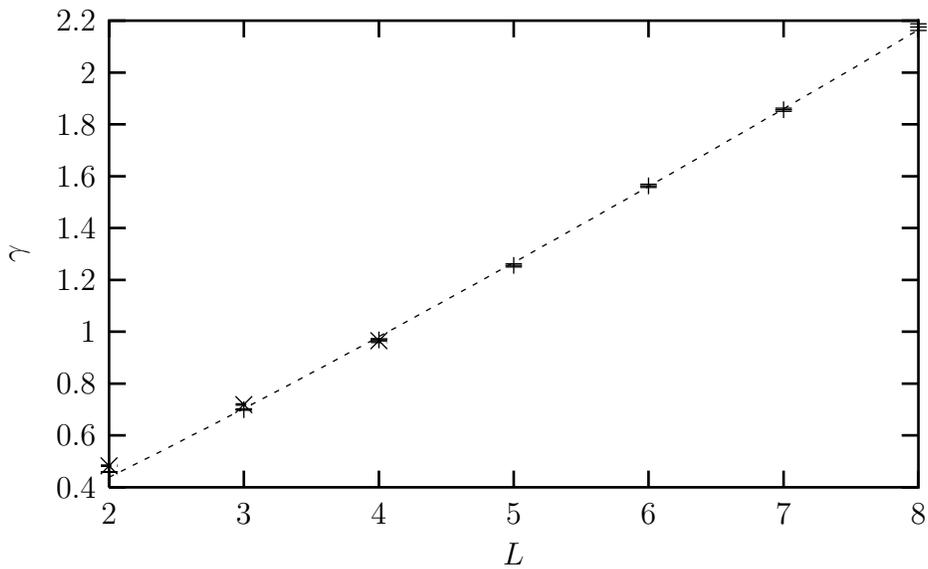

\begin	{figure}[p]
\begin	{center}
\leavevmode
\input	{polyakovwilsonmatchgraph}

\end	{center}
\vskip 0.15in
\caption{Polyakov loop eigenvalue density in SU(12) in 2+1 dimensions
in the confined phase at $\gamma=0.764$ (solid line)
and in the deconfined phase at $\gamma=0.833$ (long dashes),
and the $3\times 3$ Wilson loop in 1+1 dimensions at $\gamma=1.684$
(dashes).}
\label{fig_polyakovwilsonmatch}
\end 	{figure}
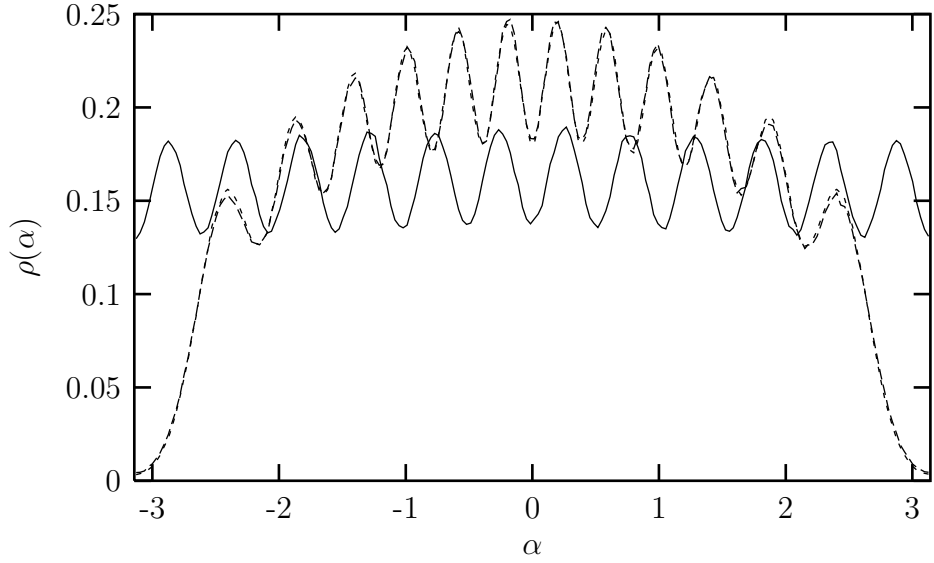

\begin	{figure}[p]
\begin	{center}
\leavevmode
\input	{plot_l1_40}
\end	{center}
\vskip 0.015in
\caption{Values of the shorter `spatial' Polyakov loop
$\langle |{\bar l}_{\mu=1}|\rangle$, $\bullet$, and
the plaquette difference 
$500\times\langle(u_{01}-u_{02})\rangle$,
$\circ$, on a $2\times 4\times 40$ lattice in SU(12)
versus the bare inverse 't Hooft coupling,
$\gamma=\beta/2N^2$.}
\label{fig_l1_40}
\end 	{figure}
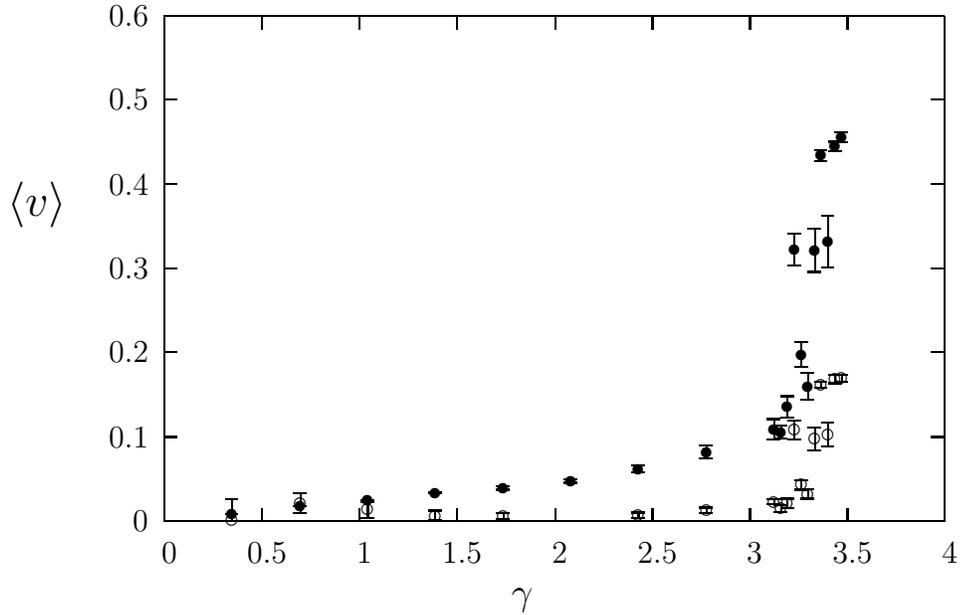

\begin	{figure}[p]
\begin	{center}
\leavevmode
\input	{plot_histl1b}
\end	{center}
\vskip 0.015in
\caption{Histogram of values of the spatial Polyakov loop, 
$|{\bar l}_{\mu=1}|$, in SU(12)
on a  $2\times 4\times 80$ lattice at $\gamma = 3.368$.}
\label{fig_histl1b}
\end 	{figure}
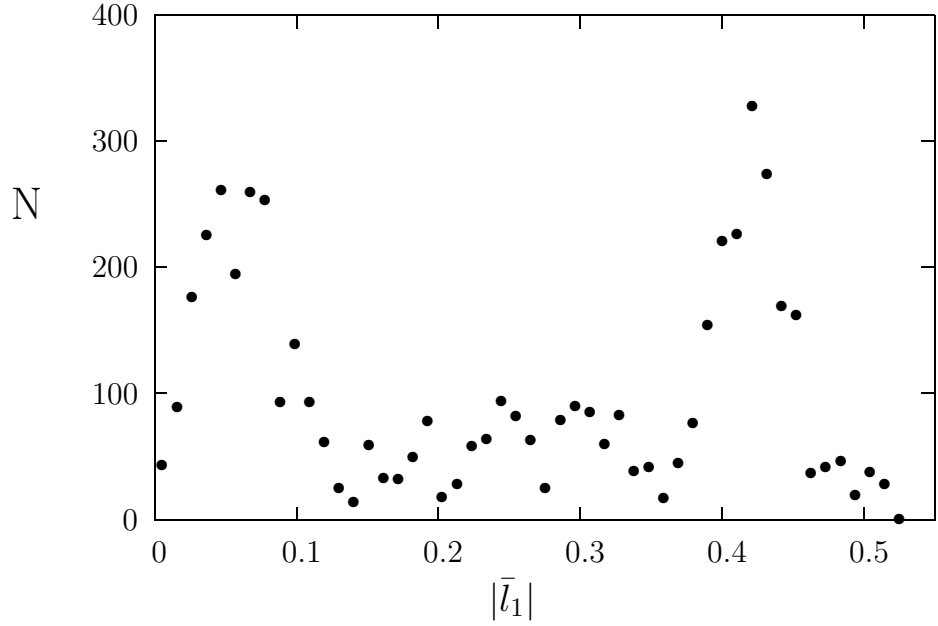

\begin	{figure}[p]
\begin	{center}
\leavevmode
\input	{plot_l2}
\end	{center}
\vskip 0.015in
\caption{The average $\mu=2$ Polyakov loop 
in $SU(12)$, $\circ$, and in  $SU(24)$, $\bullet$, on a 
$2\times 4\times 10$ lattice
versus the inverse bare 't Hooft coupling $\gamma=\beta/2N^2$.}
\label{fig_l2}
\end 	{figure}
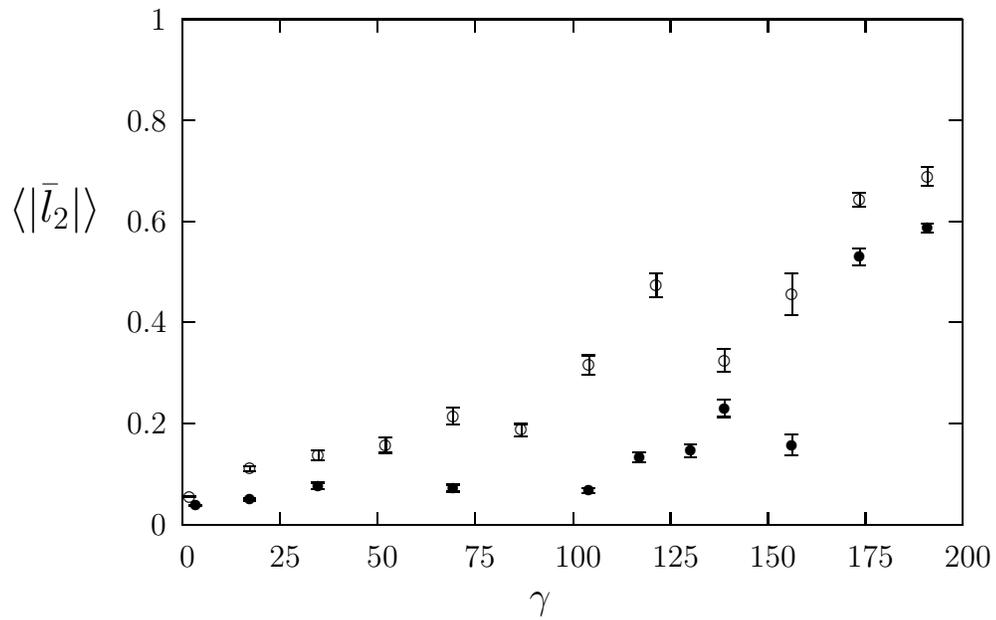

\end{document}

%% file: wilson48graph.tex
\begingroup%
  \makeatletter%
  \newcommand{\GNUPLOTspecial}{%
    \@sanitize\catcode`\%=14\relax\special}%
  \setlength{\unitlength}{0.1bp}%
{\GNUPLOTspecial{!
/gnudict 256 dict def
gnudict begin
/Color false def
/Solid false def
/gnulinewidth 5.000 def
/userlinewidth gnulinewidth def
/vshift -33 def
/dl {10 mul} def
/hpt_ 31.5 def
/vpt_ 31.5 def
/hpt hpt_ def
/vpt vpt_ def
/M {moveto} bind def
/L {lineto} bind def
/R {rmoveto} bind def
/V {rlineto} bind def
/vpt2 vpt 2 mul def
/hpt2 hpt 2 mul def
/Lshow { currentpoint stroke M
  0 vshift R show } def
/Rshow { currentpoint stroke M
  dup stringwidth pop neg vshift R show } def
/Cshow { currentpoint stroke M
  dup stringwidth pop -2 div vshift R show } def
/UP { dup vpt_ mul /vpt exch def hpt_ mul /hpt exch def
  /hpt2 hpt 2 mul def /vpt2 vpt 2 mul def } def
/DL { Color {setrgbcolor Solid {pop []} if 0 setdash }
 {pop pop pop Solid {pop []} if 0 setdash} ifelse } def
/BL { stroke userlinewidth 2 mul setlinewidth } def
/AL { stroke userlinewidth 2 div setlinewidth } def
/UL { dup gnulinewidth mul /userlinewidth exch def
      10 mul /udl exch def } def
/PL { stroke userlinewidth setlinewidth } def
/LTb { BL [] 0 0 0 DL } def
/LTa { AL [1 udl mul 2 udl mul] 0 setdash 0 0 0 setrgbcolor } def
/LT0 { PL [] 1 0 0 DL } def
/LT1 { PL [4 dl 2 dl] 0 1 0 DL } def
/LT2 { PL [2 dl 3 dl] 0 0 1 DL } def
/LT3 { PL [1 dl 1.5 dl] 1 0 1 DL } def
/LT4 { PL [5 dl 2 dl 1 dl 2 dl] 0 1 1 DL } def
/LT5 { PL [4 dl 3 dl 1 dl 3 dl] 1 1 0 DL } def
/LT6 { PL [2 dl 2 dl 2 dl 4 dl] 0 0 0 DL } def
/LT7 { PL [2 dl 2 dl 2 dl 2 dl 2 dl 4 dl] 1 0.3 0 DL } def
/LT8 { PL [2 dl 2 dl 2 dl 2 dl 2 dl 2 dl 2 dl 4 dl] 0.5 0.5 0.5 DL } def
/Pnt { stroke [] 0 setdash
   gsave 1 setlinecap M 0 0 V stroke grestore } def
/Dia { stroke [] 0 setdash 2 copy vpt add M
  hpt neg vpt neg V hpt vpt neg V
  hpt vpt V hpt neg vpt V closepath stroke
  Pnt } def
/Pls { stroke [] 0 setdash vpt sub M 0 vpt2 V
  currentpoint stroke M
  hpt neg vpt neg R hpt2 0 V stroke
  } def
/Box { stroke [] 0 setdash 2 copy exch hpt sub exch vpt add M
  0 vpt2 neg V hpt2 0 V 0 vpt2 V
  hpt2 neg 0 V closepath stroke
  Pnt } def
/Crs { stroke [] 0 setdash exch hpt sub exch vpt add M
  hpt2 vpt2 neg V currentpoint stroke M
  hpt2 neg 0 R hpt2 vpt2 V stroke } def
/TriU { stroke [] 0 setdash 2 copy vpt 1.12 mul add M
  hpt neg vpt -1.62 mul V
  hpt 2 mul 0 V
  hpt neg vpt 1.62 mul V closepath stroke
  Pnt  } def
/Star { 2 copy Pls Crs } def
/BoxF { stroke [] 0 setdash exch hpt sub exch vpt add M
  0 vpt2 neg V  hpt2 0 V  0 vpt2 V
  hpt2 neg 0 V  closepath fill } def
/TriUF { stroke [] 0 setdash vpt 1.12 mul add M
  hpt neg vpt -1.62 mul V
  hpt 2 mul 0 V
  hpt neg vpt 1.62 mul V closepath fill } def
/TriD { stroke [] 0 setdash 2 copy vpt 1.12 mul sub M
  hpt neg vpt 1.62 mul V
  hpt 2 mul 0 V
  hpt neg vpt -1.62 mul V closepath stroke
  Pnt  } def
/TriDF { stroke [] 0 setdash vpt 1.12 mul sub M
  hpt neg vpt 1.62 mul V
  hpt 2 mul 0 V
  hpt neg vpt -1.62 mul V closepath fill} def
/DiaF { stroke [] 0 setdash vpt add M
  hpt neg vpt neg V hpt vpt neg V
  hpt vpt V hpt neg vpt V closepath fill } def
/Pent { stroke [] 0 setdash 2 copy gsave
  translate 0 hpt M 4 {72 rotate 0 hpt L} repeat
  closepath stroke grestore Pnt } def
/PentF { stroke [] 0 setdash gsave
  translate 0 hpt M 4 {72 rotate 0 hpt L} repeat
  closepath fill grestore } def
/Circle { stroke [] 0 setdash 2 copy
  hpt 0 360 arc stroke Pnt } def
/CircleF { stroke [] 0 setdash hpt 0 360 arc fill } def
/C0 { BL [] 0 setdash 2 copy moveto vpt 90 450  arc } bind def
/C1 { BL [] 0 setdash 2 copy        moveto
       2 copy  vpt 0 90 arc closepath fill
               vpt 0 360 arc closepath } bind def
/C2 { BL [] 0 setdash 2 copy moveto
       2 copy  vpt 90 180 arc closepath fill
               vpt 0 360 arc closepath } bind def
/C3 { BL [] 0 setdash 2 copy moveto
       2 copy  vpt 0 180 arc closepath fill
               vpt 0 360 arc closepath } bind def
/C4 { BL [] 0 setdash 2 copy moveto
       2 copy  vpt 180 270 arc closepath fill
               vpt 0 360 arc closepath } bind def
/C5 { BL [] 0 setdash 2 copy moveto
       2 copy  vpt 0 90 arc
       2 copy moveto
       2 copy  vpt 180 270 arc closepath fill
               vpt 0 360 arc } bind def
/C6 { BL [] 0 setdash 2 copy moveto
      2 copy  vpt 90 270 arc closepath fill
              vpt 0 360 arc closepath } bind def
/C7 { BL [] 0 setdash 2 copy moveto
      2 copy  vpt 0 270 arc closepath fill
              vpt 0 360 arc closepath } bind def
/C8 { BL [] 0 setdash 2 copy moveto
      2 copy vpt 270 360 arc closepath fill
              vpt 0 360 arc closepath } bind def
/C9 { BL [] 0 setdash 2 copy moveto
      2 copy  vpt 270 450 arc closepath fill
              vpt 0 360 arc closepath } bind def
/C10 { BL [] 0 setdash 2 copy 2 copy moveto vpt 270 360 arc closepath fill
       2 copy moveto
       2 copy vpt 90 180 arc closepath fill
               vpt 0 360 arc closepath } bind def
/C11 { BL [] 0 setdash 2 copy moveto
       2 copy  vpt 0 180 arc closepath fill
       2 copy moveto
       2 copy  vpt 270 360 arc closepath fill
               vpt 0 360 arc closepath } bind def
/C12 { BL [] 0 setdash 2 copy moveto
       2 copy  vpt 180 360 arc closepath fill
               vpt 0 360 arc closepath } bind def
/C13 { BL [] 0 setdash  2 copy moveto
       2 copy  vpt 0 90 arc closepath fill
       2 copy moveto
       2 copy  vpt 180 360 arc closepath fill
               vpt 0 360 arc closepath } bind def
/C14 { BL [] 0 setdash 2 copy moveto
       2 copy  vpt 90 360 arc closepath fill
               vpt 0 360 arc } bind def
/C15 { BL [] 0 setdash 2 copy vpt 0 360 arc closepath fill
               vpt 0 360 arc closepath } bind def
/Rec   { newpath 4 2 roll moveto 1 index 0 rlineto 0 exch rlineto
       neg 0 rlineto closepath } bind def
/Square { dup Rec } bind def
/Bsquare { vpt sub exch vpt sub exch vpt2 Square } bind def
/S0 { BL [] 0 setdash 2 copy moveto 0 vpt rlineto BL Bsquare } bind def
/S1 { BL [] 0 setdash 2 copy vpt Square fill Bsquare } bind def
/S2 { BL [] 0 setdash 2 copy exch vpt sub exch vpt Square fill Bsquare } bind def
/S3 { BL [] 0 setdash 2 copy exch vpt sub exch vpt2 vpt Rec fill Bsquare } bind def
/S4 { BL [] 0 setdash 2 copy exch vpt sub exch vpt sub vpt Square fill Bsquare } bind def
/S5 { BL [] 0 setdash 2 copy 2 copy vpt Square fill
       exch vpt sub exch vpt sub vpt Square fill Bsquare } bind def
/S6 { BL [] 0 setdash 2 copy exch vpt sub exch vpt sub vpt vpt2 Rec fill Bsquare } bind def
/S7 { BL [] 0 setdash 2 copy exch vpt sub exch vpt sub vpt vpt2 Rec fill
       2 copy vpt Square fill
       Bsquare } bind def
/S8 { BL [] 0 setdash 2 copy vpt sub vpt Square fill Bsquare } bind def
/S9 { BL [] 0 setdash 2 copy vpt sub vpt vpt2 Rec fill Bsquare } bind def
/S10 { BL [] 0 setdash 2 copy vpt sub vpt Square fill 2 copy exch vpt sub exch vpt Square fill
       Bsquare } bind def
/S11 { BL [] 0 setdash 2 copy vpt sub vpt Square fill 2 copy exch vpt sub exch vpt2 vpt Rec fill
       Bsquare } bind def
/S12 { BL [] 0 setdash 2 copy exch vpt sub exch vpt sub vpt2 vpt Rec fill Bsquare } bind def
/S13 { BL [] 0 setdash 2 copy exch vpt sub exch vpt sub vpt2 vpt Rec fill
       2 copy vpt Square fill Bsquare } bind def
/S14 { BL [] 0 setdash 2 copy exch vpt sub exch vpt sub vpt2 vpt Rec fill
       2 copy exch vpt sub exch vpt Square fill Bsquare } bind def
/S15 { BL [] 0 setdash 2 copy Bsquare fill Bsquare } bind def
/D0 { gsave translate 45 rotate 0 0 S0 stroke grestore } bind def
/D1 { gsave translate 45 rotate 0 0 S1 stroke grestore } bind def
/D2 { gsave translate 45 rotate 0 0 S2 stroke grestore } bind def
/D3 { gsave translate 45 rotate 0 0 S3 stroke grestore } bind def
/D4 { gsave translate 45 rotate 0 0 S4 stroke grestore } bind def
/D5 { gsave translate 45 rotate 0 0 S5 stroke grestore } bind def
/D6 { gsave translate 45 rotate 0 0 S6 stroke grestore } bind def
/D7 { gsave translate 45 rotate 0 0 S7 stroke grestore } bind def
/D8 { gsave translate 45 rotate 0 0 S8 stroke grestore } bind def
/D9 { gsave translate 45 rotate 0 0 S9 stroke grestore } bind def
/D10 { gsave translate 45 rotate 0 0 S10 stroke grestore } bind def
/D11 { gsave translate 45 rotate 0 0 S11 stroke grestore } bind def
/D12 { gsave translate 45 rotate 0 0 S12 stroke grestore } bind def
/D13 { gsave translate 45 rotate 0 0 S13 stroke grestore } bind def
/D14 { gsave translate 45 rotate 0 0 S14 stroke grestore } bind def
/D15 { gsave translate 45 rotate 0 0 S15 stroke grestore } bind def
/DiaE { stroke [] 0 setdash vpt add M
  hpt neg vpt neg V hpt vpt neg V
  hpt vpt V hpt neg vpt V closepath stroke } def
/BoxE { stroke [] 0 setdash exch hpt sub exch vpt add M
  0 vpt2 neg V hpt2 0 V 0 vpt2 V
  hpt2 neg 0 V closepath stroke } def
/TriUE { stroke [] 0 setdash vpt 1.12 mul add M
  hpt neg vpt -1.62 mul V
  hpt 2 mul 0 V
  hpt neg vpt 1.62 mul V closepath stroke } def
/TriDE { stroke [] 0 setdash vpt 1.12 mul sub M
  hpt neg vpt 1.62 mul V
  hpt 2 mul 0 V
  hpt neg vpt -1.62 mul V closepath stroke } def
/PentE { stroke [] 0 setdash gsave
  translate 0 hpt M 4 {72 rotate 0 hpt L} repeat
  closepath stroke grestore } def
/CircE { stroke [] 0 setdash 
  hpt 0 360 arc stroke } def
/Opaque { gsave closepath 1 setgray fill grestore 0 setgray closepath } def
/DiaW { stroke [] 0 setdash vpt add M
  hpt neg vpt neg V hpt vpt neg V
  hpt vpt V hpt neg vpt V Opaque stroke } def
/BoxW { stroke [] 0 setdash exch hpt sub exch vpt add M
  0 vpt2 neg V hpt2 0 V 0 vpt2 V
  hpt2 neg 0 V Opaque stroke } def
/TriUW { stroke [] 0 setdash vpt 1.12 mul add M
  hpt neg vpt -1.62 mul V
  hpt 2 mul 0 V
  hpt neg vpt 1.62 mul V Opaque stroke } def
/TriDW { stroke [] 0 setdash vpt 1.12 mul sub M
  hpt neg vpt 1.62 mul V
  hpt 2 mul 0 V
  hpt neg vpt -1.62 mul V Opaque stroke } def
/PentW { stroke [] 0 setdash gsave
  translate 0 hpt M 4 {72 rotate 0 hpt L} repeat
  Opaque stroke grestore } def
/CircW { stroke [] 0 setdash 
  hpt 0 360 arc Opaque stroke } def
/BoxFill { gsave Rec 1 setgray fill grestore } def
end
}}%
\begin{picture}(3600,2160)(0,0)%
{\GNUPLOTspecial{"
gnudict begin
gsave
0 0 translate
0.100 0.100 scale
0 setgray
newpath
1.000 UL
LTb
450 300 M
63 0 V
2937 0 R
-63 0 V
450 476 M
63 0 V
2937 0 R
-63 0 V
450 652 M
63 0 V
2937 0 R
-63 0 V
450 828 M
63 0 V
2937 0 R
-63 0 V
450 1004 M
63 0 V
2937 0 R
-63 0 V
450 1180 M
63 0 V
2937 0 R
-63 0 V
450 1356 M
63 0 V
2937 0 R
-63 0 V
450 1532 M
63 0 V
2937 0 R
-63 0 V
450 1708 M
63 0 V
2937 0 R
-63 0 V
450 1884 M
63 0 V
2937 0 R
-63 0 V
450 2060 M
63 0 V
2937 0 R
-63 0 V
518 300 M
0 63 V
0 1697 R
0 -63 V
995 300 M
0 63 V
0 1697 R
0 -63 V
1473 300 M
0 63 V
0 1697 R
0 -63 V
1950 300 M
0 63 V
0 1697 R
0 -63 V
2427 300 M
0 63 V
0 1697 R
0 -63 V
2905 300 M
0 63 V
0 1697 R
0 -63 V
3382 300 M
0 63 V
0 1697 R
0 -63 V
1.000 UL
LTb
450 300 M
3000 0 V
0 1760 V
-3000 0 V
450 300 L
1.000 UL
LT0
457 623 M
15 86 V
16 145 V
15 143 V
15 135 V
14 31 V
15 4 V
16 -49 V
15 10 V
14 27 V
15 127 V
16 85 V
14 -26 V
16 -34 V
14 -15 V
16 66 V
15 79 V
14 27 V
16 -18 V
14 -49 V
15 40 V
16 86 V
14 35 V
15 -36 V
16 -16 V
14 40 V
15 37 V
16 39 V
14 -53 V
15 2 V
16 64 V
14 71 V
15 -72 V
16 -35 V
14 70 V
15 52 V
16 -6 V
14 -49 V
15 15 V
16 98 V
15 -42 V
14 -17 V
16 -3 V
15 105 V
14 -41 V
16 -52 V
15 34 V
14 83 V
16 -29 V
15 -52 V
14 31 V
15 80 V
15 -33 V
16 -57 V
15 49 V
15 78 V
14 -60 V
15 -42 V
15 82 V
16 19 V
15 -75 V
15 16 V
14 85 V
15 -12 V
15 -67 V
15 17 V
16 106 V
15 -64 V
15 -63 V
14 89 V
15 16 V
15 -72 V
16 25 V
15 65 V
15 -27 V
14 -67 V
15 70 V
15 63 V
16 -113 V
15 23 V
15 77 V
14 -11 V
15 -105 V
15 107 V
16 30 V
15 -80 V
15 -16 V
14 95 V
15 -33 V
15 -83 V
15 77 V
16 55 V
15 -80 V
15 -15 V
14 72 V
15 -14 V
15 -90 V
16 92 V
15 25 V
15 -78 V
14 1 V
15 62 V
15 -26 V
15 -54 V
15 54 V
15 45 V
16 -106 V
15 26 V
15 90 V
15 -64 V
15 -59 V
15 72 V
15 23 V
14 -97 V
15 33 V
15 67 V
15 -51 V
15 -71 V
15 88 V
16 13 V
15 -89 V
15 34 V
15 44 V
15 -32 V
15 -77 V
14 74 V
15 43 V
15 -82 V
15 -33 V
15 88 V
15 -26 V
16 -71 V
15 33 V
15 74 V
15 -99 V
15 -8 V
15 45 V
15 19 V
14 -99 V
15 3 V
15 78 V
15 -17 V
15 -110 V
15 64 V
16 71 V
15 -100 V
15 -44 V
15 52 V
15 38 V
15 -76 V
14 -17 V
15 54 V
15 -6 V
15 -66 V
15 -29 V
15 54 V
16 28 V
15 -98 V
15 -5 V
15 55 V
15 8 V
15 -84 V
15 -25 V
14 47 V
15 2 V
15 -56 V
15 -52 V
15 63 V
15 25 V
16 -75 V
15 -52 V
15 -12 V
15 64 V
15 0 V
15 -89 V
14 -37 V
15 44 V
15 16 V
15 -38 V
15 -74 V
15 -29 V
16 26 V
15 31 V
15 -24 V
15 -101 V
15 -50 V
15 0 V
15 68 V
14 1 V
15 -98 V
15 -83 V
15 -53 V
15 2 V
15 13 V
16 26 V
15 -31 V
15 -128 V
15 -160 V
15 -151 V
15 -75 V
1.000 UL
LT1
1950 1980 M
8 0 V
7 0 V
8 0 V
7 0 V
8 0 V
7 0 V
8 -1 V
7 0 V
8 0 V
7 0 V
8 -1 V
7 0 V
8 0 V
7 -1 V
8 0 V
7 -1 V
8 0 V
7 -1 V
8 0 V
7 -1 V
8 -1 V
7 0 V
8 -1 V
7 -1 V
8 0 V
7 -1 V
8 -1 V
7 -1 V
8 -1 V
7 -1 V
8 -1 V
7 -1 V
8 -1 V
7 -1 V
8 -1 V
7 -1 V
8 -1 V
7 -1 V
8 -1 V
7 -1 V
8 -2 V
7 -1 V
8 -1 V
7 -2 V
8 -1 V
7 -2 V
8 -1 V
7 -2 V
8 -1 V
7 -2 V
8 -1 V
7 -2 V
8 -2 V
7 -1 V
8 -2 V
7 -2 V
8 -2 V
7 -1 V
8 -2 V
7 -2 V
8 -2 V
7 -2 V
8 -2 V
7 -2 V
8 -2 V
7 -3 V
8 -2 V
7 -2 V
8 -2 V
7 -2 V
8 -3 V
7 -2 V
8 -3 V
7 -2 V
8 -2 V
7 -3 V
8 -3 V
7 -2 V
8 -3 V
7 -3 V
8 -2 V
7 -3 V
8 -3 V
7 -3 V
8 -3 V
7 -3 V
8 -2 V
7 -4 V
8 -3 V
7 -3 V
8 -3 V
7 -3 V
8 -3 V
7 -4 V
8 -3 V
7 -3 V
8 -4 V
7 -3 V
8 -4 V
7 -3 V
8 -4 V
7 -4 V
8 -4 V
7 -3 V
8 -4 V
7 -4 V
8 -4 V
7 -4 V
8 -4 V
7 -4 V
8 -4 V
7 -5 V
8 -4 V
7 -4 V
8 -5 V
7 -4 V
8 -5 V
7 -4 V
8 -5 V
7 -5 V
8 -5 V
7 -5 V
8 -4 V
7 -5 V
8 -6 V
7 -5 V
8 -5 V
7 -5 V
8 -6 V
7 -5 V
8 -6 V
7 -5 V
8 -6 V
7 -6 V
8 -6 V
7 -6 V
8 -6 V
7 -6 V
7 -6 V
8 -6 V
8 -7 V
7 -6 V
8 -7 V
7 -7 V
8 -7 V
7 -7 V
8 -7 V
7 -7 V
8 -7 V
7 -8 V
8 -7 V
7 -8 V
8 -8 V
7 -8 V
8 -8 V
7 -8 V
8 -9 V
7 -8 V
8 -9 V
7 -9 V
8 -9 V
7 -9 V
8 -10 V
7 -10 V
8 -10 V
7 -10 V
8 -10 V
7 -11 V
8 -11 V
7 -11 V
8 -11 V
7 -12 V
8 -12 V
7 -13 V
8 -12 V
7 -13 V
8 -14 V
7 -14 V
8 -14 V
7 -15 V
8 -16 V
7 -16 V
8 -17 V
7 -17 V
8 -18 V
7 -19 V
8 -20 V
7 -22 V
8 -22 V
7 -24 V
8 -25 V
7 -28 V
8 -30 V
7 -33 V
8 -37 V
7 -42 V
8 -51 V
7 -63 V
8 -90 V
7 -348 V
1.000 UL
LT1
1950 1980 M
-7 0 V
-8 0 V
-7 0 V
-8 0 V
-7 0 V
-8 0 V
-7 -1 V
-8 0 V
-7 0 V
-8 0 V
-7 -1 V
-8 0 V
-7 0 V
-8 -1 V
-7 0 V
-8 -1 V
-7 0 V
-8 -1 V
-7 0 V
-8 -1 V
-7 -1 V
-8 0 V
-7 -1 V
-8 -1 V
-7 0 V
-8 -1 V
-7 -1 V
-8 -1 V
-7 -1 V
-8 -1 V
-7 -1 V
-8 -1 V
-7 -1 V
-8 -1 V
-7 -1 V
-8 -1 V
-7 -1 V
-8 -1 V
-7 -1 V
-8 -1 V
-7 -2 V
-8 -1 V
-7 -1 V
-8 -2 V
-7 -1 V
-8 -2 V
-7 -1 V
-8 -2 V
-7 -1 V
-8 -2 V
-7 -1 V
-8 -2 V
-7 -2 V
-8 -1 V
-7 -2 V
-8 -2 V
-7 -2 V
-8 -1 V
-7 -2 V
-8 -2 V
-7 -2 V
-8 -2 V
-7 -2 V
-8 -2 V
-7 -2 V
-8 -3 V
-7 -2 V
-8 -2 V
-7 -2 V
-8 -2 V
-7 -3 V
-8 -2 V
-7 -3 V
-8 -2 V
-7 -2 V
-8 -3 V
-7 -3 V
-8 -2 V
-7 -3 V
-8 -3 V
-7 -2 V
-8 -3 V
-7 -3 V
-8 -3 V
-7 -3 V
-8 -3 V
-7 -2 V
-8 -4 V
-7 -3 V
-8 -3 V
-7 -3 V
-8 -3 V
-7 -3 V
-8 -4 V
-7 -3 V
-8 -3 V
-7 -4 V
-8 -3 V
-7 -4 V
-8 -3 V
-7 -4 V
-8 -4 V
-7 -4 V
-8 -3 V
-7 -4 V
-8 -4 V
-7 -4 V
-8 -4 V
-7 -4 V
-8 -4 V
-7 -4 V
-8 -5 V
-7 -4 V
-8 -4 V
-7 -5 V
-8 -4 V
-7 -5 V
-8 -4 V
-7 -5 V
-8 -5 V
-7 -5 V
-8 -5 V
-7 -4 V
-8 -5 V
-7 -6 V
-8 -5 V
-7 -5 V
-8 -5 V
-7 -6 V
-8 -5 V
-7 -6 V
-8 -5 V
-7 -6 V
-8 -6 V
-7 -6 V
-8 -6 V
-8 -6 V
-7 -6 V
-7 -6 V
-8 -6 V
-7 -7 V
-8 -6 V
-7 -7 V
-8 -7 V
-7 -7 V
-8 -7 V
-7 -7 V
-8 -7 V
-7 -7 V
-8 -8 V
-7 -7 V
-8 -8 V
-7 -8 V
-8 -8 V
-7 -8 V
-8 -8 V
-8 -9 V
-7 -8 V
-7 -9 V
-8 -9 V
-8 -9 V
-7 -9 V
-7 -10 V
-8 -10 V
-7 -10 V
-8 -10 V
-7 -10 V
-8 -11 V
-7 -11 V
-8 -11 V
-7 -11 V
-8 -12 V
-8 -12 V
-7 -13 V
-7 -12 V
-8 -13 V
-7 -14 V
-8 -14 V
-7 -14 V
-8 -15 V
-7 -16 V
-8 -16 V
-8 -17 V
-7 -17 V
-7 -18 V
-8 -19 V
-8 -20 V
-7 -22 V
-7 -22 V
-8 -24 V
-7 -25 V
-8 -28 V
-7 -30 V
-8 -33 V
-7 -37 V
-8 -42 V
-7 -51 V
-8 -63 V
-8 -90 V
450 300 L
stroke
grestore
end
showpage
}}%
\put(1950,50){\makebox(0,0){$\alpha$}}%
\put(100,1180){%
\makebox(0,0)[b]{\shortstack{$\rho(\alpha)$}}%
}%
\put(3382,200){\makebox(0,0){3}}%
\put(2905,200){\makebox(0,0){2}}%
\put(2427,200){\makebox(0,0){1}}%
\put(1950,200){\makebox(0,0){0}}%
\put(1473,200){\makebox(0,0){-1}}%
\put(995,200){\makebox(0,0){-2}}%
\put(518,200){\makebox(0,0){-3}}%
\put(400,2060){\makebox(0,0)[r]{0.2}}%
\put(400,1884){\makebox(0,0)[r]{0.18}}%
\put(400,1708){\makebox(0,0)[r]{0.16}}%
\put(400,1532){\makebox(0,0)[r]{0.14}}%
\put(400,1356){\makebox(0,0)[r]{0.12}}%
\put(400,1180){\makebox(0,0)[r]{0.1}}%
\put(400,1004){\makebox(0,0)[r]{0.08}}%
\put(400,828){\makebox(0,0)[r]{0.06}}%
\put(400,652){\makebox(0,0)[r]{0.04}}%
\put(400,476){\makebox(0,0)[r]{0.02}}%
\put(400,300){\makebox(0,0)[r]{0}}%
\end{picture}%
\endgroup

%% file: actiongraph.tex
\begingroup%
  \makeatletter%
  \newcommand{\GNUPLOTspecial}{%
    \@sanitize\catcode`\%=14\relax\special}%
  \setlength{\unitlength}{0.1bp}%
{\GNUPLOTspecial{!
/gnudict 256 dict def
gnudict begin
/Color false def
/Solid false def
/gnulinewidth 5.000 def
/userlinewidth gnulinewidth def
/vshift -33 def
/dl {10 mul} def
/hpt_ 31.5 def
/vpt_ 31.5 def
/hpt hpt_ def
/vpt vpt_ def
/M {moveto} bind def
/L {lineto} bind def
/R {rmoveto} bind def
/V {rlineto} bind def
/vpt2 vpt 2 mul def
/hpt2 hpt 2 mul def
/Lshow { currentpoint stroke M
  0 vshift R show } def
/Rshow { currentpoint stroke M
  dup stringwidth pop neg vshift R show } def
/Cshow { currentpoint stroke M
  dup stringwidth pop -2 div vshift R show } def
/UP { dup vpt_ mul /vpt exch def hpt_ mul /hpt exch def
  /hpt2 hpt 2 mul def /vpt2 vpt 2 mul def } def
/DL { Color {setrgbcolor Solid {pop []} if 0 setdash }
 {pop pop pop Solid {pop []} if 0 setdash} ifelse } def
/BL { stroke userlinewidth 2 mul setlinewidth } def
/AL { stroke userlinewidth 2 div setlinewidth } def
/UL { dup gnulinewidth mul /userlinewidth exch def
      10 mul /udl exch def } def
/PL { stroke userlinewidth setlinewidth } def
/LTb { BL [] 0 0 0 DL } def
/LTa { AL [1 udl mul 2 udl mul] 0 setdash 0 0 0 setrgbcolor } def
/LT0 { PL [] 1 0 0 DL } def
/LT1 { PL [4 dl 2 dl] 0 1 0 DL } def
/LT2 { PL [2 dl 3 dl] 0 0 1 DL } def
/LT3 { PL [1 dl 1.5 dl] 1 0 1 DL } def
/LT4 { PL [5 dl 2 dl 1 dl 2 dl] 0 1 1 DL } def
/LT5 { PL [4 dl 3 dl 1 dl 3 dl] 1 1 0 DL } def
/LT6 { PL [2 dl 2 dl 2 dl 4 dl] 0 0 0 DL } def
/LT7 { PL [2 dl 2 dl 2 dl 2 dl 2 dl 4 dl] 1 0.3 0 DL } def
/LT8 { PL [2 dl 2 dl 2 dl 2 dl 2 dl 2 dl 2 dl 4 dl] 0.5 0.5 0.5 DL } def
/Pnt { stroke [] 0 setdash
   gsave 1 setlinecap M 0 0 V stroke grestore } def
/Dia { stroke [] 0 setdash 2 copy vpt add M
  hpt neg vpt neg V hpt vpt neg V
  hpt vpt V hpt neg vpt V closepath stroke
  Pnt } def
/Pls { stroke [] 0 setdash vpt sub M 0 vpt2 V
  currentpoint stroke M
  hpt neg vpt neg R hpt2 0 V stroke
  } def
/Box { stroke [] 0 setdash 2 copy exch hpt sub exch vpt add M
  0 vpt2 neg V hpt2 0 V 0 vpt2 V
  hpt2 neg 0 V closepath stroke
  Pnt } def
/Crs { stroke [] 0 setdash exch hpt sub exch vpt add M
  hpt2 vpt2 neg V currentpoint stroke M
  hpt2 neg 0 R hpt2 vpt2 V stroke } def
/TriU { stroke [] 0 setdash 2 copy vpt 1.12 mul add M
  hpt neg vpt -1.62 mul V
  hpt 2 mul 0 V
  hpt neg vpt 1.62 mul V closepath stroke
  Pnt  } def
/Star { 2 copy Pls Crs } def
/BoxF { stroke [] 0 setdash exch hpt sub exch vpt add M
  0 vpt2 neg V  hpt2 0 V  0 vpt2 V
  hpt2 neg 0 V  closepath fill } def
/TriUF { stroke [] 0 setdash vpt 1.12 mul add M
  hpt neg vpt -1.62 mul V
  hpt 2 mul 0 V
  hpt neg vpt 1.62 mul V closepath fill } def
/TriD { stroke [] 0 setdash 2 copy vpt 1.12 mul sub M
  hpt neg vpt 1.62 mul V
  hpt 2 mul 0 V
  hpt neg vpt -1.62 mul V closepath stroke
  Pnt  } def
/TriDF { stroke [] 0 setdash vpt 1.12 mul sub M
  hpt neg vpt 1.62 mul V
  hpt 2 mul 0 V
  hpt neg vpt -1.62 mul V closepath fill} def
/DiaF { stroke [] 0 setdash vpt add M
  hpt neg vpt neg V hpt vpt neg V
  hpt vpt V hpt neg vpt V closepath fill } def
/Pent { stroke [] 0 setdash 2 copy gsave
  translate 0 hpt M 4 {72 rotate 0 hpt L} repeat
  closepath stroke grestore Pnt } def
/PentF { stroke [] 0 setdash gsave
  translate 0 hpt M 4 {72 rotate 0 hpt L} repeat
  closepath fill grestore } def
/Circle { stroke [] 0 setdash 2 copy
  hpt 0 360 arc stroke Pnt } def
/CircleF { stroke [] 0 setdash hpt 0 360 arc fill } def
/C0 { BL [] 0 setdash 2 copy moveto vpt 90 450  arc } bind def
/C1 { BL [] 0 setdash 2 copy        moveto
       2 copy  vpt 0 90 arc closepath fill
               vpt 0 360 arc closepath } bind def
/C2 { BL [] 0 setdash 2 copy moveto
       2 copy  vpt 90 180 arc closepath fill
               vpt 0 360 arc closepath } bind def
/C3 { BL [] 0 setdash 2 copy moveto
       2 copy  vpt 0 180 arc closepath fill
               vpt 0 360 arc closepath } bind def
/C4 { BL [] 0 setdash 2 copy moveto
       2 copy  vpt 180 270 arc closepath fill
               vpt 0 360 arc closepath } bind def
/C5 { BL [] 0 setdash 2 copy moveto
       2 copy  vpt 0 90 arc
       2 copy moveto
       2 copy  vpt 180 270 arc closepath fill
               vpt 0 360 arc } bind def
/C6 { BL [] 0 setdash 2 copy moveto
      2 copy  vpt 90 270 arc closepath fill
              vpt 0 360 arc closepath } bind def
/C7 { BL [] 0 setdash 2 copy moveto
      2 copy  vpt 0 270 arc closepath fill
              vpt 0 360 arc closepath } bind def
/C8 { BL [] 0 setdash 2 copy moveto
      2 copy vpt 270 360 arc closepath fill
              vpt 0 360 arc closepath } bind def
/C9 { BL [] 0 setdash 2 copy moveto
      2 copy  vpt 270 450 arc closepath fill
              vpt 0 360 arc closepath } bind def
/C10 { BL [] 0 setdash 2 copy 2 copy moveto vpt 270 360 arc closepath fill
       2 copy moveto
       2 copy vpt 90 180 arc closepath fill
               vpt 0 360 arc closepath } bind def
/C11 { BL [] 0 setdash 2 copy moveto
       2 copy  vpt 0 180 arc closepath fill
       2 copy moveto
       2 copy  vpt 270 360 arc closepath fill
               vpt 0 360 arc closepath } bind def
/C12 { BL [] 0 setdash 2 copy moveto
       2 copy  vpt 180 360 arc closepath fill
               vpt 0 360 arc closepath } bind def
/C13 { BL [] 0 setdash  2 copy moveto
       2 copy  vpt 0 90 arc closepath fill
       2 copy moveto
       2 copy  vpt 180 360 arc closepath fill
               vpt 0 360 arc closepath } bind def
/C14 { BL [] 0 setdash 2 copy moveto
       2 copy  vpt 90 360 arc closepath fill
               vpt 0 360 arc } bind def
/C15 { BL [] 0 setdash 2 copy vpt 0 360 arc closepath fill
               vpt 0 360 arc closepath } bind def
/Rec   { newpath 4 2 roll moveto 1 index 0 rlineto 0 exch rlineto
       neg 0 rlineto closepath } bind def
/Square { dup Rec } bind def
/Bsquare { vpt sub exch vpt sub exch vpt2 Square } bind def
/S0 { BL [] 0 setdash 2 copy moveto 0 vpt rlineto BL Bsquare } bind def
/S1 { BL [] 0 setdash 2 copy vpt Square fill Bsquare } bind def
/S2 { BL [] 0 setdash 2 copy exch vpt sub exch vpt Square fill Bsquare } bind def
/S3 { BL [] 0 setdash 2 copy exch vpt sub exch vpt2 vpt Rec fill Bsquare } bind def
/S4 { BL [] 0 setdash 2 copy exch vpt sub exch vpt sub vpt Square fill Bsquare } bind def
/S5 { BL [] 0 setdash 2 copy 2 copy vpt Square fill
       exch vpt sub exch vpt sub vpt Square fill Bsquare } bind def
/S6 { BL [] 0 setdash 2 copy exch vpt sub exch vpt sub vpt vpt2 Rec fill Bsquare } bind def
/S7 { BL [] 0 setdash 2 copy exch vpt sub exch vpt sub vpt vpt2 Rec fill
       2 copy vpt Square fill
       Bsquare } bind def
/S8 { BL [] 0 setdash 2 copy vpt sub vpt Square fill Bsquare } bind def
/S9 { BL [] 0 setdash 2 copy vpt sub vpt vpt2 Rec fill Bsquare } bind def
/S10 { BL [] 0 setdash 2 copy vpt sub vpt Square fill 2 copy exch vpt sub exch vpt Square fill
       Bsquare } bind def
/S11 { BL [] 0 setdash 2 copy vpt sub vpt Square fill 2 copy exch vpt sub exch vpt2 vpt Rec fill
       Bsquare } bind def
/S12 { BL [] 0 setdash 2 copy exch vpt sub exch vpt sub vpt2 vpt Rec fill Bsquare } bind def
/S13 { BL [] 0 setdash 2 copy exch vpt sub exch vpt sub vpt2 vpt Rec fill
       2 copy vpt Square fill Bsquare } bind def
/S14 { BL [] 0 setdash 2 copy exch vpt sub exch vpt sub vpt2 vpt Rec fill
       2 copy exch vpt sub exch vpt Square fill Bsquare } bind def
/S15 { BL [] 0 setdash 2 copy Bsquare fill Bsquare } bind def
/D0 { gsave translate 45 rotate 0 0 S0 stroke grestore } bind def
/D1 { gsave translate 45 rotate 0 0 S1 stroke grestore } bind def
/D2 { gsave translate 45 rotate 0 0 S2 stroke grestore } bind def
/D3 { gsave translate 45 rotate 0 0 S3 stroke grestore } bind def
/D4 { gsave translate 45 rotate 0 0 S4 stroke grestore } bind def
/D5 { gsave translate 45 rotate 0 0 S5 stroke grestore } bind def
/D6 { gsave translate 45 rotate 0 0 S6 stroke grestore } bind def
/D7 { gsave translate 45 rotate 0 0 S7 stroke grestore } bind def
/D8 { gsave translate 45 rotate 0 0 S8 stroke grestore } bind def
/D9 { gsave translate 45 rotate 0 0 S9 stroke grestore } bind def
/D10 { gsave translate 45 rotate 0 0 S10 stroke grestore } bind def
/D11 { gsave translate 45 rotate 0 0 S11 stroke grestore } bind def
/D12 { gsave translate 45 rotate 0 0 S12 stroke grestore } bind def
/D13 { gsave translate 45 rotate 0 0 S13 stroke grestore } bind def
/D14 { gsave translate 45 rotate 0 0 S14 stroke grestore } bind def
/D15 { gsave translate 45 rotate 0 0 S15 stroke grestore } bind def
/DiaE { stroke [] 0 setdash vpt add M
  hpt neg vpt neg V hpt vpt neg V
  hpt vpt V hpt neg vpt V closepath stroke } def
/BoxE { stroke [] 0 setdash exch hpt sub exch vpt add M
  0 vpt2 neg V hpt2 0 V 0 vpt2 V
  hpt2 neg 0 V closepath stroke } def
/TriUE { stroke [] 0 setdash vpt 1.12 mul add M
  hpt neg vpt -1.62 mul V
  hpt 2 mul 0 V
  hpt neg vpt 1.62 mul V closepath stroke } def
/TriDE { stroke [] 0 setdash vpt 1.12 mul sub M
  hpt neg vpt 1.62 mul V
  hpt 2 mul 0 V
  hpt neg vpt -1.62 mul V closepath stroke } def
/PentE { stroke [] 0 setdash gsave
  translate 0 hpt M 4 {72 rotate 0 hpt L} repeat
  closepath stroke grestore } def
/CircE { stroke [] 0 setdash 
  hpt 0 360 arc stroke } def
/Opaque { gsave closepath 1 setgray fill grestore 0 setgray closepath } def
/DiaW { stroke [] 0 setdash vpt add M
  hpt neg vpt neg V hpt vpt neg V
  hpt vpt V hpt neg vpt V Opaque stroke } def
/BoxW { stroke [] 0 setdash exch hpt sub exch vpt add M
  0 vpt2 neg V hpt2 0 V 0 vpt2 V
  hpt2 neg 0 V Opaque stroke } def
/TriUW { stroke [] 0 setdash vpt 1.12 mul add M
  hpt neg vpt -1.62 mul V
  hpt 2 mul 0 V
  hpt neg vpt 1.62 mul V Opaque stroke } def
/TriDW { stroke [] 0 setdash vpt 1.12 mul sub M
  hpt neg vpt 1.62 mul V
  hpt 2 mul 0 V
  hpt neg vpt -1.62 mul V Opaque stroke } def
/PentW { stroke [] 0 setdash gsave
  translate 0 hpt M 4 {72 rotate 0 hpt L} repeat
  Opaque stroke grestore } def
/CircW { stroke [] 0 setdash 
  hpt 0 360 arc Opaque stroke } def
/BoxFill { gsave Rec 1 setgray fill grestore } def
end
}}%
\begin{picture}(3600,2160)(0,0)%
{\GNUPLOTspecial{"
gnudict begin
gsave
0 0 translate
0.100 0.100 scale
0 setgray
newpath
1.000 UL
LTb
450 300 M
63 0 V
2937 0 R
-63 0 V
450 476 M
63 0 V
2937 0 R
-63 0 V
450 652 M
63 0 V
2937 0 R
-63 0 V
450 828 M
63 0 V
2937 0 R
-63 0 V
450 1004 M
63 0 V
2937 0 R
-63 0 V
450 1180 M
63 0 V
2937 0 R
-63 0 V
450 1356 M
63 0 V
2937 0 R
-63 0 V
450 1532 M
63 0 V
2937 0 R
-63 0 V
450 1708 M
63 0 V
2937 0 R
-63 0 V
450 1884 M
63 0 V
2937 0 R
-63 0 V
450 2060 M
63 0 V
2937 0 R
-63 0 V
450 300 M
0 63 V
0 1697 R
0 -63 V
825 300 M
0 63 V
0 1697 R
0 -63 V
1200 300 M
0 63 V
0 1697 R
0 -63 V
1575 300 M
0 63 V
0 1697 R
0 -63 V
1950 300 M
0 63 V
0 1697 R
0 -63 V
2325 300 M
0 63 V
0 1697 R
0 -63 V
2700 300 M
0 63 V
0 1697 R
0 -63 V
3075 300 M
0 63 V
0 1697 R
0 -63 V
3450 300 M
0 63 V
0 1697 R
0 -63 V
1.000 UL
LTb
450 300 M
3000 0 V
0 1760 V
-3000 0 V
450 300 L
0.600 UP
1.000 UL
LT0
450 337 M
659 450 L
520 332 V
261 199 V
130 104 V
130 106 V
260 191 V
261 157 V
521 222 V
521 157 V
187 42 V
659 450 Pls
1179 782 Pls
1440 981 Pls
1570 1085 Pls
1700 1191 Pls
1960 1382 Pls
2221 1539 Pls
2742 1761 Pls
3263 1918 Pls
0.600 UP
1.000 UL
LT1
450 309 M
659 411 L
520 283 V
261 170 V
260 208 V
130 124 V
131 114 V
260 177 V
521 239 V
521 162 V
187 44 V
659 411 Crs
1179 694 Crs
1440 864 Crs
1700 1072 Crs
1830 1196 Crs
1961 1310 Crs
2221 1487 Crs
2742 1726 Crs
3263 1888 Crs
0.600 UP
1.000 UL
LT2
1179 683 M
261 154 V
260 187 V
130 126 V
65 72 V
66 62 V
260 188 V
260 136 V
1179 683 Star
1440 837 Star
1700 1024 Star
1830 1150 Star
1895 1222 Star
1961 1284 Star
2221 1472 Star
2481 1608 Star
0.600 UP
1.000 UL
LT3
1440 834 M
260 178 V
65 56 V
65 68 V
33 37 V
32 38 V
66 68 V
260 191 V
1440 834 Box
1700 1012 Box
1765 1068 Box
1830 1136 Box
1863 1173 Box
1895 1211 Box
1961 1279 Box
2221 1470 Box
stroke
grestore
end
showpage
}}%
\put(1950,50){\makebox(0,0){$\gamma$}}%
\put(100,1180){%
\makebox(0,0)[b]{\shortstack{$<u_p>$}}%
}%
\put(3450,200){\makebox(0,0){0.65}}%
\put(3075,200){\makebox(0,0){0.6}}%
\put(2700,200){\makebox(0,0){0.55}}%
\put(2325,200){\makebox(0,0){0.5}}%
\put(1950,200){\makebox(0,0){0.45}}%
\put(1575,200){\makebox(0,0){0.4}}%
\put(1200,200){\makebox(0,0){0.35}}%
\put(825,200){\makebox(0,0){0.3}}%
\put(450,200){\makebox(0,0){0.25}}%
\put(400,2060){\makebox(0,0)[r]{0.75}}%
\put(400,1884){\makebox(0,0)[r]{0.7}}%
\put(400,1708){\makebox(0,0)[r]{0.65}}%
\put(400,1532){\makebox(0,0)[r]{0.6}}%
\put(400,1356){\makebox(0,0)[r]{0.55}}%
\put(400,1180){\makebox(0,0)[r]{0.5}}%
\put(400,1004){\makebox(0,0)[r]{0.45}}%
\put(400,828){\makebox(0,0)[r]{0.4}}%
\put(400,652){\makebox(0,0)[r]{0.35}}%
\put(400,476){\makebox(0,0)[r]{0.3}}%
\put(400,300){\makebox(0,0)[r]{0.25}}%
\end{picture}%
\endgroup

%% file: C_2graph.tex
\begingroup%
  \makeatletter%
  \newcommand{\GNUPLOTspecial}{%
    \@sanitize\catcode`\%=14\relax\special}%
  \setlength{\unitlength}{0.1bp}%
{\GNUPLOTspecial{!
/gnudict 256 dict def
gnudict begin
/Color false def
/Solid false def
/gnulinewidth 5.000 def
/userlinewidth gnulinewidth def
/vshift -33 def
/dl {10 mul} def
/hpt_ 31.5 def
/vpt_ 31.5 def
/hpt hpt_ def
/vpt vpt_ def
/M {moveto} bind def
/L {lineto} bind def
/R {rmoveto} bind def
/V {rlineto} bind def
/vpt2 vpt 2 mul def
/hpt2 hpt 2 mul def
/Lshow { currentpoint stroke M
  0 vshift R show } def
/Rshow { currentpoint stroke M
  dup stringwidth pop neg vshift R show } def
/Cshow { currentpoint stroke M
  dup stringwidth pop -2 div vshift R show } def
/UP { dup vpt_ mul /vpt exch def hpt_ mul /hpt exch def
  /hpt2 hpt 2 mul def /vpt2 vpt 2 mul def } def
/DL { Color {setrgbcolor Solid {pop []} if 0 setdash }
 {pop pop pop Solid {pop []} if 0 setdash} ifelse } def
/BL { stroke userlinewidth 2 mul setlinewidth } def
/AL { stroke userlinewidth 2 div setlinewidth } def
/UL { dup gnulinewidth mul /userlinewidth exch def
      10 mul /udl exch def } def
/PL { stroke userlinewidth setlinewidth } def
/LTb { BL [] 0 0 0 DL } def
/LTa { AL [1 udl mul 2 udl mul] 0 setdash 0 0 0 setrgbcolor } def
/LT0 { PL [] 1 0 0 DL } def
/LT1 { PL [4 dl 2 dl] 0 1 0 DL } def
/LT2 { PL [2 dl 3 dl] 0 0 1 DL } def
/LT3 { PL [1 dl 1.5 dl] 1 0 1 DL } def
/LT4 { PL [5 dl 2 dl 1 dl 2 dl] 0 1 1 DL } def
/LT5 { PL [4 dl 3 dl 1 dl 3 dl] 1 1 0 DL } def
/LT6 { PL [2 dl 2 dl 2 dl 4 dl] 0 0 0 DL } def
/LT7 { PL [2 dl 2 dl 2 dl 2 dl 2 dl 4 dl] 1 0.3 0 DL } def
/LT8 { PL [2 dl 2 dl 2 dl 2 dl 2 dl 2 dl 2 dl 4 dl] 0.5 0.5 0.5 DL } def
/Pnt { stroke [] 0 setdash
   gsave 1 setlinecap M 0 0 V stroke grestore } def
/Dia { stroke [] 0 setdash 2 copy vpt add M
  hpt neg vpt neg V hpt vpt neg V
  hpt vpt V hpt neg vpt V closepath stroke
  Pnt } def
/Pls { stroke [] 0 setdash vpt sub M 0 vpt2 V
  currentpoint stroke M
  hpt neg vpt neg R hpt2 0 V stroke
  } def
/Box { stroke [] 0 setdash 2 copy exch hpt sub exch vpt add M
  0 vpt2 neg V hpt2 0 V 0 vpt2 V
  hpt2 neg 0 V closepath stroke
  Pnt } def
/Crs { stroke [] 0 setdash exch hpt sub exch vpt add M
  hpt2 vpt2 neg V currentpoint stroke M
  hpt2 neg 0 R hpt2 vpt2 V stroke } def
/TriU { stroke [] 0 setdash 2 copy vpt 1.12 mul add M
  hpt neg vpt -1.62 mul V
  hpt 2 mul 0 V
  hpt neg vpt 1.62 mul V closepath stroke
  Pnt  } def
/Star { 2 copy Pls Crs } def
/BoxF { stroke [] 0 setdash exch hpt sub exch vpt add M
  0 vpt2 neg V  hpt2 0 V  0 vpt2 V
  hpt2 neg 0 V  closepath fill } def
/TriUF { stroke [] 0 setdash vpt 1.12 mul add M
  hpt neg vpt -1.62 mul V
  hpt 2 mul 0 V
  hpt neg vpt 1.62 mul V closepath fill } def
/TriD { stroke [] 0 setdash 2 copy vpt 1.12 mul sub M
  hpt neg vpt 1.62 mul V
  hpt 2 mul 0 V
  hpt neg vpt -1.62 mul V closepath stroke
  Pnt  } def
/TriDF { stroke [] 0 setdash vpt 1.12 mul sub M
  hpt neg vpt 1.62 mul V
  hpt 2 mul 0 V
  hpt neg vpt -1.62 mul V closepath fill} def
/DiaF { stroke [] 0 setdash vpt add M
  hpt neg vpt neg V hpt vpt neg V
  hpt vpt V hpt neg vpt V closepath fill } def
/Pent { stroke [] 0 setdash 2 copy gsave
  translate 0 hpt M 4 {72 rotate 0 hpt L} repeat
  closepath stroke grestore Pnt } def
/PentF { stroke [] 0 setdash gsave
  translate 0 hpt M 4 {72 rotate 0 hpt L} repeat
  closepath fill grestore } def
/Circle { stroke [] 0 setdash 2 copy
  hpt 0 360 arc stroke Pnt } def
/CircleF { stroke [] 0 setdash hpt 0 360 arc fill } def
/C0 { BL [] 0 setdash 2 copy moveto vpt 90 450  arc } bind def
/C1 { BL [] 0 setdash 2 copy        moveto
       2 copy  vpt 0 90 arc closepath fill
               vpt 0 360 arc closepath } bind def
/C2 { BL [] 0 setdash 2 copy moveto
       2 copy  vpt 90 180 arc closepath fill
               vpt 0 360 arc closepath } bind def
/C3 { BL [] 0 setdash 2 copy moveto
       2 copy  vpt 0 180 arc closepath fill
               vpt 0 360 arc closepath } bind def
/C4 { BL [] 0 setdash 2 copy moveto
       2 copy  vpt 180 270 arc closepath fill
               vpt 0 360 arc closepath } bind def
/C5 { BL [] 0 setdash 2 copy moveto
       2 copy  vpt 0 90 arc
       2 copy moveto
       2 copy  vpt 180 270 arc closepath fill
               vpt 0 360 arc } bind def
/C6 { BL [] 0 setdash 2 copy moveto
      2 copy  vpt 90 270 arc closepath fill
              vpt 0 360 arc closepath } bind def
/C7 { BL [] 0 setdash 2 copy moveto
      2 copy  vpt 0 270 arc closepath fill
              vpt 0 360 arc closepath } bind def
/C8 { BL [] 0 setdash 2 copy moveto
      2 copy vpt 270 360 arc closepath fill
              vpt 0 360 arc closepath } bind def
/C9 { BL [] 0 setdash 2 copy moveto
      2 copy  vpt 270 450 arc closepath fill
              vpt 0 360 arc closepath } bind def
/C10 { BL [] 0 setdash 2 copy 2 copy moveto vpt 270 360 arc closepath fill
       2 copy moveto
       2 copy vpt 90 180 arc closepath fill
               vpt 0 360 arc closepath } bind def
/C11 { BL [] 0 setdash 2 copy moveto
       2 copy  vpt 0 180 arc closepath fill
       2 copy moveto
       2 copy  vpt 270 360 arc closepath fill
               vpt 0 360 arc closepath } bind def
/C12 { BL [] 0 setdash 2 copy moveto
       2 copy  vpt 180 360 arc closepath fill
               vpt 0 360 arc closepath } bind def
/C13 { BL [] 0 setdash  2 copy moveto
       2 copy  vpt 0 90 arc closepath fill
       2 copy moveto
       2 copy  vpt 180 360 arc closepath fill
               vpt 0 360 arc closepath } bind def
/C14 { BL [] 0 setdash 2 copy moveto
       2 copy  vpt 90 360 arc closepath fill
               vpt 0 360 arc } bind def
/C15 { BL [] 0 setdash 2 copy vpt 0 360 arc closepath fill
               vpt 0 360 arc closepath } bind def
/Rec   { newpath 4 2 roll moveto 1 index 0 rlineto 0 exch rlineto
       neg 0 rlineto closepath } bind def
/Square { dup Rec } bind def
/Bsquare { vpt sub exch vpt sub exch vpt2 Square } bind def
/S0 { BL [] 0 setdash 2 copy moveto 0 vpt rlineto BL Bsquare } bind def
/S1 { BL [] 0 setdash 2 copy vpt Square fill Bsquare } bind def
/S2 { BL [] 0 setdash 2 copy exch vpt sub exch vpt Square fill Bsquare } bind def
/S3 { BL [] 0 setdash 2 copy exch vpt sub exch vpt2 vpt Rec fill Bsquare } bind def
/S4 { BL [] 0 setdash 2 copy exch vpt sub exch vpt sub vpt Square fill Bsquare } bind def
/S5 { BL [] 0 setdash 2 copy 2 copy vpt Square fill
       exch vpt sub exch vpt sub vpt Square fill Bsquare } bind def
/S6 { BL [] 0 setdash 2 copy exch vpt sub exch vpt sub vpt vpt2 Rec fill Bsquare } bind def
/S7 { BL [] 0 setdash 2 copy exch vpt sub exch vpt sub vpt vpt2 Rec fill
       2 copy vpt Square fill
       Bsquare } bind def
/S8 { BL [] 0 setdash 2 copy vpt sub vpt Square fill Bsquare } bind def
/S9 { BL [] 0 setdash 2 copy vpt sub vpt vpt2 Rec fill Bsquare } bind def
/S10 { BL [] 0 setdash 2 copy vpt sub vpt Square fill 2 copy exch vpt sub exch vpt Square fill
       Bsquare } bind def
/S11 { BL [] 0 setdash 2 copy vpt sub vpt Square fill 2 copy exch vpt sub exch vpt2 vpt Rec fill
       Bsquare } bind def
/S12 { BL [] 0 setdash 2 copy exch vpt sub exch vpt sub vpt2 vpt Rec fill Bsquare } bind def
/S13 { BL [] 0 setdash 2 copy exch vpt sub exch vpt sub vpt2 vpt Rec fill
       2 copy vpt Square fill Bsquare } bind def
/S14 { BL [] 0 setdash 2 copy exch vpt sub exch vpt sub vpt2 vpt Rec fill
       2 copy exch vpt sub exch vpt Square fill Bsquare } bind def
/S15 { BL [] 0 setdash 2 copy Bsquare fill Bsquare } bind def
/D0 { gsave translate 45 rotate 0 0 S0 stroke grestore } bind def
/D1 { gsave translate 45 rotate 0 0 S1 stroke grestore } bind def
/D2 { gsave translate 45 rotate 0 0 S2 stroke grestore } bind def
/D3 { gsave translate 45 rotate 0 0 S3 stroke grestore } bind def
/D4 { gsave translate 45 rotate 0 0 S4 stroke grestore } bind def
/D5 { gsave translate 45 rotate 0 0 S5 stroke grestore } bind def
/D6 { gsave translate 45 rotate 0 0 S6 stroke grestore } bind def
/D7 { gsave translate 45 rotate 0 0 S7 stroke grestore } bind def
/D8 { gsave translate 45 rotate 0 0 S8 stroke grestore } bind def
/D9 { gsave translate 45 rotate 0 0 S9 stroke grestore } bind def
/D10 { gsave translate 45 rotate 0 0 S10 stroke grestore } bind def
/D11 { gsave translate 45 rotate 0 0 S11 stroke grestore } bind def
/D12 { gsave translate 45 rotate 0 0 S12 stroke grestore } bind def
/D13 { gsave translate 45 rotate 0 0 S13 stroke grestore } bind def
/D14 { gsave translate 45 rotate 0 0 S14 stroke grestore } bind def
/D15 { gsave translate 45 rotate 0 0 S15 stroke grestore } bind def
/DiaE { stroke [] 0 setdash vpt add M
  hpt neg vpt neg V hpt vpt neg V
  hpt vpt V hpt neg vpt V closepath stroke } def
/BoxE { stroke [] 0 setdash exch hpt sub exch vpt add M
  0 vpt2 neg V hpt2 0 V 0 vpt2 V
  hpt2 neg 0 V closepath stroke } def
/TriUE { stroke [] 0 setdash vpt 1.12 mul add M
  hpt neg vpt -1.62 mul V
  hpt 2 mul 0 V
  hpt neg vpt 1.62 mul V closepath stroke } def
/TriDE { stroke [] 0 setdash vpt 1.12 mul sub M
  hpt neg vpt 1.62 mul V
  hpt 2 mul 0 V
  hpt neg vpt -1.62 mul V closepath stroke } def
/PentE { stroke [] 0 setdash gsave
  translate 0 hpt M 4 {72 rotate 0 hpt L} repeat
  closepath stroke grestore } def
/CircE { stroke [] 0 setdash 
  hpt 0 360 arc stroke } def
/Opaque { gsave closepath 1 setgray fill grestore 0 setgray closepath } def
/DiaW { stroke [] 0 setdash vpt add M
  hpt neg vpt neg V hpt vpt neg V
  hpt vpt V hpt neg vpt V Opaque stroke } def
/BoxW { stroke [] 0 setdash exch hpt sub exch vpt add M
  0 vpt2 neg V hpt2 0 V 0 vpt2 V
  hpt2 neg 0 V Opaque stroke } def
/TriUW { stroke [] 0 setdash vpt 1.12 mul add M
  hpt neg vpt -1.62 mul V
  hpt 2 mul 0 V
  hpt neg vpt 1.62 mul V Opaque stroke } def
/TriDW { stroke [] 0 setdash vpt 1.12 mul sub M
  hpt neg vpt 1.62 mul V
  hpt 2 mul 0 V
  hpt neg vpt -1.62 mul V Opaque stroke } def
/PentW { stroke [] 0 setdash gsave
  translate 0 hpt M 4 {72 rotate 0 hpt L} repeat
  Opaque stroke grestore } def
/CircW { stroke [] 0 setdash 
  hpt 0 360 arc Opaque stroke } def
/BoxFill { gsave Rec 1 setgray fill grestore } def
end
}}%
\begin{picture}(3600,2160)(0,0)%
{\GNUPLOTspecial{"
gnudict begin
gsave
0 0 translate
0.100 0.100 scale
0 setgray
newpath
1.000 UL
LTb
400 300 M
63 0 V
2987 0 R
-63 0 V
400 593 M
63 0 V
2987 0 R
-63 0 V
400 887 M
63 0 V
2987 0 R
-63 0 V
400 1180 M
63 0 V
2987 0 R
-63 0 V
400 1473 M
63 0 V
2987 0 R
-63 0 V
400 1767 M
63 0 V
2987 0 R
-63 0 V
400 2060 M
63 0 V
2987 0 R
-63 0 V
400 300 M
0 63 V
0 1697 R
0 -63 V
781 300 M
0 63 V
0 1697 R
0 -63 V
1163 300 M
0 63 V
0 1697 R
0 -63 V
1544 300 M
0 63 V
0 1697 R
0 -63 V
1925 300 M
0 63 V
0 1697 R
0 -63 V
2306 300 M
0 63 V
0 1697 R
0 -63 V
2687 300 M
0 63 V
0 1697 R
0 -63 V
3069 300 M
0 63 V
0 1697 R
0 -63 V
3450 300 M
0 63 V
0 1697 R
0 -63 V
1.000 UL
LTb
400 300 M
3050 0 V
0 1760 V
-3050 0 V
400 300 L
1.000 UL
LT0
400 1134 M
212 56 V
529 209 V
265 170 V
132 13 V
133 -19 V
264 -228 V
265 -209 V
2730 856 L
529 -54 V
191 -60 V
0.600 UP
1.000 UL
LT1
612 1180 M
0 21 V
-31 -21 R
62 0 V
-62 21 R
62 0 V
498 183 R
0 29 V
-31 -29 R
62 0 V
-62 29 R
62 0 V
234 140 R
0 32 V
-31 -32 R
62 0 V
-62 32 R
62 0 V
101 -52 R
0 97 V
-31 -97 R
62 0 V
-62 97 R
62 0 V
102 -85 R
0 35 V
-31 -35 R
62 0 V
-62 35 R
62 0 V
233 -256 R
0 23 V
-31 -23 R
62 0 V
-62 23 R
62 0 V
234 -230 R
0 18 V
-31 -18 R
62 0 V
-62 18 R
62 0 V
2730 849 M
0 14 V
-31 -14 R
62 0 V
-62 14 R
62 0 V
3259 737 M
0 129 V
3228 737 M
62 0 V
-62 129 R
62 0 V
612 1190 Pls
1141 1399 Pls
1406 1569 Pls
1538 1582 Pls
1671 1563 Pls
1935 1335 Pls
2200 1126 Pls
2730 856 Pls
3259 802 Pls
1.000 UL
LT2
400 1036 M
212 27 V
529 48 V
265 339 V
265 375 V
132 -92 V
133 -189 V
264 -314 V
2730 932 L
3259 783 L
191 -59 V
0.600 UP
1.000 UL
LT3
612 986 M
0 153 V
581 986 M
62 0 V
-62 153 R
62 0 V
498 -122 R
0 188 V
-31 -188 R
62 0 V
-62 188 R
62 0 V
234 210 R
0 70 V
-31 -70 R
62 0 V
-62 70 R
62 0 V
234 268 R
0 144 V
-31 -144 R
62 0 V
-62 144 R
62 0 V
101 -236 R
0 144 V
-31 -144 R
62 0 V
-62 144 R
62 0 V
102 -321 R
0 120 V
-31 -120 R
62 0 V
-62 120 R
62 0 V
233 -420 R
0 91 V
-31 -91 R
62 0 V
-62 91 R
62 0 V
2730 897 M
0 70 V
-31 -70 R
62 0 V
-62 70 R
62 0 V
3259 736 M
0 93 V
-31 -93 R
62 0 V
-62 93 R
62 0 V
612 1063 Crs
1141 1111 Crs
1406 1450 Crs
1671 1825 Crs
1803 1733 Crs
1936 1544 Crs
2200 1230 Crs
2730 932 Crs
3259 783 Crs
stroke
grestore
end
showpage
}}%
\put(1925,50){\makebox(0,0){$\gamma$}}%
\put(100,1180){%
\makebox(0,0)[b]{\shortstack{$C_2$}}%
}%
\put(3450,200){\makebox(0,0){0.65}}%
\put(3069,200){\makebox(0,0){0.6}}%
\put(2687,200){\makebox(0,0){0.55}}%
\put(2306,200){\makebox(0,0){0.5}}%
\put(1925,200){\makebox(0,0){0.45}}%
\put(1544,200){\makebox(0,0){0.4}}%
\put(1163,200){\makebox(0,0){0.35}}%
\put(781,200){\makebox(0,0){0.3}}%
\put(400,200){\makebox(0,0){0.25}}%
\put(350,2060){\makebox(0,0)[r]{1.2}}%
\put(350,1767){\makebox(0,0)[r]{1}}%
\put(350,1473){\makebox(0,0)[r]{0.8}}%
\put(350,1180){\makebox(0,0)[r]{0.6}}%
\put(350,887){\makebox(0,0)[r]{0.4}}%
\put(350,593){\makebox(0,0)[r]{0.2}}%
\put(350,300){\makebox(0,0)[r]{0}}%
\end{picture}%
\endgroup

%% file: P_2graph.tex
\begingroup%
  \makeatletter%
  \newcommand{\GNUPLOTspecial}{%
    \@sanitize\catcode`\%=14\relax\special}%
  \setlength{\unitlength}{0.1bp}%
{\GNUPLOTspecial{!
/gnudict 256 dict def
gnudict begin
/Color false def
/Solid false def
/gnulinewidth 5.000 def
/userlinewidth gnulinewidth def
/vshift -33 def
/dl {10 mul} def
/hpt_ 31.5 def
/vpt_ 31.5 def
/hpt hpt_ def
/vpt vpt_ def
/M {moveto} bind def
/L {lineto} bind def
/R {rmoveto} bind def
/V {rlineto} bind def
/vpt2 vpt 2 mul def
/hpt2 hpt 2 mul def
/Lshow { currentpoint stroke M
  0 vshift R show } def
/Rshow { currentpoint stroke M
  dup stringwidth pop neg vshift R show } def
/Cshow { currentpoint stroke M
  dup stringwidth pop -2 div vshift R show } def
/UP { dup vpt_ mul /vpt exch def hpt_ mul /hpt exch def
  /hpt2 hpt 2 mul def /vpt2 vpt 2 mul def } def
/DL { Color {setrgbcolor Solid {pop []} if 0 setdash }
 {pop pop pop Solid {pop []} if 0 setdash} ifelse } def
/BL { stroke userlinewidth 2 mul setlinewidth } def
/AL { stroke userlinewidth 2 div setlinewidth } def
/UL { dup gnulinewidth mul /userlinewidth exch def
      10 mul /udl exch def } def
/PL { stroke userlinewidth setlinewidth } def
/LTb { BL [] 0 0 0 DL } def
/LTa { AL [1 udl mul 2 udl mul] 0 setdash 0 0 0 setrgbcolor } def
/LT0 { PL [] 1 0 0 DL } def
/LT1 { PL [4 dl 2 dl] 0 1 0 DL } def
/LT2 { PL [2 dl 3 dl] 0 0 1 DL } def
/LT3 { PL [1 dl 1.5 dl] 1 0 1 DL } def
/LT4 { PL [5 dl 2 dl 1 dl 2 dl] 0 1 1 DL } def
/LT5 { PL [4 dl 3 dl 1 dl 3 dl] 1 1 0 DL } def
/LT6 { PL [2 dl 2 dl 2 dl 4 dl] 0 0 0 DL } def
/LT7 { PL [2 dl 2 dl 2 dl 2 dl 2 dl 4 dl] 1 0.3 0 DL } def
/LT8 { PL [2 dl 2 dl 2 dl 2 dl 2 dl 2 dl 2 dl 4 dl] 0.5 0.5 0.5 DL } def
/Pnt { stroke [] 0 setdash
   gsave 1 setlinecap M 0 0 V stroke grestore } def
/Dia { stroke [] 0 setdash 2 copy vpt add M
  hpt neg vpt neg V hpt vpt neg V
  hpt vpt V hpt neg vpt V closepath stroke
  Pnt } def
/Pls { stroke [] 0 setdash vpt sub M 0 vpt2 V
  currentpoint stroke M
  hpt neg vpt neg R hpt2 0 V stroke
  } def
/Box { stroke [] 0 setdash 2 copy exch hpt sub exch vpt add M
  0 vpt2 neg V hpt2 0 V 0 vpt2 V
  hpt2 neg 0 V closepath stroke
  Pnt } def
/Crs { stroke [] 0 setdash exch hpt sub exch vpt add M
  hpt2 vpt2 neg V currentpoint stroke M
  hpt2 neg 0 R hpt2 vpt2 V stroke } def
/TriU { stroke [] 0 setdash 2 copy vpt 1.12 mul add M
  hpt neg vpt -1.62 mul V
  hpt 2 mul 0 V
  hpt neg vpt 1.62 mul V closepath stroke
  Pnt  } def
/Star { 2 copy Pls Crs } def
/BoxF { stroke [] 0 setdash exch hpt sub exch vpt add M
  0 vpt2 neg V  hpt2 0 V  0 vpt2 V
  hpt2 neg 0 V  closepath fill } def
/TriUF { stroke [] 0 setdash vpt 1.12 mul add M
  hpt neg vpt -1.62 mul V
  hpt 2 mul 0 V
  hpt neg vpt 1.62 mul V closepath fill } def
/TriD { stroke [] 0 setdash 2 copy vpt 1.12 mul sub M
  hpt neg vpt 1.62 mul V
  hpt 2 mul 0 V
  hpt neg vpt -1.62 mul V closepath stroke
  Pnt  } def
/TriDF { stroke [] 0 setdash vpt 1.12 mul sub M
  hpt neg vpt 1.62 mul V
  hpt 2 mul 0 V
  hpt neg vpt -1.62 mul V closepath fill} def
/DiaF { stroke [] 0 setdash vpt add M
  hpt neg vpt neg V hpt vpt neg V
  hpt vpt V hpt neg vpt V closepath fill } def
/Pent { stroke [] 0 setdash 2 copy gsave
  translate 0 hpt M 4 {72 rotate 0 hpt L} repeat
  closepath stroke grestore Pnt } def
/PentF { stroke [] 0 setdash gsave
  translate 0 hpt M 4 {72 rotate 0 hpt L} repeat
  closepath fill grestore } def
/Circle { stroke [] 0 setdash 2 copy
  hpt 0 360 arc stroke Pnt } def
/CircleF { stroke [] 0 setdash hpt 0 360 arc fill } def
/C0 { BL [] 0 setdash 2 copy moveto vpt 90 450  arc } bind def
/C1 { BL [] 0 setdash 2 copy        moveto
       2 copy  vpt 0 90 arc closepath fill
               vpt 0 360 arc closepath } bind def
/C2 { BL [] 0 setdash 2 copy moveto
       2 copy  vpt 90 180 arc closepath fill
               vpt 0 360 arc closepath } bind def
/C3 { BL [] 0 setdash 2 copy moveto
       2 copy  vpt 0 180 arc closepath fill
               vpt 0 360 arc closepath } bind def
/C4 { BL [] 0 setdash 2 copy moveto
       2 copy  vpt 180 270 arc closepath fill
               vpt 0 360 arc closepath } bind def
/C5 { BL [] 0 setdash 2 copy moveto
       2 copy  vpt 0 90 arc
       2 copy moveto
       2 copy  vpt 180 270 arc closepath fill
               vpt 0 360 arc } bind def
/C6 { BL [] 0 setdash 2 copy moveto
      2 copy  vpt 90 270 arc closepath fill
              vpt 0 360 arc closepath } bind def
/C7 { BL [] 0 setdash 2 copy moveto
      2 copy  vpt 0 270 arc closepath fill
              vpt 0 360 arc closepath } bind def
/C8 { BL [] 0 setdash 2 copy moveto
      2 copy vpt 270 360 arc closepath fill
              vpt 0 360 arc closepath } bind def
/C9 { BL [] 0 setdash 2 copy moveto
      2 copy  vpt 270 450 arc closepath fill
              vpt 0 360 arc closepath } bind def
/C10 { BL [] 0 setdash 2 copy 2 copy moveto vpt 270 360 arc closepath fill
       2 copy moveto
       2 copy vpt 90 180 arc closepath fill
               vpt 0 360 arc closepath } bind def
/C11 { BL [] 0 setdash 2 copy moveto
       2 copy  vpt 0 180 arc closepath fill
       2 copy moveto
       2 copy  vpt 270 360 arc closepath fill
               vpt 0 360 arc closepath } bind def
/C12 { BL [] 0 setdash 2 copy moveto
       2 copy  vpt 180 360 arc closepath fill
               vpt 0 360 arc closepath } bind def
/C13 { BL [] 0 setdash  2 copy moveto
       2 copy  vpt 0 90 arc closepath fill
       2 copy moveto
       2 copy  vpt 180 360 arc closepath fill
               vpt 0 360 arc closepath } bind def
/C14 { BL [] 0 setdash 2 copy moveto
       2 copy  vpt 90 360 arc closepath fill
               vpt 0 360 arc } bind def
/C15 { BL [] 0 setdash 2 copy vpt 0 360 arc closepath fill
               vpt 0 360 arc closepath } bind def
/Rec   { newpath 4 2 roll moveto 1 index 0 rlineto 0 exch rlineto
       neg 0 rlineto closepath } bind def
/Square { dup Rec } bind def
/Bsquare { vpt sub exch vpt sub exch vpt2 Square } bind def
/S0 { BL [] 0 setdash 2 copy moveto 0 vpt rlineto BL Bsquare } bind def
/S1 { BL [] 0 setdash 2 copy vpt Square fill Bsquare } bind def
/S2 { BL [] 0 setdash 2 copy exch vpt sub exch vpt Square fill Bsquare } bind def
/S3 { BL [] 0 setdash 2 copy exch vpt sub exch vpt2 vpt Rec fill Bsquare } bind def
/S4 { BL [] 0 setdash 2 copy exch vpt sub exch vpt sub vpt Square fill Bsquare } bind def
/S5 { BL [] 0 setdash 2 copy 2 copy vpt Square fill
       exch vpt sub exch vpt sub vpt Square fill Bsquare } bind def
/S6 { BL [] 0 setdash 2 copy exch vpt sub exch vpt sub vpt vpt2 Rec fill Bsquare } bind def
/S7 { BL [] 0 setdash 2 copy exch vpt sub exch vpt sub vpt vpt2 Rec fill
       2 copy vpt Square fill
       Bsquare } bind def
/S8 { BL [] 0 setdash 2 copy vpt sub vpt Square fill Bsquare } bind def
/S9 { BL [] 0 setdash 2 copy vpt sub vpt vpt2 Rec fill Bsquare } bind def
/S10 { BL [] 0 setdash 2 copy vpt sub vpt Square fill 2 copy exch vpt sub exch vpt Square fill
       Bsquare } bind def
/S11 { BL [] 0 setdash 2 copy vpt sub vpt Square fill 2 copy exch vpt sub exch vpt2 vpt Rec fill
       Bsquare } bind def
/S12 { BL [] 0 setdash 2 copy exch vpt sub exch vpt sub vpt2 vpt Rec fill Bsquare } bind def
/S13 { BL [] 0 setdash 2 copy exch vpt sub exch vpt sub vpt2 vpt Rec fill
       2 copy vpt Square fill Bsquare } bind def
/S14 { BL [] 0 setdash 2 copy exch vpt sub exch vpt sub vpt2 vpt Rec fill
       2 copy exch vpt sub exch vpt Square fill Bsquare } bind def
/S15 { BL [] 0 setdash 2 copy Bsquare fill Bsquare } bind def
/D0 { gsave translate 45 rotate 0 0 S0 stroke grestore } bind def
/D1 { gsave translate 45 rotate 0 0 S1 stroke grestore } bind def
/D2 { gsave translate 45 rotate 0 0 S2 stroke grestore } bind def
/D3 { gsave translate 45 rotate 0 0 S3 stroke grestore } bind def
/D4 { gsave translate 45 rotate 0 0 S4 stroke grestore } bind def
/D5 { gsave translate 45 rotate 0 0 S5 stroke grestore } bind def
/D6 { gsave translate 45 rotate 0 0 S6 stroke grestore } bind def
/D7 { gsave translate 45 rotate 0 0 S7 stroke grestore } bind def
/D8 { gsave translate 45 rotate 0 0 S8 stroke grestore } bind def
/D9 { gsave translate 45 rotate 0 0 S9 stroke grestore } bind def
/D10 { gsave translate 45 rotate 0 0 S10 stroke grestore } bind def
/D11 { gsave translate 45 rotate 0 0 S11 stroke grestore } bind def
/D12 { gsave translate 45 rotate 0 0 S12 stroke grestore } bind def
/D13 { gsave translate 45 rotate 0 0 S13 stroke grestore } bind def
/D14 { gsave translate 45 rotate 0 0 S14 stroke grestore } bind def
/D15 { gsave translate 45 rotate 0 0 S15 stroke grestore } bind def
/DiaE { stroke [] 0 setdash vpt add M
  hpt neg vpt neg V hpt vpt neg V
  hpt vpt V hpt neg vpt V closepath stroke } def
/BoxE { stroke [] 0 setdash exch hpt sub exch vpt add M
  0 vpt2 neg V hpt2 0 V 0 vpt2 V
  hpt2 neg 0 V closepath stroke } def
/TriUE { stroke [] 0 setdash vpt 1.12 mul add M
  hpt neg vpt -1.62 mul V
  hpt 2 mul 0 V
  hpt neg vpt 1.62 mul V closepath stroke } def
/TriDE { stroke [] 0 setdash vpt 1.12 mul sub M
  hpt neg vpt 1.62 mul V
  hpt 2 mul 0 V
  hpt neg vpt -1.62 mul V closepath stroke } def
/PentE { stroke [] 0 setdash gsave
  translate 0 hpt M 4 {72 rotate 0 hpt L} repeat
  closepath stroke grestore } def
/CircE { stroke [] 0 setdash 
  hpt 0 360 arc stroke } def
/Opaque { gsave closepath 1 setgray fill grestore 0 setgray closepath } def
/DiaW { stroke [] 0 setdash vpt add M
  hpt neg vpt neg V hpt vpt neg V
  hpt vpt V hpt neg vpt V Opaque stroke } def
/BoxW { stroke [] 0 setdash exch hpt sub exch vpt add M
  0 vpt2 neg V hpt2 0 V 0 vpt2 V
  hpt2 neg 0 V Opaque stroke } def
/TriUW { stroke [] 0 setdash vpt 1.12 mul add M
  hpt neg vpt -1.62 mul V
  hpt 2 mul 0 V
  hpt neg vpt 1.62 mul V Opaque stroke } def
/TriDW { stroke [] 0 setdash vpt 1.12 mul sub M
  hpt neg vpt 1.62 mul V
  hpt 2 mul 0 V
  hpt neg vpt -1.62 mul V Opaque stroke } def
/PentW { stroke [] 0 setdash gsave
  translate 0 hpt M 4 {72 rotate 0 hpt L} repeat
  Opaque stroke grestore } def
/CircW { stroke [] 0 setdash 
  hpt 0 360 arc Opaque stroke } def
/BoxFill { gsave Rec 1 setgray fill grestore } def
end
}}%
\begin{picture}(3600,2160)(0,0)%
{\GNUPLOTspecial{"
gnudict begin
gsave
0 0 translate
0.100 0.100 scale
0 setgray
newpath
1.000 UL
LTb
400 300 M
63 0 V
2987 0 R
-63 0 V
400 593 M
63 0 V
2987 0 R
-63 0 V
400 887 M
63 0 V
2987 0 R
-63 0 V
400 1180 M
63 0 V
2987 0 R
-63 0 V
400 1473 M
63 0 V
2987 0 R
-63 0 V
400 1767 M
63 0 V
2987 0 R
-63 0 V
400 2060 M
63 0 V
2987 0 R
-63 0 V
400 300 M
0 63 V
0 1697 R
0 -63 V
781 300 M
0 63 V
0 1697 R
0 -63 V
1163 300 M
0 63 V
0 1697 R
0 -63 V
1544 300 M
0 63 V
0 1697 R
0 -63 V
1925 300 M
0 63 V
0 1697 R
0 -63 V
2306 300 M
0 63 V
0 1697 R
0 -63 V
2687 300 M
0 63 V
0 1697 R
0 -63 V
3069 300 M
0 63 V
0 1697 R
0 -63 V
3450 300 M
0 63 V
0 1697 R
0 -63 V
1.000 UL
LTb
400 300 M
3050 0 V
0 1760 V
-3050 0 V
400 300 L
0.600 UP
1.000 UL
LT0
400 1922 M
212 40 V
529 8 V
265 -106 V
132 -97 V
133 -119 V
264 -255 V
265 -203 V
2730 933 L
3259 778 L
191 -36 V
612 1962 Pls
1141 1970 Pls
1406 1864 Pls
1538 1767 Pls
1671 1648 Pls
1935 1393 Pls
2200 1190 Pls
2730 933 Pls
3259 778 Pls
0.600 UP
1.000 UL
LT1
400 1779 M
212 3 V
529 63 V
265 33 V
265 -73 V
132 -142 V
133 -164 V
264 -259 V
2730 957 L
3259 797 L
191 -39 V
612 1782 Crs
1141 1845 Crs
1406 1878 Crs
1671 1805 Crs
1803 1663 Crs
1936 1499 Crs
2200 1240 Crs
2730 957 Crs
3259 797 Crs
0.600 UP
1.000 UL
LT2
1141 1778 M
265 25 V
265 29 V
132 -68 V
66 -104 V
67 -108 V
264 -296 V
265 -173 V
1141 1778 Star
1406 1803 Star
1671 1832 Star
1803 1764 Star
1869 1660 Star
1936 1552 Star
2200 1256 Star
2465 1083 Star
1.000 UL
LT3
1406 1771 M
265 46 V
66 10 V
66 -15 V
33 -50 V
33 -63 V
67 -134 V
264 -312 V
0.600 UP
1.000 UL
LT4
1406 1765 M
0 13 V
-31 -13 R
62 0 V
-62 13 R
62 0 V
234 34 R
0 11 V
-31 -11 R
62 0 V
-62 11 R
62 0 V
35 -1 R
0 11 V
-31 -11 R
62 0 V
-62 11 R
62 0 V
35 -27 R
0 13 V
-31 -13 R
62 0 V
-62 13 R
62 0 V
2 -64 R
0 14 V
-31 -14 R
62 0 V
-62 14 R
62 0 V
2 -76 R
0 13 V
-31 -13 R
62 0 V
-62 13 R
62 0 V
36 -146 R
0 10 V
-31 -10 R
62 0 V
-62 10 R
62 0 V
233 -322 R
0 11 V
-31 -11 R
62 0 V
-62 11 R
62 0 V
1406 1771 Box
1671 1817 Box
1737 1827 Box
1803 1812 Box
1836 1762 Box
1869 1699 Box
1936 1565 Box
2200 1253 Box
stroke
grestore
end
showpage
}}%
\put(1925,50){\makebox(0,0){$\gamma$}}%
\put(100,1180){%
\makebox(0,0)[b]{\shortstack{$P_2$}}%
}%
\put(3450,200){\makebox(0,0){0.65}}%
\put(3069,200){\makebox(0,0){0.6}}%
\put(2687,200){\makebox(0,0){0.55}}%
\put(2306,200){\makebox(0,0){0.5}}%
\put(1925,200){\makebox(0,0){0.45}}%
\put(1544,200){\makebox(0,0){0.4}}%
\put(1163,200){\makebox(0,0){0.35}}%
\put(781,200){\makebox(0,0){0.3}}%
\put(400,200){\makebox(0,0){0.25}}%
\put(350,2060){\makebox(0,0)[r]{0.6}}%
\put(350,1767){\makebox(0,0)[r]{0.5}}%
\put(350,1473){\makebox(0,0)[r]{0.4}}%
\put(350,1180){\makebox(0,0)[r]{0.3}}%
\put(350,887){\makebox(0,0)[r]{0.2}}%
\put(350,593){\makebox(0,0)[r]{0.1}}%
\put(350,300){\makebox(0,0)[r]{0}}%
\end{picture}%
\endgroup

%% file: 1+1d_C_2graph.tex
\begingroup%
  \makeatletter%
  \newcommand{\GNUPLOTspecial}{%
    \@sanitize\catcode`\%=14\relax\special}%
  \setlength{\unitlength}{0.1bp}%
{\GNUPLOTspecial{!
/gnudict 256 dict def
gnudict begin
/Color false def
/Solid false def
/gnulinewidth 5.000 def
/userlinewidth gnulinewidth def
/vshift -33 def
/dl {10 mul} def
/hpt_ 31.5 def
/vpt_ 31.5 def
/hpt hpt_ def
/vpt vpt_ def
/M {moveto} bind def
/L {lineto} bind def
/R {rmoveto} bind def
/V {rlineto} bind def
/vpt2 vpt 2 mul def
/hpt2 hpt 2 mul def
/Lshow { currentpoint stroke M
  0 vshift R show } def
/Rshow { currentpoint stroke M
  dup stringwidth pop neg vshift R show } def
/Cshow { currentpoint stroke M
  dup stringwidth pop -2 div vshift R show } def
/UP { dup vpt_ mul /vpt exch def hpt_ mul /hpt exch def
  /hpt2 hpt 2 mul def /vpt2 vpt 2 mul def } def
/DL { Color {setrgbcolor Solid {pop []} if 0 setdash }
 {pop pop pop Solid {pop []} if 0 setdash} ifelse } def
/BL { stroke userlinewidth 2 mul setlinewidth } def
/AL { stroke userlinewidth 2 div setlinewidth } def
/UL { dup gnulinewidth mul /userlinewidth exch def
      10 mul /udl exch def } def
/PL { stroke userlinewidth setlinewidth } def
/LTb { BL [] 0 0 0 DL } def
/LTa { AL [1 udl mul 2 udl mul] 0 setdash 0 0 0 setrgbcolor } def
/LT0 { PL [] 1 0 0 DL } def
/LT1 { PL [4 dl 2 dl] 0 1 0 DL } def
/LT2 { PL [2 dl 3 dl] 0 0 1 DL } def
/LT3 { PL [1 dl 1.5 dl] 1 0 1 DL } def
/LT4 { PL [5 dl 2 dl 1 dl 2 dl] 0 1 1 DL } def
/LT5 { PL [4 dl 3 dl 1 dl 3 dl] 1 1 0 DL } def
/LT6 { PL [2 dl 2 dl 2 dl 4 dl] 0 0 0 DL } def
/LT7 { PL [2 dl 2 dl 2 dl 2 dl 2 dl 4 dl] 1 0.3 0 DL } def
/LT8 { PL [2 dl 2 dl 2 dl 2 dl 2 dl 2 dl 2 dl 4 dl] 0.5 0.5 0.5 DL } def
/Pnt { stroke [] 0 setdash
   gsave 1 setlinecap M 0 0 V stroke grestore } def
/Dia { stroke [] 0 setdash 2 copy vpt add M
  hpt neg vpt neg V hpt vpt neg V
  hpt vpt V hpt neg vpt V closepath stroke
  Pnt } def
/Pls { stroke [] 0 setdash vpt sub M 0 vpt2 V
  currentpoint stroke M
  hpt neg vpt neg R hpt2 0 V stroke
  } def
/Box { stroke [] 0 setdash 2 copy exch hpt sub exch vpt add M
  0 vpt2 neg V hpt2 0 V 0 vpt2 V
  hpt2 neg 0 V closepath stroke
  Pnt } def
/Crs { stroke [] 0 setdash exch hpt sub exch vpt add M
  hpt2 vpt2 neg V currentpoint stroke M
  hpt2 neg 0 R hpt2 vpt2 V stroke } def
/TriU { stroke [] 0 setdash 2 copy vpt 1.12 mul add M
  hpt neg vpt -1.62 mul V
  hpt 2 mul 0 V
  hpt neg vpt 1.62 mul V closepath stroke
  Pnt  } def
/Star { 2 copy Pls Crs } def
/BoxF { stroke [] 0 setdash exch hpt sub exch vpt add M
  0 vpt2 neg V  hpt2 0 V  0 vpt2 V
  hpt2 neg 0 V  closepath fill } def
/TriUF { stroke [] 0 setdash vpt 1.12 mul add M
  hpt neg vpt -1.62 mul V
  hpt 2 mul 0 V
  hpt neg vpt 1.62 mul V closepath fill } def
/TriD { stroke [] 0 setdash 2 copy vpt 1.12 mul sub M
  hpt neg vpt 1.62 mul V
  hpt 2 mul 0 V
  hpt neg vpt -1.62 mul V closepath stroke
  Pnt  } def
/TriDF { stroke [] 0 setdash vpt 1.12 mul sub M
  hpt neg vpt 1.62 mul V
  hpt 2 mul 0 V
  hpt neg vpt -1.62 mul V closepath fill} def
/DiaF { stroke [] 0 setdash vpt add M
  hpt neg vpt neg V hpt vpt neg V
  hpt vpt V hpt neg vpt V closepath fill } def
/Pent { stroke [] 0 setdash 2 copy gsave
  translate 0 hpt M 4 {72 rotate 0 hpt L} repeat
  closepath stroke grestore Pnt } def
/PentF { stroke [] 0 setdash gsave
  translate 0 hpt M 4 {72 rotate 0 hpt L} repeat
  closepath fill grestore } def
/Circle { stroke [] 0 setdash 2 copy
  hpt 0 360 arc stroke Pnt } def
/CircleF { stroke [] 0 setdash hpt 0 360 arc fill } def
/C0 { BL [] 0 setdash 2 copy moveto vpt 90 450  arc } bind def
/C1 { BL [] 0 setdash 2 copy        moveto
       2 copy  vpt 0 90 arc closepath fill
               vpt 0 360 arc closepath } bind def
/C2 { BL [] 0 setdash 2 copy moveto
       2 copy  vpt 90 180 arc closepath fill
               vpt 0 360 arc closepath } bind def
/C3 { BL [] 0 setdash 2 copy moveto
       2 copy  vpt 0 180 arc closepath fill
               vpt 0 360 arc closepath } bind def
/C4 { BL [] 0 setdash 2 copy moveto
       2 copy  vpt 180 270 arc closepath fill
               vpt 0 360 arc closepath } bind def
/C5 { BL [] 0 setdash 2 copy moveto
       2 copy  vpt 0 90 arc
       2 copy moveto
       2 copy  vpt 180 270 arc closepath fill
               vpt 0 360 arc } bind def
/C6 { BL [] 0 setdash 2 copy moveto
      2 copy  vpt 90 270 arc closepath fill
              vpt 0 360 arc closepath } bind def
/C7 { BL [] 0 setdash 2 copy moveto
      2 copy  vpt 0 270 arc closepath fill
              vpt 0 360 arc closepath } bind def
/C8 { BL [] 0 setdash 2 copy moveto
      2 copy vpt 270 360 arc closepath fill
              vpt 0 360 arc closepath } bind def
/C9 { BL [] 0 setdash 2 copy moveto
      2 copy  vpt 270 450 arc closepath fill
              vpt 0 360 arc closepath } bind def
/C10 { BL [] 0 setdash 2 copy 2 copy moveto vpt 270 360 arc closepath fill
       2 copy moveto
       2 copy vpt 90 180 arc closepath fill
               vpt 0 360 arc closepath } bind def
/C11 { BL [] 0 setdash 2 copy moveto
       2 copy  vpt 0 180 arc closepath fill
       2 copy moveto
       2 copy  vpt 270 360 arc closepath fill
               vpt 0 360 arc closepath } bind def
/C12 { BL [] 0 setdash 2 copy moveto
       2 copy  vpt 180 360 arc closepath fill
               vpt 0 360 arc closepath } bind def
/C13 { BL [] 0 setdash  2 copy moveto
       2 copy  vpt 0 90 arc closepath fill
       2 copy moveto
       2 copy  vpt 180 360 arc closepath fill
               vpt 0 360 arc closepath } bind def
/C14 { BL [] 0 setdash 2 copy moveto
       2 copy  vpt 90 360 arc closepath fill
               vpt 0 360 arc } bind def
/C15 { BL [] 0 setdash 2 copy vpt 0 360 arc closepath fill
               vpt 0 360 arc closepath } bind def
/Rec   { newpath 4 2 roll moveto 1 index 0 rlineto 0 exch rlineto
       neg 0 rlineto closepath } bind def
/Square { dup Rec } bind def
/Bsquare { vpt sub exch vpt sub exch vpt2 Square } bind def
/S0 { BL [] 0 setdash 2 copy moveto 0 vpt rlineto BL Bsquare } bind def
/S1 { BL [] 0 setdash 2 copy vpt Square fill Bsquare } bind def
/S2 { BL [] 0 setdash 2 copy exch vpt sub exch vpt Square fill Bsquare } bind def
/S3 { BL [] 0 setdash 2 copy exch vpt sub exch vpt2 vpt Rec fill Bsquare } bind def
/S4 { BL [] 0 setdash 2 copy exch vpt sub exch vpt sub vpt Square fill Bsquare } bind def
/S5 { BL [] 0 setdash 2 copy 2 copy vpt Square fill
       exch vpt sub exch vpt sub vpt Square fill Bsquare } bind def
/S6 { BL [] 0 setdash 2 copy exch vpt sub exch vpt sub vpt vpt2 Rec fill Bsquare } bind def
/S7 { BL [] 0 setdash 2 copy exch vpt sub exch vpt sub vpt vpt2 Rec fill
       2 copy vpt Square fill
       Bsquare } bind def
/S8 { BL [] 0 setdash 2 copy vpt sub vpt Square fill Bsquare } bind def
/S9 { BL [] 0 setdash 2 copy vpt sub vpt vpt2 Rec fill Bsquare } bind def
/S10 { BL [] 0 setdash 2 copy vpt sub vpt Square fill 2 copy exch vpt sub exch vpt Square fill
       Bsquare } bind def
/S11 { BL [] 0 setdash 2 copy vpt sub vpt Square fill 2 copy exch vpt sub exch vpt2 vpt Rec fill
       Bsquare } bind def
/S12 { BL [] 0 setdash 2 copy exch vpt sub exch vpt sub vpt2 vpt Rec fill Bsquare } bind def
/S13 { BL [] 0 setdash 2 copy exch vpt sub exch vpt sub vpt2 vpt Rec fill
       2 copy vpt Square fill Bsquare } bind def
/S14 { BL [] 0 setdash 2 copy exch vpt sub exch vpt sub vpt2 vpt Rec fill
       2 copy exch vpt sub exch vpt Square fill Bsquare } bind def
/S15 { BL [] 0 setdash 2 copy Bsquare fill Bsquare } bind def
/D0 { gsave translate 45 rotate 0 0 S0 stroke grestore } bind def
/D1 { gsave translate 45 rotate 0 0 S1 stroke grestore } bind def
/D2 { gsave translate 45 rotate 0 0 S2 stroke grestore } bind def
/D3 { gsave translate 45 rotate 0 0 S3 stroke grestore } bind def
/D4 { gsave translate 45 rotate 0 0 S4 stroke grestore } bind def
/D5 { gsave translate 45 rotate 0 0 S5 stroke grestore } bind def
/D6 { gsave translate 45 rotate 0 0 S6 stroke grestore } bind def
/D7 { gsave translate 45 rotate 0 0 S7 stroke grestore } bind def
/D8 { gsave translate 45 rotate 0 0 S8 stroke grestore } bind def
/D9 { gsave translate 45 rotate 0 0 S9 stroke grestore } bind def
/D10 { gsave translate 45 rotate 0 0 S10 stroke grestore } bind def
/D11 { gsave translate 45 rotate 0 0 S11 stroke grestore } bind def
/D12 { gsave translate 45 rotate 0 0 S12 stroke grestore } bind def
/D13 { gsave translate 45 rotate 0 0 S13 stroke grestore } bind def
/D14 { gsave translate 45 rotate 0 0 S14 stroke grestore } bind def
/D15 { gsave translate 45 rotate 0 0 S15 stroke grestore } bind def
/DiaE { stroke [] 0 setdash vpt add M
  hpt neg vpt neg V hpt vpt neg V
  hpt vpt V hpt neg vpt V closepath stroke } def
/BoxE { stroke [] 0 setdash exch hpt sub exch vpt add M
  0 vpt2 neg V hpt2 0 V 0 vpt2 V
  hpt2 neg 0 V closepath stroke } def
/TriUE { stroke [] 0 setdash vpt 1.12 mul add M
  hpt neg vpt -1.62 mul V
  hpt 2 mul 0 V
  hpt neg vpt 1.62 mul V closepath stroke } def
/TriDE { stroke [] 0 setdash vpt 1.12 mul sub M
  hpt neg vpt 1.62 mul V
  hpt 2 mul 0 V
  hpt neg vpt -1.62 mul V closepath stroke } def
/PentE { stroke [] 0 setdash gsave
  translate 0 hpt M 4 {72 rotate 0 hpt L} repeat
  closepath stroke grestore } def
/CircE { stroke [] 0 setdash 
  hpt 0 360 arc stroke } def
/Opaque { gsave closepath 1 setgray fill grestore 0 setgray closepath } def
/DiaW { stroke [] 0 setdash vpt add M
  hpt neg vpt neg V hpt vpt neg V
  hpt vpt V hpt neg vpt V Opaque stroke } def
/BoxW { stroke [] 0 setdash exch hpt sub exch vpt add M
  0 vpt2 neg V hpt2 0 V 0 vpt2 V
  hpt2 neg 0 V Opaque stroke } def
/TriUW { stroke [] 0 setdash vpt 1.12 mul add M
  hpt neg vpt -1.62 mul V
  hpt 2 mul 0 V
  hpt neg vpt 1.62 mul V Opaque stroke } def
/TriDW { stroke [] 0 setdash vpt 1.12 mul sub M
  hpt neg vpt 1.62 mul V
  hpt 2 mul 0 V
  hpt neg vpt -1.62 mul V Opaque stroke } def
/PentW { stroke [] 0 setdash gsave
  translate 0 hpt M 4 {72 rotate 0 hpt L} repeat
  Opaque stroke grestore } def
/CircW { stroke [] 0 setdash 
  hpt 0 360 arc Opaque stroke } def
/BoxFill { gsave Rec 1 setgray fill grestore } def
end
}}%
\begin{picture}(3600,2160)(0,0)%
{\GNUPLOTspecial{"
gnudict begin
gsave
0 0 translate
0.100 0.100 scale
0 setgray
newpath
1.000 UL
LTb
400 300 M
63 0 V
2987 0 R
-63 0 V
400 593 M
63 0 V
2987 0 R
-63 0 V
400 887 M
63 0 V
2987 0 R
-63 0 V
400 1180 M
63 0 V
2987 0 R
-63 0 V
400 1473 M
63 0 V
2987 0 R
-63 0 V
400 1767 M
63 0 V
2987 0 R
-63 0 V
400 2060 M
63 0 V
2987 0 R
-63 0 V
400 300 M
0 63 V
0 1697 R
0 -63 V
705 300 M
0 63 V
0 1697 R
0 -63 V
1010 300 M
0 63 V
0 1697 R
0 -63 V
1315 300 M
0 63 V
0 1697 R
0 -63 V
1620 300 M
0 63 V
0 1697 R
0 -63 V
1925 300 M
0 63 V
0 1697 R
0 -63 V
2230 300 M
0 63 V
0 1697 R
0 -63 V
2535 300 M
0 63 V
0 1697 R
0 -63 V
2840 300 M
0 63 V
0 1697 R
0 -63 V
3145 300 M
0 63 V
0 1697 R
0 -63 V
3450 300 M
0 63 V
0 1697 R
0 -63 V
1.000 UL
LTb
400 300 M
3050 0 V
0 1760 V
-3050 0 V
400 300 L
1.000 UL
LT0
400 1767 M
1525 0 V
31 -29 V
30 -28 V
31 -28 V
30 -26 V
31 -26 V
30 -25 V
31 -24 V
30 -24 V
31 -23 V
30 -22 V
31 -22 V
30 -21 V
31 -20 V
30 -20 V
31 -20 V
30 -19 V
31 -19 V
30 -18 V
31 -17 V
30 -17 V
31 -17 V
30 -17 V
31 -16 V
30 -15 V
31 -15 V
30 -15 V
31 -15 V
30 -14 V
31 -14 V
30 -13 V
31 -13 V
30 -13 V
31 -13 V
30 -12 V
31 -12 V
30 -12 V
31 -12 V
30 -11 V
30 -11 V
31 -11 V
30 -10 V
31 -11 V
31 -10 V
30 -10 V
31 -9 V
30 -10 V
31 -9 V
30 -9 V
31 -9 V
30 -9 V
0.600 UP
1.000 UL
LT1
400 1924 M
847 0 V
339 -158 V
169 -93 V
170 -108 V
339 -183 V
3450 918 L
400 1924 Pls
1247 1924 Pls
1586 1766 Pls
1755 1673 Pls
1925 1565 Pls
2264 1382 Pls
3450 918 Pls
1.000 UL
LT2
400 1784 M
847 81 V
339 -43 V
169 -68 V
170 -97 V
339 -204 V
3450 946 L
0.600 UP
1.000 UL
LT3
400 1782 M
0 5 V
-31 -5 R
62 0 V
-62 5 R
62 0 V
816 74 R
0 7 V
-31 -7 R
62 0 V
-62 7 R
62 0 V
308 -49 R
0 7 V
-31 -7 R
62 0 V
-62 7 R
62 0 V
138 -76 R
0 8 V
-31 -8 R
62 0 V
-62 8 R
62 0 V
139 -104 R
0 6 V
-31 -6 R
62 0 V
-62 6 R
62 0 V
308 -210 R
0 5 V
-31 -5 R
62 0 V
-62 5 R
62 0 V
3450 944 M
0 4 V
-31 -4 R
62 0 V
-62 4 R
62 0 V
400 1784 Crs
1247 1865 Crs
1586 1822 Crs
1755 1754 Crs
1925 1657 Crs
2264 1453 Crs
3450 946 Crs
1.000 UL
LT4
400 1775 M
847 18 V
339 21 V
169 -36 V
170 -75 V
171 -115 V
168 -106 V
3450 954 L
0.600 UP
1.000 UL
LT5
400 1768 M
0 15 V
-31 -15 R
62 0 V
-62 15 R
62 0 V
816 2 R
0 15 V
-31 -15 R
62 0 V
-62 15 R
62 0 V
308 9 R
0 11 V
-31 -11 R
62 0 V
-62 11 R
62 0 V
138 -48 R
0 11 V
-31 -11 R
62 0 V
-62 11 R
62 0 V
139 -86 R
0 11 V
-31 -11 R
62 0 V
-62 11 R
62 0 V
140 -144 R
0 47 V
-31 -47 R
62 0 V
-62 47 R
62 0 V
137 -135 R
0 12 V
-31 -12 R
62 0 V
-62 12 R
62 0 V
3450 949 M
0 10 V
-31 -10 R
62 0 V
-62 10 R
62 0 V
400 1775 Star
1247 1793 Star
1586 1814 Star
1755 1778 Star
1925 1703 Star
2096 1588 Star
2264 1482 Star
3450 954 Star
stroke
grestore
end
showpage
}}%
\put(1925,50){\makebox(0,0){$\gamma$}}%
\put(100,1180){%
\makebox(0,0)[b]{\shortstack{$C_2$}}%
}%
\put(3450,200){\makebox(0,0){0.75}}%
\put(3145,200){\makebox(0,0){0.7}}%
\put(2840,200){\makebox(0,0){0.65}}%
\put(2535,200){\makebox(0,0){0.6}}%
\put(2230,200){\makebox(0,0){0.55}}%
\put(1925,200){\makebox(0,0){0.5}}%
\put(1620,200){\makebox(0,0){0.45}}%
\put(1315,200){\makebox(0,0){0.4}}%
\put(1010,200){\makebox(0,0){0.35}}%
\put(705,200){\makebox(0,0){0.3}}%
\put(400,200){\makebox(0,0){0.25}}%
\put(350,2060){\makebox(0,0)[r]{0.6}}%
\put(350,1767){\makebox(0,0)[r]{0.5}}%
\put(350,1473){\makebox(0,0)[r]{0.4}}%
\put(350,1180){\makebox(0,0)[r]{0.3}}%
\put(350,887){\makebox(0,0)[r]{0.2}}%
\put(350,593){\makebox(0,0)[r]{0.1}}%
\put(350,300){\makebox(0,0)[r]{0}}%
\end{picture}%
\endgroup

%% file: P_3graph.tex
\begingroup%
  \makeatletter%
  \newcommand{\GNUPLOTspecial}{%
    \@sanitize\catcode`\%=14\relax\special}%
  \setlength{\unitlength}{0.1bp}%
{\GNUPLOTspecial{!
/gnudict 256 dict def
gnudict begin
/Color false def
/Solid false def
/gnulinewidth 5.000 def
/userlinewidth gnulinewidth def
/vshift -33 def
/dl {10 mul} def
/hpt_ 31.5 def
/vpt_ 31.5 def
/hpt hpt_ def
/vpt vpt_ def
/M {moveto} bind def
/L {lineto} bind def
/R {rmoveto} bind def
/V {rlineto} bind def
/vpt2 vpt 2 mul def
/hpt2 hpt 2 mul def
/Lshow { currentpoint stroke M
  0 vshift R show } def
/Rshow { currentpoint stroke M
  dup stringwidth pop neg vshift R show } def
/Cshow { currentpoint stroke M
  dup stringwidth pop -2 div vshift R show } def
/UP { dup vpt_ mul /vpt exch def hpt_ mul /hpt exch def
  /hpt2 hpt 2 mul def /vpt2 vpt 2 mul def } def
/DL { Color {setrgbcolor Solid {pop []} if 0 setdash }
 {pop pop pop Solid {pop []} if 0 setdash} ifelse } def
/BL { stroke userlinewidth 2 mul setlinewidth } def
/AL { stroke userlinewidth 2 div setlinewidth } def
/UL { dup gnulinewidth mul /userlinewidth exch def
      10 mul /udl exch def } def
/PL { stroke userlinewidth setlinewidth } def
/LTb { BL [] 0 0 0 DL } def
/LTa { AL [1 udl mul 2 udl mul] 0 setdash 0 0 0 setrgbcolor } def
/LT0 { PL [] 1 0 0 DL } def
/LT1 { PL [4 dl 2 dl] 0 1 0 DL } def
/LT2 { PL [2 dl 3 dl] 0 0 1 DL } def
/LT3 { PL [1 dl 1.5 dl] 1 0 1 DL } def
/LT4 { PL [5 dl 2 dl 1 dl 2 dl] 0 1 1 DL } def
/LT5 { PL [4 dl 3 dl 1 dl 3 dl] 1 1 0 DL } def
/LT6 { PL [2 dl 2 dl 2 dl 4 dl] 0 0 0 DL } def
/LT7 { PL [2 dl 2 dl 2 dl 2 dl 2 dl 4 dl] 1 0.3 0 DL } def
/LT8 { PL [2 dl 2 dl 2 dl 2 dl 2 dl 2 dl 2 dl 4 dl] 0.5 0.5 0.5 DL } def
/Pnt { stroke [] 0 setdash
   gsave 1 setlinecap M 0 0 V stroke grestore } def
/Dia { stroke [] 0 setdash 2 copy vpt add M
  hpt neg vpt neg V hpt vpt neg V
  hpt vpt V hpt neg vpt V closepath stroke
  Pnt } def
/Pls { stroke [] 0 setdash vpt sub M 0 vpt2 V
  currentpoint stroke M
  hpt neg vpt neg R hpt2 0 V stroke
  } def
/Box { stroke [] 0 setdash 2 copy exch hpt sub exch vpt add M
  0 vpt2 neg V hpt2 0 V 0 vpt2 V
  hpt2 neg 0 V closepath stroke
  Pnt } def
/Crs { stroke [] 0 setdash exch hpt sub exch vpt add M
  hpt2 vpt2 neg V currentpoint stroke M
  hpt2 neg 0 R hpt2 vpt2 V stroke } def
/TriU { stroke [] 0 setdash 2 copy vpt 1.12 mul add M
  hpt neg vpt -1.62 mul V
  hpt 2 mul 0 V
  hpt neg vpt 1.62 mul V closepath stroke
  Pnt  } def
/Star { 2 copy Pls Crs } def
/BoxF { stroke [] 0 setdash exch hpt sub exch vpt add M
  0 vpt2 neg V  hpt2 0 V  0 vpt2 V
  hpt2 neg 0 V  closepath fill } def
/TriUF { stroke [] 0 setdash vpt 1.12 mul add M
  hpt neg vpt -1.62 mul V
  hpt 2 mul 0 V
  hpt neg vpt 1.62 mul V closepath fill } def
/TriD { stroke [] 0 setdash 2 copy vpt 1.12 mul sub M
  hpt neg vpt 1.62 mul V
  hpt 2 mul 0 V
  hpt neg vpt -1.62 mul V closepath stroke
  Pnt  } def
/TriDF { stroke [] 0 setdash vpt 1.12 mul sub M
  hpt neg vpt 1.62 mul V
  hpt 2 mul 0 V
  hpt neg vpt -1.62 mul V closepath fill} def
/DiaF { stroke [] 0 setdash vpt add M
  hpt neg vpt neg V hpt vpt neg V
  hpt vpt V hpt neg vpt V closepath fill } def
/Pent { stroke [] 0 setdash 2 copy gsave
  translate 0 hpt M 4 {72 rotate 0 hpt L} repeat
  closepath stroke grestore Pnt } def
/PentF { stroke [] 0 setdash gsave
  translate 0 hpt M 4 {72 rotate 0 hpt L} repeat
  closepath fill grestore } def
/Circle { stroke [] 0 setdash 2 copy
  hpt 0 360 arc stroke Pnt } def
/CircleF { stroke [] 0 setdash hpt 0 360 arc fill } def
/C0 { BL [] 0 setdash 2 copy moveto vpt 90 450  arc } bind def
/C1 { BL [] 0 setdash 2 copy        moveto
       2 copy  vpt 0 90 arc closepath fill
               vpt 0 360 arc closepath } bind def
/C2 { BL [] 0 setdash 2 copy moveto
       2 copy  vpt 90 180 arc closepath fill
               vpt 0 360 arc closepath } bind def
/C3 { BL [] 0 setdash 2 copy moveto
       2 copy  vpt 0 180 arc closepath fill
               vpt 0 360 arc closepath } bind def
/C4 { BL [] 0 setdash 2 copy moveto
       2 copy  vpt 180 270 arc closepath fill
               vpt 0 360 arc closepath } bind def
/C5 { BL [] 0 setdash 2 copy moveto
       2 copy  vpt 0 90 arc
       2 copy moveto
       2 copy  vpt 180 270 arc closepath fill
               vpt 0 360 arc } bind def
/C6 { BL [] 0 setdash 2 copy moveto
      2 copy  vpt 90 270 arc closepath fill
              vpt 0 360 arc closepath } bind def
/C7 { BL [] 0 setdash 2 copy moveto
      2 copy  vpt 0 270 arc closepath fill
              vpt 0 360 arc closepath } bind def
/C8 { BL [] 0 setdash 2 copy moveto
      2 copy vpt 270 360 arc closepath fill
              vpt 0 360 arc closepath } bind def
/C9 { BL [] 0 setdash 2 copy moveto
      2 copy  vpt 270 450 arc closepath fill
              vpt 0 360 arc closepath } bind def
/C10 { BL [] 0 setdash 2 copy 2 copy moveto vpt 270 360 arc closepath fill
       2 copy moveto
       2 copy vpt 90 180 arc closepath fill
               vpt 0 360 arc closepath } bind def
/C11 { BL [] 0 setdash 2 copy moveto
       2 copy  vpt 0 180 arc closepath fill
       2 copy moveto
       2 copy  vpt 270 360 arc closepath fill
               vpt 0 360 arc closepath } bind def
/C12 { BL [] 0 setdash 2 copy moveto
       2 copy  vpt 180 360 arc closepath fill
               vpt 0 360 arc closepath } bind def
/C13 { BL [] 0 setdash  2 copy moveto
       2 copy  vpt 0 90 arc closepath fill
       2 copy moveto
       2 copy  vpt 180 360 arc closepath fill
               vpt 0 360 arc closepath } bind def
/C14 { BL [] 0 setdash 2 copy moveto
       2 copy  vpt 90 360 arc closepath fill
               vpt 0 360 arc } bind def
/C15 { BL [] 0 setdash 2 copy vpt 0 360 arc closepath fill
               vpt 0 360 arc closepath } bind def
/Rec   { newpath 4 2 roll moveto 1 index 0 rlineto 0 exch rlineto
       neg 0 rlineto closepath } bind def
/Square { dup Rec } bind def
/Bsquare { vpt sub exch vpt sub exch vpt2 Square } bind def
/S0 { BL [] 0 setdash 2 copy moveto 0 vpt rlineto BL Bsquare } bind def
/S1 { BL [] 0 setdash 2 copy vpt Square fill Bsquare } bind def
/S2 { BL [] 0 setdash 2 copy exch vpt sub exch vpt Square fill Bsquare } bind def
/S3 { BL [] 0 setdash 2 copy exch vpt sub exch vpt2 vpt Rec fill Bsquare } bind def
/S4 { BL [] 0 setdash 2 copy exch vpt sub exch vpt sub vpt Square fill Bsquare } bind def
/S5 { BL [] 0 setdash 2 copy 2 copy vpt Square fill
       exch vpt sub exch vpt sub vpt Square fill Bsquare } bind def
/S6 { BL [] 0 setdash 2 copy exch vpt sub exch vpt sub vpt vpt2 Rec fill Bsquare } bind def
/S7 { BL [] 0 setdash 2 copy exch vpt sub exch vpt sub vpt vpt2 Rec fill
       2 copy vpt Square fill
       Bsquare } bind def
/S8 { BL [] 0 setdash 2 copy vpt sub vpt Square fill Bsquare } bind def
/S9 { BL [] 0 setdash 2 copy vpt sub vpt vpt2 Rec fill Bsquare } bind def
/S10 { BL [] 0 setdash 2 copy vpt sub vpt Square fill 2 copy exch vpt sub exch vpt Square fill
       Bsquare } bind def
/S11 { BL [] 0 setdash 2 copy vpt sub vpt Square fill 2 copy exch vpt sub exch vpt2 vpt Rec fill
       Bsquare } bind def
/S12 { BL [] 0 setdash 2 copy exch vpt sub exch vpt sub vpt2 vpt Rec fill Bsquare } bind def
/S13 { BL [] 0 setdash 2 copy exch vpt sub exch vpt sub vpt2 vpt Rec fill
       2 copy vpt Square fill Bsquare } bind def
/S14 { BL [] 0 setdash 2 copy exch vpt sub exch vpt sub vpt2 vpt Rec fill
       2 copy exch vpt sub exch vpt Square fill Bsquare } bind def
/S15 { BL [] 0 setdash 2 copy Bsquare fill Bsquare } bind def
/D0 { gsave translate 45 rotate 0 0 S0 stroke grestore } bind def
/D1 { gsave translate 45 rotate 0 0 S1 stroke grestore } bind def
/D2 { gsave translate 45 rotate 0 0 S2 stroke grestore } bind def
/D3 { gsave translate 45 rotate 0 0 S3 stroke grestore } bind def
/D4 { gsave translate 45 rotate 0 0 S4 stroke grestore } bind def
/D5 { gsave translate 45 rotate 0 0 S5 stroke grestore } bind def
/D6 { gsave translate 45 rotate 0 0 S6 stroke grestore } bind def
/D7 { gsave translate 45 rotate 0 0 S7 stroke grestore } bind def
/D8 { gsave translate 45 rotate 0 0 S8 stroke grestore } bind def
/D9 { gsave translate 45 rotate 0 0 S9 stroke grestore } bind def
/D10 { gsave translate 45 rotate 0 0 S10 stroke grestore } bind def
/D11 { gsave translate 45 rotate 0 0 S11 stroke grestore } bind def
/D12 { gsave translate 45 rotate 0 0 S12 stroke grestore } bind def
/D13 { gsave translate 45 rotate 0 0 S13 stroke grestore } bind def
/D14 { gsave translate 45 rotate 0 0 S14 stroke grestore } bind def
/D15 { gsave translate 45 rotate 0 0 S15 stroke grestore } bind def
/DiaE { stroke [] 0 setdash vpt add M
  hpt neg vpt neg V hpt vpt neg V
  hpt vpt V hpt neg vpt V closepath stroke } def
/BoxE { stroke [] 0 setdash exch hpt sub exch vpt add M
  0 vpt2 neg V hpt2 0 V 0 vpt2 V
  hpt2 neg 0 V closepath stroke } def
/TriUE { stroke [] 0 setdash vpt 1.12 mul add M
  hpt neg vpt -1.62 mul V
  hpt 2 mul 0 V
  hpt neg vpt 1.62 mul V closepath stroke } def
/TriDE { stroke [] 0 setdash vpt 1.12 mul sub M
  hpt neg vpt 1.62 mul V
  hpt 2 mul 0 V
  hpt neg vpt -1.62 mul V closepath stroke } def
/PentE { stroke [] 0 setdash gsave
  translate 0 hpt M 4 {72 rotate 0 hpt L} repeat
  closepath stroke grestore } def
/CircE { stroke [] 0 setdash 
  hpt 0 360 arc stroke } def
/Opaque { gsave closepath 1 setgray fill grestore 0 setgray closepath } def
/DiaW { stroke [] 0 setdash vpt add M
  hpt neg vpt neg V hpt vpt neg V
  hpt vpt V hpt neg vpt V Opaque stroke } def
/BoxW { stroke [] 0 setdash exch hpt sub exch vpt add M
  0 vpt2 neg V hpt2 0 V 0 vpt2 V
  hpt2 neg 0 V Opaque stroke } def
/TriUW { stroke [] 0 setdash vpt 1.12 mul add M
  hpt neg vpt -1.62 mul V
  hpt 2 mul 0 V
  hpt neg vpt 1.62 mul V Opaque stroke } def
/TriDW { stroke [] 0 setdash vpt 1.12 mul sub M
  hpt neg vpt 1.62 mul V
  hpt 2 mul 0 V
  hpt neg vpt -1.62 mul V Opaque stroke } def
/PentW { stroke [] 0 setdash gsave
  translate 0 hpt M 4 {72 rotate 0 hpt L} repeat
  Opaque stroke grestore } def
/CircW { stroke [] 0 setdash 
  hpt 0 360 arc Opaque stroke } def
/BoxFill { gsave Rec 1 setgray fill grestore } def
end
}}%
\begin{picture}(3600,2160)(0,0)%
{\GNUPLOTspecial{"
gnudict begin
gsave
0 0 translate
0.100 0.100 scale
0 setgray
newpath
1.000 UL
LTb
450 300 M
63 0 V
2937 0 R
-63 0 V
450 520 M
63 0 V
2937 0 R
-63 0 V
450 740 M
63 0 V
2937 0 R
-63 0 V
450 960 M
63 0 V
2937 0 R
-63 0 V
450 1180 M
63 0 V
2937 0 R
-63 0 V
450 1400 M
63 0 V
2937 0 R
-63 0 V
450 1620 M
63 0 V
2937 0 R
-63 0 V
450 1840 M
63 0 V
2937 0 R
-63 0 V
450 2060 M
63 0 V
2937 0 R
-63 0 V
450 300 M
0 63 V
0 1697 R
0 -63 V
825 300 M
0 63 V
0 1697 R
0 -63 V
1200 300 M
0 63 V
0 1697 R
0 -63 V
1575 300 M
0 63 V
0 1697 R
0 -63 V
1950 300 M
0 63 V
0 1697 R
0 -63 V
2325 300 M
0 63 V
0 1697 R
0 -63 V
2700 300 M
0 63 V
0 1697 R
0 -63 V
3075 300 M
0 63 V
0 1697 R
0 -63 V
3450 300 M
0 63 V
0 1697 R
0 -63 V
1.000 UL
LTb
450 300 M
3000 0 V
0 1760 V
-3000 0 V
450 300 L
0.600 UP
1.000 UL
LT0
450 1623 M
209 -21 V
520 -448 V
1440 864 L
1570 759 L
130 -34 V
260 96 V
261 129 V
521 172 V
521 94 V
187 18 V
659 1602 Pls
1179 1154 Pls
1440 864 Pls
1570 759 Pls
1700 725 Pls
1960 821 Pls
2221 950 Pls
2742 1122 Pls
3263 1216 Pls
1.000 UL
LT1
450 1475 M
209 37 V
520 50 V
261 -217 V
1700 841 L
1830 674 L
131 35 V
260 205 V
521 198 V
521 94 V
187 21 V
1.000 UL
LT2
1179 1456 M
261 52 V
260 -273 V
1830 706 L
65 -92 V
66 14 V
260 300 V
260 95 V
1.000 UL
LT3
1440 1666 M
260 -147 V
65 -495 V
65 -344 V
33 99 V
32 -233 V
66 53 V
260 183 V
0.600 UP
1.000 UL
LT4
659 1497 M
0 31 V
-31 -31 R
62 0 V
-62 31 R
62 0 V
489 21 R
0 26 V
-31 -26 R
62 0 V
-62 26 R
62 0 V
230 -233 R
0 6 V
-31 -6 R
62 0 V
-62 6 R
62 0 V
1700 838 M
0 7 V
-31 -7 R
62 0 V
-62 7 R
62 0 V
99 -174 R
0 6 V
-31 -6 R
62 0 V
-62 6 R
62 0 V
100 31 R
0 2 V
-31 -2 R
62 0 V
-62 2 R
62 0 V
229 202 R
0 4 V
-31 -4 R
62 0 V
-62 4 R
62 0 V
490 188 R
0 16 V
-31 -16 R
62 0 V
-62 16 R
62 0 V
490 83 R
0 7 V
-31 -7 R
62 0 V
-62 7 R
62 0 V
659 1512 Crs
1179 1562 Crs
1440 1345 Crs
1700 841 Crs
1830 674 Crs
1961 709 Crs
2221 914 Crs
2742 1112 Crs
3263 1206 Crs
0.600 UP
1.000 UL
LT5
1179 1436 M
0 40 V
-31 -40 R
62 0 V
-62 40 R
62 0 V
230 4 R
0 55 V
-31 -55 R
62 0 V
-62 55 R
62 0 V
229 -351 R
0 102 V
-31 -102 R
62 0 V
-62 102 R
62 0 V
99 -598 R
0 36 V
-31 -36 R
62 0 V
-62 36 R
62 0 V
34 -160 R
0 99 V
-31 -99 R
62 0 V
-62 99 R
62 0 V
35 -64 R
0 57 V
-31 -57 R
62 0 V
-62 57 R
62 0 V
229 243 R
0 58 V
-31 -58 R
62 0 V
-62 58 R
62 0 V
229 53 R
0 26 V
-31 -26 R
62 0 V
-62 26 R
62 0 V
1179 1456 Star
1440 1508 Star
1700 1235 Star
1830 706 Star
1895 614 Star
1961 628 Star
2221 928 Star
2481 1023 Star
0.600 UP
1.000 UL
LT6
1440 1490 M
0 352 V
-31 -352 R
62 0 V
-62 352 R
62 0 V
229 -475 R
0 304 V
-31 -304 R
62 0 V
-62 304 R
62 0 V
34 -741 R
0 187 V
1734 930 M
62 0 V
-62 187 R
62 0 V
34 -596 R
0 317 V
1799 521 M
62 0 V
-62 317 R
62 0 V
2 -149 R
0 179 V
1832 689 M
62 0 V
-62 179 R
62 0 V
1 -454 R
0 264 V
1864 414 M
62 0 V
-62 264 R
62 0 V
35 -201 R
0 244 V
1930 477 M
62 0 V
-62 244 R
62 0 V
229 -32 R
0 185 V
2190 689 M
62 0 V
-62 185 R
62 0 V
1440 1666 Box
1700 1519 Box
1765 1024 Box
1830 680 Box
1863 779 Box
1895 546 Box
1961 599 Box
2221 782 Box
stroke
grestore
end
showpage
}}%
\put(1950,50){\makebox(0,0){$\gamma$}}%
\put(100,1180){%
\makebox(0,0)[b]{\shortstack{$P_3$}}%
}%
\put(3450,200){\makebox(0,0){0.65}}%
\put(3075,200){\makebox(0,0){0.6}}%
\put(2700,200){\makebox(0,0){0.55}}%
\put(2325,200){\makebox(0,0){0.5}}%
\put(1950,200){\makebox(0,0){0.45}}%
\put(1575,200){\makebox(0,0){0.4}}%
\put(1200,200){\makebox(0,0){0.35}}%
\put(825,200){\makebox(0,0){0.3}}%
\put(450,200){\makebox(0,0){0.25}}%
\put(400,2060){\makebox(0,0)[r]{0.6}}%
\put(400,1840){\makebox(0,0)[r]{0.4}}%
\put(400,1620){\makebox(0,0)[r]{0.2}}%
\put(400,1400){\makebox(0,0)[r]{0}}%
\put(400,1180){\makebox(0,0)[r]{-0.2}}%
\put(400,960){\makebox(0,0)[r]{-0.4}}%
\put(400,740){\makebox(0,0)[r]{-0.6}}%
\put(400,520){\makebox(0,0)[r]{-0.8}}%
\put(400,300){\makebox(0,0)[r]{-1}}%
\end{picture}%
\endgroup

%% file: 1+1d_C_3graph.tex
\begingroup%
  \makeatletter%
  \newcommand{\GNUPLOTspecial}{%
    \@sanitize\catcode`\%=14\relax\special}%
  \setlength{\unitlength}{0.1bp}%
{\GNUPLOTspecial{!
/gnudict 256 dict def
gnudict begin
/Color false def
/Solid false def
/gnulinewidth 5.000 def
/userlinewidth gnulinewidth def
/vshift -33 def
/dl {10 mul} def
/hpt_ 31.5 def
/vpt_ 31.5 def
/hpt hpt_ def
/vpt vpt_ def
/M {moveto} bind def
/L {lineto} bind def
/R {rmoveto} bind def
/V {rlineto} bind def
/vpt2 vpt 2 mul def
/hpt2 hpt 2 mul def
/Lshow { currentpoint stroke M
  0 vshift R show } def
/Rshow { currentpoint stroke M
  dup stringwidth pop neg vshift R show } def
/Cshow { currentpoint stroke M
  dup stringwidth pop -2 div vshift R show } def
/UP { dup vpt_ mul /vpt exch def hpt_ mul /hpt exch def
  /hpt2 hpt 2 mul def /vpt2 vpt 2 mul def } def
/DL { Color {setrgbcolor Solid {pop []} if 0 setdash }
 {pop pop pop Solid {pop []} if 0 setdash} ifelse } def
/BL { stroke userlinewidth 2 mul setlinewidth } def
/AL { stroke userlinewidth 2 div setlinewidth } def
/UL { dup gnulinewidth mul /userlinewidth exch def
      10 mul /udl exch def } def
/PL { stroke userlinewidth setlinewidth } def
/LTb { BL [] 0 0 0 DL } def
/LTa { AL [1 udl mul 2 udl mul] 0 setdash 0 0 0 setrgbcolor } def
/LT0 { PL [] 1 0 0 DL } def
/LT1 { PL [4 dl 2 dl] 0 1 0 DL } def
/LT2 { PL [2 dl 3 dl] 0 0 1 DL } def
/LT3 { PL [1 dl 1.5 dl] 1 0 1 DL } def
/LT4 { PL [5 dl 2 dl 1 dl 2 dl] 0 1 1 DL } def
/LT5 { PL [4 dl 3 dl 1 dl 3 dl] 1 1 0 DL } def
/LT6 { PL [2 dl 2 dl 2 dl 4 dl] 0 0 0 DL } def
/LT7 { PL [2 dl 2 dl 2 dl 2 dl 2 dl 4 dl] 1 0.3 0 DL } def
/LT8 { PL [2 dl 2 dl 2 dl 2 dl 2 dl 2 dl 2 dl 4 dl] 0.5 0.5 0.5 DL } def
/Pnt { stroke [] 0 setdash
   gsave 1 setlinecap M 0 0 V stroke grestore } def
/Dia { stroke [] 0 setdash 2 copy vpt add M
  hpt neg vpt neg V hpt vpt neg V
  hpt vpt V hpt neg vpt V closepath stroke
  Pnt } def
/Pls { stroke [] 0 setdash vpt sub M 0 vpt2 V
  currentpoint stroke M
  hpt neg vpt neg R hpt2 0 V stroke
  } def
/Box { stroke [] 0 setdash 2 copy exch hpt sub exch vpt add M
  0 vpt2 neg V hpt2 0 V 0 vpt2 V
  hpt2 neg 0 V closepath stroke
  Pnt } def
/Crs { stroke [] 0 setdash exch hpt sub exch vpt add M
  hpt2 vpt2 neg V currentpoint stroke M
  hpt2 neg 0 R hpt2 vpt2 V stroke } def
/TriU { stroke [] 0 setdash 2 copy vpt 1.12 mul add M
  hpt neg vpt -1.62 mul V
  hpt 2 mul 0 V
  hpt neg vpt 1.62 mul V closepath stroke
  Pnt  } def
/Star { 2 copy Pls Crs } def
/BoxF { stroke [] 0 setdash exch hpt sub exch vpt add M
  0 vpt2 neg V  hpt2 0 V  0 vpt2 V
  hpt2 neg 0 V  closepath fill } def
/TriUF { stroke [] 0 setdash vpt 1.12 mul add M
  hpt neg vpt -1.62 mul V
  hpt 2 mul 0 V
  hpt neg vpt 1.62 mul V closepath fill } def
/TriD { stroke [] 0 setdash 2 copy vpt 1.12 mul sub M
  hpt neg vpt 1.62 mul V
  hpt 2 mul 0 V
  hpt neg vpt -1.62 mul V closepath stroke
  Pnt  } def
/TriDF { stroke [] 0 setdash vpt 1.12 mul sub M
  hpt neg vpt 1.62 mul V
  hpt 2 mul 0 V
  hpt neg vpt -1.62 mul V closepath fill} def
/DiaF { stroke [] 0 setdash vpt add M
  hpt neg vpt neg V hpt vpt neg V
  hpt vpt V hpt neg vpt V closepath fill } def
/Pent { stroke [] 0 setdash 2 copy gsave
  translate 0 hpt M 4 {72 rotate 0 hpt L} repeat
  closepath stroke grestore Pnt } def
/PentF { stroke [] 0 setdash gsave
  translate 0 hpt M 4 {72 rotate 0 hpt L} repeat
  closepath fill grestore } def
/Circle { stroke [] 0 setdash 2 copy
  hpt 0 360 arc stroke Pnt } def
/CircleF { stroke [] 0 setdash hpt 0 360 arc fill } def
/C0 { BL [] 0 setdash 2 copy moveto vpt 90 450  arc } bind def
/C1 { BL [] 0 setdash 2 copy        moveto
       2 copy  vpt 0 90 arc closepath fill
               vpt 0 360 arc closepath } bind def
/C2 { BL [] 0 setdash 2 copy moveto
       2 copy  vpt 90 180 arc closepath fill
               vpt 0 360 arc closepath } bind def
/C3 { BL [] 0 setdash 2 copy moveto
       2 copy  vpt 0 180 arc closepath fill
               vpt 0 360 arc closepath } bind def
/C4 { BL [] 0 setdash 2 copy moveto
       2 copy  vpt 180 270 arc closepath fill
               vpt 0 360 arc closepath } bind def
/C5 { BL [] 0 setdash 2 copy moveto
       2 copy  vpt 0 90 arc
       2 copy moveto
       2 copy  vpt 180 270 arc closepath fill
               vpt 0 360 arc } bind def
/C6 { BL [] 0 setdash 2 copy moveto
      2 copy  vpt 90 270 arc closepath fill
              vpt 0 360 arc closepath } bind def
/C7 { BL [] 0 setdash 2 copy moveto
      2 copy  vpt 0 270 arc closepath fill
              vpt 0 360 arc closepath } bind def
/C8 { BL [] 0 setdash 2 copy moveto
      2 copy vpt 270 360 arc closepath fill
              vpt 0 360 arc closepath } bind def
/C9 { BL [] 0 setdash 2 copy moveto
      2 copy  vpt 270 450 arc closepath fill
              vpt 0 360 arc closepath } bind def
/C10 { BL [] 0 setdash 2 copy 2 copy moveto vpt 270 360 arc closepath fill
       2 copy moveto
       2 copy vpt 90 180 arc closepath fill
               vpt 0 360 arc closepath } bind def
/C11 { BL [] 0 setdash 2 copy moveto
       2 copy  vpt 0 180 arc closepath fill
       2 copy moveto
       2 copy  vpt 270 360 arc closepath fill
               vpt 0 360 arc closepath } bind def
/C12 { BL [] 0 setdash 2 copy moveto
       2 copy  vpt 180 360 arc closepath fill
               vpt 0 360 arc closepath } bind def
/C13 { BL [] 0 setdash  2 copy moveto
       2 copy  vpt 0 90 arc closepath fill
       2 copy moveto
       2 copy  vpt 180 360 arc closepath fill
               vpt 0 360 arc closepath } bind def
/C14 { BL [] 0 setdash 2 copy moveto
       2 copy  vpt 90 360 arc closepath fill
               vpt 0 360 arc } bind def
/C15 { BL [] 0 setdash 2 copy vpt 0 360 arc closepath fill
               vpt 0 360 arc closepath } bind def
/Rec   { newpath 4 2 roll moveto 1 index 0 rlineto 0 exch rlineto
       neg 0 rlineto closepath } bind def
/Square { dup Rec } bind def
/Bsquare { vpt sub exch vpt sub exch vpt2 Square } bind def
/S0 { BL [] 0 setdash 2 copy moveto 0 vpt rlineto BL Bsquare } bind def
/S1 { BL [] 0 setdash 2 copy vpt Square fill Bsquare } bind def
/S2 { BL [] 0 setdash 2 copy exch vpt sub exch vpt Square fill Bsquare } bind def
/S3 { BL [] 0 setdash 2 copy exch vpt sub exch vpt2 vpt Rec fill Bsquare } bind def
/S4 { BL [] 0 setdash 2 copy exch vpt sub exch vpt sub vpt Square fill Bsquare } bind def
/S5 { BL [] 0 setdash 2 copy 2 copy vpt Square fill
       exch vpt sub exch vpt sub vpt Square fill Bsquare } bind def
/S6 { BL [] 0 setdash 2 copy exch vpt sub exch vpt sub vpt vpt2 Rec fill Bsquare } bind def
/S7 { BL [] 0 setdash 2 copy exch vpt sub exch vpt sub vpt vpt2 Rec fill
       2 copy vpt Square fill
       Bsquare } bind def
/S8 { BL [] 0 setdash 2 copy vpt sub vpt Square fill Bsquare } bind def
/S9 { BL [] 0 setdash 2 copy vpt sub vpt vpt2 Rec fill Bsquare } bind def
/S10 { BL [] 0 setdash 2 copy vpt sub vpt Square fill 2 copy exch vpt sub exch vpt Square fill
       Bsquare } bind def
/S11 { BL [] 0 setdash 2 copy vpt sub vpt Square fill 2 copy exch vpt sub exch vpt2 vpt Rec fill
       Bsquare } bind def
/S12 { BL [] 0 setdash 2 copy exch vpt sub exch vpt sub vpt2 vpt Rec fill Bsquare } bind def
/S13 { BL [] 0 setdash 2 copy exch vpt sub exch vpt sub vpt2 vpt Rec fill
       2 copy vpt Square fill Bsquare } bind def
/S14 { BL [] 0 setdash 2 copy exch vpt sub exch vpt sub vpt2 vpt Rec fill
       2 copy exch vpt sub exch vpt Square fill Bsquare } bind def
/S15 { BL [] 0 setdash 2 copy Bsquare fill Bsquare } bind def
/D0 { gsave translate 45 rotate 0 0 S0 stroke grestore } bind def
/D1 { gsave translate 45 rotate 0 0 S1 stroke grestore } bind def
/D2 { gsave translate 45 rotate 0 0 S2 stroke grestore } bind def
/D3 { gsave translate 45 rotate 0 0 S3 stroke grestore } bind def
/D4 { gsave translate 45 rotate 0 0 S4 stroke grestore } bind def
/D5 { gsave translate 45 rotate 0 0 S5 stroke grestore } bind def
/D6 { gsave translate 45 rotate 0 0 S6 stroke grestore } bind def
/D7 { gsave translate 45 rotate 0 0 S7 stroke grestore } bind def
/D8 { gsave translate 45 rotate 0 0 S8 stroke grestore } bind def
/D9 { gsave translate 45 rotate 0 0 S9 stroke grestore } bind def
/D10 { gsave translate 45 rotate 0 0 S10 stroke grestore } bind def
/D11 { gsave translate 45 rotate 0 0 S11 stroke grestore } bind def
/D12 { gsave translate 45 rotate 0 0 S12 stroke grestore } bind def
/D13 { gsave translate 45 rotate 0 0 S13 stroke grestore } bind def
/D14 { gsave translate 45 rotate 0 0 S14 stroke grestore } bind def
/D15 { gsave translate 45 rotate 0 0 S15 stroke grestore } bind def
/DiaE { stroke [] 0 setdash vpt add M
  hpt neg vpt neg V hpt vpt neg V
  hpt vpt V hpt neg vpt V closepath stroke } def
/BoxE { stroke [] 0 setdash exch hpt sub exch vpt add M
  0 vpt2 neg V hpt2 0 V 0 vpt2 V
  hpt2 neg 0 V closepath stroke } def
/TriUE { stroke [] 0 setdash vpt 1.12 mul add M
  hpt neg vpt -1.62 mul V
  hpt 2 mul 0 V
  hpt neg vpt 1.62 mul V closepath stroke } def
/TriDE { stroke [] 0 setdash vpt 1.12 mul sub M
  hpt neg vpt 1.62 mul V
  hpt 2 mul 0 V
  hpt neg vpt -1.62 mul V closepath stroke } def
/PentE { stroke [] 0 setdash gsave
  translate 0 hpt M 4 {72 rotate 0 hpt L} repeat
  closepath stroke grestore } def
/CircE { stroke [] 0 setdash 
  hpt 0 360 arc stroke } def
/Opaque { gsave closepath 1 setgray fill grestore 0 setgray closepath } def
/DiaW { stroke [] 0 setdash vpt add M
  hpt neg vpt neg V hpt vpt neg V
  hpt vpt V hpt neg vpt V Opaque stroke } def
/BoxW { stroke [] 0 setdash exch hpt sub exch vpt add M
  0 vpt2 neg V hpt2 0 V 0 vpt2 V
  hpt2 neg 0 V Opaque stroke } def
/TriUW { stroke [] 0 setdash vpt 1.12 mul add M
  hpt neg vpt -1.62 mul V
  hpt 2 mul 0 V
  hpt neg vpt 1.62 mul V Opaque stroke } def
/TriDW { stroke [] 0 setdash vpt 1.12 mul sub M
  hpt neg vpt 1.62 mul V
  hpt 2 mul 0 V
  hpt neg vpt -1.62 mul V Opaque stroke } def
/PentW { stroke [] 0 setdash gsave
  translate 0 hpt M 4 {72 rotate 0 hpt L} repeat
  Opaque stroke grestore } def
/CircW { stroke [] 0 setdash 
  hpt 0 360 arc Opaque stroke } def
/BoxFill { gsave Rec 1 setgray fill grestore } def
end
}}%
\begin{picture}(3600,2160)(0,0)%
{\GNUPLOTspecial{"
gnudict begin
gsave
0 0 translate
0.100 0.100 scale
0 setgray
newpath
1.000 UL
LTb
450 300 M
63 0 V
2937 0 R
-63 0 V
450 551 M
63 0 V
2937 0 R
-63 0 V
450 803 M
63 0 V
2937 0 R
-63 0 V
450 1054 M
63 0 V
2937 0 R
-63 0 V
450 1306 M
63 0 V
2937 0 R
-63 0 V
450 1557 M
63 0 V
2937 0 R
-63 0 V
450 1809 M
63 0 V
2937 0 R
-63 0 V
450 2060 M
63 0 V
2937 0 R
-63 0 V
450 300 M
0 63 V
0 1697 R
0 -63 V
750 300 M
0 63 V
0 1697 R
0 -63 V
1050 300 M
0 63 V
0 1697 R
0 -63 V
1350 300 M
0 63 V
0 1697 R
0 -63 V
1650 300 M
0 63 V
0 1697 R
0 -63 V
1950 300 M
0 63 V
0 1697 R
0 -63 V
2250 300 M
0 63 V
0 1697 R
0 -63 V
2550 300 M
0 63 V
0 1697 R
0 -63 V
2850 300 M
0 63 V
0 1697 R
0 -63 V
3150 300 M
0 63 V
0 1697 R
0 -63 V
3450 300 M
0 63 V
0 1697 R
0 -63 V
1.000 UL
LTb
450 300 M
3000 0 V
0 1760 V
-3000 0 V
450 300 L
1.000 UL
LT0
450 1557 M
1500 0 V
0 -1257 V
30 37 V
30 36 V
30 34 V
30 33 V
30 31 V
30 31 V
30 29 V
30 28 V
30 27 V
30 27 V
30 25 V
30 24 V
30 24 V
30 23 V
30 22 V
30 21 V
30 20 V
30 20 V
30 19 V
30 19 V
30 18 V
30 17 V
30 17 V
30 16 V
30 15 V
30 16 V
30 14 V
30 15 V
30 14 V
30 13 V
30 13 V
30 13 V
30 12 V
30 12 V
30 11 V
30 11 V
30 11 V
30 11 V
30 10 V
30 10 V
30 10 V
30 9 V
30 9 V
30 9 V
30 9 V
30 8 V
30 8 V
30 8 V
30 8 V
30 8 V
0.600 UP
1.000 UL
LT1
450 1871 M
833 -751 V
1617 836 L
166 -49 V
167 17 V
333 50 V
1167 360 V
450 1871 Pls
1283 1120 Pls
1617 836 Pls
1783 787 Pls
1950 804 Pls
2283 854 Pls
3450 1214 Pls
1.000 UL
LT2
450 1607 M
833 -41 V
334 -429 V
1783 900 L
1950 781 L
333 -28 V
1167 435 V
1.000 UL
LT3
450 1489 M
833 78 V
333 -36 V
167 -400 V
1950 844 L
2283 667 L
1167 533 V
0.600 UP
1.000 UL
LT4
450 1584 M
0 47 V
-31 -47 R
62 0 V
-62 47 R
62 0 V
802 -75 R
0 20 V
-31 -20 R
62 0 V
-62 20 R
62 0 V
303 -453 R
0 28 V
-31 -28 R
62 0 V
-62 28 R
62 0 V
1783 878 M
0 43 V
-31 -43 R
62 0 V
-62 43 R
62 0 V
1950 766 M
0 31 V
-31 -31 R
62 0 V
-62 31 R
62 0 V
302 -57 R
0 25 V
-31 -25 R
62 0 V
-62 25 R
62 0 V
1136 412 R
0 21 V
-31 -21 R
62 0 V
-62 21 R
62 0 V
450 1607 Crs
1283 1566 Crs
1617 1137 Crs
1783 900 Crs
1950 781 Crs
2283 753 Crs
3450 1188 Crs
0.600 UP
1.000 UL
LT5
450 1395 M
0 189 V
419 1395 M
62 0 V
-62 189 R
62 0 V
802 -124 R
0 214 V
-31 -214 R
62 0 V
-62 214 R
62 0 V
302 -246 R
0 206 V
-31 -206 R
62 0 V
-62 206 R
62 0 V
136 -578 R
0 150 V
-31 -150 R
62 0 V
-62 150 R
62 0 V
1950 793 M
0 103 V
1919 793 M
62 0 V
-62 103 R
62 0 V
2283 607 M
0 120 V
2252 607 M
62 0 V
-62 120 R
62 0 V
1136 440 R
0 66 V
-31 -66 R
62 0 V
-62 66 R
62 0 V
450 1489 Star
1283 1567 Star
1616 1531 Star
1783 1131 Star
1950 844 Star
2283 667 Star
3450 1200 Star
stroke
grestore
end
showpage
}}%
\put(1950,50){\makebox(0,0){$\gamma$}}%
\put(100,1180){%
\makebox(0,0)[b]{\shortstack{$C_3$}}%
}%
\put(3450,200){\makebox(0,0){0.75}}%
\put(3150,200){\makebox(0,0){0.7}}%
\put(2850,200){\makebox(0,0){0.65}}%
\put(2550,200){\makebox(0,0){0.6}}%
\put(2250,200){\makebox(0,0){0.55}}%
\put(1950,200){\makebox(0,0){0.5}}%
\put(1650,200){\makebox(0,0){0.45}}%
\put(1350,200){\makebox(0,0){0.4}}%
\put(1050,200){\makebox(0,0){0.35}}%
\put(750,200){\makebox(0,0){0.3}}%
\put(450,200){\makebox(0,0){0.25}}%
\put(400,2060){\makebox(0,0)[r]{0.4}}%
\put(400,1809){\makebox(0,0)[r]{0.2}}%
\put(400,1557){\makebox(0,0)[r]{0}}%
\put(400,1306){\makebox(0,0)[r]{-0.2}}%
\put(400,1054){\makebox(0,0)[r]{-0.4}}%
\put(400,803){\makebox(0,0)[r]{-0.6}}%
\put(400,551){\makebox(0,0)[r]{-0.8}}%
\put(400,300){\makebox(0,0)[r]{-1}}%
\end{picture}%
\endgroup

%% file: ratiograph.tex
\begingroup%
  \makeatletter%
  \newcommand{\GNUPLOTspecial}{%
    \@sanitize\catcode`\%=14\relax\special}%
  \setlength{\unitlength}{0.1bp}%
{\GNUPLOTspecial{!
/gnudict 256 dict def
gnudict begin
/Color false def
/Solid false def
/gnulinewidth 5.000 def
/userlinewidth gnulinewidth def
/vshift -33 def
/dl {10 mul} def
/hpt_ 31.5 def
/vpt_ 31.5 def
/hpt hpt_ def
/vpt vpt_ def
/M {moveto} bind def
/L {lineto} bind def
/R {rmoveto} bind def
/V {rlineto} bind def
/vpt2 vpt 2 mul def
/hpt2 hpt 2 mul def
/Lshow { currentpoint stroke M
  0 vshift R show } def
/Rshow { currentpoint stroke M
  dup stringwidth pop neg vshift R show } def
/Cshow { currentpoint stroke M
  dup stringwidth pop -2 div vshift R show } def
/UP { dup vpt_ mul /vpt exch def hpt_ mul /hpt exch def
  /hpt2 hpt 2 mul def /vpt2 vpt 2 mul def } def
/DL { Color {setrgbcolor Solid {pop []} if 0 setdash }
 {pop pop pop Solid {pop []} if 0 setdash} ifelse } def
/BL { stroke userlinewidth 2 mul setlinewidth } def
/AL { stroke userlinewidth 2 div setlinewidth } def
/UL { dup gnulinewidth mul /userlinewidth exch def
      10 mul /udl exch def } def
/PL { stroke userlinewidth setlinewidth } def
/LTb { BL [] 0 0 0 DL } def
/LTa { AL [1 udl mul 2 udl mul] 0 setdash 0 0 0 setrgbcolor } def
/LT0 { PL [] 1 0 0 DL } def
/LT1 { PL [4 dl 2 dl] 0 1 0 DL } def
/LT2 { PL [2 dl 3 dl] 0 0 1 DL } def
/LT3 { PL [1 dl 1.5 dl] 1 0 1 DL } def
/LT4 { PL [5 dl 2 dl 1 dl 2 dl] 0 1 1 DL } def
/LT5 { PL [4 dl 3 dl 1 dl 3 dl] 1 1 0 DL } def
/LT6 { PL [2 dl 2 dl 2 dl 4 dl] 0 0 0 DL } def
/LT7 { PL [2 dl 2 dl 2 dl 2 dl 2 dl 4 dl] 1 0.3 0 DL } def
/LT8 { PL [2 dl 2 dl 2 dl 2 dl 2 dl 2 dl 2 dl 4 dl] 0.5 0.5 0.5 DL } def
/Pnt { stroke [] 0 setdash
   gsave 1 setlinecap M 0 0 V stroke grestore } def
/Dia { stroke [] 0 setdash 2 copy vpt add M
  hpt neg vpt neg V hpt vpt neg V
  hpt vpt V hpt neg vpt V closepath stroke
  Pnt } def
/Pls { stroke [] 0 setdash vpt sub M 0 vpt2 V
  currentpoint stroke M
  hpt neg vpt neg R hpt2 0 V stroke
  } def
/Box { stroke [] 0 setdash 2 copy exch hpt sub exch vpt add M
  0 vpt2 neg V hpt2 0 V 0 vpt2 V
  hpt2 neg 0 V closepath stroke
  Pnt } def
/Crs { stroke [] 0 setdash exch hpt sub exch vpt add M
  hpt2 vpt2 neg V currentpoint stroke M
  hpt2 neg 0 R hpt2 vpt2 V stroke } def
/TriU { stroke [] 0 setdash 2 copy vpt 1.12 mul add M
  hpt neg vpt -1.62 mul V
  hpt 2 mul 0 V
  hpt neg vpt 1.62 mul V closepath stroke
  Pnt  } def
/Star { 2 copy Pls Crs } def
/BoxF { stroke [] 0 setdash exch hpt sub exch vpt add M
  0 vpt2 neg V  hpt2 0 V  0 vpt2 V
  hpt2 neg 0 V  closepath fill } def
/TriUF { stroke [] 0 setdash vpt 1.12 mul add M
  hpt neg vpt -1.62 mul V
  hpt 2 mul 0 V
  hpt neg vpt 1.62 mul V closepath fill } def
/TriD { stroke [] 0 setdash 2 copy vpt 1.12 mul sub M
  hpt neg vpt 1.62 mul V
  hpt 2 mul 0 V
  hpt neg vpt -1.62 mul V closepath stroke
  Pnt  } def
/TriDF { stroke [] 0 setdash vpt 1.12 mul sub M
  hpt neg vpt 1.62 mul V
  hpt 2 mul 0 V
  hpt neg vpt -1.62 mul V closepath fill} def
/DiaF { stroke [] 0 setdash vpt add M
  hpt neg vpt neg V hpt vpt neg V
  hpt vpt V hpt neg vpt V closepath fill } def
/Pent { stroke [] 0 setdash 2 copy gsave
  translate 0 hpt M 4 {72 rotate 0 hpt L} repeat
  closepath stroke grestore Pnt } def
/PentF { stroke [] 0 setdash gsave
  translate 0 hpt M 4 {72 rotate 0 hpt L} repeat
  closepath fill grestore } def
/Circle { stroke [] 0 setdash 2 copy
  hpt 0 360 arc stroke Pnt } def
/CircleF { stroke [] 0 setdash hpt 0 360 arc fill } def
/C0 { BL [] 0 setdash 2 copy moveto vpt 90 450  arc } bind def
/C1 { BL [] 0 setdash 2 copy        moveto
       2 copy  vpt 0 90 arc closepath fill
               vpt 0 360 arc closepath } bind def
/C2 { BL [] 0 setdash 2 copy moveto
       2 copy  vpt 90 180 arc closepath fill
               vpt 0 360 arc closepath } bind def
/C3 { BL [] 0 setdash 2 copy moveto
       2 copy  vpt 0 180 arc closepath fill
               vpt 0 360 arc closepath } bind def
/C4 { BL [] 0 setdash 2 copy moveto
       2 copy  vpt 180 270 arc closepath fill
               vpt 0 360 arc closepath } bind def
/C5 { BL [] 0 setdash 2 copy moveto
       2 copy  vpt 0 90 arc
       2 copy moveto
       2 copy  vpt 180 270 arc closepath fill
               vpt 0 360 arc } bind def
/C6 { BL [] 0 setdash 2 copy moveto
      2 copy  vpt 90 270 arc closepath fill
              vpt 0 360 arc closepath } bind def
/C7 { BL [] 0 setdash 2 copy moveto
      2 copy  vpt 0 270 arc closepath fill
              vpt 0 360 arc closepath } bind def
/C8 { BL [] 0 setdash 2 copy moveto
      2 copy vpt 270 360 arc closepath fill
              vpt 0 360 arc closepath } bind def
/C9 { BL [] 0 setdash 2 copy moveto
      2 copy  vpt 270 450 arc closepath fill
              vpt 0 360 arc closepath } bind def
/C10 { BL [] 0 setdash 2 copy 2 copy moveto vpt 270 360 arc closepath fill
       2 copy moveto
       2 copy vpt 90 180 arc closepath fill
               vpt 0 360 arc closepath } bind def
/C11 { BL [] 0 setdash 2 copy moveto
       2 copy  vpt 0 180 arc closepath fill
       2 copy moveto
       2 copy  vpt 270 360 arc closepath fill
               vpt 0 360 arc closepath } bind def
/C12 { BL [] 0 setdash 2 copy moveto
       2 copy  vpt 180 360 arc closepath fill
               vpt 0 360 arc closepath } bind def
/C13 { BL [] 0 setdash  2 copy moveto
       2 copy  vpt 0 90 arc closepath fill
       2 copy moveto
       2 copy  vpt 180 360 arc closepath fill
               vpt 0 360 arc closepath } bind def
/C14 { BL [] 0 setdash 2 copy moveto
       2 copy  vpt 90 360 arc closepath fill
               vpt 0 360 arc } bind def
/C15 { BL [] 0 setdash 2 copy vpt 0 360 arc closepath fill
               vpt 0 360 arc closepath } bind def
/Rec   { newpath 4 2 roll moveto 1 index 0 rlineto 0 exch rlineto
       neg 0 rlineto closepath } bind def
/Square { dup Rec } bind def
/Bsquare { vpt sub exch vpt sub exch vpt2 Square } bind def
/S0 { BL [] 0 setdash 2 copy moveto 0 vpt rlineto BL Bsquare } bind def
/S1 { BL [] 0 setdash 2 copy vpt Square fill Bsquare } bind def
/S2 { BL [] 0 setdash 2 copy exch vpt sub exch vpt Square fill Bsquare } bind def
/S3 { BL [] 0 setdash 2 copy exch vpt sub exch vpt2 vpt Rec fill Bsquare } bind def
/S4 { BL [] 0 setdash 2 copy exch vpt sub exch vpt sub vpt Square fill Bsquare } bind def
/S5 { BL [] 0 setdash 2 copy 2 copy vpt Square fill
       exch vpt sub exch vpt sub vpt Square fill Bsquare } bind def
/S6 { BL [] 0 setdash 2 copy exch vpt sub exch vpt sub vpt vpt2 Rec fill Bsquare } bind def
/S7 { BL [] 0 setdash 2 copy exch vpt sub exch vpt sub vpt vpt2 Rec fill
       2 copy vpt Square fill
       Bsquare } bind def
/S8 { BL [] 0 setdash 2 copy vpt sub vpt Square fill Bsquare } bind def
/S9 { BL [] 0 setdash 2 copy vpt sub vpt vpt2 Rec fill Bsquare } bind def
/S10 { BL [] 0 setdash 2 copy vpt sub vpt Square fill 2 copy exch vpt sub exch vpt Square fill
       Bsquare } bind def
/S11 { BL [] 0 setdash 2 copy vpt sub vpt Square fill 2 copy exch vpt sub exch vpt2 vpt Rec fill
       Bsquare } bind def
/S12 { BL [] 0 setdash 2 copy exch vpt sub exch vpt sub vpt2 vpt Rec fill Bsquare } bind def
/S13 { BL [] 0 setdash 2 copy exch vpt sub exch vpt sub vpt2 vpt Rec fill
       2 copy vpt Square fill Bsquare } bind def
/S14 { BL [] 0 setdash 2 copy exch vpt sub exch vpt sub vpt2 vpt Rec fill
       2 copy exch vpt sub exch vpt Square fill Bsquare } bind def
/S15 { BL [] 0 setdash 2 copy Bsquare fill Bsquare } bind def
/D0 { gsave translate 45 rotate 0 0 S0 stroke grestore } bind def
/D1 { gsave translate 45 rotate 0 0 S1 stroke grestore } bind def
/D2 { gsave translate 45 rotate 0 0 S2 stroke grestore } bind def
/D3 { gsave translate 45 rotate 0 0 S3 stroke grestore } bind def
/D4 { gsave translate 45 rotate 0 0 S4 stroke grestore } bind def
/D5 { gsave translate 45 rotate 0 0 S5 stroke grestore } bind def
/D6 { gsave translate 45 rotate 0 0 S6 stroke grestore } bind def
/D7 { gsave translate 45 rotate 0 0 S7 stroke grestore } bind def
/D8 { gsave translate 45 rotate 0 0 S8 stroke grestore } bind def
/D9 { gsave translate 45 rotate 0 0 S9 stroke grestore } bind def
/D10 { gsave translate 45 rotate 0 0 S10 stroke grestore } bind def
/D11 { gsave translate 45 rotate 0 0 S11 stroke grestore } bind def
/D12 { gsave translate 45 rotate 0 0 S12 stroke grestore } bind def
/D13 { gsave translate 45 rotate 0 0 S13 stroke grestore } bind def
/D14 { gsave translate 45 rotate 0 0 S14 stroke grestore } bind def
/D15 { gsave translate 45 rotate 0 0 S15 stroke grestore } bind def
/DiaE { stroke [] 0 setdash vpt add M
  hpt neg vpt neg V hpt vpt neg V
  hpt vpt V hpt neg vpt V closepath stroke } def
/BoxE { stroke [] 0 setdash exch hpt sub exch vpt add M
  0 vpt2 neg V hpt2 0 V 0 vpt2 V
  hpt2 neg 0 V closepath stroke } def
/TriUE { stroke [] 0 setdash vpt 1.12 mul add M
  hpt neg vpt -1.62 mul V
  hpt 2 mul 0 V
  hpt neg vpt 1.62 mul V closepath stroke } def
/TriDE { stroke [] 0 setdash vpt 1.12 mul sub M
  hpt neg vpt 1.62 mul V
  hpt 2 mul 0 V
  hpt neg vpt -1.62 mul V closepath stroke } def
/PentE { stroke [] 0 setdash gsave
  translate 0 hpt M 4 {72 rotate 0 hpt L} repeat
  closepath stroke grestore } def
/CircE { stroke [] 0 setdash 
  hpt 0 360 arc stroke } def
/Opaque { gsave closepath 1 setgray fill grestore 0 setgray closepath } def
/DiaW { stroke [] 0 setdash vpt add M
  hpt neg vpt neg V hpt vpt neg V
  hpt vpt V hpt neg vpt V Opaque stroke } def
/BoxW { stroke [] 0 setdash exch hpt sub exch vpt add M
  0 vpt2 neg V hpt2 0 V 0 vpt2 V
  hpt2 neg 0 V Opaque stroke } def
/TriUW { stroke [] 0 setdash vpt 1.12 mul add M
  hpt neg vpt -1.62 mul V
  hpt 2 mul 0 V
  hpt neg vpt 1.62 mul V Opaque stroke } def
/TriDW { stroke [] 0 setdash vpt 1.12 mul sub M
  hpt neg vpt 1.62 mul V
  hpt 2 mul 0 V
  hpt neg vpt -1.62 mul V Opaque stroke } def
/PentW { stroke [] 0 setdash gsave
  translate 0 hpt M 4 {72 rotate 0 hpt L} repeat
  Opaque stroke grestore } def
/CircW { stroke [] 0 setdash 
  hpt 0 360 arc Opaque stroke } def
/BoxFill { gsave Rec 1 setgray fill grestore } def
end
}}%
\begin{picture}(3600,2160)(0,0)%
{\GNUPLOTspecial{"
gnudict begin
gsave
0 0 translate
0.100 0.100 scale
0 setgray
newpath
1.000 UL
LTb
350 300 M
63 0 V
3037 0 R
-63 0 V
350 551 M
63 0 V
3037 0 R
-63 0 V
350 803 M
63 0 V
3037 0 R
-63 0 V
350 1054 M
63 0 V
3037 0 R
-63 0 V
350 1306 M
63 0 V
3037 0 R
-63 0 V
350 1557 M
63 0 V
3037 0 R
-63 0 V
350 1809 M
63 0 V
3037 0 R
-63 0 V
350 2060 M
63 0 V
3037 0 R
-63 0 V
350 300 M
0 63 V
0 1697 R
0 -63 V
737 300 M
0 63 V
0 1697 R
0 -63 V
1125 300 M
0 63 V
0 1697 R
0 -63 V
1512 300 M
0 63 V
0 1697 R
0 -63 V
1900 300 M
0 63 V
0 1697 R
0 -63 V
2287 300 M
0 63 V
0 1697 R
0 -63 V
2675 300 M
0 63 V
0 1697 R
0 -63 V
3063 300 M
0 63 V
0 1697 R
0 -63 V
3450 300 M
0 63 V
0 1697 R
0 -63 V
1.000 UL
LTb
350 300 M
3100 0 V
0 1760 V
-3100 0 V
350 300 L
1.000 UL
LT0
350 411 M
215 10 V
538 37 V
270 11 V
134 2 V
135 -4 V
134 -5 V
135 -6 V
269 -11 V
538 -14 V
538 -7 V
194 -1 V
1.000 UL
LT1
565 466 M
538 100 V
270 60 V
269 35 V
134 -21 V
135 -37 V
269 -52 V
538 -40 V
538 -19 V
194 -4 V
1.000 UL
LT2
565 502 M
538 194 V
270 165 V
269 199 V
134 -15 V
67 -85 V
68 -89 V
2180 712 L
269 -56 V
269 -32 V
538 -35 V
194 -4 V
1.000 UL
LT3
1373 1154 M
269 651 V
67 163 V
67 -9 V
34 -123 V
33 -200 V
68 -333 V
2180 946 L
0.600 UP
1.000 UL
LT4
565 421 M
0 1 V
-31 -1 R
62 0 V
-62 1 R
62 0 V
507 36 R
0 1 V
-31 -1 R
62 0 V
-62 1 R
62 0 V
239 9 R
0 2 V
-31 -2 R
62 0 V
-62 2 R
62 0 V
103 1 R
-31 0 R
62 0 V
-62 0 R
62 0 V
104 -4 R
0 1 V
-31 -1 R
62 0 V
-62 1 R
62 0 V
103 -7 R
0 2 V
-31 -2 R
62 0 V
-62 2 R
62 0 V
104 -8 R
0 2 V
-31 -2 R
62 0 V
-62 2 R
62 0 V
238 -13 R
0 1 V
-31 -1 R
62 0 V
-62 1 R
62 0 V
507 -15 R
0 1 V
-31 -1 R
62 0 V
-62 1 R
62 0 V
507 -7 R
0 1 V
-31 -1 R
62 0 V
-62 1 R
62 0 V
565 421 Pls
1103 458 Pls
1373 469 Pls
1507 471 Pls
1642 467 Pls
1776 462 Pls
1911 456 Pls
2180 445 Pls
2718 431 Pls
3256 424 Pls
0.600 UP
1.000 UL
LT5
565 465 M
0 1 V
-31 -1 R
62 0 V
-62 1 R
62 0 V
507 99 R
0 1 V
-31 -1 R
62 0 V
-62 1 R
62 0 V
239 59 R
0 2 V
-31 -2 R
62 0 V
-62 2 R
62 0 V
238 33 R
0 2 V
-31 -2 R
62 0 V
-62 2 R
62 0 V
103 -24 R
0 3 V
-31 -3 R
62 0 V
-62 3 R
62 0 V
104 -40 R
0 4 V
-31 -4 R
62 0 V
-62 4 R
62 0 V
238 -55 R
0 2 V
-31 -2 R
62 0 V
-62 2 R
62 0 V
507 -42 R
0 1 V
-31 -1 R
62 0 V
-62 1 R
62 0 V
507 -19 R
0 1 V
-31 -1 R
62 0 V
-62 1 R
62 0 V
565 466 Crs
1103 566 Crs
1373 626 Crs
1642 661 Crs
1776 640 Crs
1911 603 Crs
2180 551 Crs
2718 511 Crs
3256 492 Crs
0.600 UP
1.000 UL
LT6
565 501 M
0 3 V
-31 -3 R
62 0 V
-62 3 R
62 0 V
507 191 R
0 3 V
-31 -3 R
62 0 V
-62 3 R
62 0 V
239 160 R
0 6 V
-31 -6 R
62 0 V
-62 6 R
62 0 V
238 192 R
0 8 V
-31 -8 R
62 0 V
-62 8 R
62 0 V
103 -24 R
0 10 V
-31 -10 R
62 0 V
-62 10 R
62 0 V
36 -94 R
0 7 V
-31 -7 R
62 0 V
-62 7 R
62 0 V
37 -94 R
0 4 V
-31 -4 R
62 0 V
-62 4 R
62 0 V
2180 710 M
0 4 V
-31 -4 R
62 0 V
-62 4 R
62 0 V
238 -59 R
0 2 V
-31 -2 R
62 0 V
-62 2 R
62 0 V
238 -35 R
0 4 V
-31 -4 R
62 0 V
-62 4 R
62 0 V
507 -38 R
0 1 V
-31 -1 R
62 0 V
-62 1 R
62 0 V
565 502 Star
1103 696 Star
1373 861 Star
1642 1060 Star
1776 1045 Star
1843 960 Star
1911 871 Star
2180 712 Star
2449 656 Star
2718 624 Star
3256 589 Star
0.600 UP
1.000 UL
LT7
1373 1148 M
0 12 V
-31 -12 R
62 0 V
-62 12 R
62 0 V
238 638 R
0 15 V
-31 -15 R
62 0 V
-62 15 R
62 0 V
36 145 R
0 21 V
-31 -21 R
62 0 V
-62 21 R
62 0 V
36 -29 R
0 18 V
-31 -18 R
62 0 V
-62 18 R
62 0 V
3 -144 R
0 24 V
-31 -24 R
62 0 V
-62 24 R
62 0 V
2 -219 R
0 14 V
-31 -14 R
62 0 V
-62 14 R
62 0 V
37 -348 R
0 16 V
-31 -16 R
62 0 V
-62 16 R
62 0 V
2180 943 M
0 6 V
-31 -6 R
62 0 V
-62 6 R
62 0 V
1373 1154 Box
1642 1805 Box
1709 1968 Box
1776 1959 Box
1810 1836 Box
1843 1636 Box
1911 1303 Box
2180 946 Box
stroke
grestore
end
showpage
}}%
\put(1900,50){\makebox(0,0){$\gamma$}}%
\put(100,1180){%
\makebox(0,0)[b]{\shortstack{$R_p$}}%
}%
\put(3450,200){\makebox(0,0){0.65}}%
\put(3063,200){\makebox(0,0){0.6}}%
\put(2675,200){\makebox(0,0){0.55}}%
\put(2287,200){\makebox(0,0){0.5}}%
\put(1900,200){\makebox(0,0){0.45}}%
\put(1512,200){\makebox(0,0){0.4}}%
\put(1125,200){\makebox(0,0){0.35}}%
\put(737,200){\makebox(0,0){0.3}}%
\put(350,200){\makebox(0,0){0.25}}%
\put(300,2060){\makebox(0,0)[r]{35}}%
\put(300,1809){\makebox(0,0)[r]{30}}%
\put(300,1557){\makebox(0,0)[r]{25}}%
\put(300,1306){\makebox(0,0)[r]{20}}%
\put(300,1054){\makebox(0,0)[r]{15}}%
\put(300,803){\makebox(0,0)[r]{10}}%
\put(300,551){\makebox(0,0)[r]{5}}%
\put(300,300){\makebox(0,0)[r]{0}}%
\end{picture}%
\endgroup

%% file: 1+1d_ratiograph.tex
\begingroup%
  \makeatletter%
  \newcommand{\GNUPLOTspecial}{%
    \@sanitize\catcode`\%=14\relax\special}%
  \setlength{\unitlength}{0.1bp}%
{\GNUPLOTspecial{!
/gnudict 256 dict def
gnudict begin
/Color false def
/Solid false def
/gnulinewidth 5.000 def
/userlinewidth gnulinewidth def
/vshift -33 def
/dl {10 mul} def
/hpt_ 31.5 def
/vpt_ 31.5 def
/hpt hpt_ def
/vpt vpt_ def
/M {moveto} bind def
/L {lineto} bind def
/R {rmoveto} bind def
/V {rlineto} bind def
/vpt2 vpt 2 mul def
/hpt2 hpt 2 mul def
/Lshow { currentpoint stroke M
  0 vshift R show } def
/Rshow { currentpoint stroke M
  dup stringwidth pop neg vshift R show } def
/Cshow { currentpoint stroke M
  dup stringwidth pop -2 div vshift R show } def
/UP { dup vpt_ mul /vpt exch def hpt_ mul /hpt exch def
  /hpt2 hpt 2 mul def /vpt2 vpt 2 mul def } def
/DL { Color {setrgbcolor Solid {pop []} if 0 setdash }
 {pop pop pop Solid {pop []} if 0 setdash} ifelse } def
/BL { stroke userlinewidth 2 mul setlinewidth } def
/AL { stroke userlinewidth 2 div setlinewidth } def
/UL { dup gnulinewidth mul /userlinewidth exch def
      10 mul /udl exch def } def
/PL { stroke userlinewidth setlinewidth } def
/LTb { BL [] 0 0 0 DL } def
/LTa { AL [1 udl mul 2 udl mul] 0 setdash 0 0 0 setrgbcolor } def
/LT0 { PL [] 1 0 0 DL } def
/LT1 { PL [4 dl 2 dl] 0 1 0 DL } def
/LT2 { PL [2 dl 3 dl] 0 0 1 DL } def
/LT3 { PL [1 dl 1.5 dl] 1 0 1 DL } def
/LT4 { PL [5 dl 2 dl 1 dl 2 dl] 0 1 1 DL } def
/LT5 { PL [4 dl 3 dl 1 dl 3 dl] 1 1 0 DL } def
/LT6 { PL [2 dl 2 dl 2 dl 4 dl] 0 0 0 DL } def
/LT7 { PL [2 dl 2 dl 2 dl 2 dl 2 dl 4 dl] 1 0.3 0 DL } def
/LT8 { PL [2 dl 2 dl 2 dl 2 dl 2 dl 2 dl 2 dl 4 dl] 0.5 0.5 0.5 DL } def
/Pnt { stroke [] 0 setdash
   gsave 1 setlinecap M 0 0 V stroke grestore } def
/Dia { stroke [] 0 setdash 2 copy vpt add M
  hpt neg vpt neg V hpt vpt neg V
  hpt vpt V hpt neg vpt V closepath stroke
  Pnt } def
/Pls { stroke [] 0 setdash vpt sub M 0 vpt2 V
  currentpoint stroke M
  hpt neg vpt neg R hpt2 0 V stroke
  } def
/Box { stroke [] 0 setdash 2 copy exch hpt sub exch vpt add M
  0 vpt2 neg V hpt2 0 V 0 vpt2 V
  hpt2 neg 0 V closepath stroke
  Pnt } def
/Crs { stroke [] 0 setdash exch hpt sub exch vpt add M
  hpt2 vpt2 neg V currentpoint stroke M
  hpt2 neg 0 R hpt2 vpt2 V stroke } def
/TriU { stroke [] 0 setdash 2 copy vpt 1.12 mul add M
  hpt neg vpt -1.62 mul V
  hpt 2 mul 0 V
  hpt neg vpt 1.62 mul V closepath stroke
  Pnt  } def
/Star { 2 copy Pls Crs } def
/BoxF { stroke [] 0 setdash exch hpt sub exch vpt add M
  0 vpt2 neg V  hpt2 0 V  0 vpt2 V
  hpt2 neg 0 V  closepath fill } def
/TriUF { stroke [] 0 setdash vpt 1.12 mul add M
  hpt neg vpt -1.62 mul V
  hpt 2 mul 0 V
  hpt neg vpt 1.62 mul V closepath fill } def
/TriD { stroke [] 0 setdash 2 copy vpt 1.12 mul sub M
  hpt neg vpt 1.62 mul V
  hpt 2 mul 0 V
  hpt neg vpt -1.62 mul V closepath stroke
  Pnt  } def
/TriDF { stroke [] 0 setdash vpt 1.12 mul sub M
  hpt neg vpt 1.62 mul V
  hpt 2 mul 0 V
  hpt neg vpt -1.62 mul V closepath fill} def
/DiaF { stroke [] 0 setdash vpt add M
  hpt neg vpt neg V hpt vpt neg V
  hpt vpt V hpt neg vpt V closepath fill } def
/Pent { stroke [] 0 setdash 2 copy gsave
  translate 0 hpt M 4 {72 rotate 0 hpt L} repeat
  closepath stroke grestore Pnt } def
/PentF { stroke [] 0 setdash gsave
  translate 0 hpt M 4 {72 rotate 0 hpt L} repeat
  closepath fill grestore } def
/Circle { stroke [] 0 setdash 2 copy
  hpt 0 360 arc stroke Pnt } def
/CircleF { stroke [] 0 setdash hpt 0 360 arc fill } def
/C0 { BL [] 0 setdash 2 copy moveto vpt 90 450  arc } bind def
/C1 { BL [] 0 setdash 2 copy        moveto
       2 copy  vpt 0 90 arc closepath fill
               vpt 0 360 arc closepath } bind def
/C2 { BL [] 0 setdash 2 copy moveto
       2 copy  vpt 90 180 arc closepath fill
               vpt 0 360 arc closepath } bind def
/C3 { BL [] 0 setdash 2 copy moveto
       2 copy  vpt 0 180 arc closepath fill
               vpt 0 360 arc closepath } bind def
/C4 { BL [] 0 setdash 2 copy moveto
       2 copy  vpt 180 270 arc closepath fill
               vpt 0 360 arc closepath } bind def
/C5 { BL [] 0 setdash 2 copy moveto
       2 copy  vpt 0 90 arc
       2 copy moveto
       2 copy  vpt 180 270 arc closepath fill
               vpt 0 360 arc } bind def
/C6 { BL [] 0 setdash 2 copy moveto
      2 copy  vpt 90 270 arc closepath fill
              vpt 0 360 arc closepath } bind def
/C7 { BL [] 0 setdash 2 copy moveto
      2 copy  vpt 0 270 arc closepath fill
              vpt 0 360 arc closepath } bind def
/C8 { BL [] 0 setdash 2 copy moveto
      2 copy vpt 270 360 arc closepath fill
              vpt 0 360 arc closepath } bind def
/C9 { BL [] 0 setdash 2 copy moveto
      2 copy  vpt 270 450 arc closepath fill
              vpt 0 360 arc closepath } bind def
/C10 { BL [] 0 setdash 2 copy 2 copy moveto vpt 270 360 arc closepath fill
       2 copy moveto
       2 copy vpt 90 180 arc closepath fill
               vpt 0 360 arc closepath } bind def
/C11 { BL [] 0 setdash 2 copy moveto
       2 copy  vpt 0 180 arc closepath fill
       2 copy moveto
       2 copy  vpt 270 360 arc closepath fill
               vpt 0 360 arc closepath } bind def
/C12 { BL [] 0 setdash 2 copy moveto
       2 copy  vpt 180 360 arc closepath fill
               vpt 0 360 arc closepath } bind def
/C13 { BL [] 0 setdash  2 copy moveto
       2 copy  vpt 0 90 arc closepath fill
       2 copy moveto
       2 copy  vpt 180 360 arc closepath fill
               vpt 0 360 arc closepath } bind def
/C14 { BL [] 0 setdash 2 copy moveto
       2 copy  vpt 90 360 arc closepath fill
               vpt 0 360 arc } bind def
/C15 { BL [] 0 setdash 2 copy vpt 0 360 arc closepath fill
               vpt 0 360 arc closepath } bind def
/Rec   { newpath 4 2 roll moveto 1 index 0 rlineto 0 exch rlineto
       neg 0 rlineto closepath } bind def
/Square { dup Rec } bind def
/Bsquare { vpt sub exch vpt sub exch vpt2 Square } bind def
/S0 { BL [] 0 setdash 2 copy moveto 0 vpt rlineto BL Bsquare } bind def
/S1 { BL [] 0 setdash 2 copy vpt Square fill Bsquare } bind def
/S2 { BL [] 0 setdash 2 copy exch vpt sub exch vpt Square fill Bsquare } bind def
/S3 { BL [] 0 setdash 2 copy exch vpt sub exch vpt2 vpt Rec fill Bsquare } bind def
/S4 { BL [] 0 setdash 2 copy exch vpt sub exch vpt sub vpt Square fill Bsquare } bind def
/S5 { BL [] 0 setdash 2 copy 2 copy vpt Square fill
       exch vpt sub exch vpt sub vpt Square fill Bsquare } bind def
/S6 { BL [] 0 setdash 2 copy exch vpt sub exch vpt sub vpt vpt2 Rec fill Bsquare } bind def
/S7 { BL [] 0 setdash 2 copy exch vpt sub exch vpt sub vpt vpt2 Rec fill
       2 copy vpt Square fill
       Bsquare } bind def
/S8 { BL [] 0 setdash 2 copy vpt sub vpt Square fill Bsquare } bind def
/S9 { BL [] 0 setdash 2 copy vpt sub vpt vpt2 Rec fill Bsquare } bind def
/S10 { BL [] 0 setdash 2 copy vpt sub vpt Square fill 2 copy exch vpt sub exch vpt Square fill
       Bsquare } bind def
/S11 { BL [] 0 setdash 2 copy vpt sub vpt Square fill 2 copy exch vpt sub exch vpt2 vpt Rec fill
       Bsquare } bind def
/S12 { BL [] 0 setdash 2 copy exch vpt sub exch vpt sub vpt2 vpt Rec fill Bsquare } bind def
/S13 { BL [] 0 setdash 2 copy exch vpt sub exch vpt sub vpt2 vpt Rec fill
       2 copy vpt Square fill Bsquare } bind def
/S14 { BL [] 0 setdash 2 copy exch vpt sub exch vpt sub vpt2 vpt Rec fill
       2 copy exch vpt sub exch vpt Square fill Bsquare } bind def
/S15 { BL [] 0 setdash 2 copy Bsquare fill Bsquare } bind def
/D0 { gsave translate 45 rotate 0 0 S0 stroke grestore } bind def
/D1 { gsave translate 45 rotate 0 0 S1 stroke grestore } bind def
/D2 { gsave translate 45 rotate 0 0 S2 stroke grestore } bind def
/D3 { gsave translate 45 rotate 0 0 S3 stroke grestore } bind def
/D4 { gsave translate 45 rotate 0 0 S4 stroke grestore } bind def
/D5 { gsave translate 45 rotate 0 0 S5 stroke grestore } bind def
/D6 { gsave translate 45 rotate 0 0 S6 stroke grestore } bind def
/D7 { gsave translate 45 rotate 0 0 S7 stroke grestore } bind def
/D8 { gsave translate 45 rotate 0 0 S8 stroke grestore } bind def
/D9 { gsave translate 45 rotate 0 0 S9 stroke grestore } bind def
/D10 { gsave translate 45 rotate 0 0 S10 stroke grestore } bind def
/D11 { gsave translate 45 rotate 0 0 S11 stroke grestore } bind def
/D12 { gsave translate 45 rotate 0 0 S12 stroke grestore } bind def
/D13 { gsave translate 45 rotate 0 0 S13 stroke grestore } bind def
/D14 { gsave translate 45 rotate 0 0 S14 stroke grestore } bind def
/D15 { gsave translate 45 rotate 0 0 S15 stroke grestore } bind def
/DiaE { stroke [] 0 setdash vpt add M
  hpt neg vpt neg V hpt vpt neg V
  hpt vpt V hpt neg vpt V closepath stroke } def
/BoxE { stroke [] 0 setdash exch hpt sub exch vpt add M
  0 vpt2 neg V hpt2 0 V 0 vpt2 V
  hpt2 neg 0 V closepath stroke } def
/TriUE { stroke [] 0 setdash vpt 1.12 mul add M
  hpt neg vpt -1.62 mul V
  hpt 2 mul 0 V
  hpt neg vpt 1.62 mul V closepath stroke } def
/TriDE { stroke [] 0 setdash vpt 1.12 mul sub M
  hpt neg vpt 1.62 mul V
  hpt 2 mul 0 V
  hpt neg vpt -1.62 mul V closepath stroke } def
/PentE { stroke [] 0 setdash gsave
  translate 0 hpt M 4 {72 rotate 0 hpt L} repeat
  closepath stroke grestore } def
/CircE { stroke [] 0 setdash 
  hpt 0 360 arc stroke } def
/Opaque { gsave closepath 1 setgray fill grestore 0 setgray closepath } def
/DiaW { stroke [] 0 setdash vpt add M
  hpt neg vpt neg V hpt vpt neg V
  hpt vpt V hpt neg vpt V Opaque stroke } def
/BoxW { stroke [] 0 setdash exch hpt sub exch vpt add M
  0 vpt2 neg V hpt2 0 V 0 vpt2 V
  hpt2 neg 0 V Opaque stroke } def
/TriUW { stroke [] 0 setdash vpt 1.12 mul add M
  hpt neg vpt -1.62 mul V
  hpt 2 mul 0 V
  hpt neg vpt 1.62 mul V Opaque stroke } def
/TriDW { stroke [] 0 setdash vpt 1.12 mul sub M
  hpt neg vpt 1.62 mul V
  hpt 2 mul 0 V
  hpt neg vpt -1.62 mul V Opaque stroke } def
/PentW { stroke [] 0 setdash gsave
  translate 0 hpt M 4 {72 rotate 0 hpt L} repeat
  Opaque stroke grestore } def
/CircW { stroke [] 0 setdash 
  hpt 0 360 arc Opaque stroke } def
/BoxFill { gsave Rec 1 setgray fill grestore } def
end
}}%
\begin{picture}(3600,2160)(0,0)%
{\GNUPLOTspecial{"
gnudict begin
gsave
0 0 translate
0.100 0.100 scale
0 setgray
newpath
1.000 UL
LTb
350 300 M
63 0 V
3037 0 R
-63 0 V
350 544 M
63 0 V
3037 0 R
-63 0 V
350 789 M
63 0 V
3037 0 R
-63 0 V
350 1033 M
63 0 V
3037 0 R
-63 0 V
350 1278 M
63 0 V
3037 0 R
-63 0 V
350 1522 M
63 0 V
3037 0 R
-63 0 V
350 1767 M
63 0 V
3037 0 R
-63 0 V
350 2011 M
63 0 V
3037 0 R
-63 0 V
350 300 M
0 63 V
0 1697 R
0 -63 V
660 300 M
0 63 V
0 1697 R
0 -63 V
970 300 M
0 63 V
0 1697 R
0 -63 V
1280 300 M
0 63 V
0 1697 R
0 -63 V
1590 300 M
0 63 V
0 1697 R
0 -63 V
1900 300 M
0 63 V
0 1697 R
0 -63 V
2210 300 M
0 63 V
0 1697 R
0 -63 V
2520 300 M
0 63 V
0 1697 R
0 -63 V
2830 300 M
0 63 V
0 1697 R
0 -63 V
3140 300 M
0 63 V
0 1697 R
0 -63 V
3450 300 M
0 63 V
0 1697 R
0 -63 V
1.000 UL
LTb
350 300 M
3100 0 V
0 1760 V
-3100 0 V
350 300 L
1.000 UL
LT0
350 401 M
861 60 V
345 8 V
172 -1 V
172 -1 V
344 -8 V
3450 436 L
1.000 UL
LT1
350 426 M
861 173 V
345 54 V
172 7 V
172 -8 V
344 -36 V
3450 529 L
1.000 UL
LT2
350 449 M
861 364 V
345 201 V
172 67 V
172 -16 V
174 -82 V
170 -69 V
3450 666 L
1.000 UL
LT3
1211 985 M
345 646 V
172 297 V
172 24 V
172 -234 V
172 -289 V
0.600 UP
1.000 UL
LT4
350 401 M
0 1 V
-31 -1 R
62 0 V
-62 1 R
62 0 V
830 58 R
0 2 V
-31 -2 R
62 0 V
-62 2 R
62 0 V
314 6 R
0 1 V
-31 -1 R
62 0 V
-62 1 R
62 0 V
141 -1 R
0 1 V
-31 -1 R
62 0 V
-62 1 R
62 0 V
141 -3 R
0 1 V
-31 -1 R
62 0 V
-62 1 R
62 0 V
313 -8 R
0 1 V
-31 -1 R
62 0 V
-62 1 R
62 0 V
3450 436 M
0 1 V
-31 -1 R
62 0 V
-62 1 R
62 0 V
350 401 Pls
1211 461 Pls
1556 469 Pls
1728 468 Pls
1900 467 Pls
2244 459 Pls
3450 436 Pls
0.600 UP
1.000 UL
LT5
350 426 M
0 1 V
-31 -1 R
62 0 V
-62 1 R
62 0 V
830 171 R
0 2 V
-31 -2 R
62 0 V
-62 2 R
62 0 V
314 52 R
0 3 V
-31 -3 R
62 0 V
-62 3 R
62 0 V
141 4 R
0 1 V
-31 -1 R
62 0 V
-62 1 R
62 0 V
141 -9 R
0 2 V
-31 -2 R
62 0 V
-62 2 R
62 0 V
313 -38 R
0 3 V
-31 -3 R
62 0 V
-62 3 R
62 0 V
3450 528 M
0 2 V
-31 -2 R
62 0 V
-62 2 R
62 0 V
350 426 Crs
1211 599 Crs
1556 653 Crs
1728 660 Crs
1900 652 Crs
2244 616 Crs
3450 529 Crs
0.600 UP
1.000 UL
LT6
350 449 M
0 1 V
-31 -1 R
62 0 V
-62 1 R
62 0 V
830 351 R
0 25 V
-31 -25 R
62 0 V
-62 25 R
62 0 V
314 186 R
0 4 V
-31 -4 R
62 0 V
-62 4 R
62 0 V
141 63 R
0 5 V
-31 -5 R
62 0 V
-62 5 R
62 0 V
141 -21 R
0 4 V
-31 -4 R
62 0 V
-62 4 R
62 0 V
2074 958 M
0 50 V
-31 -50 R
62 0 V
-62 50 R
62 0 V
139 -96 R
0 4 V
-31 -4 R
62 0 V
-62 4 R
62 0 V
3450 665 M
0 1 V
-31 -1 R
62 0 V
-62 1 R
62 0 V
350 449 Star
1211 813 Star
1556 1014 Star
1728 1081 Star
1900 1065 Star
2074 983 Star
2244 914 Star
3450 666 Star
0.600 UP
1.000 UL
LT7
1211 968 M
0 34 V
-31 -34 R
62 0 V
-62 34 R
62 0 V
314 594 R
0 69 V
-31 -69 R
62 0 V
-62 69 R
62 0 V
141 199 R
0 128 V
-31 -128 R
62 0 V
-62 128 R
62 0 V
141 -128 R
0 176 V
-31 -176 R
62 0 V
-62 176 R
62 0 V
141 -400 R
0 156 V
-31 -156 R
62 0 V
-62 156 R
62 0 V
141 -474 R
0 215 V
-31 -215 R
62 0 V
-62 215 R
62 0 V
1211 985 Box
1556 1631 Box
1728 1928 Box
1900 1952 Box
2072 1718 Box
2244 1429 Box
stroke
grestore
end
showpage
}}%
\put(1900,50){\makebox(0,0){$\gamma$}}%
\put(100,1180){%
\makebox(0,0)[b]{\shortstack{$R_p$}}%
}%
\put(3450,200){\makebox(0,0){0.75}}%
\put(3140,200){\makebox(0,0){0.7}}%
\put(2830,200){\makebox(0,0){0.65}}%
\put(2520,200){\makebox(0,0){0.6}}%
\put(2210,200){\makebox(0,0){0.55}}%
\put(1900,200){\makebox(0,0){0.5}}%
\put(1590,200){\makebox(0,0){0.45}}%
\put(1280,200){\makebox(0,0){0.4}}%
\put(970,200){\makebox(0,0){0.35}}%
\put(660,200){\makebox(0,0){0.3}}%
\put(350,200){\makebox(0,0){0.25}}%
\put(300,2011){\makebox(0,0)[r]{35}}%
\put(300,1767){\makebox(0,0)[r]{30}}%
\put(300,1522){\makebox(0,0)[r]{25}}%
\put(300,1278){\makebox(0,0)[r]{20}}%
\put(300,1033){\makebox(0,0)[r]{15}}%
\put(300,789){\makebox(0,0)[r]{10}}%
\put(300,544){\makebox(0,0)[r]{5}}%
\put(300,300){\makebox(0,0)[r]{0}}%
\end{picture}%
\endgroup

%% file: eigenvaluedensitygraph.tex
\begingroup%
  \makeatletter%
  \newcommand{\GNUPLOTspecial}{%
    \@sanitize\catcode`\%=14\relax\special}%
  \setlength{\unitlength}{0.1bp}%
{\GNUPLOTspecial{!
/gnudict 256 dict def
gnudict begin
/Color false def
/Solid false def
/gnulinewidth 5.000 def
/userlinewidth gnulinewidth def
/vshift -33 def
/dl {10 mul} def
/hpt_ 31.5 def
/vpt_ 31.5 def
/hpt hpt_ def
/vpt vpt_ def
/M {moveto} bind def
/L {lineto} bind def
/R {rmoveto} bind def
/V {rlineto} bind def
/vpt2 vpt 2 mul def
/hpt2 hpt 2 mul def
/Lshow { currentpoint stroke M
  0 vshift R show } def
/Rshow { currentpoint stroke M
  dup stringwidth pop neg vshift R show } def
/Cshow { currentpoint stroke M
  dup stringwidth pop -2 div vshift R show } def
/UP { dup vpt_ mul /vpt exch def hpt_ mul /hpt exch def
  /hpt2 hpt 2 mul def /vpt2 vpt 2 mul def } def
/DL { Color {setrgbcolor Solid {pop []} if 0 setdash }
 {pop pop pop Solid {pop []} if 0 setdash} ifelse } def
/BL { stroke userlinewidth 2 mul setlinewidth } def
/AL { stroke userlinewidth 2 div setlinewidth } def
/UL { dup gnulinewidth mul /userlinewidth exch def
      10 mul /udl exch def } def
/PL { stroke userlinewidth setlinewidth } def
/LTb { BL [] 0 0 0 DL } def
/LTa { AL [1 udl mul 2 udl mul] 0 setdash 0 0 0 setrgbcolor } def
/LT0 { PL [] 1 0 0 DL } def
/LT1 { PL [4 dl 2 dl] 0 1 0 DL } def
/LT2 { PL [2 dl 3 dl] 0 0 1 DL } def
/LT3 { PL [1 dl 1.5 dl] 1 0 1 DL } def
/LT4 { PL [5 dl 2 dl 1 dl 2 dl] 0 1 1 DL } def
/LT5 { PL [4 dl 3 dl 1 dl 3 dl] 1 1 0 DL } def
/LT6 { PL [2 dl 2 dl 2 dl 4 dl] 0 0 0 DL } def
/LT7 { PL [2 dl 2 dl 2 dl 2 dl 2 dl 4 dl] 1 0.3 0 DL } def
/LT8 { PL [2 dl 2 dl 2 dl 2 dl 2 dl 2 dl 2 dl 4 dl] 0.5 0.5 0.5 DL } def
/Pnt { stroke [] 0 setdash
   gsave 1 setlinecap M 0 0 V stroke grestore } def
/Dia { stroke [] 0 setdash 2 copy vpt add M
  hpt neg vpt neg V hpt vpt neg V
  hpt vpt V hpt neg vpt V closepath stroke
  Pnt } def
/Pls { stroke [] 0 setdash vpt sub M 0 vpt2 V
  currentpoint stroke M
  hpt neg vpt neg R hpt2 0 V stroke
  } def
/Box { stroke [] 0 setdash 2 copy exch hpt sub exch vpt add M
  0 vpt2 neg V hpt2 0 V 0 vpt2 V
  hpt2 neg 0 V closepath stroke
  Pnt } def
/Crs { stroke [] 0 setdash exch hpt sub exch vpt add M
  hpt2 vpt2 neg V currentpoint stroke M
  hpt2 neg 0 R hpt2 vpt2 V stroke } def
/TriU { stroke [] 0 setdash 2 copy vpt 1.12 mul add M
  hpt neg vpt -1.62 mul V
  hpt 2 mul 0 V
  hpt neg vpt 1.62 mul V closepath stroke
  Pnt  } def
/Star { 2 copy Pls Crs } def
/BoxF { stroke [] 0 setdash exch hpt sub exch vpt add M
  0 vpt2 neg V  hpt2 0 V  0 vpt2 V
  hpt2 neg 0 V  closepath fill } def
/TriUF { stroke [] 0 setdash vpt 1.12 mul add M
  hpt neg vpt -1.62 mul V
  hpt 2 mul 0 V
  hpt neg vpt 1.62 mul V closepath fill } def
/TriD { stroke [] 0 setdash 2 copy vpt 1.12 mul sub M
  hpt neg vpt 1.62 mul V
  hpt 2 mul 0 V
  hpt neg vpt -1.62 mul V closepath stroke
  Pnt  } def
/TriDF { stroke [] 0 setdash vpt 1.12 mul sub M
  hpt neg vpt 1.62 mul V
  hpt 2 mul 0 V
  hpt neg vpt -1.62 mul V closepath fill} def
/DiaF { stroke [] 0 setdash vpt add M
  hpt neg vpt neg V hpt vpt neg V
  hpt vpt V hpt neg vpt V closepath fill } def
/Pent { stroke [] 0 setdash 2 copy gsave
  translate 0 hpt M 4 {72 rotate 0 hpt L} repeat
  closepath stroke grestore Pnt } def
/PentF { stroke [] 0 setdash gsave
  translate 0 hpt M 4 {72 rotate 0 hpt L} repeat
  closepath fill grestore } def
/Circle { stroke [] 0 setdash 2 copy
  hpt 0 360 arc stroke Pnt } def
/CircleF { stroke [] 0 setdash hpt 0 360 arc fill } def
/C0 { BL [] 0 setdash 2 copy moveto vpt 90 450  arc } bind def
/C1 { BL [] 0 setdash 2 copy        moveto
       2 copy  vpt 0 90 arc closepath fill
               vpt 0 360 arc closepath } bind def
/C2 { BL [] 0 setdash 2 copy moveto
       2 copy  vpt 90 180 arc closepath fill
               vpt 0 360 arc closepath } bind def
/C3 { BL [] 0 setdash 2 copy moveto
       2 copy  vpt 0 180 arc closepath fill
               vpt 0 360 arc closepath } bind def
/C4 { BL [] 0 setdash 2 copy moveto
       2 copy  vpt 180 270 arc closepath fill
               vpt 0 360 arc closepath } bind def
/C5 { BL [] 0 setdash 2 copy moveto
       2 copy  vpt 0 90 arc
       2 copy moveto
       2 copy  vpt 180 270 arc closepath fill
               vpt 0 360 arc } bind def
/C6 { BL [] 0 setdash 2 copy moveto
      2 copy  vpt 90 270 arc closepath fill
              vpt 0 360 arc closepath } bind def
/C7 { BL [] 0 setdash 2 copy moveto
      2 copy  vpt 0 270 arc closepath fill
              vpt 0 360 arc closepath } bind def
/C8 { BL [] 0 setdash 2 copy moveto
      2 copy vpt 270 360 arc closepath fill
              vpt 0 360 arc closepath } bind def
/C9 { BL [] 0 setdash 2 copy moveto
      2 copy  vpt 270 450 arc closepath fill
              vpt 0 360 arc closepath } bind def
/C10 { BL [] 0 setdash 2 copy 2 copy moveto vpt 270 360 arc closepath fill
       2 copy moveto
       2 copy vpt 90 180 arc closepath fill
               vpt 0 360 arc closepath } bind def
/C11 { BL [] 0 setdash 2 copy moveto
       2 copy  vpt 0 180 arc closepath fill
       2 copy moveto
       2 copy  vpt 270 360 arc closepath fill
               vpt 0 360 arc closepath } bind def
/C12 { BL [] 0 setdash 2 copy moveto
       2 copy  vpt 180 360 arc closepath fill
               vpt 0 360 arc closepath } bind def
/C13 { BL [] 0 setdash  2 copy moveto
       2 copy  vpt 0 90 arc closepath fill
       2 copy moveto
       2 copy  vpt 180 360 arc closepath fill
               vpt 0 360 arc closepath } bind def
/C14 { BL [] 0 setdash 2 copy moveto
       2 copy  vpt 90 360 arc closepath fill
               vpt 0 360 arc } bind def
/C15 { BL [] 0 setdash 2 copy vpt 0 360 arc closepath fill
               vpt 0 360 arc closepath } bind def
/Rec   { newpath 4 2 roll moveto 1 index 0 rlineto 0 exch rlineto
       neg 0 rlineto closepath } bind def
/Square { dup Rec } bind def
/Bsquare { vpt sub exch vpt sub exch vpt2 Square } bind def
/S0 { BL [] 0 setdash 2 copy moveto 0 vpt rlineto BL Bsquare } bind def
/S1 { BL [] 0 setdash 2 copy vpt Square fill Bsquare } bind def
/S2 { BL [] 0 setdash 2 copy exch vpt sub exch vpt Square fill Bsquare } bind def
/S3 { BL [] 0 setdash 2 copy exch vpt sub exch vpt2 vpt Rec fill Bsquare } bind def
/S4 { BL [] 0 setdash 2 copy exch vpt sub exch vpt sub vpt Square fill Bsquare } bind def
/S5 { BL [] 0 setdash 2 copy 2 copy vpt Square fill
       exch vpt sub exch vpt sub vpt Square fill Bsquare } bind def
/S6 { BL [] 0 setdash 2 copy exch vpt sub exch vpt sub vpt vpt2 Rec fill Bsquare } bind def
/S7 { BL [] 0 setdash 2 copy exch vpt sub exch vpt sub vpt vpt2 Rec fill
       2 copy vpt Square fill
       Bsquare } bind def
/S8 { BL [] 0 setdash 2 copy vpt sub vpt Square fill Bsquare } bind def
/S9 { BL [] 0 setdash 2 copy vpt sub vpt vpt2 Rec fill Bsquare } bind def
/S10 { BL [] 0 setdash 2 copy vpt sub vpt Square fill 2 copy exch vpt sub exch vpt Square fill
       Bsquare } bind def
/S11 { BL [] 0 setdash 2 copy vpt sub vpt Square fill 2 copy exch vpt sub exch vpt2 vpt Rec fill
       Bsquare } bind def
/S12 { BL [] 0 setdash 2 copy exch vpt sub exch vpt sub vpt2 vpt Rec fill Bsquare } bind def
/S13 { BL [] 0 setdash 2 copy exch vpt sub exch vpt sub vpt2 vpt Rec fill
       2 copy vpt Square fill Bsquare } bind def
/S14 { BL [] 0 setdash 2 copy exch vpt sub exch vpt sub vpt2 vpt Rec fill
       2 copy exch vpt sub exch vpt Square fill Bsquare } bind def
/S15 { BL [] 0 setdash 2 copy Bsquare fill Bsquare } bind def
/D0 { gsave translate 45 rotate 0 0 S0 stroke grestore } bind def
/D1 { gsave translate 45 rotate 0 0 S1 stroke grestore } bind def
/D2 { gsave translate 45 rotate 0 0 S2 stroke grestore } bind def
/D3 { gsave translate 45 rotate 0 0 S3 stroke grestore } bind def
/D4 { gsave translate 45 rotate 0 0 S4 stroke grestore } bind def
/D5 { gsave translate 45 rotate 0 0 S5 stroke grestore } bind def
/D6 { gsave translate 45 rotate 0 0 S6 stroke grestore } bind def
/D7 { gsave translate 45 rotate 0 0 S7 stroke grestore } bind def
/D8 { gsave translate 45 rotate 0 0 S8 stroke grestore } bind def
/D9 { gsave translate 45 rotate 0 0 S9 stroke grestore } bind def
/D10 { gsave translate 45 rotate 0 0 S10 stroke grestore } bind def
/D11 { gsave translate 45 rotate 0 0 S11 stroke grestore } bind def
/D12 { gsave translate 45 rotate 0 0 S12 stroke grestore } bind def
/D13 { gsave translate 45 rotate 0 0 S13 stroke grestore } bind def
/D14 { gsave translate 45 rotate 0 0 S14 stroke grestore } bind def
/D15 { gsave translate 45 rotate 0 0 S15 stroke grestore } bind def
/DiaE { stroke [] 0 setdash vpt add M
  hpt neg vpt neg V hpt vpt neg V
  hpt vpt V hpt neg vpt V closepath stroke } def
/BoxE { stroke [] 0 setdash exch hpt sub exch vpt add M
  0 vpt2 neg V hpt2 0 V 0 vpt2 V
  hpt2 neg 0 V closepath stroke } def
/TriUE { stroke [] 0 setdash vpt 1.12 mul add M
  hpt neg vpt -1.62 mul V
  hpt 2 mul 0 V
  hpt neg vpt 1.62 mul V closepath stroke } def
/TriDE { stroke [] 0 setdash vpt 1.12 mul sub M
  hpt neg vpt 1.62 mul V
  hpt 2 mul 0 V
  hpt neg vpt -1.62 mul V closepath stroke } def
/PentE { stroke [] 0 setdash gsave
  translate 0 hpt M 4 {72 rotate 0 hpt L} repeat
  closepath stroke grestore } def
/CircE { stroke [] 0 setdash 
  hpt 0 360 arc stroke } def
/Opaque { gsave closepath 1 setgray fill grestore 0 setgray closepath } def
/DiaW { stroke [] 0 setdash vpt add M
  hpt neg vpt neg V hpt vpt neg V
  hpt vpt V hpt neg vpt V Opaque stroke } def
/BoxW { stroke [] 0 setdash exch hpt sub exch vpt add M
  0 vpt2 neg V hpt2 0 V 0 vpt2 V
  hpt2 neg 0 V Opaque stroke } def
/TriUW { stroke [] 0 setdash vpt 1.12 mul add M
  hpt neg vpt -1.62 mul V
  hpt 2 mul 0 V
  hpt neg vpt 1.62 mul V Opaque stroke } def
/TriDW { stroke [] 0 setdash vpt 1.12 mul sub M
  hpt neg vpt 1.62 mul V
  hpt 2 mul 0 V
  hpt neg vpt -1.62 mul V Opaque stroke } def
/PentW { stroke [] 0 setdash gsave
  translate 0 hpt M 4 {72 rotate 0 hpt L} repeat
  Opaque stroke grestore } def
/CircW { stroke [] 0 setdash 
  hpt 0 360 arc Opaque stroke } def
/BoxFill { gsave Rec 1 setgray fill grestore } def
end
}}%
\begin{picture}(3600,2160)(0,0)%
{\GNUPLOTspecial{"
gnudict begin
gsave
0 0 translate
0.100 0.100 scale
0 setgray
newpath
1.000 UL
LTb
450 300 M
63 0 V
2937 0 R
-63 0 V
450 520 M
63 0 V
2937 0 R
-63 0 V
450 740 M
63 0 V
2937 0 R
-63 0 V
450 960 M
63 0 V
2937 0 R
-63 0 V
450 1180 M
63 0 V
2937 0 R
-63 0 V
450 1400 M
63 0 V
2937 0 R
-63 0 V
450 1620 M
63 0 V
2937 0 R
-63 0 V
450 1840 M
63 0 V
2937 0 R
-63 0 V
450 2060 M
63 0 V
2937 0 R
-63 0 V
518 300 M
0 63 V
0 1697 R
0 -63 V
995 300 M
0 63 V
0 1697 R
0 -63 V
1473 300 M
0 63 V
0 1697 R
0 -63 V
1950 300 M
0 63 V
0 1697 R
0 -63 V
2427 300 M
0 63 V
0 1697 R
0 -63 V
2905 300 M
0 63 V
0 1697 R
0 -63 V
3382 300 M
0 63 V
0 1697 R
0 -63 V
1.000 UL
LTb
450 300 M
3000 0 V
0 1760 V
-3000 0 V
450 300 L
1.000 UL
LT0
457 314 M
15 1 V
16 1 V
15 0 V
15 1 V
14 2 V
15 3 V
16 2 V
15 2 V
14 4 V
15 6 V
16 4 V
14 5 V
16 6 V
14 7 V
16 10 V
15 11 V
14 13 V
16 11 V
14 16 V
15 17 V
16 18 V
14 25 V
15 20 V
16 24 V
14 30 V
15 25 V
16 33 V
14 37 V
15 29 V
16 33 V
14 34 V
15 28 V
16 31 V
14 20 V
15 24 V
16 7 V
14 11 V
15 6 V
16 -11 V
15 -12 V
14 -15 V
16 -6 V
15 -16 V
14 1 V
16 2 V
15 17 V
14 33 V
16 38 V
15 58 V
14 63 V
15 67 V
15 64 V
16 43 V
15 26 V
15 11 V
14 -10 V
15 -49 V
15 -30 V
16 -31 V
15 -25 V
15 20 V
14 50 V
15 74 V
15 85 V
15 78 V
16 52 V
15 31 V
15 -29 V
14 -49 V
15 -61 V
15 -54 V
16 -2 V
15 41 V
15 86 V
14 106 V
15 75 V
15 49 V
16 -27 V
15 -68 V
15 -76 V
14 -61 V
15 -3 V
15 62 V
16 91 V
15 109 V
15 71 V
14 -17 V
15 -62 V
15 -105 V
15 -64 V
16 -20 V
15 65 V
15 104 V
14 87 V
15 60 V
15 -26 V
16 -83 V
15 -113 V
15 -60 V
14 -4 V
15 70 V
15 105 V
15 75 V
15 30 V
15 -41 V
16 -105 V
15 -97 V
15 -63 V
15 9 V
15 74 V
15 98 V
15 77 V
14 -12 V
15 -61 V
15 -85 V
15 -106 V
15 -64 V
15 10 V
16 51 V
15 83 V
15 66 V
15 17 V
15 -32 V
15 -81 V
14 -114 V
15 -78 V
15 -38 V
15 5 V
15 35 V
15 69 V
16 49 V
15 22 V
15 -13 V
15 -63 V
15 -67 V
15 -91 V
15 -70 V
14 -52 V
15 -14 V
15 7 V
15 31 V
15 51 V
15 32 V
16 12 V
15 0 V
15 -34 V
15 -54 V
15 -62 V
15 -53 V
14 -67 V
15 -61 V
15 -49 V
15 -23 V
15 -16 V
15 -4 V
16 5 V
15 11 V
15 11 V
15 11 V
15 11 V
15 5 V
15 5 V
14 -10 V
15 -13 V
15 -14 V
15 -26 V
15 -28 V
15 -30 V
16 -41 V
15 -29 V
15 -31 V
15 -39 V
15 -30 V
15 -29 V
14 -26 V
15 -21 V
15 -24 V
15 -20 V
15 -20 V
15 -19 V
16 -15 V
15 -13 V
15 -11 V
15 -9 V
15 -12 V
15 -7 V
15 -5 V
14 -7 V
15 -3 V
15 -6 V
15 -3 V
15 -4 V
15 -2 V
16 -2 V
15 -1 V
15 -2 V
15 -2 V
15 0 V
15 1 V
1.000 UL
LT1
457 316 M
15 1 V
15 0 V
15 2 V
15 0 V
15 3 V
15 1 V
15 4 V
15 2 V
15 3 V
15 5 V
15 6 V
15 6 V
15 5 V
15 9 V
15 9 V
15 13 V
15 11 V
15 11 V
15 15 V
15 19 V
15 19 V
15 18 V
15 25 V
15 24 V
15 27 V
15 30 V
15 31 V
15 29 V
15 33 V
15 32 V
15 30 V
15 30 V
15 29 V
15 21 V
15 17 V
15 15 V
15 7 V
15 -3 V
15 -1 V
15 -13 V
15 -9 V
15 -15 V
15 -16 V
15 -1 V
15 2 V
15 15 V
15 29 V
15 48 V
15 56 V
15 67 V
15 64 V
15 60 V
15 42 V
15 38 V
15 6 V
15 -13 V
15 -30 V
15 -40 V
15 -31 V
15 -17 V
15 6 V
15 46 V
15 70 V
15 89 V
15 81 V
15 63 V
15 22 V
15 -19 V
15 -49 V
15 -59 V
15 -46 V
15 -9 V
15 42 V
15 66 V
15 104 V
15 85 V
15 44 V
15 -10 V
15 -58 V
15 -75 V
15 -63 V
15 -9 V
15 53 V
15 91 V
15 108 V
15 60 V
15 1 V
15 -59 V
15 -96 V
15 -75 V
15 -20 V
15 55 V
15 97 V
15 110 V
15 49 V
15 -24 V
15 -86 V
15 -96 V
15 -70 V
16 -1 V
15 73 V
15 97 V
15 76 V
15 27 V
15 -42 V
15 -111 V
15 -99 V
15 -55 V
15 22 V
15 74 V
15 91 V
15 67 V
15 -5 V
15 -68 V
15 -97 V
15 -98 V
15 -46 V
15 4 V
15 61 V
15 76 V
15 66 V
15 10 V
15 -45 V
15 -85 V
15 -105 V
15 -77 V
15 -41 V
15 11 V
15 54 V
15 57 V
15 50 V
15 24 V
15 -27 V
15 -63 V
15 -84 V
15 -84 V
15 -73 V
15 -47 V
15 -5 V
15 14 V
15 29 V
15 42 V
15 34 V
15 13 V
15 -2 V
15 -35 V
15 -54 V
15 -58 V
15 -65 V
15 -65 V
15 -56 V
15 -52 V
15 -24 V
15 -14 V
15 -7 V
15 4 V
15 16 V
15 8 V
15 16 V
15 12 V
15 3 V
15 1 V
15 -5 V
15 -18 V
15 -14 V
15 -20 V
15 -30 V
15 -31 V
15 -33 V
15 -28 V
15 -34 V
15 -29 V
15 -34 V
15 -28 V
15 -27 V
15 -22 V
15 -24 V
15 -21 V
15 -19 V
15 -17 V
15 -15 V
15 -12 V
15 -14 V
15 -9 V
15 -10 V
15 -7 V
15 -7 V
15 -7 V
15 -4 V
15 -5 V
15 -5 V
15 -2 V
15 -3 V
15 -2 V
15 -1 V
15 -2 V
15 -1 V
15 0 V
15 -1 V
1.000 UL
LT2
457 301 M
15 0 V
16 0 V
15 0 V
15 1 V
14 0 V
15 0 V
16 0 V
15 1 V
14 0 V
15 1 V
16 0 V
14 1 V
16 1 V
14 2 V
16 2 V
15 2 V
14 3 V
16 5 V
14 2 V
15 6 V
16 6 V
14 7 V
15 12 V
16 9 V
14 14 V
15 15 V
16 19 V
14 21 V
15 29 V
16 31 V
14 31 V
15 33 V
16 44 V
14 53 V
15 48 V
16 45 V
14 47 V
15 52 V
16 45 V
15 37 V
14 30 V
16 20 V
15 7 V
14 3 V
16 -9 V
15 -22 V
14 -22 V
16 -17 V
15 -17 V
14 13 V
15 25 V
15 47 V
16 63 V
15 86 V
15 78 V
14 65 V
15 46 V
15 26 V
16 -8 V
15 -41 V
15 -50 V
14 -48 V
15 -26 V
15 27 V
15 64 V
16 101 V
15 95 V
15 93 V
14 39 V
15 -18 V
15 -72 V
16 -83 V
15 -49 V
15 -5 V
14 63 V
15 111 V
15 113 V
16 77 V
15 17 V
15 -70 V
14 -101 V
15 -78 V
15 -29 V
16 66 V
15 129 V
15 110 V
14 74 V
15 -30 V
15 -95 V
15 -110 V
16 -70 V
15 26 V
15 101 V
14 133 V
15 91 V
15 -9 V
16 -102 V
15 -117 V
15 -94 V
14 -4 V
15 97 V
15 127 V
15 92 V
15 -4 V
15 -74 V
16 -149 V
15 -89 V
15 -27 V
15 68 V
15 124 V
15 87 V
15 33 V
14 -78 V
15 -115 V
15 -121 V
15 -67 V
15 25 V
15 82 V
16 101 V
15 67 V
15 -21 V
15 -72 V
15 -113 V
15 -111 V
14 -71 V
15 17 V
15 46 V
15 84 V
15 64 V
15 13 V
16 -33 V
15 -82 V
15 -95 V
15 -107 V
15 -57 V
15 -33 V
15 21 V
14 48 V
15 47 V
15 42 V
15 14 V
15 -22 V
15 -48 V
16 -71 V
15 -84 V
15 -77 V
15 -62 V
15 -47 V
15 -25 V
14 -6 V
15 6 V
15 23 V
15 13 V
15 27 V
15 12 V
16 -8 V
15 -7 V
15 -16 V
15 -34 V
15 -37 V
15 -37 V
15 -53 V
14 -52 V
15 -48 V
15 -50 V
15 -44 V
15 -41 V
15 -38 V
16 -32 V
15 -28 V
15 -28 V
15 -23 V
15 -19 V
15 -19 V
14 -12 V
15 -11 V
15 -9 V
15 -7 V
15 -7 V
15 -7 V
16 -3 V
15 -4 V
15 -2 V
15 -2 V
15 -2 V
15 -1 V
15 -1 V
14 -1 V
15 -1 V
15 -1 V
15 -1 V
15 0 V
15 0 V
16 0 V
15 -1 V
15 0 V
15 0 V
15 0 V
15 0 V
1.000 UL
LT3
457 302 M
15 0 V
15 0 V
15 0 V
15 0 V
15 1 V
15 0 V
15 0 V
15 1 V
15 1 V
15 1 V
15 0 V
15 2 V
15 1 V
15 1 V
15 3 V
15 2 V
15 4 V
15 2 V
15 7 V
15 4 V
15 6 V
15 9 V
15 8 V
15 11 V
15 13 V
15 15 V
15 20 V
15 19 V
15 24 V
15 28 V
15 34 V
15 34 V
15 38 V
15 38 V
15 49 V
15 40 V
15 53 V
15 43 V
15 48 V
15 33 V
15 35 V
15 25 V
15 11 V
15 2 V
15 -6 V
15 -18 V
15 -20 V
15 -12 V
15 -20 V
15 0 V
15 15 V
15 37 V
15 60 V
15 72 V
15 86 V
15 75 V
15 71 V
15 30 V
15 12 V
15 -27 V
15 -46 V
15 -52 V
15 -32 V
15 -2 V
15 44 V
15 78 V
15 116 V
15 91 V
15 65 V
15 14 V
15 -46 V
15 -74 V
15 -73 V
15 -32 V
15 36 V
15 94 V
15 122 V
15 95 V
15 47 V
15 -33 V
15 -94 V
15 -95 V
15 -51 V
15 40 V
15 113 V
15 139 V
15 79 V
15 -6 V
15 -90 V
15 -114 V
15 -84 V
15 18 V
15 93 V
15 126 V
15 103 V
15 0 V
15 -85 V
15 -133 V
15 -90 V
16 -1 V
15 88 V
15 129 V
15 94 V
15 -4 V
15 -96 V
15 -136 V
15 -95 V
15 -3 V
15 75 V
15 124 V
15 84 V
15 2 V
15 -92 V
15 -127 V
15 -108 V
15 -38 V
15 31 V
15 105 V
15 102 V
15 32 V
15 -49 V
15 -86 V
15 -132 V
15 -88 V
15 -40 V
15 28 V
15 75 V
15 79 V
15 40 V
15 -14 V
15 -62 V
15 -94 V
15 -110 V
15 -77 V
15 -52 V
15 1 V
15 35 V
15 51 V
15 42 V
15 34 V
15 -5 V
15 -35 V
15 -70 V
15 -82 V
15 -83 V
15 -75 V
15 -58 V
15 -33 V
15 -25 V
15 6 V
15 15 V
15 17 V
15 21 V
15 22 V
15 1 V
15 3 V
15 -18 V
15 -20 V
15 -34 V
15 -39 V
15 -44 V
15 -44 V
15 -47 V
15 -46 V
15 -47 V
15 -42 V
15 -38 V
15 -33 V
15 -34 V
15 -25 V
15 -26 V
15 -20 V
15 -19 V
15 -14 V
15 -15 V
15 -10 V
15 -10 V
15 -6 V
15 -8 V
15 -5 V
15 -4 V
15 -4 V
15 -3 V
15 -2 V
15 -2 V
15 -2 V
15 -1 V
15 -2 V
15 -1 V
15 0 V
15 -1 V
15 -1 V
15 0 V
15 0 V
15 0 V
15 -1 V
15 0 V
15 0 V
15 0 V
stroke
grestore
end
showpage
}}%
\put(1950,50){\makebox(0,0){$\alpha$}}%
\put(100,1180){%
\makebox(0,0)[b]{\shortstack{$\rho(\alpha)$}}%
}%
\put(3382,200){\makebox(0,0){3}}%
\put(2905,200){\makebox(0,0){2}}%
\put(2427,200){\makebox(0,0){1}}%
\put(1950,200){\makebox(0,0){0}}%
\put(1473,200){\makebox(0,0){-1}}%
\put(995,200){\makebox(0,0){-2}}%
\put(518,200){\makebox(0,0){-3}}%
\put(400,2060){\makebox(0,0)[r]{0.4}}%
\put(400,1840){\makebox(0,0)[r]{0.35}}%
\put(400,1620){\makebox(0,0)[r]{0.3}}%
\put(400,1400){\makebox(0,0)[r]{0.25}}%
\put(400,1180){\makebox(0,0)[r]{0.2}}%
\put(400,960){\makebox(0,0)[r]{0.15}}%
\put(400,740){\makebox(0,0)[r]{0.1}}%
\put(400,520){\makebox(0,0)[r]{0.05}}%
\put(400,300){\makebox(0,0)[r]{0}}%
\end{picture}%
\endgroup

%% file: outplanegraph.tex
\begingroup%
  \makeatletter%
  \newcommand{\GNUPLOTspecial}{%
    \@sanitize\catcode`\%=14\relax\special}%
  \setlength{\unitlength}{0.1bp}%
{\GNUPLOTspecial{!
/gnudict 256 dict def
gnudict begin
/Color false def
/Solid false def
/gnulinewidth 5.000 def
/userlinewidth gnulinewidth def
/vshift -33 def
/dl {10 mul} def
/hpt_ 31.5 def
/vpt_ 31.5 def
/hpt hpt_ def
/vpt vpt_ def
/M {moveto} bind def
/L {lineto} bind def
/R {rmoveto} bind def
/V {rlineto} bind def
/vpt2 vpt 2 mul def
/hpt2 hpt 2 mul def
/Lshow { currentpoint stroke M
  0 vshift R show } def
/Rshow { currentpoint stroke M
  dup stringwidth pop neg vshift R show } def
/Cshow { currentpoint stroke M
  dup stringwidth pop -2 div vshift R show } def
/UP { dup vpt_ mul /vpt exch def hpt_ mul /hpt exch def
  /hpt2 hpt 2 mul def /vpt2 vpt 2 mul def } def
/DL { Color {setrgbcolor Solid {pop []} if 0 setdash }
 {pop pop pop Solid {pop []} if 0 setdash} ifelse } def
/BL { stroke userlinewidth 2 mul setlinewidth } def
/AL { stroke userlinewidth 2 div setlinewidth } def
/UL { dup gnulinewidth mul /userlinewidth exch def
      10 mul /udl exch def } def
/PL { stroke userlinewidth setlinewidth } def
/LTb { BL [] 0 0 0 DL } def
/LTa { AL [1 udl mul 2 udl mul] 0 setdash 0 0 0 setrgbcolor } def
/LT0 { PL [] 1 0 0 DL } def
/LT1 { PL [4 dl 2 dl] 0 1 0 DL } def
/LT2 { PL [2 dl 3 dl] 0 0 1 DL } def
/LT3 { PL [1 dl 1.5 dl] 1 0 1 DL } def
/LT4 { PL [5 dl 2 dl 1 dl 2 dl] 0 1 1 DL } def
/LT5 { PL [4 dl 3 dl 1 dl 3 dl] 1 1 0 DL } def
/LT6 { PL [2 dl 2 dl 2 dl 4 dl] 0 0 0 DL } def
/LT7 { PL [2 dl 2 dl 2 dl 2 dl 2 dl 4 dl] 1 0.3 0 DL } def
/LT8 { PL [2 dl 2 dl 2 dl 2 dl 2 dl 2 dl 2 dl 4 dl] 0.5 0.5 0.5 DL } def
/Pnt { stroke [] 0 setdash
   gsave 1 setlinecap M 0 0 V stroke grestore } def
/Dia { stroke [] 0 setdash 2 copy vpt add M
  hpt neg vpt neg V hpt vpt neg V
  hpt vpt V hpt neg vpt V closepath stroke
  Pnt } def
/Pls { stroke [] 0 setdash vpt sub M 0 vpt2 V
  currentpoint stroke M
  hpt neg vpt neg R hpt2 0 V stroke
  } def
/Box { stroke [] 0 setdash 2 copy exch hpt sub exch vpt add M
  0 vpt2 neg V hpt2 0 V 0 vpt2 V
  hpt2 neg 0 V closepath stroke
  Pnt } def
/Crs { stroke [] 0 setdash exch hpt sub exch vpt add M
  hpt2 vpt2 neg V currentpoint stroke M
  hpt2 neg 0 R hpt2 vpt2 V stroke } def
/TriU { stroke [] 0 setdash 2 copy vpt 1.12 mul add M
  hpt neg vpt -1.62 mul V
  hpt 2 mul 0 V
  hpt neg vpt 1.62 mul V closepath stroke
  Pnt  } def
/Star { 2 copy Pls Crs } def
/BoxF { stroke [] 0 setdash exch hpt sub exch vpt add M
  0 vpt2 neg V  hpt2 0 V  0 vpt2 V
  hpt2 neg 0 V  closepath fill } def
/TriUF { stroke [] 0 setdash vpt 1.12 mul add M
  hpt neg vpt -1.62 mul V
  hpt 2 mul 0 V
  hpt neg vpt 1.62 mul V closepath fill } def
/TriD { stroke [] 0 setdash 2 copy vpt 1.12 mul sub M
  hpt neg vpt 1.62 mul V
  hpt 2 mul 0 V
  hpt neg vpt -1.62 mul V closepath stroke
  Pnt  } def
/TriDF { stroke [] 0 setdash vpt 1.12 mul sub M
  hpt neg vpt 1.62 mul V
  hpt 2 mul 0 V
  hpt neg vpt -1.62 mul V closepath fill} def
/DiaF { stroke [] 0 setdash vpt add M
  hpt neg vpt neg V hpt vpt neg V
  hpt vpt V hpt neg vpt V closepath fill } def
/Pent { stroke [] 0 setdash 2 copy gsave
  translate 0 hpt M 4 {72 rotate 0 hpt L} repeat
  closepath stroke grestore Pnt } def
/PentF { stroke [] 0 setdash gsave
  translate 0 hpt M 4 {72 rotate 0 hpt L} repeat
  closepath fill grestore } def
/Circle { stroke [] 0 setdash 2 copy
  hpt 0 360 arc stroke Pnt } def
/CircleF { stroke [] 0 setdash hpt 0 360 arc fill } def
/C0 { BL [] 0 setdash 2 copy moveto vpt 90 450  arc } bind def
/C1 { BL [] 0 setdash 2 copy        moveto
       2 copy  vpt 0 90 arc closepath fill
               vpt 0 360 arc closepath } bind def
/C2 { BL [] 0 setdash 2 copy moveto
       2 copy  vpt 90 180 arc closepath fill
               vpt 0 360 arc closepath } bind def
/C3 { BL [] 0 setdash 2 copy moveto
       2 copy  vpt 0 180 arc closepath fill
               vpt 0 360 arc closepath } bind def
/C4 { BL [] 0 setdash 2 copy moveto
       2 copy  vpt 180 270 arc closepath fill
               vpt 0 360 arc closepath } bind def
/C5 { BL [] 0 setdash 2 copy moveto
       2 copy  vpt 0 90 arc
       2 copy moveto
       2 copy  vpt 180 270 arc closepath fill
               vpt 0 360 arc } bind def
/C6 { BL [] 0 setdash 2 copy moveto
      2 copy  vpt 90 270 arc closepath fill
              vpt 0 360 arc closepath } bind def
/C7 { BL [] 0 setdash 2 copy moveto
      2 copy  vpt 0 270 arc closepath fill
              vpt 0 360 arc closepath } bind def
/C8 { BL [] 0 setdash 2 copy moveto
      2 copy vpt 270 360 arc closepath fill
              vpt 0 360 arc closepath } bind def
/C9 { BL [] 0 setdash 2 copy moveto
      2 copy  vpt 270 450 arc closepath fill
              vpt 0 360 arc closepath } bind def
/C10 { BL [] 0 setdash 2 copy 2 copy moveto vpt 270 360 arc closepath fill
       2 copy moveto
       2 copy vpt 90 180 arc closepath fill
               vpt 0 360 arc closepath } bind def
/C11 { BL [] 0 setdash 2 copy moveto
       2 copy  vpt 0 180 arc closepath fill
       2 copy moveto
       2 copy  vpt 270 360 arc closepath fill
               vpt 0 360 arc closepath } bind def
/C12 { BL [] 0 setdash 2 copy moveto
       2 copy  vpt 180 360 arc closepath fill
               vpt 0 360 arc closepath } bind def
/C13 { BL [] 0 setdash  2 copy moveto
       2 copy  vpt 0 90 arc closepath fill
       2 copy moveto
       2 copy  vpt 180 360 arc closepath fill
               vpt 0 360 arc closepath } bind def
/C14 { BL [] 0 setdash 2 copy moveto
       2 copy  vpt 90 360 arc closepath fill
               vpt 0 360 arc } bind def
/C15 { BL [] 0 setdash 2 copy vpt 0 360 arc closepath fill
               vpt 0 360 arc closepath } bind def
/Rec   { newpath 4 2 roll moveto 1 index 0 rlineto 0 exch rlineto
       neg 0 rlineto closepath } bind def
/Square { dup Rec } bind def
/Bsquare { vpt sub exch vpt sub exch vpt2 Square } bind def
/S0 { BL [] 0 setdash 2 copy moveto 0 vpt rlineto BL Bsquare } bind def
/S1 { BL [] 0 setdash 2 copy vpt Square fill Bsquare } bind def
/S2 { BL [] 0 setdash 2 copy exch vpt sub exch vpt Square fill Bsquare } bind def
/S3 { BL [] 0 setdash 2 copy exch vpt sub exch vpt2 vpt Rec fill Bsquare } bind def
/S4 { BL [] 0 setdash 2 copy exch vpt sub exch vpt sub vpt Square fill Bsquare } bind def
/S5 { BL [] 0 setdash 2 copy 2 copy vpt Square fill
       exch vpt sub exch vpt sub vpt Square fill Bsquare } bind def
/S6 { BL [] 0 setdash 2 copy exch vpt sub exch vpt sub vpt vpt2 Rec fill Bsquare } bind def
/S7 { BL [] 0 setdash 2 copy exch vpt sub exch vpt sub vpt vpt2 Rec fill
       2 copy vpt Square fill
       Bsquare } bind def
/S8 { BL [] 0 setdash 2 copy vpt sub vpt Square fill Bsquare } bind def
/S9 { BL [] 0 setdash 2 copy vpt sub vpt vpt2 Rec fill Bsquare } bind def
/S10 { BL [] 0 setdash 2 copy vpt sub vpt Square fill 2 copy exch vpt sub exch vpt Square fill
       Bsquare } bind def
/S11 { BL [] 0 setdash 2 copy vpt sub vpt Square fill 2 copy exch vpt sub exch vpt2 vpt Rec fill
       Bsquare } bind def
/S12 { BL [] 0 setdash 2 copy exch vpt sub exch vpt sub vpt2 vpt Rec fill Bsquare } bind def
/S13 { BL [] 0 setdash 2 copy exch vpt sub exch vpt sub vpt2 vpt Rec fill
       2 copy vpt Square fill Bsquare } bind def
/S14 { BL [] 0 setdash 2 copy exch vpt sub exch vpt sub vpt2 vpt Rec fill
       2 copy exch vpt sub exch vpt Square fill Bsquare } bind def
/S15 { BL [] 0 setdash 2 copy Bsquare fill Bsquare } bind def
/D0 { gsave translate 45 rotate 0 0 S0 stroke grestore } bind def
/D1 { gsave translate 45 rotate 0 0 S1 stroke grestore } bind def
/D2 { gsave translate 45 rotate 0 0 S2 stroke grestore } bind def
/D3 { gsave translate 45 rotate 0 0 S3 stroke grestore } bind def
/D4 { gsave translate 45 rotate 0 0 S4 stroke grestore } bind def
/D5 { gsave translate 45 rotate 0 0 S5 stroke grestore } bind def
/D6 { gsave translate 45 rotate 0 0 S6 stroke grestore } bind def
/D7 { gsave translate 45 rotate 0 0 S7 stroke grestore } bind def
/D8 { gsave translate 45 rotate 0 0 S8 stroke grestore } bind def
/D9 { gsave translate 45 rotate 0 0 S9 stroke grestore } bind def
/D10 { gsave translate 45 rotate 0 0 S10 stroke grestore } bind def
/D11 { gsave translate 45 rotate 0 0 S11 stroke grestore } bind def
/D12 { gsave translate 45 rotate 0 0 S12 stroke grestore } bind def
/D13 { gsave translate 45 rotate 0 0 S13 stroke grestore } bind def
/D14 { gsave translate 45 rotate 0 0 S14 stroke grestore } bind def
/D15 { gsave translate 45 rotate 0 0 S15 stroke grestore } bind def
/DiaE { stroke [] 0 setdash vpt add M
  hpt neg vpt neg V hpt vpt neg V
  hpt vpt V hpt neg vpt V closepath stroke } def
/BoxE { stroke [] 0 setdash exch hpt sub exch vpt add M
  0 vpt2 neg V hpt2 0 V 0 vpt2 V
  hpt2 neg 0 V closepath stroke } def
/TriUE { stroke [] 0 setdash vpt 1.12 mul add M
  hpt neg vpt -1.62 mul V
  hpt 2 mul 0 V
  hpt neg vpt 1.62 mul V closepath stroke } def
/TriDE { stroke [] 0 setdash vpt 1.12 mul sub M
  hpt neg vpt 1.62 mul V
  hpt 2 mul 0 V
  hpt neg vpt -1.62 mul V closepath stroke } def
/PentE { stroke [] 0 setdash gsave
  translate 0 hpt M 4 {72 rotate 0 hpt L} repeat
  closepath stroke grestore } def
/CircE { stroke [] 0 setdash 
  hpt 0 360 arc stroke } def
/Opaque { gsave closepath 1 setgray fill grestore 0 setgray closepath } def
/DiaW { stroke [] 0 setdash vpt add M
  hpt neg vpt neg V hpt vpt neg V
  hpt vpt V hpt neg vpt V Opaque stroke } def
/BoxW { stroke [] 0 setdash exch hpt sub exch vpt add M
  0 vpt2 neg V hpt2 0 V 0 vpt2 V
  hpt2 neg 0 V Opaque stroke } def
/TriUW { stroke [] 0 setdash vpt 1.12 mul add M
  hpt neg vpt -1.62 mul V
  hpt 2 mul 0 V
  hpt neg vpt 1.62 mul V Opaque stroke } def
/TriDW { stroke [] 0 setdash vpt 1.12 mul sub M
  hpt neg vpt 1.62 mul V
  hpt 2 mul 0 V
  hpt neg vpt -1.62 mul V Opaque stroke } def
/PentW { stroke [] 0 setdash gsave
  translate 0 hpt M 4 {72 rotate 0 hpt L} repeat
  Opaque stroke grestore } def
/CircW { stroke [] 0 setdash 
  hpt 0 360 arc Opaque stroke } def
/BoxFill { gsave Rec 1 setgray fill grestore } def
end
}}%
\begin{picture}(3600,2160)(0,0)%
{\GNUPLOTspecial{"
gnudict begin
gsave
0 0 translate
0.100 0.100 scale
0 setgray
newpath
1.000 UL
LTb
450 300 M
63 0 V
2937 0 R
-63 0 V
450 520 M
63 0 V
2937 0 R
-63 0 V
450 740 M
63 0 V
2937 0 R
-63 0 V
450 960 M
63 0 V
2937 0 R
-63 0 V
450 1180 M
63 0 V
2937 0 R
-63 0 V
450 1400 M
63 0 V
2937 0 R
-63 0 V
450 1620 M
63 0 V
2937 0 R
-63 0 V
450 1840 M
63 0 V
2937 0 R
-63 0 V
450 2060 M
63 0 V
2937 0 R
-63 0 V
450 300 M
0 63 V
0 1697 R
0 -63 V
825 300 M
0 63 V
0 1697 R
0 -63 V
1200 300 M
0 63 V
0 1697 R
0 -63 V
1575 300 M
0 63 V
0 1697 R
0 -63 V
1950 300 M
0 63 V
0 1697 R
0 -63 V
2325 300 M
0 63 V
0 1697 R
0 -63 V
2700 300 M
0 63 V
0 1697 R
0 -63 V
3075 300 M
0 63 V
0 1697 R
0 -63 V
3450 300 M
0 63 V
0 1697 R
0 -63 V
1.000 UL
LTb
450 300 M
3000 0 V
0 1760 V
-3000 0 V
450 300 L
1.000 UL
LT0
450 418 M
209 52 V
520 399 V
261 288 V
130 93 V
130 10 V
261 -158 V
2221 914 L
2742 692 L
3263 582 L
187 -23 V
1.000 UL
LT1
1179 621 M
261 321 V
260 498 V
130 17 V
131 -127 V
2221 982 L
2742 718 L
3263 595 L
187 -24 V
1.000 UL
LT2
1179 612 M
261 176 V
260 480 V
130 383 V
65 -27 V
66 -189 V
2221 995 L
2481 832 L
2742 718 L
3263 617 L
187 -32 V
1.000 UL
LT3
1440 775 M
260 449 V
65 176 V
65 295 V
33 105 V
32 -114 V
66 -229 V
2221 995 L
0.600 UP
1.000 UL
LT4
659 469 M
0 2 V
-31 -2 R
62 0 V
-62 2 R
62 0 V
489 397 R
0 2 V
-31 -2 R
62 0 V
-62 2 R
62 0 V
230 286 R
0 2 V
-31 -2 R
62 0 V
-62 2 R
62 0 V
99 88 R
0 9 V
-31 -9 R
62 0 V
-62 9 R
62 0 V
99 4 R
0 2 V
-31 -2 R
62 0 V
-62 2 R
62 0 V
230 -160 R
0 2 V
-31 -2 R
62 0 V
-62 2 R
62 0 V
2221 913 M
0 1 V
-31 -1 R
62 0 V
-62 1 R
62 0 V
2742 683 M
0 17 V
-31 -17 R
62 0 V
-62 17 R
62 0 V
3263 573 M
0 17 V
-31 -17 R
62 0 V
-62 17 R
62 0 V
659 470 Pls
1179 869 Pls
1440 1157 Pls
1570 1250 Pls
1700 1260 Pls
1961 1102 Pls
2221 914 Pls
2742 692 Pls
3263 582 Pls
0.600 UP
1.000 UL
LT5
1179 612 M
0 18 V
-31 -18 R
62 0 V
-62 18 R
62 0 V
230 295 R
0 35 V
-31 -35 R
62 0 V
-62 35 R
62 0 V
229 462 R
0 35 V
-31 -35 R
62 0 V
-62 35 R
62 0 V
99 -22 R
0 44 V
-31 -44 R
62 0 V
-62 44 R
62 0 V
100 -158 R
0 17 V
-31 -17 R
62 0 V
-62 17 R
62 0 V
2221 973 M
0 18 V
-31 -18 R
62 0 V
-62 18 R
62 0 V
2742 714 M
0 8 V
-31 -8 R
62 0 V
-62 8 R
62 0 V
3263 590 M
0 9 V
-31 -9 R
62 0 V
-62 9 R
62 0 V
1179 621 Crs
1440 942 Crs
1700 1440 Crs
1830 1457 Crs
1961 1330 Crs
2221 982 Crs
2742 718 Crs
3263 595 Crs
0.600 UP
1.000 UL
LT6
1179 590 M
0 44 V
-31 -44 R
62 0 V
-62 44 R
62 0 V
230 128 R
0 53 V
-31 -53 R
62 0 V
-62 53 R
62 0 V
229 431 R
0 44 V
-31 -44 R
62 0 V
-62 44 R
62 0 V
99 343 R
0 35 V
-31 -35 R
62 0 V
-62 35 R
62 0 V
34 -88 R
0 88 V
-31 -88 R
62 0 V
-62 88 R
62 0 V
35 -255 R
0 44 V
-31 -44 R
62 0 V
-62 44 R
62 0 V
2221 982 M
0 26 V
-31 -26 R
62 0 V
-62 26 R
62 0 V
2481 824 M
0 17 V
-31 -17 R
62 0 V
-62 17 R
62 0 V
2742 709 M
0 18 V
-31 -18 R
62 0 V
-62 18 R
62 0 V
3263 612 M
0 9 V
-31 -9 R
62 0 V
-62 9 R
62 0 V
1179 612 Star
1440 788 Star
1700 1268 Star
1830 1651 Star
1895 1624 Star
1961 1435 Star
2221 995 Star
2481 832 Star
2742 718 Star
3263 617 Star
0.600 UP
1.000 UL
LT7
1440 749 M
0 53 V
-31 -53 R
62 0 V
-62 53 R
62 0 V
229 391 R
0 62 V
-31 -62 R
62 0 V
-62 62 R
62 0 V
34 105 R
0 80 V
-31 -80 R
62 0 V
-62 80 R
62 0 V
34 224 R
0 62 V
-31 -62 R
62 0 V
-62 62 R
62 0 V
2 26 R
0 97 V
-31 -97 R
62 0 V
-62 97 R
62 0 V
1 -220 R
0 114 V
-31 -114 R
62 0 V
-62 114 R
62 0 V
35 -321 R
0 70 V
-31 -70 R
62 0 V
-62 70 R
62 0 V
2221 978 M
0 35 V
-31 -35 R
62 0 V
-62 35 R
62 0 V
1440 775 Box
1700 1224 Box
1765 1400 Box
1830 1695 Box
1863 1800 Box
1895 1686 Box
1961 1457 Box
2221 995 Box
stroke
grestore
end
showpage
}}%
\put(1950,50){\makebox(0,0){$\gamma$}}%
\put(100,1180){%
\makebox(0,0)[b]{\shortstack{$C_o$}}%
}%
\put(3450,200){\makebox(0,0){0.65}}%
\put(3075,200){\makebox(0,0){0.6}}%
\put(2700,200){\makebox(0,0){0.55}}%
\put(2325,200){\makebox(0,0){0.5}}%
\put(1950,200){\makebox(0,0){0.45}}%
\put(1575,200){\makebox(0,0){0.4}}%
\put(1200,200){\makebox(0,0){0.35}}%
\put(825,200){\makebox(0,0){0.3}}%
\put(450,200){\makebox(0,0){0.25}}%
\put(400,2060){\makebox(0,0)[r]{0.4}}%
\put(400,1840){\makebox(0,0)[r]{0.35}}%
\put(400,1620){\makebox(0,0)[r]{0.3}}%
\put(400,1400){\makebox(0,0)[r]{0.25}}%
\put(400,1180){\makebox(0,0)[r]{0.2}}%
\put(400,960){\makebox(0,0)[r]{0.15}}%
\put(400,740){\makebox(0,0)[r]{0.1}}%
\put(400,520){\makebox(0,0)[r]{0.05}}%
\put(400,300){\makebox(0,0)[r]{0}}%
\end{picture}%
\endgroup

%% file: massgraph.tex
\begingroup%
  \makeatletter%
  \newcommand{\GNUPLOTspecial}{%
    \@sanitize\catcode`\%=14\relax\special}%
  \setlength{\unitlength}{0.1bp}%
{\GNUPLOTspecial{!
/gnudict 256 dict def
gnudict begin
/Color false def
/Solid false def
/gnulinewidth 5.000 def
/userlinewidth gnulinewidth def
/vshift -33 def
/dl {10 mul} def
/hpt_ 31.5 def
/vpt_ 31.5 def
/hpt hpt_ def
/vpt vpt_ def
/M {moveto} bind def
/L {lineto} bind def
/R {rmoveto} bind def
/V {rlineto} bind def
/vpt2 vpt 2 mul def
/hpt2 hpt 2 mul def
/Lshow { currentpoint stroke M
  0 vshift R show } def
/Rshow { currentpoint stroke M
  dup stringwidth pop neg vshift R show } def
/Cshow { currentpoint stroke M
  dup stringwidth pop -2 div vshift R show } def
/UP { dup vpt_ mul /vpt exch def hpt_ mul /hpt exch def
  /hpt2 hpt 2 mul def /vpt2 vpt 2 mul def } def
/DL { Color {setrgbcolor Solid {pop []} if 0 setdash }
 {pop pop pop Solid {pop []} if 0 setdash} ifelse } def
/BL { stroke userlinewidth 2 mul setlinewidth } def
/AL { stroke userlinewidth 2 div setlinewidth } def
/UL { dup gnulinewidth mul /userlinewidth exch def
      10 mul /udl exch def } def
/PL { stroke userlinewidth setlinewidth } def
/LTb { BL [] 0 0 0 DL } def
/LTa { AL [1 udl mul 2 udl mul] 0 setdash 0 0 0 setrgbcolor } def
/LT0 { PL [] 1 0 0 DL } def
/LT1 { PL [4 dl 2 dl] 0 1 0 DL } def
/LT2 { PL [2 dl 3 dl] 0 0 1 DL } def
/LT3 { PL [1 dl 1.5 dl] 1 0 1 DL } def
/LT4 { PL [5 dl 2 dl 1 dl 2 dl] 0 1 1 DL } def
/LT5 { PL [4 dl 3 dl 1 dl 3 dl] 1 1 0 DL } def
/LT6 { PL [2 dl 2 dl 2 dl 4 dl] 0 0 0 DL } def
/LT7 { PL [2 dl 2 dl 2 dl 2 dl 2 dl 4 dl] 1 0.3 0 DL } def
/LT8 { PL [2 dl 2 dl 2 dl 2 dl 2 dl 2 dl 2 dl 4 dl] 0.5 0.5 0.5 DL } def
/Pnt { stroke [] 0 setdash
   gsave 1 setlinecap M 0 0 V stroke grestore } def
/Dia { stroke [] 0 setdash 2 copy vpt add M
  hpt neg vpt neg V hpt vpt neg V
  hpt vpt V hpt neg vpt V closepath stroke
  Pnt } def
/Pls { stroke [] 0 setdash vpt sub M 0 vpt2 V
  currentpoint stroke M
  hpt neg vpt neg R hpt2 0 V stroke
  } def
/Box { stroke [] 0 setdash 2 copy exch hpt sub exch vpt add M
  0 vpt2 neg V hpt2 0 V 0 vpt2 V
  hpt2 neg 0 V closepath stroke
  Pnt } def
/Crs { stroke [] 0 setdash exch hpt sub exch vpt add M
  hpt2 vpt2 neg V currentpoint stroke M
  hpt2 neg 0 R hpt2 vpt2 V stroke } def
/TriU { stroke [] 0 setdash 2 copy vpt 1.12 mul add M
  hpt neg vpt -1.62 mul V
  hpt 2 mul 0 V
  hpt neg vpt 1.62 mul V closepath stroke
  Pnt  } def
/Star { 2 copy Pls Crs } def
/BoxF { stroke [] 0 setdash exch hpt sub exch vpt add M
  0 vpt2 neg V  hpt2 0 V  0 vpt2 V
  hpt2 neg 0 V  closepath fill } def
/TriUF { stroke [] 0 setdash vpt 1.12 mul add M
  hpt neg vpt -1.62 mul V
  hpt 2 mul 0 V
  hpt neg vpt 1.62 mul V closepath fill } def
/TriD { stroke [] 0 setdash 2 copy vpt 1.12 mul sub M
  hpt neg vpt 1.62 mul V
  hpt 2 mul 0 V
  hpt neg vpt -1.62 mul V closepath stroke
  Pnt  } def
/TriDF { stroke [] 0 setdash vpt 1.12 mul sub M
  hpt neg vpt 1.62 mul V
  hpt 2 mul 0 V
  hpt neg vpt -1.62 mul V closepath fill} def
/DiaF { stroke [] 0 setdash vpt add M
  hpt neg vpt neg V hpt vpt neg V
  hpt vpt V hpt neg vpt V closepath fill } def
/Pent { stroke [] 0 setdash 2 copy gsave
  translate 0 hpt M 4 {72 rotate 0 hpt L} repeat
  closepath stroke grestore Pnt } def
/PentF { stroke [] 0 setdash gsave
  translate 0 hpt M 4 {72 rotate 0 hpt L} repeat
  closepath fill grestore } def
/Circle { stroke [] 0 setdash 2 copy
  hpt 0 360 arc stroke Pnt } def
/CircleF { stroke [] 0 setdash hpt 0 360 arc fill } def
/C0 { BL [] 0 setdash 2 copy moveto vpt 90 450  arc } bind def
/C1 { BL [] 0 setdash 2 copy        moveto
       2 copy  vpt 0 90 arc closepath fill
               vpt 0 360 arc closepath } bind def
/C2 { BL [] 0 setdash 2 copy moveto
       2 copy  vpt 90 180 arc closepath fill
               vpt 0 360 arc closepath } bind def
/C3 { BL [] 0 setdash 2 copy moveto
       2 copy  vpt 0 180 arc closepath fill
               vpt 0 360 arc closepath } bind def
/C4 { BL [] 0 setdash 2 copy moveto
       2 copy  vpt 180 270 arc closepath fill
               vpt 0 360 arc closepath } bind def
/C5 { BL [] 0 setdash 2 copy moveto
       2 copy  vpt 0 90 arc
       2 copy moveto
       2 copy  vpt 180 270 arc closepath fill
               vpt 0 360 arc } bind def
/C6 { BL [] 0 setdash 2 copy moveto
      2 copy  vpt 90 270 arc closepath fill
              vpt 0 360 arc closepath } bind def
/C7 { BL [] 0 setdash 2 copy moveto
      2 copy  vpt 0 270 arc closepath fill
              vpt 0 360 arc closepath } bind def
/C8 { BL [] 0 setdash 2 copy moveto
      2 copy vpt 270 360 arc closepath fill
              vpt 0 360 arc closepath } bind def
/C9 { BL [] 0 setdash 2 copy moveto
      2 copy  vpt 270 450 arc closepath fill
              vpt 0 360 arc closepath } bind def
/C10 { BL [] 0 setdash 2 copy 2 copy moveto vpt 270 360 arc closepath fill
       2 copy moveto
       2 copy vpt 90 180 arc closepath fill
               vpt 0 360 arc closepath } bind def
/C11 { BL [] 0 setdash 2 copy moveto
       2 copy  vpt 0 180 arc closepath fill
       2 copy moveto
       2 copy  vpt 270 360 arc closepath fill
               vpt 0 360 arc closepath } bind def
/C12 { BL [] 0 setdash 2 copy moveto
       2 copy  vpt 180 360 arc closepath fill
               vpt 0 360 arc closepath } bind def
/C13 { BL [] 0 setdash  2 copy moveto
       2 copy  vpt 0 90 arc closepath fill
       2 copy moveto
       2 copy  vpt 180 360 arc closepath fill
               vpt 0 360 arc closepath } bind def
/C14 { BL [] 0 setdash 2 copy moveto
       2 copy  vpt 90 360 arc closepath fill
               vpt 0 360 arc } bind def
/C15 { BL [] 0 setdash 2 copy vpt 0 360 arc closepath fill
               vpt 0 360 arc closepath } bind def
/Rec   { newpath 4 2 roll moveto 1 index 0 rlineto 0 exch rlineto
       neg 0 rlineto closepath } bind def
/Square { dup Rec } bind def
/Bsquare { vpt sub exch vpt sub exch vpt2 Square } bind def
/S0 { BL [] 0 setdash 2 copy moveto 0 vpt rlineto BL Bsquare } bind def
/S1 { BL [] 0 setdash 2 copy vpt Square fill Bsquare } bind def
/S2 { BL [] 0 setdash 2 copy exch vpt sub exch vpt Square fill Bsquare } bind def
/S3 { BL [] 0 setdash 2 copy exch vpt sub exch vpt2 vpt Rec fill Bsquare } bind def
/S4 { BL [] 0 setdash 2 copy exch vpt sub exch vpt sub vpt Square fill Bsquare } bind def
/S5 { BL [] 0 setdash 2 copy 2 copy vpt Square fill
       exch vpt sub exch vpt sub vpt Square fill Bsquare } bind def
/S6 { BL [] 0 setdash 2 copy exch vpt sub exch vpt sub vpt vpt2 Rec fill Bsquare } bind def
/S7 { BL [] 0 setdash 2 copy exch vpt sub exch vpt sub vpt vpt2 Rec fill
       2 copy vpt Square fill
       Bsquare } bind def
/S8 { BL [] 0 setdash 2 copy vpt sub vpt Square fill Bsquare } bind def
/S9 { BL [] 0 setdash 2 copy vpt sub vpt vpt2 Rec fill Bsquare } bind def
/S10 { BL [] 0 setdash 2 copy vpt sub vpt Square fill 2 copy exch vpt sub exch vpt Square fill
       Bsquare } bind def
/S11 { BL [] 0 setdash 2 copy vpt sub vpt Square fill 2 copy exch vpt sub exch vpt2 vpt Rec fill
       Bsquare } bind def
/S12 { BL [] 0 setdash 2 copy exch vpt sub exch vpt sub vpt2 vpt Rec fill Bsquare } bind def
/S13 { BL [] 0 setdash 2 copy exch vpt sub exch vpt sub vpt2 vpt Rec fill
       2 copy vpt Square fill Bsquare } bind def
/S14 { BL [] 0 setdash 2 copy exch vpt sub exch vpt sub vpt2 vpt Rec fill
       2 copy exch vpt sub exch vpt Square fill Bsquare } bind def
/S15 { BL [] 0 setdash 2 copy Bsquare fill Bsquare } bind def
/D0 { gsave translate 45 rotate 0 0 S0 stroke grestore } bind def
/D1 { gsave translate 45 rotate 0 0 S1 stroke grestore } bind def
/D2 { gsave translate 45 rotate 0 0 S2 stroke grestore } bind def
/D3 { gsave translate 45 rotate 0 0 S3 stroke grestore } bind def
/D4 { gsave translate 45 rotate 0 0 S4 stroke grestore } bind def
/D5 { gsave translate 45 rotate 0 0 S5 stroke grestore } bind def
/D6 { gsave translate 45 rotate 0 0 S6 stroke grestore } bind def
/D7 { gsave translate 45 rotate 0 0 S7 stroke grestore } bind def
/D8 { gsave translate 45 rotate 0 0 S8 stroke grestore } bind def
/D9 { gsave translate 45 rotate 0 0 S9 stroke grestore } bind def
/D10 { gsave translate 45 rotate 0 0 S10 stroke grestore } bind def
/D11 { gsave translate 45 rotate 0 0 S11 stroke grestore } bind def
/D12 { gsave translate 45 rotate 0 0 S12 stroke grestore } bind def
/D13 { gsave translate 45 rotate 0 0 S13 stroke grestore } bind def
/D14 { gsave translate 45 rotate 0 0 S14 stroke grestore } bind def
/D15 { gsave translate 45 rotate 0 0 S15 stroke grestore } bind def
/DiaE { stroke [] 0 setdash vpt add M
  hpt neg vpt neg V hpt vpt neg V
  hpt vpt V hpt neg vpt V closepath stroke } def
/BoxE { stroke [] 0 setdash exch hpt sub exch vpt add M
  0 vpt2 neg V hpt2 0 V 0 vpt2 V
  hpt2 neg 0 V closepath stroke } def
/TriUE { stroke [] 0 setdash vpt 1.12 mul add M
  hpt neg vpt -1.62 mul V
  hpt 2 mul 0 V
  hpt neg vpt 1.62 mul V closepath stroke } def
/TriDE { stroke [] 0 setdash vpt 1.12 mul sub M
  hpt neg vpt 1.62 mul V
  hpt 2 mul 0 V
  hpt neg vpt -1.62 mul V closepath stroke } def
/PentE { stroke [] 0 setdash gsave
  translate 0 hpt M 4 {72 rotate 0 hpt L} repeat
  closepath stroke grestore } def
/CircE { stroke [] 0 setdash 
  hpt 0 360 arc stroke } def
/Opaque { gsave closepath 1 setgray fill grestore 0 setgray closepath } def
/DiaW { stroke [] 0 setdash vpt add M
  hpt neg vpt neg V hpt vpt neg V
  hpt vpt V hpt neg vpt V Opaque stroke } def
/BoxW { stroke [] 0 setdash exch hpt sub exch vpt add M
  0 vpt2 neg V hpt2 0 V 0 vpt2 V
  hpt2 neg 0 V Opaque stroke } def
/TriUW { stroke [] 0 setdash vpt 1.12 mul add M
  hpt neg vpt -1.62 mul V
  hpt 2 mul 0 V
  hpt neg vpt 1.62 mul V Opaque stroke } def
/TriDW { stroke [] 0 setdash vpt 1.12 mul sub M
  hpt neg vpt 1.62 mul V
  hpt 2 mul 0 V
  hpt neg vpt -1.62 mul V Opaque stroke } def
/PentW { stroke [] 0 setdash gsave
  translate 0 hpt M 4 {72 rotate 0 hpt L} repeat
  Opaque stroke grestore } def
/CircW { stroke [] 0 setdash 
  hpt 0 360 arc Opaque stroke } def
/BoxFill { gsave Rec 1 setgray fill grestore } def
end
}}%
\begin{picture}(3600,2160)(0,0)%
{\GNUPLOTspecial{"
gnudict begin
gsave
0 0 translate
0.100 0.100 scale
0 setgray
newpath
1.000 UL
LTb
400 300 M
63 0 V
2987 0 R
-63 0 V
400 652 M
63 0 V
2987 0 R
-63 0 V
400 1004 M
63 0 V
2987 0 R
-63 0 V
400 1356 M
63 0 V
2987 0 R
-63 0 V
400 1708 M
63 0 V
2987 0 R
-63 0 V
400 2060 M
63 0 V
2987 0 R
-63 0 V
400 300 M
0 63 V
0 1697 R
0 -63 V
781 300 M
0 63 V
0 1697 R
0 -63 V
1163 300 M
0 63 V
0 1697 R
0 -63 V
1544 300 M
0 63 V
0 1697 R
0 -63 V
1925 300 M
0 63 V
0 1697 R
0 -63 V
2306 300 M
0 63 V
0 1697 R
0 -63 V
2687 300 M
0 63 V
0 1697 R
0 -63 V
3069 300 M
0 63 V
0 1697 R
0 -63 V
3450 300 M
0 63 V
0 1697 R
0 -63 V
1.000 UL
LTb
400 300 M
3050 0 V
0 1760 V
-3050 0 V
400 300 L
0.600 UP
1.000 UL
LT0
612 1892 M
529 -844 V
1406 715 L
1671 531 L
265 -18 V
264 47 V
612 1892 Pls
1141 1048 Pls
1406 715 Pls
1671 531 Pls
1936 513 Pls
2200 560 Pls
0.600 UP
1.000 UL
LT1
1406 938 M
1671 513 L
1803 408 L
133 18 V
264 92 V
1406 938 Crs
1671 513 Crs
1803 408 Crs
1936 426 Crs
2200 518 Crs
stroke
grestore
end
showpage
}}%
\put(1925,50){\makebox(0,0){$\gamma$}}%
\put(100,1180){%
\makebox(0,0)[b]{\shortstack{$a m_\mathrm{eff}(1)$}}%
}%
\put(3450,200){\makebox(0,0){0.65}}%
\put(3069,200){\makebox(0,0){0.6}}%
\put(2687,200){\makebox(0,0){0.55}}%
\put(2306,200){\makebox(0,0){0.5}}%
\put(1925,200){\makebox(0,0){0.45}}%
\put(1544,200){\makebox(0,0){0.4}}%
\put(1163,200){\makebox(0,0){0.35}}%
\put(781,200){\makebox(0,0){0.3}}%
\put(400,200){\makebox(0,0){0.25}}%
\put(350,2060){\makebox(0,0)[r]{5}}%
\put(350,1708){\makebox(0,0)[r]{4.5}}%
\put(350,1356){\makebox(0,0)[r]{4}}%
\put(350,1004){\makebox(0,0)[r]{3.5}}%
\put(350,652){\makebox(0,0)[r]{3}}%
\put(350,300){\makebox(0,0)[r]{2.5}}%
\end{picture}%
\endgroup

%% file: wilsongraph.tex
\begingroup%
  \makeatletter%
  \newcommand{\GNUPLOTspecial}{%
    \@sanitize\catcode`\%=14\relax\special}%
  \setlength{\unitlength}{0.1bp}%
{\GNUPLOTspecial{!
/gnudict 256 dict def
gnudict begin
/Color false def
/Solid false def
/gnulinewidth 5.000 def
/userlinewidth gnulinewidth def
/vshift -33 def
/dl {10 mul} def
/hpt_ 31.5 def
/vpt_ 31.5 def
/hpt hpt_ def
/vpt vpt_ def
/M {moveto} bind def
/L {lineto} bind def
/R {rmoveto} bind def
/V {rlineto} bind def
/vpt2 vpt 2 mul def
/hpt2 hpt 2 mul def
/Lshow { currentpoint stroke M
  0 vshift R show } def
/Rshow { currentpoint stroke M
  dup stringwidth pop neg vshift R show } def
/Cshow { currentpoint stroke M
  dup stringwidth pop -2 div vshift R show } def
/UP { dup vpt_ mul /vpt exch def hpt_ mul /hpt exch def
  /hpt2 hpt 2 mul def /vpt2 vpt 2 mul def } def
/DL { Color {setrgbcolor Solid {pop []} if 0 setdash }
 {pop pop pop Solid {pop []} if 0 setdash} ifelse } def
/BL { stroke userlinewidth 2 mul setlinewidth } def
/AL { stroke userlinewidth 2 div setlinewidth } def
/UL { dup gnulinewidth mul /userlinewidth exch def
      10 mul /udl exch def } def
/PL { stroke userlinewidth setlinewidth } def
/LTb { BL [] 0 0 0 DL } def
/LTa { AL [1 udl mul 2 udl mul] 0 setdash 0 0 0 setrgbcolor } def
/LT0 { PL [] 1 0 0 DL } def
/LT1 { PL [4 dl 2 dl] 0 1 0 DL } def
/LT2 { PL [2 dl 3 dl] 0 0 1 DL } def
/LT3 { PL [1 dl 1.5 dl] 1 0 1 DL } def
/LT4 { PL [5 dl 2 dl 1 dl 2 dl] 0 1 1 DL } def
/LT5 { PL [4 dl 3 dl 1 dl 3 dl] 1 1 0 DL } def
/LT6 { PL [2 dl 2 dl 2 dl 4 dl] 0 0 0 DL } def
/LT7 { PL [2 dl 2 dl 2 dl 2 dl 2 dl 4 dl] 1 0.3 0 DL } def
/LT8 { PL [2 dl 2 dl 2 dl 2 dl 2 dl 2 dl 2 dl 4 dl] 0.5 0.5 0.5 DL } def
/Pnt { stroke [] 0 setdash
   gsave 1 setlinecap M 0 0 V stroke grestore } def
/Dia { stroke [] 0 setdash 2 copy vpt add M
  hpt neg vpt neg V hpt vpt neg V
  hpt vpt V hpt neg vpt V closepath stroke
  Pnt } def
/Pls { stroke [] 0 setdash vpt sub M 0 vpt2 V
  currentpoint stroke M
  hpt neg vpt neg R hpt2 0 V stroke
  } def
/Box { stroke [] 0 setdash 2 copy exch hpt sub exch vpt add M
  0 vpt2 neg V hpt2 0 V 0 vpt2 V
  hpt2 neg 0 V closepath stroke
  Pnt } def
/Crs { stroke [] 0 setdash exch hpt sub exch vpt add M
  hpt2 vpt2 neg V currentpoint stroke M
  hpt2 neg 0 R hpt2 vpt2 V stroke } def
/TriU { stroke [] 0 setdash 2 copy vpt 1.12 mul add M
  hpt neg vpt -1.62 mul V
  hpt 2 mul 0 V
  hpt neg vpt 1.62 mul V closepath stroke
  Pnt  } def
/Star { 2 copy Pls Crs } def
/BoxF { stroke [] 0 setdash exch hpt sub exch vpt add M
  0 vpt2 neg V  hpt2 0 V  0 vpt2 V
  hpt2 neg 0 V  closepath fill } def
/TriUF { stroke [] 0 setdash vpt 1.12 mul add M
  hpt neg vpt -1.62 mul V
  hpt 2 mul 0 V
  hpt neg vpt 1.62 mul V closepath fill } def
/TriD { stroke [] 0 setdash 2 copy vpt 1.12 mul sub M
  hpt neg vpt 1.62 mul V
  hpt 2 mul 0 V
  hpt neg vpt -1.62 mul V closepath stroke
  Pnt  } def
/TriDF { stroke [] 0 setdash vpt 1.12 mul sub M
  hpt neg vpt 1.62 mul V
  hpt 2 mul 0 V
  hpt neg vpt -1.62 mul V closepath fill} def
/DiaF { stroke [] 0 setdash vpt add M
  hpt neg vpt neg V hpt vpt neg V
  hpt vpt V hpt neg vpt V closepath fill } def
/Pent { stroke [] 0 setdash 2 copy gsave
  translate 0 hpt M 4 {72 rotate 0 hpt L} repeat
  closepath stroke grestore Pnt } def
/PentF { stroke [] 0 setdash gsave
  translate 0 hpt M 4 {72 rotate 0 hpt L} repeat
  closepath fill grestore } def
/Circle { stroke [] 0 setdash 2 copy
  hpt 0 360 arc stroke Pnt } def
/CircleF { stroke [] 0 setdash hpt 0 360 arc fill } def
/C0 { BL [] 0 setdash 2 copy moveto vpt 90 450  arc } bind def
/C1 { BL [] 0 setdash 2 copy        moveto
       2 copy  vpt 0 90 arc closepath fill
               vpt 0 360 arc closepath } bind def
/C2 { BL [] 0 setdash 2 copy moveto
       2 copy  vpt 90 180 arc closepath fill
               vpt 0 360 arc closepath } bind def
/C3 { BL [] 0 setdash 2 copy moveto
       2 copy  vpt 0 180 arc closepath fill
               vpt 0 360 arc closepath } bind def
/C4 { BL [] 0 setdash 2 copy moveto
       2 copy  vpt 180 270 arc closepath fill
               vpt 0 360 arc closepath } bind def
/C5 { BL [] 0 setdash 2 copy moveto
       2 copy  vpt 0 90 arc
       2 copy moveto
       2 copy  vpt 180 270 arc closepath fill
               vpt 0 360 arc } bind def
/C6 { BL [] 0 setdash 2 copy moveto
      2 copy  vpt 90 270 arc closepath fill
              vpt 0 360 arc closepath } bind def
/C7 { BL [] 0 setdash 2 copy moveto
      2 copy  vpt 0 270 arc closepath fill
              vpt 0 360 arc closepath } bind def
/C8 { BL [] 0 setdash 2 copy moveto
      2 copy vpt 270 360 arc closepath fill
              vpt 0 360 arc closepath } bind def
/C9 { BL [] 0 setdash 2 copy moveto
      2 copy  vpt 270 450 arc closepath fill
              vpt 0 360 arc closepath } bind def
/C10 { BL [] 0 setdash 2 copy 2 copy moveto vpt 270 360 arc closepath fill
       2 copy moveto
       2 copy vpt 90 180 arc closepath fill
               vpt 0 360 arc closepath } bind def
/C11 { BL [] 0 setdash 2 copy moveto
       2 copy  vpt 0 180 arc closepath fill
       2 copy moveto
       2 copy  vpt 270 360 arc closepath fill
               vpt 0 360 arc closepath } bind def
/C12 { BL [] 0 setdash 2 copy moveto
       2 copy  vpt 180 360 arc closepath fill
               vpt 0 360 arc closepath } bind def
/C13 { BL [] 0 setdash  2 copy moveto
       2 copy  vpt 0 90 arc closepath fill
       2 copy moveto
       2 copy  vpt 180 360 arc closepath fill
               vpt 0 360 arc closepath } bind def
/C14 { BL [] 0 setdash 2 copy moveto
       2 copy  vpt 90 360 arc closepath fill
               vpt 0 360 arc } bind def
/C15 { BL [] 0 setdash 2 copy vpt 0 360 arc closepath fill
               vpt 0 360 arc closepath } bind def
/Rec   { newpath 4 2 roll moveto 1 index 0 rlineto 0 exch rlineto
       neg 0 rlineto closepath } bind def
/Square { dup Rec } bind def
/Bsquare { vpt sub exch vpt sub exch vpt2 Square } bind def
/S0 { BL [] 0 setdash 2 copy moveto 0 vpt rlineto BL Bsquare } bind def
/S1 { BL [] 0 setdash 2 copy vpt Square fill Bsquare } bind def
/S2 { BL [] 0 setdash 2 copy exch vpt sub exch vpt Square fill Bsquare } bind def
/S3 { BL [] 0 setdash 2 copy exch vpt sub exch vpt2 vpt Rec fill Bsquare } bind def
/S4 { BL [] 0 setdash 2 copy exch vpt sub exch vpt sub vpt Square fill Bsquare } bind def
/S5 { BL [] 0 setdash 2 copy 2 copy vpt Square fill
       exch vpt sub exch vpt sub vpt Square fill Bsquare } bind def
/S6 { BL [] 0 setdash 2 copy exch vpt sub exch vpt sub vpt vpt2 Rec fill Bsquare } bind def
/S7 { BL [] 0 setdash 2 copy exch vpt sub exch vpt sub vpt vpt2 Rec fill
       2 copy vpt Square fill
       Bsquare } bind def
/S8 { BL [] 0 setdash 2 copy vpt sub vpt Square fill Bsquare } bind def
/S9 { BL [] 0 setdash 2 copy vpt sub vpt vpt2 Rec fill Bsquare } bind def
/S10 { BL [] 0 setdash 2 copy vpt sub vpt Square fill 2 copy exch vpt sub exch vpt Square fill
       Bsquare } bind def
/S11 { BL [] 0 setdash 2 copy vpt sub vpt Square fill 2 copy exch vpt sub exch vpt2 vpt Rec fill
       Bsquare } bind def
/S12 { BL [] 0 setdash 2 copy exch vpt sub exch vpt sub vpt2 vpt Rec fill Bsquare } bind def
/S13 { BL [] 0 setdash 2 copy exch vpt sub exch vpt sub vpt2 vpt Rec fill
       2 copy vpt Square fill Bsquare } bind def
/S14 { BL [] 0 setdash 2 copy exch vpt sub exch vpt sub vpt2 vpt Rec fill
       2 copy exch vpt sub exch vpt Square fill Bsquare } bind def
/S15 { BL [] 0 setdash 2 copy Bsquare fill Bsquare } bind def
/D0 { gsave translate 45 rotate 0 0 S0 stroke grestore } bind def
/D1 { gsave translate 45 rotate 0 0 S1 stroke grestore } bind def
/D2 { gsave translate 45 rotate 0 0 S2 stroke grestore } bind def
/D3 { gsave translate 45 rotate 0 0 S3 stroke grestore } bind def
/D4 { gsave translate 45 rotate 0 0 S4 stroke grestore } bind def
/D5 { gsave translate 45 rotate 0 0 S5 stroke grestore } bind def
/D6 { gsave translate 45 rotate 0 0 S6 stroke grestore } bind def
/D7 { gsave translate 45 rotate 0 0 S7 stroke grestore } bind def
/D8 { gsave translate 45 rotate 0 0 S8 stroke grestore } bind def
/D9 { gsave translate 45 rotate 0 0 S9 stroke grestore } bind def
/D10 { gsave translate 45 rotate 0 0 S10 stroke grestore } bind def
/D11 { gsave translate 45 rotate 0 0 S11 stroke grestore } bind def
/D12 { gsave translate 45 rotate 0 0 S12 stroke grestore } bind def
/D13 { gsave translate 45 rotate 0 0 S13 stroke grestore } bind def
/D14 { gsave translate 45 rotate 0 0 S14 stroke grestore } bind def
/D15 { gsave translate 45 rotate 0 0 S15 stroke grestore } bind def
/DiaE { stroke [] 0 setdash vpt add M
  hpt neg vpt neg V hpt vpt neg V
  hpt vpt V hpt neg vpt V closepath stroke } def
/BoxE { stroke [] 0 setdash exch hpt sub exch vpt add M
  0 vpt2 neg V hpt2 0 V 0 vpt2 V
  hpt2 neg 0 V closepath stroke } def
/TriUE { stroke [] 0 setdash vpt 1.12 mul add M
  hpt neg vpt -1.62 mul V
  hpt 2 mul 0 V
  hpt neg vpt 1.62 mul V closepath stroke } def
/TriDE { stroke [] 0 setdash vpt 1.12 mul sub M
  hpt neg vpt 1.62 mul V
  hpt 2 mul 0 V
  hpt neg vpt -1.62 mul V closepath stroke } def
/PentE { stroke [] 0 setdash gsave
  translate 0 hpt M 4 {72 rotate 0 hpt L} repeat
  closepath stroke grestore } def
/CircE { stroke [] 0 setdash 
  hpt 0 360 arc stroke } def
/Opaque { gsave closepath 1 setgray fill grestore 0 setgray closepath } def
/DiaW { stroke [] 0 setdash vpt add M
  hpt neg vpt neg V hpt vpt neg V
  hpt vpt V hpt neg vpt V Opaque stroke } def
/BoxW { stroke [] 0 setdash exch hpt sub exch vpt add M
  0 vpt2 neg V hpt2 0 V 0 vpt2 V
  hpt2 neg 0 V Opaque stroke } def
/TriUW { stroke [] 0 setdash vpt 1.12 mul add M
  hpt neg vpt -1.62 mul V
  hpt 2 mul 0 V
  hpt neg vpt 1.62 mul V Opaque stroke } def
/TriDW { stroke [] 0 setdash vpt 1.12 mul sub M
  hpt neg vpt 1.62 mul V
  hpt 2 mul 0 V
  hpt neg vpt -1.62 mul V Opaque stroke } def
/PentW { stroke [] 0 setdash gsave
  translate 0 hpt M 4 {72 rotate 0 hpt L} repeat
  Opaque stroke grestore } def
/CircW { stroke [] 0 setdash 
  hpt 0 360 arc Opaque stroke } def
/BoxFill { gsave Rec 1 setgray fill grestore } def
end
}}%
\begin{picture}(3600,2160)(0,0)%
{\GNUPLOTspecial{"
gnudict begin
gsave
0 0 translate
0.100 0.100 scale
0 setgray
newpath
1.000 UL
LTb
400 300 M
63 0 V
2987 0 R
-63 0 V
400 551 M
63 0 V
2987 0 R
-63 0 V
400 803 M
63 0 V
2987 0 R
-63 0 V
400 1054 M
63 0 V
2987 0 R
-63 0 V
400 1306 M
63 0 V
2987 0 R
-63 0 V
400 1557 M
63 0 V
2987 0 R
-63 0 V
400 1809 M
63 0 V
2987 0 R
-63 0 V
400 2060 M
63 0 V
2987 0 R
-63 0 V
654 300 M
0 63 V
0 1697 R
0 -63 V
1163 300 M
0 63 V
0 1697 R
0 -63 V
1671 300 M
0 63 V
0 1697 R
0 -63 V
2179 300 M
0 63 V
0 1697 R
0 -63 V
2688 300 M
0 63 V
0 1697 R
0 -63 V
3196 300 M
0 63 V
0 1697 R
0 -63 V
1.000 UL
LTb
400 300 M
3050 0 V
0 1760 V
-3050 0 V
400 300 L
0.600 UP
1.000 UL
LT0
400 331 M
120 26 V
88 45 V
88 74 V
89 99 V
88 110 V
177 200 V
176 176 V
177 144 V
353 240 V
353 180 V
353 135 V
706 197 V
282 52 V
520 357 Pls
608 402 Pls
696 476 Pls
785 575 Pls
873 685 Pls
1050 885 Pls
1226 1061 Pls
1403 1205 Pls
1756 1445 Pls
2109 1625 Pls
2462 1760 Pls
3168 1957 Pls
0.600 UP
1.000 UL
LT1
400 325 M
120 19 V
88 29 V
88 56 V
45 49 V
44 56 V
88 113 V
177 205 V
176 174 V
177 150 V
353 240 V
706 321 V
706 202 V
282 53 V
520 344 Crs
608 373 Crs
696 429 Crs
741 478 Crs
785 534 Crs
873 647 Crs
1050 852 Crs
1226 1026 Crs
1403 1176 Crs
1756 1416 Crs
2462 1737 Crs
3168 1939 Crs
0.600 UP
1.000 UL
LT2
400 300 M
120 1 V
88 1 V
88 5 V
89 13 V
88 27 V
177 76 V
176 100 V
177 103 V
353 216 V
353 194 V
353 159 V
706 252 V
282 74 V
520 301 Star
608 302 Star
696 307 Star
785 320 Star
873 347 Star
1050 423 Star
1226 523 Star
1403 626 Star
1756 842 Star
2109 1036 Star
2462 1195 Star
3168 1447 Star
0.600 UP
1.000 UL
LT3
400 300 M
120 0 V
88 1 V
88 3 V
45 3 V
44 7 V
88 23 V
177 71 V
176 93 V
177 102 V
353 207 V
706 351 V
706 260 V
282 74 V
520 300 Box
608 301 Box
696 304 Box
741 307 Box
785 314 Box
873 337 Box
1050 408 Box
1226 501 Box
1403 603 Box
1756 810 Box
2462 1161 Box
3168 1421 Box
stroke
grestore
end
showpage
}}%
\put(1925,50){\makebox(0,0){$\gamma$}}%
\put(100,1180){%
\makebox(0,0)[b]{\shortstack{$<u_w>$}}%
}%
\put(3196,200){\makebox(0,0){1.4}}%
\put(2688,200){\makebox(0,0){1.2}}%
\put(2179,200){\makebox(0,0){1}}%
\put(1671,200){\makebox(0,0){0.8}}%
\put(1163,200){\makebox(0,0){0.6}}%
\put(654,200){\makebox(0,0){0.4}}%
\put(350,2060){\makebox(0,0)[r]{0.7}}%
\put(350,1809){\makebox(0,0)[r]{0.6}}%
\put(350,1557){\makebox(0,0)[r]{0.5}}%
\put(350,1306){\makebox(0,0)[r]{0.4}}%
\put(350,1054){\makebox(0,0)[r]{0.3}}%
\put(350,803){\makebox(0,0)[r]{0.2}}%
\put(350,551){\makebox(0,0)[r]{0.1}}%
\put(350,300){\makebox(0,0)[r]{0}}%
\end{picture}%
\endgroup

%% file: P_2wgraph.tex
\begingroup%
  \makeatletter%
  \newcommand{\GNUPLOTspecial}{%
    \@sanitize\catcode`\%=14\relax\special}%
  \setlength{\unitlength}{0.1bp}%
{\GNUPLOTspecial{!
/gnudict 256 dict def
gnudict begin
/Color false def
/Solid false def
/gnulinewidth 5.000 def
/userlinewidth gnulinewidth def
/vshift -33 def
/dl {10 mul} def
/hpt_ 31.5 def
/vpt_ 31.5 def
/hpt hpt_ def
/vpt vpt_ def
/M {moveto} bind def
/L {lineto} bind def
/R {rmoveto} bind def
/V {rlineto} bind def
/vpt2 vpt 2 mul def
/hpt2 hpt 2 mul def
/Lshow { currentpoint stroke M
  0 vshift R show } def
/Rshow { currentpoint stroke M
  dup stringwidth pop neg vshift R show } def
/Cshow { currentpoint stroke M
  dup stringwidth pop -2 div vshift R show } def
/UP { dup vpt_ mul /vpt exch def hpt_ mul /hpt exch def
  /hpt2 hpt 2 mul def /vpt2 vpt 2 mul def } def
/DL { Color {setrgbcolor Solid {pop []} if 0 setdash }
 {pop pop pop Solid {pop []} if 0 setdash} ifelse } def
/BL { stroke userlinewidth 2 mul setlinewidth } def
/AL { stroke userlinewidth 2 div setlinewidth } def
/UL { dup gnulinewidth mul /userlinewidth exch def
      10 mul /udl exch def } def
/PL { stroke userlinewidth setlinewidth } def
/LTb { BL [] 0 0 0 DL } def
/LTa { AL [1 udl mul 2 udl mul] 0 setdash 0 0 0 setrgbcolor } def
/LT0 { PL [] 1 0 0 DL } def
/LT1 { PL [4 dl 2 dl] 0 1 0 DL } def
/LT2 { PL [2 dl 3 dl] 0 0 1 DL } def
/LT3 { PL [1 dl 1.5 dl] 1 0 1 DL } def
/LT4 { PL [5 dl 2 dl 1 dl 2 dl] 0 1 1 DL } def
/LT5 { PL [4 dl 3 dl 1 dl 3 dl] 1 1 0 DL } def
/LT6 { PL [2 dl 2 dl 2 dl 4 dl] 0 0 0 DL } def
/LT7 { PL [2 dl 2 dl 2 dl 2 dl 2 dl 4 dl] 1 0.3 0 DL } def
/LT8 { PL [2 dl 2 dl 2 dl 2 dl 2 dl 2 dl 2 dl 4 dl] 0.5 0.5 0.5 DL } def
/Pnt { stroke [] 0 setdash
   gsave 1 setlinecap M 0 0 V stroke grestore } def
/Dia { stroke [] 0 setdash 2 copy vpt add M
  hpt neg vpt neg V hpt vpt neg V
  hpt vpt V hpt neg vpt V closepath stroke
  Pnt } def
/Pls { stroke [] 0 setdash vpt sub M 0 vpt2 V
  currentpoint stroke M
  hpt neg vpt neg R hpt2 0 V stroke
  } def
/Box { stroke [] 0 setdash 2 copy exch hpt sub exch vpt add M
  0 vpt2 neg V hpt2 0 V 0 vpt2 V
  hpt2 neg 0 V closepath stroke
  Pnt } def
/Crs { stroke [] 0 setdash exch hpt sub exch vpt add M
  hpt2 vpt2 neg V currentpoint stroke M
  hpt2 neg 0 R hpt2 vpt2 V stroke } def
/TriU { stroke [] 0 setdash 2 copy vpt 1.12 mul add M
  hpt neg vpt -1.62 mul V
  hpt 2 mul 0 V
  hpt neg vpt 1.62 mul V closepath stroke
  Pnt  } def
/Star { 2 copy Pls Crs } def
/BoxF { stroke [] 0 setdash exch hpt sub exch vpt add M
  0 vpt2 neg V  hpt2 0 V  0 vpt2 V
  hpt2 neg 0 V  closepath fill } def
/TriUF { stroke [] 0 setdash vpt 1.12 mul add M
  hpt neg vpt -1.62 mul V
  hpt 2 mul 0 V
  hpt neg vpt 1.62 mul V closepath fill } def
/TriD { stroke [] 0 setdash 2 copy vpt 1.12 mul sub M
  hpt neg vpt 1.62 mul V
  hpt 2 mul 0 V
  hpt neg vpt -1.62 mul V closepath stroke
  Pnt  } def
/TriDF { stroke [] 0 setdash vpt 1.12 mul sub M
  hpt neg vpt 1.62 mul V
  hpt 2 mul 0 V
  hpt neg vpt -1.62 mul V closepath fill} def
/DiaF { stroke [] 0 setdash vpt add M
  hpt neg vpt neg V hpt vpt neg V
  hpt vpt V hpt neg vpt V closepath fill } def
/Pent { stroke [] 0 setdash 2 copy gsave
  translate 0 hpt M 4 {72 rotate 0 hpt L} repeat
  closepath stroke grestore Pnt } def
/PentF { stroke [] 0 setdash gsave
  translate 0 hpt M 4 {72 rotate 0 hpt L} repeat
  closepath fill grestore } def
/Circle { stroke [] 0 setdash 2 copy
  hpt 0 360 arc stroke Pnt } def
/CircleF { stroke [] 0 setdash hpt 0 360 arc fill } def
/C0 { BL [] 0 setdash 2 copy moveto vpt 90 450  arc } bind def
/C1 { BL [] 0 setdash 2 copy        moveto
       2 copy  vpt 0 90 arc closepath fill
               vpt 0 360 arc closepath } bind def
/C2 { BL [] 0 setdash 2 copy moveto
       2 copy  vpt 90 180 arc closepath fill
               vpt 0 360 arc closepath } bind def
/C3 { BL [] 0 setdash 2 copy moveto
       2 copy  vpt 0 180 arc closepath fill
               vpt 0 360 arc closepath } bind def
/C4 { BL [] 0 setdash 2 copy moveto
       2 copy  vpt 180 270 arc closepath fill
               vpt 0 360 arc closepath } bind def
/C5 { BL [] 0 setdash 2 copy moveto
       2 copy  vpt 0 90 arc
       2 copy moveto
       2 copy  vpt 180 270 arc closepath fill
               vpt 0 360 arc } bind def
/C6 { BL [] 0 setdash 2 copy moveto
      2 copy  vpt 90 270 arc closepath fill
              vpt 0 360 arc closepath } bind def
/C7 { BL [] 0 setdash 2 copy moveto
      2 copy  vpt 0 270 arc closepath fill
              vpt 0 360 arc closepath } bind def
/C8 { BL [] 0 setdash 2 copy moveto
      2 copy vpt 270 360 arc closepath fill
              vpt 0 360 arc closepath } bind def
/C9 { BL [] 0 setdash 2 copy moveto
      2 copy  vpt 270 450 arc closepath fill
              vpt 0 360 arc closepath } bind def
/C10 { BL [] 0 setdash 2 copy 2 copy moveto vpt 270 360 arc closepath fill
       2 copy moveto
       2 copy vpt 90 180 arc closepath fill
               vpt 0 360 arc closepath } bind def
/C11 { BL [] 0 setdash 2 copy moveto
       2 copy  vpt 0 180 arc closepath fill
       2 copy moveto
       2 copy  vpt 270 360 arc closepath fill
               vpt 0 360 arc closepath } bind def
/C12 { BL [] 0 setdash 2 copy moveto
       2 copy  vpt 180 360 arc closepath fill
               vpt 0 360 arc closepath } bind def
/C13 { BL [] 0 setdash  2 copy moveto
       2 copy  vpt 0 90 arc closepath fill
       2 copy moveto
       2 copy  vpt 180 360 arc closepath fill
               vpt 0 360 arc closepath } bind def
/C14 { BL [] 0 setdash 2 copy moveto
       2 copy  vpt 90 360 arc closepath fill
               vpt 0 360 arc } bind def
/C15 { BL [] 0 setdash 2 copy vpt 0 360 arc closepath fill
               vpt 0 360 arc closepath } bind def
/Rec   { newpath 4 2 roll moveto 1 index 0 rlineto 0 exch rlineto
       neg 0 rlineto closepath } bind def
/Square { dup Rec } bind def
/Bsquare { vpt sub exch vpt sub exch vpt2 Square } bind def
/S0 { BL [] 0 setdash 2 copy moveto 0 vpt rlineto BL Bsquare } bind def
/S1 { BL [] 0 setdash 2 copy vpt Square fill Bsquare } bind def
/S2 { BL [] 0 setdash 2 copy exch vpt sub exch vpt Square fill Bsquare } bind def
/S3 { BL [] 0 setdash 2 copy exch vpt sub exch vpt2 vpt Rec fill Bsquare } bind def
/S4 { BL [] 0 setdash 2 copy exch vpt sub exch vpt sub vpt Square fill Bsquare } bind def
/S5 { BL [] 0 setdash 2 copy 2 copy vpt Square fill
       exch vpt sub exch vpt sub vpt Square fill Bsquare } bind def
/S6 { BL [] 0 setdash 2 copy exch vpt sub exch vpt sub vpt vpt2 Rec fill Bsquare } bind def
/S7 { BL [] 0 setdash 2 copy exch vpt sub exch vpt sub vpt vpt2 Rec fill
       2 copy vpt Square fill
       Bsquare } bind def
/S8 { BL [] 0 setdash 2 copy vpt sub vpt Square fill Bsquare } bind def
/S9 { BL [] 0 setdash 2 copy vpt sub vpt vpt2 Rec fill Bsquare } bind def
/S10 { BL [] 0 setdash 2 copy vpt sub vpt Square fill 2 copy exch vpt sub exch vpt Square fill
       Bsquare } bind def
/S11 { BL [] 0 setdash 2 copy vpt sub vpt Square fill 2 copy exch vpt sub exch vpt2 vpt Rec fill
       Bsquare } bind def
/S12 { BL [] 0 setdash 2 copy exch vpt sub exch vpt sub vpt2 vpt Rec fill Bsquare } bind def
/S13 { BL [] 0 setdash 2 copy exch vpt sub exch vpt sub vpt2 vpt Rec fill
       2 copy vpt Square fill Bsquare } bind def
/S14 { BL [] 0 setdash 2 copy exch vpt sub exch vpt sub vpt2 vpt Rec fill
       2 copy exch vpt sub exch vpt Square fill Bsquare } bind def
/S15 { BL [] 0 setdash 2 copy Bsquare fill Bsquare } bind def
/D0 { gsave translate 45 rotate 0 0 S0 stroke grestore } bind def
/D1 { gsave translate 45 rotate 0 0 S1 stroke grestore } bind def
/D2 { gsave translate 45 rotate 0 0 S2 stroke grestore } bind def
/D3 { gsave translate 45 rotate 0 0 S3 stroke grestore } bind def
/D4 { gsave translate 45 rotate 0 0 S4 stroke grestore } bind def
/D5 { gsave translate 45 rotate 0 0 S5 stroke grestore } bind def
/D6 { gsave translate 45 rotate 0 0 S6 stroke grestore } bind def
/D7 { gsave translate 45 rotate 0 0 S7 stroke grestore } bind def
/D8 { gsave translate 45 rotate 0 0 S8 stroke grestore } bind def
/D9 { gsave translate 45 rotate 0 0 S9 stroke grestore } bind def
/D10 { gsave translate 45 rotate 0 0 S10 stroke grestore } bind def
/D11 { gsave translate 45 rotate 0 0 S11 stroke grestore } bind def
/D12 { gsave translate 45 rotate 0 0 S12 stroke grestore } bind def
/D13 { gsave translate 45 rotate 0 0 S13 stroke grestore } bind def
/D14 { gsave translate 45 rotate 0 0 S14 stroke grestore } bind def
/D15 { gsave translate 45 rotate 0 0 S15 stroke grestore } bind def
/DiaE { stroke [] 0 setdash vpt add M
  hpt neg vpt neg V hpt vpt neg V
  hpt vpt V hpt neg vpt V closepath stroke } def
/BoxE { stroke [] 0 setdash exch hpt sub exch vpt add M
  0 vpt2 neg V hpt2 0 V 0 vpt2 V
  hpt2 neg 0 V closepath stroke } def
/TriUE { stroke [] 0 setdash vpt 1.12 mul add M
  hpt neg vpt -1.62 mul V
  hpt 2 mul 0 V
  hpt neg vpt 1.62 mul V closepath stroke } def
/TriDE { stroke [] 0 setdash vpt 1.12 mul sub M
  hpt neg vpt 1.62 mul V
  hpt 2 mul 0 V
  hpt neg vpt -1.62 mul V closepath stroke } def
/PentE { stroke [] 0 setdash gsave
  translate 0 hpt M 4 {72 rotate 0 hpt L} repeat
  closepath stroke grestore } def
/CircE { stroke [] 0 setdash 
  hpt 0 360 arc stroke } def
/Opaque { gsave closepath 1 setgray fill grestore 0 setgray closepath } def
/DiaW { stroke [] 0 setdash vpt add M
  hpt neg vpt neg V hpt vpt neg V
  hpt vpt V hpt neg vpt V Opaque stroke } def
/BoxW { stroke [] 0 setdash exch hpt sub exch vpt add M
  0 vpt2 neg V hpt2 0 V 0 vpt2 V
  hpt2 neg 0 V Opaque stroke } def
/TriUW { stroke [] 0 setdash vpt 1.12 mul add M
  hpt neg vpt -1.62 mul V
  hpt 2 mul 0 V
  hpt neg vpt 1.62 mul V Opaque stroke } def
/TriDW { stroke [] 0 setdash vpt 1.12 mul sub M
  hpt neg vpt 1.62 mul V
  hpt 2 mul 0 V
  hpt neg vpt -1.62 mul V Opaque stroke } def
/PentW { stroke [] 0 setdash gsave
  translate 0 hpt M 4 {72 rotate 0 hpt L} repeat
  Opaque stroke grestore } def
/CircW { stroke [] 0 setdash 
  hpt 0 360 arc Opaque stroke } def
/BoxFill { gsave Rec 1 setgray fill grestore } def
end
}}%
\begin{picture}(3600,2160)(0,0)%
{\GNUPLOTspecial{"
gnudict begin
gsave
0 0 translate
0.100 0.100 scale
0 setgray
newpath
1.000 UL
LTb
450 300 M
63 0 V
2937 0 R
-63 0 V
450 551 M
63 0 V
2937 0 R
-63 0 V
450 803 M
63 0 V
2937 0 R
-63 0 V
450 1054 M
63 0 V
2937 0 R
-63 0 V
450 1306 M
63 0 V
2937 0 R
-63 0 V
450 1557 M
63 0 V
2937 0 R
-63 0 V
450 1809 M
63 0 V
2937 0 R
-63 0 V
450 2060 M
63 0 V
2937 0 R
-63 0 V
450 300 M
0 63 V
0 1697 R
0 -63 V
723 300 M
0 63 V
0 1697 R
0 -63 V
995 300 M
0 63 V
0 1697 R
0 -63 V
1268 300 M
0 63 V
0 1697 R
0 -63 V
1541 300 M
0 63 V
0 1697 R
0 -63 V
1814 300 M
0 63 V
0 1697 R
0 -63 V
2086 300 M
0 63 V
0 1697 R
0 -63 V
2359 300 M
0 63 V
0 1697 R
0 -63 V
2632 300 M
0 63 V
0 1697 R
0 -63 V
2905 300 M
0 63 V
0 1697 R
0 -63 V
3177 300 M
0 63 V
0 1697 R
0 -63 V
3450 300 M
0 63 V
0 1697 R
0 -63 V
1.000 UL
LTb
450 300 M
3000 0 V
0 1760 V
-3000 0 V
450 300 L
1.000 UL
LT0
707 1137 M
379 704 V
190 141 V
174 13 V
394 -174 V
379 -186 V
515 -241 V
621 -227 V
1.000 UL
LT1
1086 1693 M
190 289 V
174 -2 V
394 -154 V
379 -181 V
515 -239 V
621 -221 V
1.000 UL
LT2
707 308 M
379 55 V
190 59 V
174 54 V
394 91 V
379 50 V
515 12 V
621 -25 V
1.000 UL
LT3
1086 333 M
190 67 V
174 58 V
394 94 V
379 52 V
515 15 V
621 -14 V
0.600 UP
1.000 UL
LT4
707 1132 M
0 10 V
-31 -10 R
62 0 V
-62 10 R
62 0 V
348 692 R
0 15 V
-31 -15 R
62 0 V
-62 15 R
62 0 V
159 128 R
0 10 V
-31 -10 R
62 0 V
-62 10 R
62 0 V
143 3 R
0 10 V
-31 -10 R
62 0 V
-62 10 R
62 0 V
363 -184 R
0 10 V
-31 -10 R
62 0 V
-62 10 R
62 0 V
348 -196 R
0 10 V
-31 -10 R
62 0 V
-62 10 R
62 0 V
484 -251 R
0 10 V
-31 -10 R
62 0 V
-62 10 R
62 0 V
590 -234 R
0 5 V
-31 -5 R
62 0 V
-62 5 R
62 0 V
707 1137 Pls
1086 1841 Pls
1276 1982 Pls
1450 1995 Pls
1844 1821 Pls
2223 1635 Pls
2738 1394 Pls
3359 1167 Pls
0.600 UP
1.000 UL
LT5
1086 1688 M
0 10 V
-31 -10 R
62 0 V
-62 10 R
62 0 V
159 277 R
0 15 V
-31 -15 R
62 0 V
-62 15 R
62 0 V
143 -15 R
0 10 V
-31 -10 R
62 0 V
-62 10 R
62 0 V
363 -164 R
0 10 V
-31 -10 R
62 0 V
-62 10 R
62 0 V
348 -191 R
0 10 V
-31 -10 R
62 0 V
-62 10 R
62 0 V
484 -246 R
0 5 V
-31 -5 R
62 0 V
-62 5 R
62 0 V
590 -226 R
0 5 V
-31 -5 R
62 0 V
-62 5 R
62 0 V
1086 1693 Crs
1276 1982 Crs
1450 1980 Crs
1844 1826 Crs
2223 1645 Crs
2738 1406 Crs
3359 1185 Crs
0.600 UP
1.000 UL
LT6
707 305 M
0 6 V
-31 -6 R
62 0 V
-62 6 R
62 0 V
348 50 R
0 4 V
-31 -4 R
62 0 V
-62 4 R
62 0 V
159 54 R
0 6 V
-31 -6 R
62 0 V
-62 6 R
62 0 V
143 49 R
0 4 V
-31 -4 R
62 0 V
-62 4 R
62 0 V
363 86 R
0 5 V
-31 -5 R
62 0 V
-62 5 R
62 0 V
348 45 R
0 5 V
-31 -5 R
62 0 V
-62 5 R
62 0 V
484 8 R
0 5 V
-31 -5 R
62 0 V
-62 5 R
62 0 V
590 -30 R
0 5 V
-31 -5 R
62 0 V
-62 5 R
62 0 V
707 308 Star
1086 363 Star
1276 422 Star
1450 476 Star
1844 567 Star
2223 617 Star
2738 629 Star
3359 604 Star
0.600 UP
1.000 UL
LT7
1086 329 M
0 8 V
-31 -8 R
62 0 V
-62 8 R
62 0 V
159 59 R
0 8 V
-31 -8 R
62 0 V
-62 8 R
62 0 V
143 51 R
0 6 V
-31 -6 R
62 0 V
-62 6 R
62 0 V
363 88 R
0 5 V
-31 -5 R
62 0 V
-62 5 R
62 0 V
348 48 R
0 5 V
-31 -5 R
62 0 V
-62 5 R
62 0 V
484 10 R
0 5 V
-31 -5 R
62 0 V
-62 5 R
62 0 V
590 -19 R
0 4 V
-31 -4 R
62 0 V
-62 4 R
62 0 V
1086 333 Box
1276 400 Box
1450 458 Box
1844 552 Box
2223 604 Box
2738 619 Box
3359 605 Box
stroke
grestore
end
showpage
}}%
\put(1950,50){\makebox(0,0){$\gamma$}}%
\put(100,1180){%
\makebox(0,0)[b]{\shortstack{$P_2^w$}}%
}%
\put(3450,200){\makebox(0,0){0.85}}%
\put(3177,200){\makebox(0,0){0.8}}%
\put(2905,200){\makebox(0,0){0.75}}%
\put(2632,200){\makebox(0,0){0.7}}%
\put(2359,200){\makebox(0,0){0.65}}%
\put(2086,200){\makebox(0,0){0.6}}%
\put(1814,200){\makebox(0,0){0.55}}%
\put(1541,200){\makebox(0,0){0.5}}%
\put(1268,200){\makebox(0,0){0.45}}%
\put(995,200){\makebox(0,0){0.4}}%
\put(723,200){\makebox(0,0){0.35}}%
\put(450,200){\makebox(0,0){0.3}}%
\put(400,2060){\makebox(0,0)[r]{0.07}}%
\put(400,1809){\makebox(0,0)[r]{0.06}}%
\put(400,1557){\makebox(0,0)[r]{0.05}}%
\put(400,1306){\makebox(0,0)[r]{0.04}}%
\put(400,1054){\makebox(0,0)[r]{0.03}}%
\put(400,803){\makebox(0,0)[r]{0.02}}%
\put(400,551){\makebox(0,0)[r]{0.01}}%
\put(400,300){\makebox(0,0)[r]{0}}%
\end{picture}%
\endgroup

%% file: W_2graph.tex
\begingroup%
  \makeatletter%
  \newcommand{\GNUPLOTspecial}{%
    \@sanitize\catcode`\%=14\relax\special}%
  \setlength{\unitlength}{0.1bp}%
{\GNUPLOTspecial{!
/gnudict 256 dict def
gnudict begin
/Color false def
/Solid false def
/gnulinewidth 5.000 def
/userlinewidth gnulinewidth def
/vshift -33 def
/dl {10 mul} def
/hpt_ 31.5 def
/vpt_ 31.5 def
/hpt hpt_ def
/vpt vpt_ def
/M {moveto} bind def
/L {lineto} bind def
/R {rmoveto} bind def
/V {rlineto} bind def
/vpt2 vpt 2 mul def
/hpt2 hpt 2 mul def
/Lshow { currentpoint stroke M
  0 vshift R show } def
/Rshow { currentpoint stroke M
  dup stringwidth pop neg vshift R show } def
/Cshow { currentpoint stroke M
  dup stringwidth pop -2 div vshift R show } def
/UP { dup vpt_ mul /vpt exch def hpt_ mul /hpt exch def
  /hpt2 hpt 2 mul def /vpt2 vpt 2 mul def } def
/DL { Color {setrgbcolor Solid {pop []} if 0 setdash }
 {pop pop pop Solid {pop []} if 0 setdash} ifelse } def
/BL { stroke userlinewidth 2 mul setlinewidth } def
/AL { stroke userlinewidth 2 div setlinewidth } def
/UL { dup gnulinewidth mul /userlinewidth exch def
      10 mul /udl exch def } def
/PL { stroke userlinewidth setlinewidth } def
/LTb { BL [] 0 0 0 DL } def
/LTa { AL [1 udl mul 2 udl mul] 0 setdash 0 0 0 setrgbcolor } def
/LT0 { PL [] 1 0 0 DL } def
/LT1 { PL [4 dl 2 dl] 0 1 0 DL } def
/LT2 { PL [2 dl 3 dl] 0 0 1 DL } def
/LT3 { PL [1 dl 1.5 dl] 1 0 1 DL } def
/LT4 { PL [5 dl 2 dl 1 dl 2 dl] 0 1 1 DL } def
/LT5 { PL [4 dl 3 dl 1 dl 3 dl] 1 1 0 DL } def
/LT6 { PL [2 dl 2 dl 2 dl 4 dl] 0 0 0 DL } def
/LT7 { PL [2 dl 2 dl 2 dl 2 dl 2 dl 4 dl] 1 0.3 0 DL } def
/LT8 { PL [2 dl 2 dl 2 dl 2 dl 2 dl 2 dl 2 dl 4 dl] 0.5 0.5 0.5 DL } def
/Pnt { stroke [] 0 setdash
   gsave 1 setlinecap M 0 0 V stroke grestore } def
/Dia { stroke [] 0 setdash 2 copy vpt add M
  hpt neg vpt neg V hpt vpt neg V
  hpt vpt V hpt neg vpt V closepath stroke
  Pnt } def
/Pls { stroke [] 0 setdash vpt sub M 0 vpt2 V
  currentpoint stroke M
  hpt neg vpt neg R hpt2 0 V stroke
  } def
/Box { stroke [] 0 setdash 2 copy exch hpt sub exch vpt add M
  0 vpt2 neg V hpt2 0 V 0 vpt2 V
  hpt2 neg 0 V closepath stroke
  Pnt } def
/Crs { stroke [] 0 setdash exch hpt sub exch vpt add M
  hpt2 vpt2 neg V currentpoint stroke M
  hpt2 neg 0 R hpt2 vpt2 V stroke } def
/TriU { stroke [] 0 setdash 2 copy vpt 1.12 mul add M
  hpt neg vpt -1.62 mul V
  hpt 2 mul 0 V
  hpt neg vpt 1.62 mul V closepath stroke
  Pnt  } def
/Star { 2 copy Pls Crs } def
/BoxF { stroke [] 0 setdash exch hpt sub exch vpt add M
  0 vpt2 neg V  hpt2 0 V  0 vpt2 V
  hpt2 neg 0 V  closepath fill } def
/TriUF { stroke [] 0 setdash vpt 1.12 mul add M
  hpt neg vpt -1.62 mul V
  hpt 2 mul 0 V
  hpt neg vpt 1.62 mul V closepath fill } def
/TriD { stroke [] 0 setdash 2 copy vpt 1.12 mul sub M
  hpt neg vpt 1.62 mul V
  hpt 2 mul 0 V
  hpt neg vpt -1.62 mul V closepath stroke
  Pnt  } def
/TriDF { stroke [] 0 setdash vpt 1.12 mul sub M
  hpt neg vpt 1.62 mul V
  hpt 2 mul 0 V
  hpt neg vpt -1.62 mul V closepath fill} def
/DiaF { stroke [] 0 setdash vpt add M
  hpt neg vpt neg V hpt vpt neg V
  hpt vpt V hpt neg vpt V closepath fill } def
/Pent { stroke [] 0 setdash 2 copy gsave
  translate 0 hpt M 4 {72 rotate 0 hpt L} repeat
  closepath stroke grestore Pnt } def
/PentF { stroke [] 0 setdash gsave
  translate 0 hpt M 4 {72 rotate 0 hpt L} repeat
  closepath fill grestore } def
/Circle { stroke [] 0 setdash 2 copy
  hpt 0 360 arc stroke Pnt } def
/CircleF { stroke [] 0 setdash hpt 0 360 arc fill } def
/C0 { BL [] 0 setdash 2 copy moveto vpt 90 450  arc } bind def
/C1 { BL [] 0 setdash 2 copy        moveto
       2 copy  vpt 0 90 arc closepath fill
               vpt 0 360 arc closepath } bind def
/C2 { BL [] 0 setdash 2 copy moveto
       2 copy  vpt 90 180 arc closepath fill
               vpt 0 360 arc closepath } bind def
/C3 { BL [] 0 setdash 2 copy moveto
       2 copy  vpt 0 180 arc closepath fill
               vpt 0 360 arc closepath } bind def
/C4 { BL [] 0 setdash 2 copy moveto
       2 copy  vpt 180 270 arc closepath fill
               vpt 0 360 arc closepath } bind def
/C5 { BL [] 0 setdash 2 copy moveto
       2 copy  vpt 0 90 arc
       2 copy moveto
       2 copy  vpt 180 270 arc closepath fill
               vpt 0 360 arc } bind def
/C6 { BL [] 0 setdash 2 copy moveto
      2 copy  vpt 90 270 arc closepath fill
              vpt 0 360 arc closepath } bind def
/C7 { BL [] 0 setdash 2 copy moveto
      2 copy  vpt 0 270 arc closepath fill
              vpt 0 360 arc closepath } bind def
/C8 { BL [] 0 setdash 2 copy moveto
      2 copy vpt 270 360 arc closepath fill
              vpt 0 360 arc closepath } bind def
/C9 { BL [] 0 setdash 2 copy moveto
      2 copy  vpt 270 450 arc closepath fill
              vpt 0 360 arc closepath } bind def
/C10 { BL [] 0 setdash 2 copy 2 copy moveto vpt 270 360 arc closepath fill
       2 copy moveto
       2 copy vpt 90 180 arc closepath fill
               vpt 0 360 arc closepath } bind def
/C11 { BL [] 0 setdash 2 copy moveto
       2 copy  vpt 0 180 arc closepath fill
       2 copy moveto
       2 copy  vpt 270 360 arc closepath fill
               vpt 0 360 arc closepath } bind def
/C12 { BL [] 0 setdash 2 copy moveto
       2 copy  vpt 180 360 arc closepath fill
               vpt 0 360 arc closepath } bind def
/C13 { BL [] 0 setdash  2 copy moveto
       2 copy  vpt 0 90 arc closepath fill
       2 copy moveto
       2 copy  vpt 180 360 arc closepath fill
               vpt 0 360 arc closepath } bind def
/C14 { BL [] 0 setdash 2 copy moveto
       2 copy  vpt 90 360 arc closepath fill
               vpt 0 360 arc } bind def
/C15 { BL [] 0 setdash 2 copy vpt 0 360 arc closepath fill
               vpt 0 360 arc closepath } bind def
/Rec   { newpath 4 2 roll moveto 1 index 0 rlineto 0 exch rlineto
       neg 0 rlineto closepath } bind def
/Square { dup Rec } bind def
/Bsquare { vpt sub exch vpt sub exch vpt2 Square } bind def
/S0 { BL [] 0 setdash 2 copy moveto 0 vpt rlineto BL Bsquare } bind def
/S1 { BL [] 0 setdash 2 copy vpt Square fill Bsquare } bind def
/S2 { BL [] 0 setdash 2 copy exch vpt sub exch vpt Square fill Bsquare } bind def
/S3 { BL [] 0 setdash 2 copy exch vpt sub exch vpt2 vpt Rec fill Bsquare } bind def
/S4 { BL [] 0 setdash 2 copy exch vpt sub exch vpt sub vpt Square fill Bsquare } bind def
/S5 { BL [] 0 setdash 2 copy 2 copy vpt Square fill
       exch vpt sub exch vpt sub vpt Square fill Bsquare } bind def
/S6 { BL [] 0 setdash 2 copy exch vpt sub exch vpt sub vpt vpt2 Rec fill Bsquare } bind def
/S7 { BL [] 0 setdash 2 copy exch vpt sub exch vpt sub vpt vpt2 Rec fill
       2 copy vpt Square fill
       Bsquare } bind def
/S8 { BL [] 0 setdash 2 copy vpt sub vpt Square fill Bsquare } bind def
/S9 { BL [] 0 setdash 2 copy vpt sub vpt vpt2 Rec fill Bsquare } bind def
/S10 { BL [] 0 setdash 2 copy vpt sub vpt Square fill 2 copy exch vpt sub exch vpt Square fill
       Bsquare } bind def
/S11 { BL [] 0 setdash 2 copy vpt sub vpt Square fill 2 copy exch vpt sub exch vpt2 vpt Rec fill
       Bsquare } bind def
/S12 { BL [] 0 setdash 2 copy exch vpt sub exch vpt sub vpt2 vpt Rec fill Bsquare } bind def
/S13 { BL [] 0 setdash 2 copy exch vpt sub exch vpt sub vpt2 vpt Rec fill
       2 copy vpt Square fill Bsquare } bind def
/S14 { BL [] 0 setdash 2 copy exch vpt sub exch vpt sub vpt2 vpt Rec fill
       2 copy exch vpt sub exch vpt Square fill Bsquare } bind def
/S15 { BL [] 0 setdash 2 copy Bsquare fill Bsquare } bind def
/D0 { gsave translate 45 rotate 0 0 S0 stroke grestore } bind def
/D1 { gsave translate 45 rotate 0 0 S1 stroke grestore } bind def
/D2 { gsave translate 45 rotate 0 0 S2 stroke grestore } bind def
/D3 { gsave translate 45 rotate 0 0 S3 stroke grestore } bind def
/D4 { gsave translate 45 rotate 0 0 S4 stroke grestore } bind def
/D5 { gsave translate 45 rotate 0 0 S5 stroke grestore } bind def
/D6 { gsave translate 45 rotate 0 0 S6 stroke grestore } bind def
/D7 { gsave translate 45 rotate 0 0 S7 stroke grestore } bind def
/D8 { gsave translate 45 rotate 0 0 S8 stroke grestore } bind def
/D9 { gsave translate 45 rotate 0 0 S9 stroke grestore } bind def
/D10 { gsave translate 45 rotate 0 0 S10 stroke grestore } bind def
/D11 { gsave translate 45 rotate 0 0 S11 stroke grestore } bind def
/D12 { gsave translate 45 rotate 0 0 S12 stroke grestore } bind def
/D13 { gsave translate 45 rotate 0 0 S13 stroke grestore } bind def
/D14 { gsave translate 45 rotate 0 0 S14 stroke grestore } bind def
/D15 { gsave translate 45 rotate 0 0 S15 stroke grestore } bind def
/DiaE { stroke [] 0 setdash vpt add M
  hpt neg vpt neg V hpt vpt neg V
  hpt vpt V hpt neg vpt V closepath stroke } def
/BoxE { stroke [] 0 setdash exch hpt sub exch vpt add M
  0 vpt2 neg V hpt2 0 V 0 vpt2 V
  hpt2 neg 0 V closepath stroke } def
/TriUE { stroke [] 0 setdash vpt 1.12 mul add M
  hpt neg vpt -1.62 mul V
  hpt 2 mul 0 V
  hpt neg vpt 1.62 mul V closepath stroke } def
/TriDE { stroke [] 0 setdash vpt 1.12 mul sub M
  hpt neg vpt 1.62 mul V
  hpt 2 mul 0 V
  hpt neg vpt -1.62 mul V closepath stroke } def
/PentE { stroke [] 0 setdash gsave
  translate 0 hpt M 4 {72 rotate 0 hpt L} repeat
  closepath stroke grestore } def
/CircE { stroke [] 0 setdash 
  hpt 0 360 arc stroke } def
/Opaque { gsave closepath 1 setgray fill grestore 0 setgray closepath } def
/DiaW { stroke [] 0 setdash vpt add M
  hpt neg vpt neg V hpt vpt neg V
  hpt vpt V hpt neg vpt V Opaque stroke } def
/BoxW { stroke [] 0 setdash exch hpt sub exch vpt add M
  0 vpt2 neg V hpt2 0 V 0 vpt2 V
  hpt2 neg 0 V Opaque stroke } def
/TriUW { stroke [] 0 setdash vpt 1.12 mul add M
  hpt neg vpt -1.62 mul V
  hpt 2 mul 0 V
  hpt neg vpt 1.62 mul V Opaque stroke } def
/TriDW { stroke [] 0 setdash vpt 1.12 mul sub M
  hpt neg vpt 1.62 mul V
  hpt 2 mul 0 V
  hpt neg vpt -1.62 mul V Opaque stroke } def
/PentW { stroke [] 0 setdash gsave
  translate 0 hpt M 4 {72 rotate 0 hpt L} repeat
  Opaque stroke grestore } def
/CircW { stroke [] 0 setdash 
  hpt 0 360 arc Opaque stroke } def
/BoxFill { gsave Rec 1 setgray fill grestore } def
end
}}%
\begin{picture}(3600,2160)(0,0)%
{\GNUPLOTspecial{"
gnudict begin
gsave
0 0 translate
0.100 0.100 scale
0 setgray
newpath
1.000 UL
LTb
400 300 M
63 0 V
2987 0 R
-63 0 V
400 593 M
63 0 V
2987 0 R
-63 0 V
400 887 M
63 0 V
2987 0 R
-63 0 V
400 1180 M
63 0 V
2987 0 R
-63 0 V
400 1473 M
63 0 V
2987 0 R
-63 0 V
400 1767 M
63 0 V
2987 0 R
-63 0 V
400 2060 M
63 0 V
2987 0 R
-63 0 V
400 300 M
0 63 V
0 1697 R
0 -63 V
705 300 M
0 63 V
0 1697 R
0 -63 V
1010 300 M
0 63 V
0 1697 R
0 -63 V
1315 300 M
0 63 V
0 1697 R
0 -63 V
1620 300 M
0 63 V
0 1697 R
0 -63 V
1925 300 M
0 63 V
0 1697 R
0 -63 V
2230 300 M
0 63 V
0 1697 R
0 -63 V
2535 300 M
0 63 V
0 1697 R
0 -63 V
2840 300 M
0 63 V
0 1697 R
0 -63 V
3145 300 M
0 63 V
0 1697 R
0 -63 V
3450 300 M
0 63 V
0 1697 R
0 -63 V
1.000 UL
LTb
400 300 M
3050 0 V
0 1760 V
-3050 0 V
400 300 L
0.600 UP
1.000 UL
LT0
400 1773 M
288 9 V
212 26 V
106 17 V
106 25 V
211 45 V
212 12 V
424 -41 V
424 -99 V
423 -119 V
644 -181 V
688 1782 Pls
900 1808 Pls
1006 1825 Pls
1112 1850 Pls
1323 1895 Pls
1535 1907 Pls
1959 1866 Pls
2383 1767 Pls
2806 1648 Pls
0.600 UP
1.000 UL
LT1
400 1769 M
288 2 V
212 9 V
212 30 V
106 24 V
106 27 V
211 12 V
424 -35 V
424 -83 V
423 -118 V
644 -171 V
688 1771 Crs
900 1780 Crs
1112 1810 Crs
1218 1834 Crs
1324 1861 Crs
1535 1873 Crs
1959 1838 Crs
2383 1755 Crs
2806 1637 Crs
0.600 UP
1.000 UL
LT2
688 1775 M
212 0 V
212 20 V
106 29 V
52 17 V
54 9 V
211 17 V
212 -8 V
688 1775 Star
900 1775 Star
1112 1795 Star
1218 1824 Star
1270 1841 Star
1324 1850 Star
1535 1867 Star
1747 1859 Star
0.600 UP
1.000 UL
LT3
900 1774 M
212 19 V
106 21 V
52 40 V
54 -2 V
211 9 V
900 1774 Box
1112 1793 Box
1218 1814 Box
1270 1854 Box
1324 1852 Box
1535 1861 Box
0.600 UP
1.000 UL
LT4
900 1767 M
0 15 V
-31 -15 R
62 0 V
-62 15 R
62 0 V
181 4 R
0 14 V
-31 -14 R
62 0 V
-62 14 R
62 0 V
75 9 R
0 10 V
-31 -10 R
62 0 V
-62 10 R
62 0 V
21 32 R
0 7 V
-31 -7 R
62 0 V
-62 7 R
62 0 V
23 -15 R
0 17 V
-31 -17 R
62 0 V
-62 17 R
62 0 V
180 -8 R
0 18 V
-31 -18 R
62 0 V
-62 18 R
62 0 V
900 1774 BoxF
1112 1793 BoxF
1218 1814 BoxF
1270 1854 BoxF
1324 1852 BoxF
1535 1861 BoxF
stroke
grestore
end
showpage
}}%
\put(1925,50){\makebox(0,0){$\gamma$}}%
\put(100,1180){%
\makebox(0,0)[b]{\shortstack{$W_2^{2\times 2}$}}%
}%
\put(3450,200){\makebox(0,0){0.8}}%
\put(3145,200){\makebox(0,0){0.75}}%
\put(2840,200){\makebox(0,0){0.7}}%
\put(2535,200){\makebox(0,0){0.65}}%
\put(2230,200){\makebox(0,0){0.6}}%
\put(1925,200){\makebox(0,0){0.55}}%
\put(1620,200){\makebox(0,0){0.5}}%
\put(1315,200){\makebox(0,0){0.45}}%
\put(1010,200){\makebox(0,0){0.4}}%
\put(705,200){\makebox(0,0){0.35}}%
\put(400,200){\makebox(0,0){0.3}}%
\put(350,2060){\makebox(0,0)[r]{0.6}}%
\put(350,1767){\makebox(0,0)[r]{0.5}}%
\put(350,1473){\makebox(0,0)[r]{0.4}}%
\put(350,1180){\makebox(0,0)[r]{0.3}}%
\put(350,887){\makebox(0,0)[r]{0.2}}%
\put(350,593){\makebox(0,0)[r]{0.1}}%
\put(350,300){\makebox(0,0)[r]{0}}%
\end{picture}%
\endgroup

%% file: W_3graph.tex
\begingroup%
  \makeatletter%
  \newcommand{\GNUPLOTspecial}{%
    \@sanitize\catcode`\%=14\relax\special}%
  \setlength{\unitlength}{0.1bp}%
{\GNUPLOTspecial{!
/gnudict 256 dict def
gnudict begin
/Color false def
/Solid false def
/gnulinewidth 5.000 def
/userlinewidth gnulinewidth def
/vshift -33 def
/dl {10 mul} def
/hpt_ 31.5 def
/vpt_ 31.5 def
/hpt hpt_ def
/vpt vpt_ def
/M {moveto} bind def
/L {lineto} bind def
/R {rmoveto} bind def
/V {rlineto} bind def
/vpt2 vpt 2 mul def
/hpt2 hpt 2 mul def
/Lshow { currentpoint stroke M
  0 vshift R show } def
/Rshow { currentpoint stroke M
  dup stringwidth pop neg vshift R show } def
/Cshow { currentpoint stroke M
  dup stringwidth pop -2 div vshift R show } def
/UP { dup vpt_ mul /vpt exch def hpt_ mul /hpt exch def
  /hpt2 hpt 2 mul def /vpt2 vpt 2 mul def } def
/DL { Color {setrgbcolor Solid {pop []} if 0 setdash }
 {pop pop pop Solid {pop []} if 0 setdash} ifelse } def
/BL { stroke userlinewidth 2 mul setlinewidth } def
/AL { stroke userlinewidth 2 div setlinewidth } def
/UL { dup gnulinewidth mul /userlinewidth exch def
      10 mul /udl exch def } def
/PL { stroke userlinewidth setlinewidth } def
/LTb { BL [] 0 0 0 DL } def
/LTa { AL [1 udl mul 2 udl mul] 0 setdash 0 0 0 setrgbcolor } def
/LT0 { PL [] 1 0 0 DL } def
/LT1 { PL [4 dl 2 dl] 0 1 0 DL } def
/LT2 { PL [2 dl 3 dl] 0 0 1 DL } def
/LT3 { PL [1 dl 1.5 dl] 1 0 1 DL } def
/LT4 { PL [5 dl 2 dl 1 dl 2 dl] 0 1 1 DL } def
/LT5 { PL [4 dl 3 dl 1 dl 3 dl] 1 1 0 DL } def
/LT6 { PL [2 dl 2 dl 2 dl 4 dl] 0 0 0 DL } def
/LT7 { PL [2 dl 2 dl 2 dl 2 dl 2 dl 4 dl] 1 0.3 0 DL } def
/LT8 { PL [2 dl 2 dl 2 dl 2 dl 2 dl 2 dl 2 dl 4 dl] 0.5 0.5 0.5 DL } def
/Pnt { stroke [] 0 setdash
   gsave 1 setlinecap M 0 0 V stroke grestore } def
/Dia { stroke [] 0 setdash 2 copy vpt add M
  hpt neg vpt neg V hpt vpt neg V
  hpt vpt V hpt neg vpt V closepath stroke
  Pnt } def
/Pls { stroke [] 0 setdash vpt sub M 0 vpt2 V
  currentpoint stroke M
  hpt neg vpt neg R hpt2 0 V stroke
  } def
/Box { stroke [] 0 setdash 2 copy exch hpt sub exch vpt add M
  0 vpt2 neg V hpt2 0 V 0 vpt2 V
  hpt2 neg 0 V closepath stroke
  Pnt } def
/Crs { stroke [] 0 setdash exch hpt sub exch vpt add M
  hpt2 vpt2 neg V currentpoint stroke M
  hpt2 neg 0 R hpt2 vpt2 V stroke } def
/TriU { stroke [] 0 setdash 2 copy vpt 1.12 mul add M
  hpt neg vpt -1.62 mul V
  hpt 2 mul 0 V
  hpt neg vpt 1.62 mul V closepath stroke
  Pnt  } def
/Star { 2 copy Pls Crs } def
/BoxF { stroke [] 0 setdash exch hpt sub exch vpt add M
  0 vpt2 neg V  hpt2 0 V  0 vpt2 V
  hpt2 neg 0 V  closepath fill } def
/TriUF { stroke [] 0 setdash vpt 1.12 mul add M
  hpt neg vpt -1.62 mul V
  hpt 2 mul 0 V
  hpt neg vpt 1.62 mul V closepath fill } def
/TriD { stroke [] 0 setdash 2 copy vpt 1.12 mul sub M
  hpt neg vpt 1.62 mul V
  hpt 2 mul 0 V
  hpt neg vpt -1.62 mul V closepath stroke
  Pnt  } def
/TriDF { stroke [] 0 setdash vpt 1.12 mul sub M
  hpt neg vpt 1.62 mul V
  hpt 2 mul 0 V
  hpt neg vpt -1.62 mul V closepath fill} def
/DiaF { stroke [] 0 setdash vpt add M
  hpt neg vpt neg V hpt vpt neg V
  hpt vpt V hpt neg vpt V closepath fill } def
/Pent { stroke [] 0 setdash 2 copy gsave
  translate 0 hpt M 4 {72 rotate 0 hpt L} repeat
  closepath stroke grestore Pnt } def
/PentF { stroke [] 0 setdash gsave
  translate 0 hpt M 4 {72 rotate 0 hpt L} repeat
  closepath fill grestore } def
/Circle { stroke [] 0 setdash 2 copy
  hpt 0 360 arc stroke Pnt } def
/CircleF { stroke [] 0 setdash hpt 0 360 arc fill } def
/C0 { BL [] 0 setdash 2 copy moveto vpt 90 450  arc } bind def
/C1 { BL [] 0 setdash 2 copy        moveto
       2 copy  vpt 0 90 arc closepath fill
               vpt 0 360 arc closepath } bind def
/C2 { BL [] 0 setdash 2 copy moveto
       2 copy  vpt 90 180 arc closepath fill
               vpt 0 360 arc closepath } bind def
/C3 { BL [] 0 setdash 2 copy moveto
       2 copy  vpt 0 180 arc closepath fill
               vpt 0 360 arc closepath } bind def
/C4 { BL [] 0 setdash 2 copy moveto
       2 copy  vpt 180 270 arc closepath fill
               vpt 0 360 arc closepath } bind def
/C5 { BL [] 0 setdash 2 copy moveto
       2 copy  vpt 0 90 arc
       2 copy moveto
       2 copy  vpt 180 270 arc closepath fill
               vpt 0 360 arc } bind def
/C6 { BL [] 0 setdash 2 copy moveto
      2 copy  vpt 90 270 arc closepath fill
              vpt 0 360 arc closepath } bind def
/C7 { BL [] 0 setdash 2 copy moveto
      2 copy  vpt 0 270 arc closepath fill
              vpt 0 360 arc closepath } bind def
/C8 { BL [] 0 setdash 2 copy moveto
      2 copy vpt 270 360 arc closepath fill
              vpt 0 360 arc closepath } bind def
/C9 { BL [] 0 setdash 2 copy moveto
      2 copy  vpt 270 450 arc closepath fill
              vpt 0 360 arc closepath } bind def
/C10 { BL [] 0 setdash 2 copy 2 copy moveto vpt 270 360 arc closepath fill
       2 copy moveto
       2 copy vpt 90 180 arc closepath fill
               vpt 0 360 arc closepath } bind def
/C11 { BL [] 0 setdash 2 copy moveto
       2 copy  vpt 0 180 arc closepath fill
       2 copy moveto
       2 copy  vpt 270 360 arc closepath fill
               vpt 0 360 arc closepath } bind def
/C12 { BL [] 0 setdash 2 copy moveto
       2 copy  vpt 180 360 arc closepath fill
               vpt 0 360 arc closepath } bind def
/C13 { BL [] 0 setdash  2 copy moveto
       2 copy  vpt 0 90 arc closepath fill
       2 copy moveto
       2 copy  vpt 180 360 arc closepath fill
               vpt 0 360 arc closepath } bind def
/C14 { BL [] 0 setdash 2 copy moveto
       2 copy  vpt 90 360 arc closepath fill
               vpt 0 360 arc } bind def
/C15 { BL [] 0 setdash 2 copy vpt 0 360 arc closepath fill
               vpt 0 360 arc closepath } bind def
/Rec   { newpath 4 2 roll moveto 1 index 0 rlineto 0 exch rlineto
       neg 0 rlineto closepath } bind def
/Square { dup Rec } bind def
/Bsquare { vpt sub exch vpt sub exch vpt2 Square } bind def
/S0 { BL [] 0 setdash 2 copy moveto 0 vpt rlineto BL Bsquare } bind def
/S1 { BL [] 0 setdash 2 copy vpt Square fill Bsquare } bind def
/S2 { BL [] 0 setdash 2 copy exch vpt sub exch vpt Square fill Bsquare } bind def
/S3 { BL [] 0 setdash 2 copy exch vpt sub exch vpt2 vpt Rec fill Bsquare } bind def
/S4 { BL [] 0 setdash 2 copy exch vpt sub exch vpt sub vpt Square fill Bsquare } bind def
/S5 { BL [] 0 setdash 2 copy 2 copy vpt Square fill
       exch vpt sub exch vpt sub vpt Square fill Bsquare } bind def
/S6 { BL [] 0 setdash 2 copy exch vpt sub exch vpt sub vpt vpt2 Rec fill Bsquare } bind def
/S7 { BL [] 0 setdash 2 copy exch vpt sub exch vpt sub vpt vpt2 Rec fill
       2 copy vpt Square fill
       Bsquare } bind def
/S8 { BL [] 0 setdash 2 copy vpt sub vpt Square fill Bsquare } bind def
/S9 { BL [] 0 setdash 2 copy vpt sub vpt vpt2 Rec fill Bsquare } bind def
/S10 { BL [] 0 setdash 2 copy vpt sub vpt Square fill 2 copy exch vpt sub exch vpt Square fill
       Bsquare } bind def
/S11 { BL [] 0 setdash 2 copy vpt sub vpt Square fill 2 copy exch vpt sub exch vpt2 vpt Rec fill
       Bsquare } bind def
/S12 { BL [] 0 setdash 2 copy exch vpt sub exch vpt sub vpt2 vpt Rec fill Bsquare } bind def
/S13 { BL [] 0 setdash 2 copy exch vpt sub exch vpt sub vpt2 vpt Rec fill
       2 copy vpt Square fill Bsquare } bind def
/S14 { BL [] 0 setdash 2 copy exch vpt sub exch vpt sub vpt2 vpt Rec fill
       2 copy exch vpt sub exch vpt Square fill Bsquare } bind def
/S15 { BL [] 0 setdash 2 copy Bsquare fill Bsquare } bind def
/D0 { gsave translate 45 rotate 0 0 S0 stroke grestore } bind def
/D1 { gsave translate 45 rotate 0 0 S1 stroke grestore } bind def
/D2 { gsave translate 45 rotate 0 0 S2 stroke grestore } bind def
/D3 { gsave translate 45 rotate 0 0 S3 stroke grestore } bind def
/D4 { gsave translate 45 rotate 0 0 S4 stroke grestore } bind def
/D5 { gsave translate 45 rotate 0 0 S5 stroke grestore } bind def
/D6 { gsave translate 45 rotate 0 0 S6 stroke grestore } bind def
/D7 { gsave translate 45 rotate 0 0 S7 stroke grestore } bind def
/D8 { gsave translate 45 rotate 0 0 S8 stroke grestore } bind def
/D9 { gsave translate 45 rotate 0 0 S9 stroke grestore } bind def
/D10 { gsave translate 45 rotate 0 0 S10 stroke grestore } bind def
/D11 { gsave translate 45 rotate 0 0 S11 stroke grestore } bind def
/D12 { gsave translate 45 rotate 0 0 S12 stroke grestore } bind def
/D13 { gsave translate 45 rotate 0 0 S13 stroke grestore } bind def
/D14 { gsave translate 45 rotate 0 0 S14 stroke grestore } bind def
/D15 { gsave translate 45 rotate 0 0 S15 stroke grestore } bind def
/DiaE { stroke [] 0 setdash vpt add M
  hpt neg vpt neg V hpt vpt neg V
  hpt vpt V hpt neg vpt V closepath stroke } def
/BoxE { stroke [] 0 setdash exch hpt sub exch vpt add M
  0 vpt2 neg V hpt2 0 V 0 vpt2 V
  hpt2 neg 0 V closepath stroke } def
/TriUE { stroke [] 0 setdash vpt 1.12 mul add M
  hpt neg vpt -1.62 mul V
  hpt 2 mul 0 V
  hpt neg vpt 1.62 mul V closepath stroke } def
/TriDE { stroke [] 0 setdash vpt 1.12 mul sub M
  hpt neg vpt 1.62 mul V
  hpt 2 mul 0 V
  hpt neg vpt -1.62 mul V closepath stroke } def
/PentE { stroke [] 0 setdash gsave
  translate 0 hpt M 4 {72 rotate 0 hpt L} repeat
  closepath stroke grestore } def
/CircE { stroke [] 0 setdash 
  hpt 0 360 arc stroke } def
/Opaque { gsave closepath 1 setgray fill grestore 0 setgray closepath } def
/DiaW { stroke [] 0 setdash vpt add M
  hpt neg vpt neg V hpt vpt neg V
  hpt vpt V hpt neg vpt V Opaque stroke } def
/BoxW { stroke [] 0 setdash exch hpt sub exch vpt add M
  0 vpt2 neg V hpt2 0 V 0 vpt2 V
  hpt2 neg 0 V Opaque stroke } def
/TriUW { stroke [] 0 setdash vpt 1.12 mul add M
  hpt neg vpt -1.62 mul V
  hpt 2 mul 0 V
  hpt neg vpt 1.62 mul V Opaque stroke } def
/TriDW { stroke [] 0 setdash vpt 1.12 mul sub M
  hpt neg vpt 1.62 mul V
  hpt 2 mul 0 V
  hpt neg vpt -1.62 mul V Opaque stroke } def
/PentW { stroke [] 0 setdash gsave
  translate 0 hpt M 4 {72 rotate 0 hpt L} repeat
  Opaque stroke grestore } def
/CircW { stroke [] 0 setdash 
  hpt 0 360 arc Opaque stroke } def
/BoxFill { gsave Rec 1 setgray fill grestore } def
end
}}%
\begin{picture}(3600,2160)(0,0)%
{\GNUPLOTspecial{"
gnudict begin
gsave
0 0 translate
0.100 0.100 scale
0 setgray
newpath
1.000 UL
LTb
500 300 M
63 0 V
2887 0 R
-63 0 V
500 551 M
63 0 V
2887 0 R
-63 0 V
500 803 M
63 0 V
2887 0 R
-63 0 V
500 1054 M
63 0 V
2887 0 R
-63 0 V
500 1306 M
63 0 V
2887 0 R
-63 0 V
500 1557 M
63 0 V
2887 0 R
-63 0 V
500 1809 M
63 0 V
2887 0 R
-63 0 V
500 2060 M
63 0 V
2887 0 R
-63 0 V
500 300 M
0 63 V
0 1697 R
0 -63 V
828 300 M
0 63 V
0 1697 R
0 -63 V
1156 300 M
0 63 V
0 1697 R
0 -63 V
1483 300 M
0 63 V
0 1697 R
0 -63 V
1811 300 M
0 63 V
0 1697 R
0 -63 V
2139 300 M
0 63 V
0 1697 R
0 -63 V
2467 300 M
0 63 V
0 1697 R
0 -63 V
2794 300 M
0 63 V
0 1697 R
0 -63 V
3122 300 M
0 63 V
0 1697 R
0 -63 V
3450 300 M
0 63 V
0 1697 R
0 -63 V
1.000 UL
LTb
500 300 M
2950 0 V
0 1760 V
-2950 0 V
500 300 L
1.000 UL
LT0
500 1579 M
155 37 V
114 93 V
56 95 V
57 52 V
114 39 V
114 -135 V
228 -495 V
1565 778 L
1793 491 L
2248 350 L
455 61 V
456 115 V
291 97 V
1.000 UL
LT1
500 1599 M
155 24 V
114 65 V
882 1582 L
57 201 V
57 -171 V
114 -20 V
228 -211 V
1565 697 L
1793 587 L
2248 406 L
911 130 V
291 92 V
0.600 UP
1.000 UL
LT2
655 1613 M
0 6 V
-31 -6 R
62 0 V
-62 6 R
62 0 V
83 87 R
0 6 V
-31 -6 R
62 0 V
-62 6 R
62 0 V
25 81 R
0 21 V
-31 -21 R
62 0 V
-62 21 R
62 0 V
26 40 R
0 5 V
-31 -5 R
62 0 V
-62 5 R
62 0 V
83 33 R
0 6 V
-31 -6 R
62 0 V
-62 6 R
62 0 V
83 -142 R
0 7 V
-31 -7 R
62 0 V
-62 7 R
62 0 V
197 -508 R
0 21 V
-31 -21 R
62 0 V
-62 21 R
62 0 V
1565 748 M
0 60 V
-31 -60 R
62 0 V
-62 60 R
62 0 V
1793 456 M
0 70 V
-31 -70 R
62 0 V
-62 70 R
62 0 V
2248 335 M
0 30 V
-31 -30 R
62 0 V
-62 30 R
62 0 V
424 26 R
0 40 V
-31 -40 R
62 0 V
-62 40 R
62 0 V
425 85 R
0 20 V
-31 -20 R
62 0 V
-62 20 R
62 0 V
655 1616 Pls
769 1709 Pls
825 1804 Pls
882 1856 Pls
996 1895 Pls
1110 1760 Pls
1338 1265 Pls
1565 778 Pls
1793 491 Pls
2248 350 Pls
2703 411 Pls
3159 526 Pls
0.600 UP
1.000 UL
LT3
655 1557 M
0 131 V
624 1557 M
62 0 V
-62 131 R
62 0 V
83 -60 R
0 120 V
738 1628 M
62 0 V
-62 120 R
62 0 V
82 -246 R
0 161 V
851 1502 M
62 0 V
-62 161 R
62 0 V
26 65 R
0 111 V
908 1728 M
62 0 V
-62 111 R
62 0 V
26 -292 R
0 131 V
965 1547 M
62 0 V
-62 131 R
62 0 V
83 -146 R
0 121 V
-31 -121 R
62 0 V
-62 121 R
62 0 V
197 -352 R
0 161 V
-31 -161 R
62 0 V
-62 161 R
62 0 V
1565 667 M
0 60 V
-31 -60 R
62 0 V
-62 60 R
62 0 V
1793 531 M
0 111 V
1762 531 M
62 0 V
-62 111 R
62 0 V
2248 375 M
0 61 V
-31 -61 R
62 0 V
-62 61 R
62 0 V
880 85 R
0 30 V
-31 -30 R
62 0 V
-62 30 R
62 0 V
655 1623 Crs
769 1688 Crs
882 1582 Crs
939 1783 Crs
996 1612 Crs
1110 1592 Crs
1338 1381 Crs
1565 697 Crs
1793 587 Crs
2248 406 Crs
3159 536 Crs
stroke
grestore
end
showpage
}}%
\put(1975,50){\makebox(0,0){$\gamma$}}%
\put(100,1180){%
\makebox(0,0)[b]{\shortstack{$W_3^{2\times 2}$}}%
}%
\put(3450,200){\makebox(0,0){1.2}}%
\put(3122,200){\makebox(0,0){1.1}}%
\put(2794,200){\makebox(0,0){1}}%
\put(2467,200){\makebox(0,0){0.9}}%
\put(2139,200){\makebox(0,0){0.8}}%
\put(1811,200){\makebox(0,0){0.7}}%
\put(1483,200){\makebox(0,0){0.6}}%
\put(1156,200){\makebox(0,0){0.5}}%
\put(828,200){\makebox(0,0){0.4}}%
\put(500,200){\makebox(0,0){0.3}}%
\put(450,2060){\makebox(0,0)[r]{0.1}}%
\put(450,1809){\makebox(0,0)[r]{0.05}}%
\put(450,1557){\makebox(0,0)[r]{0}}%
\put(450,1306){\makebox(0,0)[r]{-0.05}}%
\put(450,1054){\makebox(0,0)[r]{-0.1}}%
\put(450,803){\makebox(0,0)[r]{-0.15}}%
\put(450,551){\makebox(0,0)[r]{-0.2}}%
\put(450,300){\makebox(0,0)[r]{-0.25}}%
\end{picture}%
\endgroup

%% file: wilsonratiograph.tex
\begingroup%
  \makeatletter%
  \newcommand{\GNUPLOTspecial}{%
    \@sanitize\catcode`\%=14\relax\special}%
  \setlength{\unitlength}{0.1bp}%
{\GNUPLOTspecial{!
/gnudict 256 dict def
gnudict begin
/Color false def
/Solid false def
/gnulinewidth 5.000 def
/userlinewidth gnulinewidth def
/vshift -33 def
/dl {10 mul} def
/hpt_ 31.5 def
/vpt_ 31.5 def
/hpt hpt_ def
/vpt vpt_ def
/M {moveto} bind def
/L {lineto} bind def
/R {rmoveto} bind def
/V {rlineto} bind def
/vpt2 vpt 2 mul def
/hpt2 hpt 2 mul def
/Lshow { currentpoint stroke M
  0 vshift R show } def
/Rshow { currentpoint stroke M
  dup stringwidth pop neg vshift R show } def
/Cshow { currentpoint stroke M
  dup stringwidth pop -2 div vshift R show } def
/UP { dup vpt_ mul /vpt exch def hpt_ mul /hpt exch def
  /hpt2 hpt 2 mul def /vpt2 vpt 2 mul def } def
/DL { Color {setrgbcolor Solid {pop []} if 0 setdash }
 {pop pop pop Solid {pop []} if 0 setdash} ifelse } def
/BL { stroke userlinewidth 2 mul setlinewidth } def
/AL { stroke userlinewidth 2 div setlinewidth } def
/UL { dup gnulinewidth mul /userlinewidth exch def
      10 mul /udl exch def } def
/PL { stroke userlinewidth setlinewidth } def
/LTb { BL [] 0 0 0 DL } def
/LTa { AL [1 udl mul 2 udl mul] 0 setdash 0 0 0 setrgbcolor } def
/LT0 { PL [] 1 0 0 DL } def
/LT1 { PL [4 dl 2 dl] 0 1 0 DL } def
/LT2 { PL [2 dl 3 dl] 0 0 1 DL } def
/LT3 { PL [1 dl 1.5 dl] 1 0 1 DL } def
/LT4 { PL [5 dl 2 dl 1 dl 2 dl] 0 1 1 DL } def
/LT5 { PL [4 dl 3 dl 1 dl 3 dl] 1 1 0 DL } def
/LT6 { PL [2 dl 2 dl 2 dl 4 dl] 0 0 0 DL } def
/LT7 { PL [2 dl 2 dl 2 dl 2 dl 2 dl 4 dl] 1 0.3 0 DL } def
/LT8 { PL [2 dl 2 dl 2 dl 2 dl 2 dl 2 dl 2 dl 4 dl] 0.5 0.5 0.5 DL } def
/Pnt { stroke [] 0 setdash
   gsave 1 setlinecap M 0 0 V stroke grestore } def
/Dia { stroke [] 0 setdash 2 copy vpt add M
  hpt neg vpt neg V hpt vpt neg V
  hpt vpt V hpt neg vpt V closepath stroke
  Pnt } def
/Pls { stroke [] 0 setdash vpt sub M 0 vpt2 V
  currentpoint stroke M
  hpt neg vpt neg R hpt2 0 V stroke
  } def
/Box { stroke [] 0 setdash 2 copy exch hpt sub exch vpt add M
  0 vpt2 neg V hpt2 0 V 0 vpt2 V
  hpt2 neg 0 V closepath stroke
  Pnt } def
/Crs { stroke [] 0 setdash exch hpt sub exch vpt add M
  hpt2 vpt2 neg V currentpoint stroke M
  hpt2 neg 0 R hpt2 vpt2 V stroke } def
/TriU { stroke [] 0 setdash 2 copy vpt 1.12 mul add M
  hpt neg vpt -1.62 mul V
  hpt 2 mul 0 V
  hpt neg vpt 1.62 mul V closepath stroke
  Pnt  } def
/Star { 2 copy Pls Crs } def
/BoxF { stroke [] 0 setdash exch hpt sub exch vpt add M
  0 vpt2 neg V  hpt2 0 V  0 vpt2 V
  hpt2 neg 0 V  closepath fill } def
/TriUF { stroke [] 0 setdash vpt 1.12 mul add M
  hpt neg vpt -1.62 mul V
  hpt 2 mul 0 V
  hpt neg vpt 1.62 mul V closepath fill } def
/TriD { stroke [] 0 setdash 2 copy vpt 1.12 mul sub M
  hpt neg vpt 1.62 mul V
  hpt 2 mul 0 V
  hpt neg vpt -1.62 mul V closepath stroke
  Pnt  } def
/TriDF { stroke [] 0 setdash vpt 1.12 mul sub M
  hpt neg vpt 1.62 mul V
  hpt 2 mul 0 V
  hpt neg vpt -1.62 mul V closepath fill} def
/DiaF { stroke [] 0 setdash vpt add M
  hpt neg vpt neg V hpt vpt neg V
  hpt vpt V hpt neg vpt V closepath fill } def
/Pent { stroke [] 0 setdash 2 copy gsave
  translate 0 hpt M 4 {72 rotate 0 hpt L} repeat
  closepath stroke grestore Pnt } def
/PentF { stroke [] 0 setdash gsave
  translate 0 hpt M 4 {72 rotate 0 hpt L} repeat
  closepath fill grestore } def
/Circle { stroke [] 0 setdash 2 copy
  hpt 0 360 arc stroke Pnt } def
/CircleF { stroke [] 0 setdash hpt 0 360 arc fill } def
/C0 { BL [] 0 setdash 2 copy moveto vpt 90 450  arc } bind def
/C1 { BL [] 0 setdash 2 copy        moveto
       2 copy  vpt 0 90 arc closepath fill
               vpt 0 360 arc closepath } bind def
/C2 { BL [] 0 setdash 2 copy moveto
       2 copy  vpt 90 180 arc closepath fill
               vpt 0 360 arc closepath } bind def
/C3 { BL [] 0 setdash 2 copy moveto
       2 copy  vpt 0 180 arc closepath fill
               vpt 0 360 arc closepath } bind def
/C4 { BL [] 0 setdash 2 copy moveto
       2 copy  vpt 180 270 arc closepath fill
               vpt 0 360 arc closepath } bind def
/C5 { BL [] 0 setdash 2 copy moveto
       2 copy  vpt 0 90 arc
       2 copy moveto
       2 copy  vpt 180 270 arc closepath fill
               vpt 0 360 arc } bind def
/C6 { BL [] 0 setdash 2 copy moveto
      2 copy  vpt 90 270 arc closepath fill
              vpt 0 360 arc closepath } bind def
/C7 { BL [] 0 setdash 2 copy moveto
      2 copy  vpt 0 270 arc closepath fill
              vpt 0 360 arc closepath } bind def
/C8 { BL [] 0 setdash 2 copy moveto
      2 copy vpt 270 360 arc closepath fill
              vpt 0 360 arc closepath } bind def
/C9 { BL [] 0 setdash 2 copy moveto
      2 copy  vpt 270 450 arc closepath fill
              vpt 0 360 arc closepath } bind def
/C10 { BL [] 0 setdash 2 copy 2 copy moveto vpt 270 360 arc closepath fill
       2 copy moveto
       2 copy vpt 90 180 arc closepath fill
               vpt 0 360 arc closepath } bind def
/C11 { BL [] 0 setdash 2 copy moveto
       2 copy  vpt 0 180 arc closepath fill
       2 copy moveto
       2 copy  vpt 270 360 arc closepath fill
               vpt 0 360 arc closepath } bind def
/C12 { BL [] 0 setdash 2 copy moveto
       2 copy  vpt 180 360 arc closepath fill
               vpt 0 360 arc closepath } bind def
/C13 { BL [] 0 setdash  2 copy moveto
       2 copy  vpt 0 90 arc closepath fill
       2 copy moveto
       2 copy  vpt 180 360 arc closepath fill
               vpt 0 360 arc closepath } bind def
/C14 { BL [] 0 setdash 2 copy moveto
       2 copy  vpt 90 360 arc closepath fill
               vpt 0 360 arc } bind def
/C15 { BL [] 0 setdash 2 copy vpt 0 360 arc closepath fill
               vpt 0 360 arc closepath } bind def
/Rec   { newpath 4 2 roll moveto 1 index 0 rlineto 0 exch rlineto
       neg 0 rlineto closepath } bind def
/Square { dup Rec } bind def
/Bsquare { vpt sub exch vpt sub exch vpt2 Square } bind def
/S0 { BL [] 0 setdash 2 copy moveto 0 vpt rlineto BL Bsquare } bind def
/S1 { BL [] 0 setdash 2 copy vpt Square fill Bsquare } bind def
/S2 { BL [] 0 setdash 2 copy exch vpt sub exch vpt Square fill Bsquare } bind def
/S3 { BL [] 0 setdash 2 copy exch vpt sub exch vpt2 vpt Rec fill Bsquare } bind def
/S4 { BL [] 0 setdash 2 copy exch vpt sub exch vpt sub vpt Square fill Bsquare } bind def
/S5 { BL [] 0 setdash 2 copy 2 copy vpt Square fill
       exch vpt sub exch vpt sub vpt Square fill Bsquare } bind def
/S6 { BL [] 0 setdash 2 copy exch vpt sub exch vpt sub vpt vpt2 Rec fill Bsquare } bind def
/S7 { BL [] 0 setdash 2 copy exch vpt sub exch vpt sub vpt vpt2 Rec fill
       2 copy vpt Square fill
       Bsquare } bind def
/S8 { BL [] 0 setdash 2 copy vpt sub vpt Square fill Bsquare } bind def
/S9 { BL [] 0 setdash 2 copy vpt sub vpt vpt2 Rec fill Bsquare } bind def
/S10 { BL [] 0 setdash 2 copy vpt sub vpt Square fill 2 copy exch vpt sub exch vpt Square fill
       Bsquare } bind def
/S11 { BL [] 0 setdash 2 copy vpt sub vpt Square fill 2 copy exch vpt sub exch vpt2 vpt Rec fill
       Bsquare } bind def
/S12 { BL [] 0 setdash 2 copy exch vpt sub exch vpt sub vpt2 vpt Rec fill Bsquare } bind def
/S13 { BL [] 0 setdash 2 copy exch vpt sub exch vpt sub vpt2 vpt Rec fill
       2 copy vpt Square fill Bsquare } bind def
/S14 { BL [] 0 setdash 2 copy exch vpt sub exch vpt sub vpt2 vpt Rec fill
       2 copy exch vpt sub exch vpt Square fill Bsquare } bind def
/S15 { BL [] 0 setdash 2 copy Bsquare fill Bsquare } bind def
/D0 { gsave translate 45 rotate 0 0 S0 stroke grestore } bind def
/D1 { gsave translate 45 rotate 0 0 S1 stroke grestore } bind def
/D2 { gsave translate 45 rotate 0 0 S2 stroke grestore } bind def
/D3 { gsave translate 45 rotate 0 0 S3 stroke grestore } bind def
/D4 { gsave translate 45 rotate 0 0 S4 stroke grestore } bind def
/D5 { gsave translate 45 rotate 0 0 S5 stroke grestore } bind def
/D6 { gsave translate 45 rotate 0 0 S6 stroke grestore } bind def
/D7 { gsave translate 45 rotate 0 0 S7 stroke grestore } bind def
/D8 { gsave translate 45 rotate 0 0 S8 stroke grestore } bind def
/D9 { gsave translate 45 rotate 0 0 S9 stroke grestore } bind def
/D10 { gsave translate 45 rotate 0 0 S10 stroke grestore } bind def
/D11 { gsave translate 45 rotate 0 0 S11 stroke grestore } bind def
/D12 { gsave translate 45 rotate 0 0 S12 stroke grestore } bind def
/D13 { gsave translate 45 rotate 0 0 S13 stroke grestore } bind def
/D14 { gsave translate 45 rotate 0 0 S14 stroke grestore } bind def
/D15 { gsave translate 45 rotate 0 0 S15 stroke grestore } bind def
/DiaE { stroke [] 0 setdash vpt add M
  hpt neg vpt neg V hpt vpt neg V
  hpt vpt V hpt neg vpt V closepath stroke } def
/BoxE { stroke [] 0 setdash exch hpt sub exch vpt add M
  0 vpt2 neg V hpt2 0 V 0 vpt2 V
  hpt2 neg 0 V closepath stroke } def
/TriUE { stroke [] 0 setdash vpt 1.12 mul add M
  hpt neg vpt -1.62 mul V
  hpt 2 mul 0 V
  hpt neg vpt 1.62 mul V closepath stroke } def
/TriDE { stroke [] 0 setdash vpt 1.12 mul sub M
  hpt neg vpt 1.62 mul V
  hpt 2 mul 0 V
  hpt neg vpt -1.62 mul V closepath stroke } def
/PentE { stroke [] 0 setdash gsave
  translate 0 hpt M 4 {72 rotate 0 hpt L} repeat
  closepath stroke grestore } def
/CircE { stroke [] 0 setdash 
  hpt 0 360 arc stroke } def
/Opaque { gsave closepath 1 setgray fill grestore 0 setgray closepath } def
/DiaW { stroke [] 0 setdash vpt add M
  hpt neg vpt neg V hpt vpt neg V
  hpt vpt V hpt neg vpt V Opaque stroke } def
/BoxW { stroke [] 0 setdash exch hpt sub exch vpt add M
  0 vpt2 neg V hpt2 0 V 0 vpt2 V
  hpt2 neg 0 V Opaque stroke } def
/TriUW { stroke [] 0 setdash vpt 1.12 mul add M
  hpt neg vpt -1.62 mul V
  hpt 2 mul 0 V
  hpt neg vpt 1.62 mul V Opaque stroke } def
/TriDW { stroke [] 0 setdash vpt 1.12 mul sub M
  hpt neg vpt 1.62 mul V
  hpt 2 mul 0 V
  hpt neg vpt -1.62 mul V Opaque stroke } def
/PentW { stroke [] 0 setdash gsave
  translate 0 hpt M 4 {72 rotate 0 hpt L} repeat
  Opaque stroke grestore } def
/CircW { stroke [] 0 setdash 
  hpt 0 360 arc Opaque stroke } def
/BoxFill { gsave Rec 1 setgray fill grestore } def
end
}}%
\begin{picture}(3600,2160)(0,0)%
{\GNUPLOTspecial{"
gnudict begin
gsave
0 0 translate
0.100 0.100 scale
0 setgray
newpath
1.000 UL
LTb
400 300 M
63 0 V
2987 0 R
-63 0 V
400 551 M
63 0 V
2987 0 R
-63 0 V
400 803 M
63 0 V
2987 0 R
-63 0 V
400 1054 M
63 0 V
2987 0 R
-63 0 V
400 1306 M
63 0 V
2987 0 R
-63 0 V
400 1557 M
63 0 V
2987 0 R
-63 0 V
400 1809 M
63 0 V
2987 0 R
-63 0 V
400 2060 M
63 0 V
2987 0 R
-63 0 V
400 300 M
0 63 V
0 1697 R
0 -63 V
705 300 M
0 63 V
0 1697 R
0 -63 V
1010 300 M
0 63 V
0 1697 R
0 -63 V
1315 300 M
0 63 V
0 1697 R
0 -63 V
1620 300 M
0 63 V
0 1697 R
0 -63 V
1925 300 M
0 63 V
0 1697 R
0 -63 V
2230 300 M
0 63 V
0 1697 R
0 -63 V
2535 300 M
0 63 V
0 1697 R
0 -63 V
2840 300 M
0 63 V
0 1697 R
0 -63 V
3145 300 M
0 63 V
0 1697 R
0 -63 V
3450 300 M
0 63 V
0 1697 R
0 -63 V
1.000 UL
LTb
400 300 M
3050 0 V
0 1760 V
-3050 0 V
400 300 L
1.000 UL
LT0
400 419 M
500 18 V
106 27 V
106 34 V
212 95 V
211 117 V
424 195 V
424 91 V
423 35 V
424 11 V
220 -1 V
1.000 UL
LT1
1112 402 M
423 381 V
424 433 V
424 154 V
423 21 V
424 30 V
220 2 V
1.000 UL
LT2
1112 331 M
423 567 V
424 783 V
424 191 V
1067 30 V
0.600 UP
1.000 UL
LT3
900 435 M
0 3 V
-31 -3 R
62 0 V
-62 3 R
62 0 V
75 26 R
0 1 V
-31 -1 R
62 0 V
-62 1 R
62 0 V
75 30 R
0 5 V
-31 -5 R
62 0 V
-62 5 R
62 0 V
181 89 R
0 7 V
-31 -7 R
62 0 V
-62 7 R
62 0 V
180 110 R
0 7 V
-31 -7 R
62 0 V
-62 7 R
62 0 V
393 189 R
0 6 V
-31 -6 R
62 0 V
-62 6 R
62 0 V
393 85 R
0 5 V
-31 -5 R
62 0 V
-62 5 R
62 0 V
392 30 R
0 6 V
-31 -6 R
62 0 V
-62 6 R
62 0 V
393 5 R
0 6 V
-31 -6 R
62 0 V
-62 6 R
62 0 V
900 437 Pls
1006 464 Pls
1112 498 Pls
1324 593 Pls
1535 710 Pls
1959 905 Pls
2383 996 Pls
2806 1031 Pls
3230 1042 Pls
0.600 UP
1.000 UL
LT4
1112 401 M
0 2 V
-31 -2 R
62 0 V
-62 2 R
62 0 V
392 378 R
0 5 V
-31 -5 R
62 0 V
-62 5 R
62 0 V
393 427 R
0 6 V
-31 -6 R
62 0 V
-62 6 R
62 0 V
393 147 R
0 7 V
-31 -7 R
62 0 V
-62 7 R
62 0 V
392 15 R
0 7 V
-31 -7 R
62 0 V
-62 7 R
62 0 V
393 24 R
0 5 V
-31 -5 R
62 0 V
-62 5 R
62 0 V
1112 402 Crs
1535 783 Crs
1959 1216 Crs
2383 1370 Crs
2806 1391 Crs
3230 1421 Crs
0.600 UP
1.000 UL
LT5
1112 329 M
0 4 V
-31 -4 R
62 0 V
-62 4 R
62 0 V
392 558 R
0 13 V
-31 -13 R
62 0 V
-62 13 R
62 0 V
393 768 R
0 17 V
-31 -17 R
62 0 V
-62 17 R
62 0 V
393 172 R
0 22 V
-31 -22 R
62 0 V
-62 22 R
62 0 V
1112 331 Star
1535 898 Star
1959 1681 Star
2383 1872 Star
stroke
grestore
end
showpage
}}%
\put(1925,50){\makebox(0,0){$\gamma$}}%
\put(100,1180){%
\makebox(0,0)[b]{\shortstack{$R^{2 \times 2}$}}%
}%
\put(3450,200){\makebox(0,0){0.8}}%
\put(3145,200){\makebox(0,0){0.75}}%
\put(2840,200){\makebox(0,0){0.7}}%
\put(2535,200){\makebox(0,0){0.65}}%
\put(2230,200){\makebox(0,0){0.6}}%
\put(1925,200){\makebox(0,0){0.55}}%
\put(1620,200){\makebox(0,0){0.5}}%
\put(1315,200){\makebox(0,0){0.45}}%
\put(1010,200){\makebox(0,0){0.4}}%
\put(705,200){\makebox(0,0){0.35}}%
\put(400,200){\makebox(0,0){0.3}}%
\put(350,2060){\makebox(0,0)[r]{4}}%
\put(350,1809){\makebox(0,0)[r]{3.5}}%
\put(350,1557){\makebox(0,0)[r]{3}}%
\put(350,1306){\makebox(0,0)[r]{2.5}}%
\put(350,1054){\makebox(0,0)[r]{2}}%
\put(350,803){\makebox(0,0)[r]{1.5}}%
\put(350,551){\makebox(0,0)[r]{1}}%
\put(350,300){\makebox(0,0)[r]{0.5}}%
\end{picture}%
\endgroup

%% file: wilsoneigenvaluedensitygraph.tex
\begingroup%
  \makeatletter%
  \newcommand{\GNUPLOTspecial}{%
    \@sanitize\catcode`\%=14\relax\special}%
  \setlength{\unitlength}{0.1bp}%
{\GNUPLOTspecial{!
/gnudict 256 dict def
gnudict begin
/Color false def
/Solid false def
/gnulinewidth 5.000 def
/userlinewidth gnulinewidth def
/vshift -33 def
/dl {10 mul} def
/hpt_ 31.5 def
/vpt_ 31.5 def
/hpt hpt_ def
/vpt vpt_ def
/M {moveto} bind def
/L {lineto} bind def
/R {rmoveto} bind def
/V {rlineto} bind def
/vpt2 vpt 2 mul def
/hpt2 hpt 2 mul def
/Lshow { currentpoint stroke M
  0 vshift R show } def
/Rshow { currentpoint stroke M
  dup stringwidth pop neg vshift R show } def
/Cshow { currentpoint stroke M
  dup stringwidth pop -2 div vshift R show } def
/UP { dup vpt_ mul /vpt exch def hpt_ mul /hpt exch def
  /hpt2 hpt 2 mul def /vpt2 vpt 2 mul def } def
/DL { Color {setrgbcolor Solid {pop []} if 0 setdash }
 {pop pop pop Solid {pop []} if 0 setdash} ifelse } def
/BL { stroke userlinewidth 2 mul setlinewidth } def
/AL { stroke userlinewidth 2 div setlinewidth } def
/UL { dup gnulinewidth mul /userlinewidth exch def
      10 mul /udl exch def } def
/PL { stroke userlinewidth setlinewidth } def
/LTb { BL [] 0 0 0 DL } def
/LTa { AL [1 udl mul 2 udl mul] 0 setdash 0 0 0 setrgbcolor } def
/LT0 { PL [] 1 0 0 DL } def
/LT1 { PL [4 dl 2 dl] 0 1 0 DL } def
/LT2 { PL [2 dl 3 dl] 0 0 1 DL } def
/LT3 { PL [1 dl 1.5 dl] 1 0 1 DL } def
/LT4 { PL [5 dl 2 dl 1 dl 2 dl] 0 1 1 DL } def
/LT5 { PL [4 dl 3 dl 1 dl 3 dl] 1 1 0 DL } def
/LT6 { PL [2 dl 2 dl 2 dl 4 dl] 0 0 0 DL } def
/LT7 { PL [2 dl 2 dl 2 dl 2 dl 2 dl 4 dl] 1 0.3 0 DL } def
/LT8 { PL [2 dl 2 dl 2 dl 2 dl 2 dl 2 dl 2 dl 4 dl] 0.5 0.5 0.5 DL } def
/Pnt { stroke [] 0 setdash
   gsave 1 setlinecap M 0 0 V stroke grestore } def
/Dia { stroke [] 0 setdash 2 copy vpt add M
  hpt neg vpt neg V hpt vpt neg V
  hpt vpt V hpt neg vpt V closepath stroke
  Pnt } def
/Pls { stroke [] 0 setdash vpt sub M 0 vpt2 V
  currentpoint stroke M
  hpt neg vpt neg R hpt2 0 V stroke
  } def
/Box { stroke [] 0 setdash 2 copy exch hpt sub exch vpt add M
  0 vpt2 neg V hpt2 0 V 0 vpt2 V
  hpt2 neg 0 V closepath stroke
  Pnt } def
/Crs { stroke [] 0 setdash exch hpt sub exch vpt add M
  hpt2 vpt2 neg V currentpoint stroke M
  hpt2 neg 0 R hpt2 vpt2 V stroke } def
/TriU { stroke [] 0 setdash 2 copy vpt 1.12 mul add M
  hpt neg vpt -1.62 mul V
  hpt 2 mul 0 V
  hpt neg vpt 1.62 mul V closepath stroke
  Pnt  } def
/Star { 2 copy Pls Crs } def
/BoxF { stroke [] 0 setdash exch hpt sub exch vpt add M
  0 vpt2 neg V  hpt2 0 V  0 vpt2 V
  hpt2 neg 0 V  closepath fill } def
/TriUF { stroke [] 0 setdash vpt 1.12 mul add M
  hpt neg vpt -1.62 mul V
  hpt 2 mul 0 V
  hpt neg vpt 1.62 mul V closepath fill } def
/TriD { stroke [] 0 setdash 2 copy vpt 1.12 mul sub M
  hpt neg vpt 1.62 mul V
  hpt 2 mul 0 V
  hpt neg vpt -1.62 mul V closepath stroke
  Pnt  } def
/TriDF { stroke [] 0 setdash vpt 1.12 mul sub M
  hpt neg vpt 1.62 mul V
  hpt 2 mul 0 V
  hpt neg vpt -1.62 mul V closepath fill} def
/DiaF { stroke [] 0 setdash vpt add M
  hpt neg vpt neg V hpt vpt neg V
  hpt vpt V hpt neg vpt V closepath fill } def
/Pent { stroke [] 0 setdash 2 copy gsave
  translate 0 hpt M 4 {72 rotate 0 hpt L} repeat
  closepath stroke grestore Pnt } def
/PentF { stroke [] 0 setdash gsave
  translate 0 hpt M 4 {72 rotate 0 hpt L} repeat
  closepath fill grestore } def
/Circle { stroke [] 0 setdash 2 copy
  hpt 0 360 arc stroke Pnt } def
/CircleF { stroke [] 0 setdash hpt 0 360 arc fill } def
/C0 { BL [] 0 setdash 2 copy moveto vpt 90 450  arc } bind def
/C1 { BL [] 0 setdash 2 copy        moveto
       2 copy  vpt 0 90 arc closepath fill
               vpt 0 360 arc closepath } bind def
/C2 { BL [] 0 setdash 2 copy moveto
       2 copy  vpt 90 180 arc closepath fill
               vpt 0 360 arc closepath } bind def
/C3 { BL [] 0 setdash 2 copy moveto
       2 copy  vpt 0 180 arc closepath fill
               vpt 0 360 arc closepath } bind def
/C4 { BL [] 0 setdash 2 copy moveto
       2 copy  vpt 180 270 arc closepath fill
               vpt 0 360 arc closepath } bind def
/C5 { BL [] 0 setdash 2 copy moveto
       2 copy  vpt 0 90 arc
       2 copy moveto
       2 copy  vpt 180 270 arc closepath fill
               vpt 0 360 arc } bind def
/C6 { BL [] 0 setdash 2 copy moveto
      2 copy  vpt 90 270 arc closepath fill
              vpt 0 360 arc closepath } bind def
/C7 { BL [] 0 setdash 2 copy moveto
      2 copy  vpt 0 270 arc closepath fill
              vpt 0 360 arc closepath } bind def
/C8 { BL [] 0 setdash 2 copy moveto
      2 copy vpt 270 360 arc closepath fill
              vpt 0 360 arc closepath } bind def
/C9 { BL [] 0 setdash 2 copy moveto
      2 copy  vpt 270 450 arc closepath fill
              vpt 0 360 arc closepath } bind def
/C10 { BL [] 0 setdash 2 copy 2 copy moveto vpt 270 360 arc closepath fill
       2 copy moveto
       2 copy vpt 90 180 arc closepath fill
               vpt 0 360 arc closepath } bind def
/C11 { BL [] 0 setdash 2 copy moveto
       2 copy  vpt 0 180 arc closepath fill
       2 copy moveto
       2 copy  vpt 270 360 arc closepath fill
               vpt 0 360 arc closepath } bind def
/C12 { BL [] 0 setdash 2 copy moveto
       2 copy  vpt 180 360 arc closepath fill
               vpt 0 360 arc closepath } bind def
/C13 { BL [] 0 setdash  2 copy moveto
       2 copy  vpt 0 90 arc closepath fill
       2 copy moveto
       2 copy  vpt 180 360 arc closepath fill
               vpt 0 360 arc closepath } bind def
/C14 { BL [] 0 setdash 2 copy moveto
       2 copy  vpt 90 360 arc closepath fill
               vpt 0 360 arc } bind def
/C15 { BL [] 0 setdash 2 copy vpt 0 360 arc closepath fill
               vpt 0 360 arc closepath } bind def
/Rec   { newpath 4 2 roll moveto 1 index 0 rlineto 0 exch rlineto
       neg 0 rlineto closepath } bind def
/Square { dup Rec } bind def
/Bsquare { vpt sub exch vpt sub exch vpt2 Square } bind def
/S0 { BL [] 0 setdash 2 copy moveto 0 vpt rlineto BL Bsquare } bind def
/S1 { BL [] 0 setdash 2 copy vpt Square fill Bsquare } bind def
/S2 { BL [] 0 setdash 2 copy exch vpt sub exch vpt Square fill Bsquare } bind def
/S3 { BL [] 0 setdash 2 copy exch vpt sub exch vpt2 vpt Rec fill Bsquare } bind def
/S4 { BL [] 0 setdash 2 copy exch vpt sub exch vpt sub vpt Square fill Bsquare } bind def
/S5 { BL [] 0 setdash 2 copy 2 copy vpt Square fill
       exch vpt sub exch vpt sub vpt Square fill Bsquare } bind def
/S6 { BL [] 0 setdash 2 copy exch vpt sub exch vpt sub vpt vpt2 Rec fill Bsquare } bind def
/S7 { BL [] 0 setdash 2 copy exch vpt sub exch vpt sub vpt vpt2 Rec fill
       2 copy vpt Square fill
       Bsquare } bind def
/S8 { BL [] 0 setdash 2 copy vpt sub vpt Square fill Bsquare } bind def
/S9 { BL [] 0 setdash 2 copy vpt sub vpt vpt2 Rec fill Bsquare } bind def
/S10 { BL [] 0 setdash 2 copy vpt sub vpt Square fill 2 copy exch vpt sub exch vpt Square fill
       Bsquare } bind def
/S11 { BL [] 0 setdash 2 copy vpt sub vpt Square fill 2 copy exch vpt sub exch vpt2 vpt Rec fill
       Bsquare } bind def
/S12 { BL [] 0 setdash 2 copy exch vpt sub exch vpt sub vpt2 vpt Rec fill Bsquare } bind def
/S13 { BL [] 0 setdash 2 copy exch vpt sub exch vpt sub vpt2 vpt Rec fill
       2 copy vpt Square fill Bsquare } bind def
/S14 { BL [] 0 setdash 2 copy exch vpt sub exch vpt sub vpt2 vpt Rec fill
       2 copy exch vpt sub exch vpt Square fill Bsquare } bind def
/S15 { BL [] 0 setdash 2 copy Bsquare fill Bsquare } bind def
/D0 { gsave translate 45 rotate 0 0 S0 stroke grestore } bind def
/D1 { gsave translate 45 rotate 0 0 S1 stroke grestore } bind def
/D2 { gsave translate 45 rotate 0 0 S2 stroke grestore } bind def
/D3 { gsave translate 45 rotate 0 0 S3 stroke grestore } bind def
/D4 { gsave translate 45 rotate 0 0 S4 stroke grestore } bind def
/D5 { gsave translate 45 rotate 0 0 S5 stroke grestore } bind def
/D6 { gsave translate 45 rotate 0 0 S6 stroke grestore } bind def
/D7 { gsave translate 45 rotate 0 0 S7 stroke grestore } bind def
/D8 { gsave translate 45 rotate 0 0 S8 stroke grestore } bind def
/D9 { gsave translate 45 rotate 0 0 S9 stroke grestore } bind def
/D10 { gsave translate 45 rotate 0 0 S10 stroke grestore } bind def
/D11 { gsave translate 45 rotate 0 0 S11 stroke grestore } bind def
/D12 { gsave translate 45 rotate 0 0 S12 stroke grestore } bind def
/D13 { gsave translate 45 rotate 0 0 S13 stroke grestore } bind def
/D14 { gsave translate 45 rotate 0 0 S14 stroke grestore } bind def
/D15 { gsave translate 45 rotate 0 0 S15 stroke grestore } bind def
/DiaE { stroke [] 0 setdash vpt add M
  hpt neg vpt neg V hpt vpt neg V
  hpt vpt V hpt neg vpt V closepath stroke } def
/BoxE { stroke [] 0 setdash exch hpt sub exch vpt add M
  0 vpt2 neg V hpt2 0 V 0 vpt2 V
  hpt2 neg 0 V closepath stroke } def
/TriUE { stroke [] 0 setdash vpt 1.12 mul add M
  hpt neg vpt -1.62 mul V
  hpt 2 mul 0 V
  hpt neg vpt 1.62 mul V closepath stroke } def
/TriDE { stroke [] 0 setdash vpt 1.12 mul sub M
  hpt neg vpt 1.62 mul V
  hpt 2 mul 0 V
  hpt neg vpt -1.62 mul V closepath stroke } def
/PentE { stroke [] 0 setdash gsave
  translate 0 hpt M 4 {72 rotate 0 hpt L} repeat
  closepath stroke grestore } def
/CircE { stroke [] 0 setdash 
  hpt 0 360 arc stroke } def
/Opaque { gsave closepath 1 setgray fill grestore 0 setgray closepath } def
/DiaW { stroke [] 0 setdash vpt add M
  hpt neg vpt neg V hpt vpt neg V
  hpt vpt V hpt neg vpt V Opaque stroke } def
/BoxW { stroke [] 0 setdash exch hpt sub exch vpt add M
  0 vpt2 neg V hpt2 0 V 0 vpt2 V
  hpt2 neg 0 V Opaque stroke } def
/TriUW { stroke [] 0 setdash vpt 1.12 mul add M
  hpt neg vpt -1.62 mul V
  hpt 2 mul 0 V
  hpt neg vpt 1.62 mul V Opaque stroke } def
/TriDW { stroke [] 0 setdash vpt 1.12 mul sub M
  hpt neg vpt 1.62 mul V
  hpt 2 mul 0 V
  hpt neg vpt -1.62 mul V Opaque stroke } def
/PentW { stroke [] 0 setdash gsave
  translate 0 hpt M 4 {72 rotate 0 hpt L} repeat
  Opaque stroke grestore } def
/CircW { stroke [] 0 setdash 
  hpt 0 360 arc Opaque stroke } def
/BoxFill { gsave Rec 1 setgray fill grestore } def
end
}}%
\begin{picture}(3600,2160)(0,0)%
{\GNUPLOTspecial{"
gnudict begin
gsave
0 0 translate
0.100 0.100 scale
0 setgray
newpath
1.000 UL
LTb
450 300 M
63 0 V
2937 0 R
-63 0 V
450 652 M
63 0 V
2937 0 R
-63 0 V
450 1004 M
63 0 V
2937 0 R
-63 0 V
450 1356 M
63 0 V
2937 0 R
-63 0 V
450 1708 M
63 0 V
2937 0 R
-63 0 V
450 2060 M
63 0 V
2937 0 R
-63 0 V
518 300 M
0 63 V
0 1697 R
0 -63 V
995 300 M
0 63 V
0 1697 R
0 -63 V
1473 300 M
0 63 V
0 1697 R
0 -63 V
1950 300 M
0 63 V
0 1697 R
0 -63 V
2427 300 M
0 63 V
0 1697 R
0 -63 V
2905 300 M
0 63 V
0 1697 R
0 -63 V
3382 300 M
0 63 V
0 1697 R
0 -63 V
1.000 UL
LTb
450 300 M
3000 0 V
0 1760 V
-3000 0 V
450 300 L
1.000 UL
LT0
457 559 M
15 12 V
15 28 V
15 39 V
15 54 V
15 66 V
15 74 V
15 69 V
15 85 V
15 83 V
15 73 V
15 74 V
15 64 V
15 39 V
15 26 V
15 5 V
15 2 V
15 -17 V
15 -27 V
15 -42 V
15 -61 V
15 -16 V
15 -37 V
15 -21 V
15 -6 V
15 14 V
15 31 V
15 53 V
15 76 V
15 55 V
15 73 V
15 64 V
15 32 V
15 37 V
15 12 V
15 -21 V
15 -47 V
15 -46 V
15 -60 V
15 -57 V
15 -44 V
15 -16 V
15 12 V
15 40 V
15 46 V
15 67 V
15 83 V
15 76 V
15 44 V
15 44 V
15 -1 V
15 -24 V
15 -43 V
15 -76 V
15 -78 V
15 -60 V
15 -31 V
15 -16 V
15 17 V
15 58 V
15 75 V
15 81 V
15 83 V
15 63 V
15 38 V
15 -18 V
15 -17 V
15 -66 V
15 -86 V
15 -86 V
15 -46 V
15 -56 V
15 8 V
15 32 V
15 78 V
15 89 V
15 92 V
15 62 V
15 59 V
15 9 V
15 -42 V
15 -59 V
15 -95 V
15 -67 V
15 -85 V
15 -49 V
15 -4 V
15 38 V
15 65 V
15 119 V
15 77 V
15 75 V
15 32 V
15 23 V
15 -43 V
15 -81 V
15 -92 V
15 -86 V
15 -64 V
15 -54 V
16 -3 V
15 48 V
15 84 V
15 72 V
15 105 V
15 75 V
15 19 V
15 0 V
15 -44 V
15 -69 V
15 -83 V
15 -107 V
15 -67 V
15 -39 V
15 2 V
15 56 V
15 66 V
15 92 V
15 83 V
15 60 V
15 24 V
15 5 V
15 -45 V
15 -73 V
15 -82 V
15 -95 V
15 -75 V
15 -28 V
15 -8 V
15 30 V
15 67 V
15 78 V
15 76 V
15 67 V
15 48 V
15 -4 V
15 -31 V
15 -71 V
15 -65 V
15 -104 V
15 -67 V
15 -51 V
15 -26 V
15 14 V
15 46 V
15 62 V
15 62 V
15 70 V
15 59 V
15 10 V
15 10 V
15 -41 V
15 -58 V
15 -61 V
15 -84 V
15 -77 V
15 -44 V
15 -46 V
15 -5 V
15 25 V
15 39 V
15 56 V
15 61 V
15 61 V
15 28 V
15 17 V
15 -2 V
15 -20 V
15 -49 V
15 -68 V
15 -73 V
15 -64 V
15 -72 V
15 -45 V
15 -29 V
15 -24 V
15 6 V
15 20 V
15 35 V
15 47 V
15 43 V
15 29 V
15 24 V
15 34 V
15 11 V
15 -25 V
15 -17 V
15 -47 V
15 -77 V
15 -51 V
15 -70 V
15 -100 V
15 -80 V
15 -73 V
15 -71 V
15 -69 V
15 -47 V
15 -44 V
15 -29 V
15 -19 V
1.000 UL
LT1
457 558 M
15 13 V
16 25 V
15 41 V
15 53 V
14 67 V
15 77 V
16 76 V
15 85 V
14 77 V
15 74 V
16 74 V
14 58 V
16 42 V
14 30 V
16 11 V
15 -7 V
14 -16 V
16 -33 V
14 -42 V
15 -44 V
16 -37 V
14 -35 V
15 -19 V
16 0 V
14 15 V
15 29 V
16 55 V
14 63 V
15 72 V
16 64 V
14 73 V
15 41 V
16 26 V
14 4 V
15 -21 V
16 -51 V
14 -52 V
15 -50 V
16 -55 V
15 -41 V
14 -18 V
16 10 V
15 38 V
14 49 V
16 81 V
15 72 V
14 70 V
16 60 V
15 23 V
14 6 V
15 -30 V
15 -56 V
16 -64 V
15 -64 V
15 -63 V
14 -44 V
15 -10 V
15 27 V
16 58 V
15 72 V
15 76 V
14 85 V
15 57 V
15 43 V
15 -1 V
16 -40 V
15 -67 V
15 -76 V
14 -84 V
15 -71 V
15 -29 V
16 7 V
15 33 V
15 71 V
14 94 V
15 88 V
15 74 V
16 42 V
15 3 V
15 -30 V
14 -61 V
15 -86 V
15 -91 V
16 -73 V
15 -44 V
15 -5 V
14 38 V
15 75 V
15 86 V
15 95 V
16 84 V
15 39 V
15 -2 V
14 -38 V
15 -68 V
15 -91 V
16 -92 V
15 -79 V
15 -34 V
14 -4 V
15 45 V
15 69 V
15 100 V
15 82 V
15 82 V
16 26 V
15 3 V
15 -47 V
15 -78 V
15 -90 V
15 -96 V
15 -67 V
14 -35 V
15 4 V
15 41 V
15 75 V
15 81 V
15 94 V
16 66 V
15 29 V
15 1 V
15 -56 V
15 -70 V
15 -91 V
14 -90 V
15 -66 V
15 -31 V
15 -8 V
15 32 V
15 63 V
16 84 V
15 71 V
15 68 V
15 37 V
15 6 V
15 -37 V
15 -63 V
14 -74 V
15 -85 V
15 -78 V
15 -45 V
15 -28 V
15 4 V
16 46 V
15 62 V
15 69 V
15 66 V
15 56 V
15 23 V
14 1 V
15 -39 V
15 -50 V
15 -73 V
15 -80 V
15 -67 V
16 -63 V
15 -31 V
15 -9 V
15 19 V
15 43 V
15 55 V
15 61 V
14 47 V
15 39 V
15 25 V
15 -6 V
15 -22 V
15 -46 V
16 -65 V
15 -69 V
15 -68 V
15 -65 V
15 -51 V
15 -36 V
14 -20 V
15 5 V
15 20 V
15 30 V
15 43 V
15 51 V
16 33 V
15 38 V
15 12 V
15 7 V
15 -10 V
15 -30 V
15 -45 V
14 -56 V
15 -65 V
15 -82 V
15 -85 V
15 -78 V
15 -78 V
16 -75 V
15 -65 V
15 -48 V
15 -44 V
15 -29 V
15 -15 V
1.000 UL
LT2
1950 1644 M
8 0 V
7 0 V
8 0 V
7 0 V
8 0 V
7 0 V
8 0 V
7 -1 V
8 0 V
7 0 V
8 0 V
7 -1 V
8 0 V
7 0 V
8 -1 V
7 0 V
8 0 V
7 -1 V
8 0 V
7 -1 V
8 0 V
7 -1 V
8 0 V
7 -1 V
8 0 V
7 -1 V
8 -1 V
7 0 V
8 -1 V
7 -1 V
8 -1 V
7 0 V
8 -1 V
7 -1 V
8 -1 V
7 -1 V
8 -1 V
7 -1 V
8 -1 V
7 -1 V
8 -1 V
7 -1 V
8 -1 V
7 -1 V
8 -1 V
7 -1 V
8 -1 V
7 -1 V
8 -2 V
7 -1 V
8 -1 V
7 -2 V
8 -1 V
7 -1 V
8 -2 V
7 -1 V
8 -1 V
7 -2 V
8 -1 V
7 -2 V
8 -2 V
7 -1 V
8 -2 V
7 -1 V
8 -2 V
7 -2 V
8 -2 V
7 -1 V
8 -2 V
7 -2 V
8 -2 V
7 -2 V
8 -2 V
7 -2 V
8 -2 V
7 -2 V
8 -2 V
7 -2 V
8 -2 V
7 -2 V
8 -2 V
7 -2 V
8 -3 V
7 -2 V
8 -2 V
7 -3 V
8 -2 V
7 -2 V
8 -3 V
7 -2 V
8 -3 V
7 -2 V
8 -3 V
7 -3 V
8 -2 V
7 -3 V
8 -3 V
7 -3 V
8 -2 V
7 -3 V
8 -3 V
7 -3 V
8 -3 V
7 -3 V
8 -3 V
7 -3 V
8 -4 V
7 -3 V
8 -3 V
7 -3 V
8 -4 V
7 -3 V
8 -4 V
7 -3 V
8 -4 V
7 -3 V
8 -4 V
7 -4 V
8 -3 V
7 -4 V
8 -4 V
7 -4 V
8 -4 V
7 -4 V
8 -4 V
7 -4 V
8 -4 V
7 -4 V
8 -5 V
7 -4 V
8 -5 V
7 -4 V
8 -5 V
7 -4 V
8 -5 V
7 -5 V
8 -5 V
7 -4 V
7 -5 V
8 -5 V
8 -6 V
7 -5 V
8 -5 V
7 -6 V
8 -5 V
7 -6 V
8 -5 V
7 -6 V
8 -6 V
7 -6 V
8 -6 V
7 -6 V
8 -6 V
7 -7 V
8 -6 V
7 -7 V
8 -7 V
7 -7 V
8 -7 V
7 -7 V
8 -7 V
7 -8 V
8 -7 V
7 -8 V
8 -8 V
7 -8 V
8 -8 V
7 -9 V
8 -9 V
7 -9 V
8 -9 V
7 -9 V
8 -10 V
7 -10 V
8 -10 V
7 -10 V
8 -11 V
7 -11 V
8 -12 V
7 -12 V
8 -12 V
7 -13 V
8 -13 V
7 -14 V
8 -15 V
7 -15 V
8 -16 V
7 -17 V
8 -18 V
7 -19 V
8 -21 V
7 -22 V
8 -24 V
7 -26 V
8 -30 V
7 -34 V
8 -40 V
7 -51 V
8 -72 V
7 -278 V
1.000 UL
LT2
1950 1644 M
-7 0 V
-8 0 V
-7 0 V
-8 0 V
-7 0 V
-8 0 V
-7 0 V
-8 -1 V
-7 0 V
-8 0 V
-7 0 V
-8 -1 V
-7 0 V
-8 0 V
-7 -1 V
-8 0 V
-7 0 V
-8 -1 V
-7 0 V
-8 -1 V
-7 0 V
-8 -1 V
-7 0 V
-8 -1 V
-7 0 V
-8 -1 V
-7 -1 V
-8 0 V
-7 -1 V
-8 -1 V
-7 -1 V
-8 0 V
-7 -1 V
-8 -1 V
-7 -1 V
-8 -1 V
-7 -1 V
-8 -1 V
-7 -1 V
-8 -1 V
-7 -1 V
-8 -1 V
-7 -1 V
-8 -1 V
-7 -1 V
-8 -1 V
-7 -1 V
-8 -1 V
-7 -2 V
-8 -1 V
-7 -1 V
-8 -2 V
-7 -1 V
-8 -1 V
-7 -2 V
-8 -1 V
-7 -1 V
-8 -2 V
-7 -1 V
-8 -2 V
-7 -2 V
-8 -1 V
-7 -2 V
-8 -1 V
-7 -2 V
-8 -2 V
-7 -2 V
-8 -1 V
-7 -2 V
-8 -2 V
-7 -2 V
-8 -2 V
-7 -2 V
-8 -2 V
-7 -2 V
-8 -2 V
-7 -2 V
-8 -2 V
-7 -2 V
-8 -2 V
-7 -2 V
-8 -2 V
-7 -3 V
-8 -2 V
-7 -2 V
-8 -3 V
-7 -2 V
-8 -2 V
-7 -3 V
-8 -2 V
-7 -3 V
-8 -2 V
-7 -3 V
-8 -3 V
-7 -2 V
-8 -3 V
-7 -3 V
-8 -3 V
-7 -2 V
-8 -3 V
-7 -3 V
-8 -3 V
-7 -3 V
-8 -3 V
-7 -3 V
-8 -3 V
-7 -4 V
-8 -3 V
-7 -3 V
-8 -3 V
-7 -4 V
-8 -3 V
-7 -4 V
-8 -3 V
-7 -4 V
-8 -3 V
-7 -4 V
-8 -4 V
-7 -3 V
-8 -4 V
-7 -4 V
-8 -4 V
-7 -4 V
-8 -4 V
-7 -4 V
-8 -4 V
-7 -4 V
-8 -4 V
-7 -5 V
-8 -4 V
-7 -5 V
-8 -4 V
-7 -5 V
-8 -4 V
-7 -5 V
-8 -5 V
-8 -5 V
-7 -4 V
-7 -5 V
-8 -5 V
-7 -6 V
-8 -5 V
-7 -5 V
-8 -6 V
-7 -5 V
-8 -6 V
-7 -5 V
-8 -6 V
-7 -6 V
-8 -6 V
-7 -6 V
-8 -6 V
-7 -6 V
-8 -7 V
-7 -6 V
-8 -7 V
-8 -7 V
-7 -7 V
-7 -7 V
-8 -7 V
-8 -7 V
-7 -8 V
-7 -7 V
-8 -8 V
-7 -8 V
-8 -8 V
-7 -8 V
-8 -9 V
-7 -9 V
-8 -9 V
-7 -9 V
-8 -9 V
-8 -10 V
-7 -10 V
-7 -10 V
-8 -10 V
-7 -11 V
-8 -11 V
-7 -12 V
-8 -12 V
-7 -12 V
-8 -13 V
-8 -13 V
-7 -14 V
-7 -15 V
-8 -15 V
-8 -16 V
-7 -17 V
-7 -18 V
-8 -19 V
-7 -21 V
-8 -22 V
-7 -24 V
-8 -26 V
-7 -30 V
-8 -34 V
-7 -40 V
-8 -51 V
-8 -72 V
450 300 L
stroke
grestore
end
showpage
}}%
\put(1950,50){\makebox(0,0){$\alpha$}}%
\put(100,1180){%
\makebox(0,0)[b]{\shortstack{$\rho(\alpha)$}}%
}%
\put(3382,200){\makebox(0,0){3}}%
\put(2905,200){\makebox(0,0){2}}%
\put(2427,200){\makebox(0,0){1}}%
\put(1950,200){\makebox(0,0){0}}%
\put(1473,200){\makebox(0,0){-1}}%
\put(995,200){\makebox(0,0){-2}}%
\put(518,200){\makebox(0,0){-3}}%
\put(400,2060){\makebox(0,0)[r]{0.25}}%
\put(400,1708){\makebox(0,0)[r]{0.2}}%
\put(400,1356){\makebox(0,0)[r]{0.15}}%
\put(400,1004){\makebox(0,0)[r]{0.1}}%
\put(400,652){\makebox(0,0)[r]{0.05}}%
\put(400,300){\makebox(0,0)[r]{0}}%
\end{picture}%
\endgroup

%% file: wilsoneigenvaluedensitygraph2.tex
\begingroup%
  \makeatletter%
  \newcommand{\GNUPLOTspecial}{%
    \@sanitize\catcode`\%=14\relax\special}%
  \setlength{\unitlength}{0.1bp}%
{\GNUPLOTspecial{!
/gnudict 256 dict def
gnudict begin
/Color false def
/Solid false def
/gnulinewidth 5.000 def
/userlinewidth gnulinewidth def
/vshift -33 def
/dl {10 mul} def
/hpt_ 31.5 def
/vpt_ 31.5 def
/hpt hpt_ def
/vpt vpt_ def
/M {moveto} bind def
/L {lineto} bind def
/R {rmoveto} bind def
/V {rlineto} bind def
/vpt2 vpt 2 mul def
/hpt2 hpt 2 mul def
/Lshow { currentpoint stroke M
  0 vshift R show } def
/Rshow { currentpoint stroke M
  dup stringwidth pop neg vshift R show } def
/Cshow { currentpoint stroke M
  dup stringwidth pop -2 div vshift R show } def
/UP { dup vpt_ mul /vpt exch def hpt_ mul /hpt exch def
  /hpt2 hpt 2 mul def /vpt2 vpt 2 mul def } def
/DL { Color {setrgbcolor Solid {pop []} if 0 setdash }
 {pop pop pop Solid {pop []} if 0 setdash} ifelse } def
/BL { stroke userlinewidth 2 mul setlinewidth } def
/AL { stroke userlinewidth 2 div setlinewidth } def
/UL { dup gnulinewidth mul /userlinewidth exch def
      10 mul /udl exch def } def
/PL { stroke userlinewidth setlinewidth } def
/LTb { BL [] 0 0 0 DL } def
/LTa { AL [1 udl mul 2 udl mul] 0 setdash 0 0 0 setrgbcolor } def
/LT0 { PL [] 1 0 0 DL } def
/LT1 { PL [4 dl 2 dl] 0 1 0 DL } def
/LT2 { PL [2 dl 3 dl] 0 0 1 DL } def
/LT3 { PL [1 dl 1.5 dl] 1 0 1 DL } def
/LT4 { PL [5 dl 2 dl 1 dl 2 dl] 0 1 1 DL } def
/LT5 { PL [4 dl 3 dl 1 dl 3 dl] 1 1 0 DL } def
/LT6 { PL [2 dl 2 dl 2 dl 4 dl] 0 0 0 DL } def
/LT7 { PL [2 dl 2 dl 2 dl 2 dl 2 dl 4 dl] 1 0.3 0 DL } def
/LT8 { PL [2 dl 2 dl 2 dl 2 dl 2 dl 2 dl 2 dl 4 dl] 0.5 0.5 0.5 DL } def
/Pnt { stroke [] 0 setdash
   gsave 1 setlinecap M 0 0 V stroke grestore } def
/Dia { stroke [] 0 setdash 2 copy vpt add M
  hpt neg vpt neg V hpt vpt neg V
  hpt vpt V hpt neg vpt V closepath stroke
  Pnt } def
/Pls { stroke [] 0 setdash vpt sub M 0 vpt2 V
  currentpoint stroke M
  hpt neg vpt neg R hpt2 0 V stroke
  } def
/Box { stroke [] 0 setdash 2 copy exch hpt sub exch vpt add M
  0 vpt2 neg V hpt2 0 V 0 vpt2 V
  hpt2 neg 0 V closepath stroke
  Pnt } def
/Crs { stroke [] 0 setdash exch hpt sub exch vpt add M
  hpt2 vpt2 neg V currentpoint stroke M
  hpt2 neg 0 R hpt2 vpt2 V stroke } def
/TriU { stroke [] 0 setdash 2 copy vpt 1.12 mul add M
  hpt neg vpt -1.62 mul V
  hpt 2 mul 0 V
  hpt neg vpt 1.62 mul V closepath stroke
  Pnt  } def
/Star { 2 copy Pls Crs } def
/BoxF { stroke [] 0 setdash exch hpt sub exch vpt add M
  0 vpt2 neg V  hpt2 0 V  0 vpt2 V
  hpt2 neg 0 V  closepath fill } def
/TriUF { stroke [] 0 setdash vpt 1.12 mul add M
  hpt neg vpt -1.62 mul V
  hpt 2 mul 0 V
  hpt neg vpt 1.62 mul V closepath fill } def
/TriD { stroke [] 0 setdash 2 copy vpt 1.12 mul sub M
  hpt neg vpt 1.62 mul V
  hpt 2 mul 0 V
  hpt neg vpt -1.62 mul V closepath stroke
  Pnt  } def
/TriDF { stroke [] 0 setdash vpt 1.12 mul sub M
  hpt neg vpt 1.62 mul V
  hpt 2 mul 0 V
  hpt neg vpt -1.62 mul V closepath fill} def
/DiaF { stroke [] 0 setdash vpt add M
  hpt neg vpt neg V hpt vpt neg V
  hpt vpt V hpt neg vpt V closepath fill } def
/Pent { stroke [] 0 setdash 2 copy gsave
  translate 0 hpt M 4 {72 rotate 0 hpt L} repeat
  closepath stroke grestore Pnt } def
/PentF { stroke [] 0 setdash gsave
  translate 0 hpt M 4 {72 rotate 0 hpt L} repeat
  closepath fill grestore } def
/Circle { stroke [] 0 setdash 2 copy
  hpt 0 360 arc stroke Pnt } def
/CircleF { stroke [] 0 setdash hpt 0 360 arc fill } def
/C0 { BL [] 0 setdash 2 copy moveto vpt 90 450  arc } bind def
/C1 { BL [] 0 setdash 2 copy        moveto
       2 copy  vpt 0 90 arc closepath fill
               vpt 0 360 arc closepath } bind def
/C2 { BL [] 0 setdash 2 copy moveto
       2 copy  vpt 90 180 arc closepath fill
               vpt 0 360 arc closepath } bind def
/C3 { BL [] 0 setdash 2 copy moveto
       2 copy  vpt 0 180 arc closepath fill
               vpt 0 360 arc closepath } bind def
/C4 { BL [] 0 setdash 2 copy moveto
       2 copy  vpt 180 270 arc closepath fill
               vpt 0 360 arc closepath } bind def
/C5 { BL [] 0 setdash 2 copy moveto
       2 copy  vpt 0 90 arc
       2 copy moveto
       2 copy  vpt 180 270 arc closepath fill
               vpt 0 360 arc } bind def
/C6 { BL [] 0 setdash 2 copy moveto
      2 copy  vpt 90 270 arc closepath fill
              vpt 0 360 arc closepath } bind def
/C7 { BL [] 0 setdash 2 copy moveto
      2 copy  vpt 0 270 arc closepath fill
              vpt 0 360 arc closepath } bind def
/C8 { BL [] 0 setdash 2 copy moveto
      2 copy vpt 270 360 arc closepath fill
              vpt 0 360 arc closepath } bind def
/C9 { BL [] 0 setdash 2 copy moveto
      2 copy  vpt 270 450 arc closepath fill
              vpt 0 360 arc closepath } bind def
/C10 { BL [] 0 setdash 2 copy 2 copy moveto vpt 270 360 arc closepath fill
       2 copy moveto
       2 copy vpt 90 180 arc closepath fill
               vpt 0 360 arc closepath } bind def
/C11 { BL [] 0 setdash 2 copy moveto
       2 copy  vpt 0 180 arc closepath fill
       2 copy moveto
       2 copy  vpt 270 360 arc closepath fill
               vpt 0 360 arc closepath } bind def
/C12 { BL [] 0 setdash 2 copy moveto
       2 copy  vpt 180 360 arc closepath fill
               vpt 0 360 arc closepath } bind def
/C13 { BL [] 0 setdash  2 copy moveto
       2 copy  vpt 0 90 arc closepath fill
       2 copy moveto
       2 copy  vpt 180 360 arc closepath fill
               vpt 0 360 arc closepath } bind def
/C14 { BL [] 0 setdash 2 copy moveto
       2 copy  vpt 90 360 arc closepath fill
               vpt 0 360 arc } bind def
/C15 { BL [] 0 setdash 2 copy vpt 0 360 arc closepath fill
               vpt 0 360 arc closepath } bind def
/Rec   { newpath 4 2 roll moveto 1 index 0 rlineto 0 exch rlineto
       neg 0 rlineto closepath } bind def
/Square { dup Rec } bind def
/Bsquare { vpt sub exch vpt sub exch vpt2 Square } bind def
/S0 { BL [] 0 setdash 2 copy moveto 0 vpt rlineto BL Bsquare } bind def
/S1 { BL [] 0 setdash 2 copy vpt Square fill Bsquare } bind def
/S2 { BL [] 0 setdash 2 copy exch vpt sub exch vpt Square fill Bsquare } bind def
/S3 { BL [] 0 setdash 2 copy exch vpt sub exch vpt2 vpt Rec fill Bsquare } bind def
/S4 { BL [] 0 setdash 2 copy exch vpt sub exch vpt sub vpt Square fill Bsquare } bind def
/S5 { BL [] 0 setdash 2 copy 2 copy vpt Square fill
       exch vpt sub exch vpt sub vpt Square fill Bsquare } bind def
/S6 { BL [] 0 setdash 2 copy exch vpt sub exch vpt sub vpt vpt2 Rec fill Bsquare } bind def
/S7 { BL [] 0 setdash 2 copy exch vpt sub exch vpt sub vpt vpt2 Rec fill
       2 copy vpt Square fill
       Bsquare } bind def
/S8 { BL [] 0 setdash 2 copy vpt sub vpt Square fill Bsquare } bind def
/S9 { BL [] 0 setdash 2 copy vpt sub vpt vpt2 Rec fill Bsquare } bind def
/S10 { BL [] 0 setdash 2 copy vpt sub vpt Square fill 2 copy exch vpt sub exch vpt Square fill
       Bsquare } bind def
/S11 { BL [] 0 setdash 2 copy vpt sub vpt Square fill 2 copy exch vpt sub exch vpt2 vpt Rec fill
       Bsquare } bind def
/S12 { BL [] 0 setdash 2 copy exch vpt sub exch vpt sub vpt2 vpt Rec fill Bsquare } bind def
/S13 { BL [] 0 setdash 2 copy exch vpt sub exch vpt sub vpt2 vpt Rec fill
       2 copy vpt Square fill Bsquare } bind def
/S14 { BL [] 0 setdash 2 copy exch vpt sub exch vpt sub vpt2 vpt Rec fill
       2 copy exch vpt sub exch vpt Square fill Bsquare } bind def
/S15 { BL [] 0 setdash 2 copy Bsquare fill Bsquare } bind def
/D0 { gsave translate 45 rotate 0 0 S0 stroke grestore } bind def
/D1 { gsave translate 45 rotate 0 0 S1 stroke grestore } bind def
/D2 { gsave translate 45 rotate 0 0 S2 stroke grestore } bind def
/D3 { gsave translate 45 rotate 0 0 S3 stroke grestore } bind def
/D4 { gsave translate 45 rotate 0 0 S4 stroke grestore } bind def
/D5 { gsave translate 45 rotate 0 0 S5 stroke grestore } bind def
/D6 { gsave translate 45 rotate 0 0 S6 stroke grestore } bind def
/D7 { gsave translate 45 rotate 0 0 S7 stroke grestore } bind def
/D8 { gsave translate 45 rotate 0 0 S8 stroke grestore } bind def
/D9 { gsave translate 45 rotate 0 0 S9 stroke grestore } bind def
/D10 { gsave translate 45 rotate 0 0 S10 stroke grestore } bind def
/D11 { gsave translate 45 rotate 0 0 S11 stroke grestore } bind def
/D12 { gsave translate 45 rotate 0 0 S12 stroke grestore } bind def
/D13 { gsave translate 45 rotate 0 0 S13 stroke grestore } bind def
/D14 { gsave translate 45 rotate 0 0 S14 stroke grestore } bind def
/D15 { gsave translate 45 rotate 0 0 S15 stroke grestore } bind def
/DiaE { stroke [] 0 setdash vpt add M
  hpt neg vpt neg V hpt vpt neg V
  hpt vpt V hpt neg vpt V closepath stroke } def
/BoxE { stroke [] 0 setdash exch hpt sub exch vpt add M
  0 vpt2 neg V hpt2 0 V 0 vpt2 V
  hpt2 neg 0 V closepath stroke } def
/TriUE { stroke [] 0 setdash vpt 1.12 mul add M
  hpt neg vpt -1.62 mul V
  hpt 2 mul 0 V
  hpt neg vpt 1.62 mul V closepath stroke } def
/TriDE { stroke [] 0 setdash vpt 1.12 mul sub M
  hpt neg vpt 1.62 mul V
  hpt 2 mul 0 V
  hpt neg vpt -1.62 mul V closepath stroke } def
/PentE { stroke [] 0 setdash gsave
  translate 0 hpt M 4 {72 rotate 0 hpt L} repeat
  closepath stroke grestore } def
/CircE { stroke [] 0 setdash 
  hpt 0 360 arc stroke } def
/Opaque { gsave closepath 1 setgray fill grestore 0 setgray closepath } def
/DiaW { stroke [] 0 setdash vpt add M
  hpt neg vpt neg V hpt vpt neg V
  hpt vpt V hpt neg vpt V Opaque stroke } def
/BoxW { stroke [] 0 setdash exch hpt sub exch vpt add M
  0 vpt2 neg V hpt2 0 V 0 vpt2 V
  hpt2 neg 0 V Opaque stroke } def
/TriUW { stroke [] 0 setdash vpt 1.12 mul add M
  hpt neg vpt -1.62 mul V
  hpt 2 mul 0 V
  hpt neg vpt 1.62 mul V Opaque stroke } def
/TriDW { stroke [] 0 setdash vpt 1.12 mul sub M
  hpt neg vpt 1.62 mul V
  hpt 2 mul 0 V
  hpt neg vpt -1.62 mul V Opaque stroke } def
/PentW { stroke [] 0 setdash gsave
  translate 0 hpt M 4 {72 rotate 0 hpt L} repeat
  Opaque stroke grestore } def
/CircW { stroke [] 0 setdash 
  hpt 0 360 arc Opaque stroke } def
/BoxFill { gsave Rec 1 setgray fill grestore } def
end
}}%
\begin{picture}(3600,2160)(0,0)%
{\GNUPLOTspecial{"
gnudict begin
gsave
0 0 translate
0.100 0.100 scale
0 setgray
newpath
1.000 UL
LTb
500 300 M
63 0 V
2887 0 R
-63 0 V
500 551 M
63 0 V
2887 0 R
-63 0 V
500 803 M
63 0 V
2887 0 R
-63 0 V
500 1054 M
63 0 V
2887 0 R
-63 0 V
500 1306 M
63 0 V
2887 0 R
-63 0 V
500 1557 M
63 0 V
2887 0 R
-63 0 V
500 1809 M
63 0 V
2887 0 R
-63 0 V
500 2060 M
63 0 V
2887 0 R
-63 0 V
566 300 M
0 63 V
0 1697 R
0 -63 V
1036 300 M
0 63 V
0 1697 R
0 -63 V
1505 300 M
0 63 V
0 1697 R
0 -63 V
1975 300 M
0 63 V
0 1697 R
0 -63 V
2445 300 M
0 63 V
0 1697 R
0 -63 V
2914 300 M
0 63 V
0 1697 R
0 -63 V
3384 300 M
0 63 V
0 1697 R
0 -63 V
1.000 UL
LTb
500 300 M
2950 0 V
0 1760 V
-2950 0 V
500 300 L
1.000 UL
LT0
507 552 M
15 0 V
15 0 V
15 0 V
14 0 V
15 1 V
15 1 V
15 2 V
14 2 V
15 2 V
15 5 V
15 6 V
14 9 V
15 10 V
15 15 V
15 17 V
14 25 V
15 31 V
15 35 V
15 47 V
14 52 V
15 55 V
15 67 V
15 62 V
14 75 V
15 86 V
15 60 V
15 62 V
14 54 V
15 36 V
15 32 V
15 14 V
14 -20 V
15 -24 V
15 -22 V
15 -35 V
14 -34 V
15 -21 V
15 -3 V
15 13 V
14 30 V
15 63 V
15 69 V
15 80 V
14 55 V
15 44 V
15 27 V
15 -16 V
14 -24 V
15 -52 V
15 -56 V
15 -40 V
14 -37 V
15 5 V
15 33 V
15 68 V
14 65 V
15 91 V
15 64 V
15 21 V
14 -9 V
15 -30 V
15 -71 V
15 -66 V
14 -58 V
15 -26 V
15 10 V
15 50 V
14 82 V
15 101 V
15 71 V
15 38 V
14 -9 V
15 -38 V
15 -97 V
15 -80 V
14 -62 V
15 -25 V
15 26 V
15 76 V
14 82 V
15 97 V
15 71 V
15 21 V
14 -47 V
15 -72 V
15 -91 V
15 -89 V
14 -50 V
15 6 V
15 54 V
15 105 V
14 106 V
15 78 V
15 16 V
15 -25 V
14 -71 V
15 -114 V
15 -90 V
15 -50 V
14 -1 V
15 53 V
15 89 V
15 103 V
14 85 V
15 22 V
15 -17 V
15 -84 V
14 -96 V
15 -100 V
15 -59 V
15 -9 V
14 43 V
15 88 V
15 107 V
15 67 V
14 33 V
15 -15 V
15 -69 V
15 -86 V
14 -91 V
15 -77 V
15 -24 V
15 24 V
14 70 V
15 82 V
15 74 V
15 54 V
14 7 V
15 -49 V
15 -61 V
15 -101 V
14 -75 V
15 -69 V
15 -7 V
15 34 V
14 63 V
15 71 V
15 65 V
15 40 V
14 -4 V
15 -34 V
15 -58 V
15 -81 V
14 -76 V
15 -61 V
15 -31 V
15 0 V
14 31 V
15 41 V
15 62 V
15 48 V
14 28 V
15 1 V
15 -23 V
15 -36 V
14 -71 V
15 -64 V
15 -73 V
15 -59 V
14 -32 V
15 -18 V
15 1 V
15 22 V
14 30 V
15 36 V
15 36 V
15 16 V
14 13 V
15 -15 V
15 -14 V
15 -40 V
14 -59 V
15 -54 V
15 -72 V
15 -81 V
14 -70 V
15 -67 V
15 -67 V
15 -57 V
14 -51 V
15 -46 V
15 -36 V
15 -29 V
14 -25 V
15 -19 V
15 -14 V
15 -13 V
14 -7 V
15 -6 V
15 -4 V
15 -3 V
14 -2 V
15 -2 V
15 -1 V
15 -1 V
14 0 V
15 0 V
15 0 V
15 0 V
1.000 UL
LT1
507 552 M
15 0 V
15 0 V
15 0 V
14 1 V
15 0 V
15 1 V
15 2 V
14 2 V
15 3 V
15 6 V
15 5 V
14 8 V
15 13 V
15 13 V
15 20 V
14 25 V
15 28 V
15 41 V
15 40 V
14 49 V
15 64 V
15 64 V
15 73 V
14 65 V
15 72 V
15 77 V
15 66 V
14 51 V
15 33 V
15 27 V
15 2 V
14 -8 V
15 -27 V
15 -25 V
15 -37 V
14 -17 V
15 -39 V
15 -2 V
15 28 V
14 39 V
15 50 V
15 70 V
15 66 V
14 62 V
15 50 V
15 22 V
15 -7 V
14 -33 V
15 -54 V
15 -51 V
15 -56 V
14 -22 V
15 9 V
15 35 V
15 62 V
14 74 V
15 77 V
15 69 V
15 22 V
14 -1 V
15 -47 V
15 -58 V
15 -71 V
14 -70 V
15 -15 V
15 6 V
15 59 V
14 83 V
15 98 V
15 69 V
15 35 V
14 -3 V
15 -58 V
15 -80 V
15 -82 V
14 -66 V
15 -16 V
15 11 V
15 86 V
14 86 V
15 104 V
15 65 V
15 8 V
14 -28 V
15 -81 V
15 -109 V
15 -75 V
14 -55 V
15 17 V
15 62 V
15 103 V
14 99 V
15 66 V
15 31 V
15 -34 V
14 -74 V
15 -102 V
15 -85 V
15 -62 V
14 -10 V
15 71 V
15 79 V
15 106 V
14 86 V
15 32 V
15 -36 V
15 -75 V
14 -95 V
15 -108 V
15 -44 V
15 -21 V
14 49 V
15 83 V
15 99 V
15 67 V
14 42 V
15 -12 V
15 -71 V
15 -84 V
14 -97 V
15 -72 V
15 -24 V
15 22 V
14 57 V
15 91 V
15 84 V
15 48 V
14 7 V
15 -33 V
15 -84 V
15 -85 V
14 -89 V
15 -45 V
15 -21 V
15 24 V
14 64 V
15 67 V
15 62 V
15 51 V
14 -2 V
15 -24 V
15 -57 V
15 -81 V
14 -80 V
15 -60 V
15 -39 V
15 -5 V
14 21 V
15 54 V
15 54 V
15 48 V
14 42 V
15 -1 V
15 -24 V
15 -42 V
14 -54 V
15 -76 V
15 -69 V
15 -56 V
14 -43 V
15 -20 V
15 11 V
15 12 V
14 38 V
15 33 V
15 29 V
15 26 V
14 12 V
15 2 V
15 -31 V
15 -39 V
14 -58 V
15 -57 V
15 -69 V
15 -78 V
14 -72 V
15 -74 V
15 -58 V
15 -61 V
14 -48 V
15 -46 V
15 -39 V
15 -31 V
14 -23 V
15 -21 V
15 -15 V
15 -10 V
14 -7 V
15 -8 V
15 -4 V
15 -3 V
14 -2 V
15 -1 V
15 -2 V
15 0 V
14 -1 V
15 0 V
15 0 V
15 0 V
1.000 UL
LT2
1975 1750 M
7 0 V
8 0 V
7 0 V
8 -1 V
7 0 V
7 0 V
8 0 V
7 0 V
7 -1 V
8 0 V
7 -1 V
8 0 V
7 -1 V
7 0 V
8 -1 V
7 0 V
7 -1 V
8 -1 V
7 0 V
8 -1 V
7 -1 V
7 -1 V
8 -1 V
7 -1 V
7 -1 V
8 -1 V
7 -1 V
8 -1 V
7 -1 V
7 -1 V
8 -1 V
7 -2 V
7 -1 V
8 -1 V
7 -2 V
8 -1 V
7 -2 V
7 -1 V
8 -2 V
7 -1 V
7 -2 V
8 -2 V
7 -1 V
8 -2 V
7 -2 V
7 -2 V
8 -2 V
7 -2 V
7 -2 V
8 -2 V
7 -2 V
8 -2 V
7 -2 V
7 -3 V
8 -2 V
7 -2 V
7 -3 V
8 -2 V
7 -2 V
8 -3 V
7 -3 V
7 -2 V
8 -3 V
7 -3 V
7 -2 V
8 -3 V
7 -3 V
8 -3 V
7 -3 V
7 -3 V
8 -3 V
7 -3 V
7 -4 V
8 -3 V
7 -3 V
8 -4 V
7 -3 V
7 -3 V
8 -4 V
7 -4 V
7 -3 V
8 -4 V
7 -4 V
8 -4 V
7 -4 V
7 -4 V
8 -4 V
7 -4 V
7 -4 V
8 -4 V
7 -5 V
8 -4 V
7 -4 V
7 -5 V
8 -5 V
7 -4 V
7 -5 V
8 -5 V
7 -5 V
8 -5 V
7 -5 V
7 -5 V
8 -5 V
7 -6 V
7 -5 V
8 -6 V
7 -5 V
8 -6 V
7 -6 V
7 -6 V
8 -6 V
7 -6 V
7 -6 V
8 -6 V
7 -6 V
8 -7 V
7 -7 V
7 -6 V
8 -7 V
7 -7 V
7 -7 V
8 -8 V
7 -7 V
8 -7 V
7 -8 V
7 -8 V
8 -8 V
7 -8 V
7 -8 V
8 -8 V
7 -9 V
8 -9 V
7 -9 V
7 -9 V
8 -9 V
7 -10 V
7 -10 V
8 -10 V
7 -10 V
8 -10 V
7 -11 V
7 -11 V
8 -11 V
7 -12 V
7 -12 V
8 -12 V
7 -13 V
8 -13 V
7 -13 V
7 -14 V
8 -14 V
7 -15 V
7 -15 V
8 -16 V
7 -17 V
8 -18 V
7 -18 V
7 -19 V
8 -20 V
7 -22 V
7 -23 V
8 -25 V
7 -27 V
8 -30 V
7 -33 V
7 -40 V
8 -49 V
7 -74 V
7 -89 V
8 0 V
7 0 V
8 0 V
7 0 V
7 0 V
8 0 V
7 0 V
7 0 V
8 0 V
7 0 V
8 0 V
7 0 V
7 0 V
8 0 V
7 0 V
7 0 V
8 0 V
7 0 V
8 0 V
7 0 V
7 0 V
8 0 V
7 0 V
7 0 V
8 0 V
7 0 V
8 0 V
7 0 V
7 0 V
8 0 V
7 0 V
1.000 UL
LT2
1975 1750 M
-7 0 V
-8 0 V
-7 0 V
-7 -1 V
-8 0 V
-7 0 V
-8 0 V
-7 0 V
-7 -1 V
-8 0 V
-7 -1 V
-7 0 V
-8 -1 V
-7 0 V
-8 -1 V
-7 0 V
-7 -1 V
-8 -1 V
-7 0 V
-7 -1 V
-8 -1 V
-7 -1 V
-8 -1 V
-7 -1 V
-7 -1 V
-8 -1 V
-7 -1 V
-7 -1 V
-8 -1 V
-7 -1 V
-8 -1 V
-7 -2 V
-7 -1 V
-8 -1 V
-7 -2 V
-7 -1 V
-8 -2 V
-7 -1 V
-8 -2 V
-7 -1 V
-7 -2 V
-8 -2 V
-7 -1 V
-7 -2 V
-8 -2 V
-7 -2 V
-8 -2 V
-7 -2 V
-7 -2 V
-8 -2 V
-7 -2 V
-7 -2 V
-8 -2 V
-7 -3 V
-8 -2 V
-7 -2 V
-7 -3 V
-8 -2 V
-7 -2 V
-7 -3 V
-8 -3 V
-7 -2 V
-8 -3 V
-7 -3 V
-7 -2 V
-8 -3 V
-7 -3 V
-7 -3 V
-8 -3 V
-7 -3 V
-8 -3 V
-7 -3 V
-7 -4 V
-8 -3 V
-7 -3 V
-7 -4 V
-8 -3 V
-7 -3 V
-8 -4 V
-7 -4 V
-7 -3 V
-8 -4 V
-7 -4 V
-7 -4 V
-8 -4 V
-7 -4 V
-8 -4 V
-7 -4 V
-7 -4 V
-8 -4 V
-7 -5 V
-7 -4 V
-8 -4 V
-7 -5 V
-8 -5 V
-7 -4 V
-7 -5 V
-8 -5 V
-7 -5 V
-7 -5 V
-8 -5 V
-7 -5 V
-8 -5 V
-7 -6 V
-7 -5 V
-8 -6 V
-7 -5 V
-7 -6 V
-8 -6 V
-7 -6 V
-8 -6 V
-7 -6 V
-7 -6 V
-8 -6 V
-7 -6 V
-7 -7 V
-8 -7 V
-7 -6 V
-8 -7 V
-7 -7 V
-7 -7 V
-8 -8 V
-7 -7 V
-7 -7 V
-8 -8 V
-7 -8 V
-8 -8 V
-7 -8 V
-7 -8 V
-8 -8 V
-7 -9 V
-8 -9 V
-7 -9 V
-7 -9 V
-8 -9 V
-7 -10 V
-7 -10 V
-8 -10 V
-7 -10 V
-7 -10 V
-8 -11 V
-7 -11 V
-8 -11 V
-7 -12 V
-7 -12 V
-8 -12 V
-7 -13 V
-7 -13 V
-8 -13 V
-7 -14 V
-8 -14 V
-7 -15 V
-7 -15 V
-8 -16 V
-7 -17 V
-7 -18 V
-8 -18 V
-7 -19 V
-8 -20 V
-7 -22 V
-7 -23 V
-8 -25 V
-7 -27 V
-7 -30 V
-8 -33 V
-7 -40 V
-8 -49 V
-7 -74 V
-7 -89 V
-8 0 V
-7 0 V
-7 0 V
-8 0 V
-7 0 V
-8 0 V
-7 0 V
-7 0 V
-8 0 V
-7 0 V
-7 0 V
-8 0 V
-7 0 V
-8 0 V
-7 0 V
-7 0 V
-8 0 V
-7 0 V
-7 0 V
-8 0 V
-7 0 V
-8 0 V
-7 0 V
-7 0 V
-8 0 V
-7 0 V
-7 0 V
-8 0 V
-7 0 V
-8 0 V
-7 0 V
stroke
grestore
end
showpage
}}%
\put(1975,50){\makebox(0,0){$\alpha$}}%
\put(100,1180){%
\makebox(0,0)[b]{\shortstack{$\rho(\alpha)$}}%
}%
\put(3384,200){\makebox(0,0){3}}%
\put(2914,200){\makebox(0,0){2}}%
\put(2445,200){\makebox(0,0){1}}%
\put(1975,200){\makebox(0,0){0}}%
\put(1505,200){\makebox(0,0){-1}}%
\put(1036,200){\makebox(0,0){-2}}%
\put(566,200){\makebox(0,0){-3}}%
\put(450,2060){\makebox(0,0)[r]{0.3}}%
\put(450,1809){\makebox(0,0)[r]{0.25}}%
\put(450,1557){\makebox(0,0)[r]{0.2}}%
\put(450,1306){\makebox(0,0)[r]{0.15}}%
\put(450,1054){\makebox(0,0)[r]{0.1}}%
\put(450,803){\makebox(0,0)[r]{0.05}}%
\put(450,551){\makebox(0,0)[r]{0}}%
\put(450,300){\makebox(0,0)[r]{-0.05}}%
\end{picture}%
\endgroup

%% file: wilsonsizematchgraph.tex
\begingroup%
  \makeatletter%
  \newcommand{\GNUPLOTspecial}{%
    \@sanitize\catcode`\%=14\relax\special}%
  \setlength{\unitlength}{0.1bp}%
{\GNUPLOTspecial{!
/gnudict 256 dict def
gnudict begin
/Color false def
/Solid false def
/gnulinewidth 5.000 def
/userlinewidth gnulinewidth def
/vshift -33 def
/dl {10 mul} def
/hpt_ 31.5 def
/vpt_ 31.5 def
/hpt hpt_ def
/vpt vpt_ def
/M {moveto} bind def
/L {lineto} bind def
/R {rmoveto} bind def
/V {rlineto} bind def
/vpt2 vpt 2 mul def
/hpt2 hpt 2 mul def
/Lshow { currentpoint stroke M
  0 vshift R show } def
/Rshow { currentpoint stroke M
  dup stringwidth pop neg vshift R show } def
/Cshow { currentpoint stroke M
  dup stringwidth pop -2 div vshift R show } def
/UP { dup vpt_ mul /vpt exch def hpt_ mul /hpt exch def
  /hpt2 hpt 2 mul def /vpt2 vpt 2 mul def } def
/DL { Color {setrgbcolor Solid {pop []} if 0 setdash }
 {pop pop pop Solid {pop []} if 0 setdash} ifelse } def
/BL { stroke userlinewidth 2 mul setlinewidth } def
/AL { stroke userlinewidth 2 div setlinewidth } def
/UL { dup gnulinewidth mul /userlinewidth exch def
      10 mul /udl exch def } def
/PL { stroke userlinewidth setlinewidth } def
/LTb { BL [] 0 0 0 DL } def
/LTa { AL [1 udl mul 2 udl mul] 0 setdash 0 0 0 setrgbcolor } def
/LT0 { PL [] 1 0 0 DL } def
/LT1 { PL [4 dl 2 dl] 0 1 0 DL } def
/LT2 { PL [2 dl 3 dl] 0 0 1 DL } def
/LT3 { PL [1 dl 1.5 dl] 1 0 1 DL } def
/LT4 { PL [5 dl 2 dl 1 dl 2 dl] 0 1 1 DL } def
/LT5 { PL [4 dl 3 dl 1 dl 3 dl] 1 1 0 DL } def
/LT6 { PL [2 dl 2 dl 2 dl 4 dl] 0 0 0 DL } def
/LT7 { PL [2 dl 2 dl 2 dl 2 dl 2 dl 4 dl] 1 0.3 0 DL } def
/LT8 { PL [2 dl 2 dl 2 dl 2 dl 2 dl 2 dl 2 dl 4 dl] 0.5 0.5 0.5 DL } def
/Pnt { stroke [] 0 setdash
   gsave 1 setlinecap M 0 0 V stroke grestore } def
/Dia { stroke [] 0 setdash 2 copy vpt add M
  hpt neg vpt neg V hpt vpt neg V
  hpt vpt V hpt neg vpt V closepath stroke
  Pnt } def
/Pls { stroke [] 0 setdash vpt sub M 0 vpt2 V
  currentpoint stroke M
  hpt neg vpt neg R hpt2 0 V stroke
  } def
/Box { stroke [] 0 setdash 2 copy exch hpt sub exch vpt add M
  0 vpt2 neg V hpt2 0 V 0 vpt2 V
  hpt2 neg 0 V closepath stroke
  Pnt } def
/Crs { stroke [] 0 setdash exch hpt sub exch vpt add M
  hpt2 vpt2 neg V currentpoint stroke M
  hpt2 neg 0 R hpt2 vpt2 V stroke } def
/TriU { stroke [] 0 setdash 2 copy vpt 1.12 mul add M
  hpt neg vpt -1.62 mul V
  hpt 2 mul 0 V
  hpt neg vpt 1.62 mul V closepath stroke
  Pnt  } def
/Star { 2 copy Pls Crs } def
/BoxF { stroke [] 0 setdash exch hpt sub exch vpt add M
  0 vpt2 neg V  hpt2 0 V  0 vpt2 V
  hpt2 neg 0 V  closepath fill } def
/TriUF { stroke [] 0 setdash vpt 1.12 mul add M
  hpt neg vpt -1.62 mul V
  hpt 2 mul 0 V
  hpt neg vpt 1.62 mul V closepath fill } def
/TriD { stroke [] 0 setdash 2 copy vpt 1.12 mul sub M
  hpt neg vpt 1.62 mul V
  hpt 2 mul 0 V
  hpt neg vpt -1.62 mul V closepath stroke
  Pnt  } def
/TriDF { stroke [] 0 setdash vpt 1.12 mul sub M
  hpt neg vpt 1.62 mul V
  hpt 2 mul 0 V
  hpt neg vpt -1.62 mul V closepath fill} def
/DiaF { stroke [] 0 setdash vpt add M
  hpt neg vpt neg V hpt vpt neg V
  hpt vpt V hpt neg vpt V closepath fill } def
/Pent { stroke [] 0 setdash 2 copy gsave
  translate 0 hpt M 4 {72 rotate 0 hpt L} repeat
  closepath stroke grestore Pnt } def
/PentF { stroke [] 0 setdash gsave
  translate 0 hpt M 4 {72 rotate 0 hpt L} repeat
  closepath fill grestore } def
/Circle { stroke [] 0 setdash 2 copy
  hpt 0 360 arc stroke Pnt } def
/CircleF { stroke [] 0 setdash hpt 0 360 arc fill } def
/C0 { BL [] 0 setdash 2 copy moveto vpt 90 450  arc } bind def
/C1 { BL [] 0 setdash 2 copy        moveto
       2 copy  vpt 0 90 arc closepath fill
               vpt 0 360 arc closepath } bind def
/C2 { BL [] 0 setdash 2 copy moveto
       2 copy  vpt 90 180 arc closepath fill
               vpt 0 360 arc closepath } bind def
/C3 { BL [] 0 setdash 2 copy moveto
       2 copy  vpt 0 180 arc closepath fill
               vpt 0 360 arc closepath } bind def
/C4 { BL [] 0 setdash 2 copy moveto
       2 copy  vpt 180 270 arc closepath fill
               vpt 0 360 arc closepath } bind def
/C5 { BL [] 0 setdash 2 copy moveto
       2 copy  vpt 0 90 arc
       2 copy moveto
       2 copy  vpt 180 270 arc closepath fill
               vpt 0 360 arc } bind def
/C6 { BL [] 0 setdash 2 copy moveto
      2 copy  vpt 90 270 arc closepath fill
              vpt 0 360 arc closepath } bind def
/C7 { BL [] 0 setdash 2 copy moveto
      2 copy  vpt 0 270 arc closepath fill
              vpt 0 360 arc closepath } bind def
/C8 { BL [] 0 setdash 2 copy moveto
      2 copy vpt 270 360 arc closepath fill
              vpt 0 360 arc closepath } bind def
/C9 { BL [] 0 setdash 2 copy moveto
      2 copy  vpt 270 450 arc closepath fill
              vpt 0 360 arc closepath } bind def
/C10 { BL [] 0 setdash 2 copy 2 copy moveto vpt 270 360 arc closepath fill
       2 copy moveto
       2 copy vpt 90 180 arc closepath fill
               vpt 0 360 arc closepath } bind def
/C11 { BL [] 0 setdash 2 copy moveto
       2 copy  vpt 0 180 arc closepath fill
       2 copy moveto
       2 copy  vpt 270 360 arc closepath fill
               vpt 0 360 arc closepath } bind def
/C12 { BL [] 0 setdash 2 copy moveto
       2 copy  vpt 180 360 arc closepath fill
               vpt 0 360 arc closepath } bind def
/C13 { BL [] 0 setdash  2 copy moveto
       2 copy  vpt 0 90 arc closepath fill
       2 copy moveto
       2 copy  vpt 180 360 arc closepath fill
               vpt 0 360 arc closepath } bind def
/C14 { BL [] 0 setdash 2 copy moveto
       2 copy  vpt 90 360 arc closepath fill
               vpt 0 360 arc } bind def
/C15 { BL [] 0 setdash 2 copy vpt 0 360 arc closepath fill
               vpt 0 360 arc closepath } bind def
/Rec   { newpath 4 2 roll moveto 1 index 0 rlineto 0 exch rlineto
       neg 0 rlineto closepath } bind def
/Square { dup Rec } bind def
/Bsquare { vpt sub exch vpt sub exch vpt2 Square } bind def
/S0 { BL [] 0 setdash 2 copy moveto 0 vpt rlineto BL Bsquare } bind def
/S1 { BL [] 0 setdash 2 copy vpt Square fill Bsquare } bind def
/S2 { BL [] 0 setdash 2 copy exch vpt sub exch vpt Square fill Bsquare } bind def
/S3 { BL [] 0 setdash 2 copy exch vpt sub exch vpt2 vpt Rec fill Bsquare } bind def
/S4 { BL [] 0 setdash 2 copy exch vpt sub exch vpt sub vpt Square fill Bsquare } bind def
/S5 { BL [] 0 setdash 2 copy 2 copy vpt Square fill
       exch vpt sub exch vpt sub vpt Square fill Bsquare } bind def
/S6 { BL [] 0 setdash 2 copy exch vpt sub exch vpt sub vpt vpt2 Rec fill Bsquare } bind def
/S7 { BL [] 0 setdash 2 copy exch vpt sub exch vpt sub vpt vpt2 Rec fill
       2 copy vpt Square fill
       Bsquare } bind def
/S8 { BL [] 0 setdash 2 copy vpt sub vpt Square fill Bsquare } bind def
/S9 { BL [] 0 setdash 2 copy vpt sub vpt vpt2 Rec fill Bsquare } bind def
/S10 { BL [] 0 setdash 2 copy vpt sub vpt Square fill 2 copy exch vpt sub exch vpt Square fill
       Bsquare } bind def
/S11 { BL [] 0 setdash 2 copy vpt sub vpt Square fill 2 copy exch vpt sub exch vpt2 vpt Rec fill
       Bsquare } bind def
/S12 { BL [] 0 setdash 2 copy exch vpt sub exch vpt sub vpt2 vpt Rec fill Bsquare } bind def
/S13 { BL [] 0 setdash 2 copy exch vpt sub exch vpt sub vpt2 vpt Rec fill
       2 copy vpt Square fill Bsquare } bind def
/S14 { BL [] 0 setdash 2 copy exch vpt sub exch vpt sub vpt2 vpt Rec fill
       2 copy exch vpt sub exch vpt Square fill Bsquare } bind def
/S15 { BL [] 0 setdash 2 copy Bsquare fill Bsquare } bind def
/D0 { gsave translate 45 rotate 0 0 S0 stroke grestore } bind def
/D1 { gsave translate 45 rotate 0 0 S1 stroke grestore } bind def
/D2 { gsave translate 45 rotate 0 0 S2 stroke grestore } bind def
/D3 { gsave translate 45 rotate 0 0 S3 stroke grestore } bind def
/D4 { gsave translate 45 rotate 0 0 S4 stroke grestore } bind def
/D5 { gsave translate 45 rotate 0 0 S5 stroke grestore } bind def
/D6 { gsave translate 45 rotate 0 0 S6 stroke grestore } bind def
/D7 { gsave translate 45 rotate 0 0 S7 stroke grestore } bind def
/D8 { gsave translate 45 rotate 0 0 S8 stroke grestore } bind def
/D9 { gsave translate 45 rotate 0 0 S9 stroke grestore } bind def
/D10 { gsave translate 45 rotate 0 0 S10 stroke grestore } bind def
/D11 { gsave translate 45 rotate 0 0 S11 stroke grestore } bind def
/D12 { gsave translate 45 rotate 0 0 S12 stroke grestore } bind def
/D13 { gsave translate 45 rotate 0 0 S13 stroke grestore } bind def
/D14 { gsave translate 45 rotate 0 0 S14 stroke grestore } bind def
/D15 { gsave translate 45 rotate 0 0 S15 stroke grestore } bind def
/DiaE { stroke [] 0 setdash vpt add M
  hpt neg vpt neg V hpt vpt neg V
  hpt vpt V hpt neg vpt V closepath stroke } def
/BoxE { stroke [] 0 setdash exch hpt sub exch vpt add M
  0 vpt2 neg V hpt2 0 V 0 vpt2 V
  hpt2 neg 0 V closepath stroke } def
/TriUE { stroke [] 0 setdash vpt 1.12 mul add M
  hpt neg vpt -1.62 mul V
  hpt 2 mul 0 V
  hpt neg vpt 1.62 mul V closepath stroke } def
/TriDE { stroke [] 0 setdash vpt 1.12 mul sub M
  hpt neg vpt 1.62 mul V
  hpt 2 mul 0 V
  hpt neg vpt -1.62 mul V closepath stroke } def
/PentE { stroke [] 0 setdash gsave
  translate 0 hpt M 4 {72 rotate 0 hpt L} repeat
  closepath stroke grestore } def
/CircE { stroke [] 0 setdash 
  hpt 0 360 arc stroke } def
/Opaque { gsave closepath 1 setgray fill grestore 0 setgray closepath } def
/DiaW { stroke [] 0 setdash vpt add M
  hpt neg vpt neg V hpt vpt neg V
  hpt vpt V hpt neg vpt V Opaque stroke } def
/BoxW { stroke [] 0 setdash exch hpt sub exch vpt add M
  0 vpt2 neg V hpt2 0 V 0 vpt2 V
  hpt2 neg 0 V Opaque stroke } def
/TriUW { stroke [] 0 setdash vpt 1.12 mul add M
  hpt neg vpt -1.62 mul V
  hpt 2 mul 0 V
  hpt neg vpt 1.62 mul V Opaque stroke } def
/TriDW { stroke [] 0 setdash vpt 1.12 mul sub M
  hpt neg vpt 1.62 mul V
  hpt 2 mul 0 V
  hpt neg vpt -1.62 mul V Opaque stroke } def
/PentW { stroke [] 0 setdash gsave
  translate 0 hpt M 4 {72 rotate 0 hpt L} repeat
  Opaque stroke grestore } def
/CircW { stroke [] 0 setdash 
  hpt 0 360 arc Opaque stroke } def
/BoxFill { gsave Rec 1 setgray fill grestore } def
end
}}%
\begin{picture}(3600,2160)(0,0)%
{\GNUPLOTspecial{"
gnudict begin
gsave
0 0 translate
0.100 0.100 scale
0 setgray
newpath
1.000 UL
LTb
450 300 M
63 0 V
2937 0 R
-63 0 V
450 593 M
63 0 V
2937 0 R
-63 0 V
450 887 M
63 0 V
2937 0 R
-63 0 V
450 1180 M
63 0 V
2937 0 R
-63 0 V
450 1473 M
63 0 V
2937 0 R
-63 0 V
450 1767 M
63 0 V
2937 0 R
-63 0 V
450 2060 M
63 0 V
2937 0 R
-63 0 V
518 300 M
0 63 V
0 1697 R
0 -63 V
995 300 M
0 63 V
0 1697 R
0 -63 V
1473 300 M
0 63 V
0 1697 R
0 -63 V
1950 300 M
0 63 V
0 1697 R
0 -63 V
2427 300 M
0 63 V
0 1697 R
0 -63 V
2905 300 M
0 63 V
0 1697 R
0 -63 V
3382 300 M
0 63 V
0 1697 R
0 -63 V
1.000 UL
LTb
450 300 M
3000 0 V
0 1760 V
-3000 0 V
450 300 L
1.000 UL
LT0
457 482 M
15 1 V
16 8 V
15 16 V
15 23 V
14 26 V
15 32 V
16 36 V
15 35 V
14 49 V
15 45 V
16 49 V
14 59 V
16 56 V
14 52 V
16 53 V
15 59 V
14 54 V
16 60 V
14 45 V
15 50 V
16 40 V
14 28 V
15 30 V
16 23 V
14 17 V
15 -4 V
16 -1 V
14 -5 V
15 -22 V
16 -21 V
14 -48 V
15 -30 V
16 -48 V
14 -40 V
15 -48 V
16 -42 V
14 -43 V
15 -33 V
16 -28 V
15 -30 V
14 -5 V
16 -3 V
15 13 V
14 14 V
16 27 V
15 44 V
14 50 V
16 57 V
15 70 V
14 54 V
15 84 V
15 56 V
16 55 V
15 58 V
15 59 V
14 25 V
15 22 V
15 1 V
16 -8 V
15 -17 V
15 -36 V
14 -46 V
15 -62 V
15 -61 V
15 -61 V
16 -69 V
15 -66 V
15 -60 V
14 -49 V
15 -34 V
15 -18 V
16 -12 V
15 15 V
15 17 V
14 56 V
15 47 V
15 75 V
16 69 V
15 81 V
15 72 V
14 84 V
15 62 V
15 51 V
16 46 V
15 27 V
15 14 V
14 -16 V
15 -25 V
15 -36 V
15 -65 V
16 -70 V
15 -76 V
15 -89 V
14 -79 V
15 -62 V
15 -71 V
16 -47 V
15 -34 V
15 -24 V
14 0 V
15 25 V
15 32 V
15 48 V
15 64 V
15 73 V
16 84 V
15 77 V
15 79 V
15 67 V
15 57 V
15 46 V
15 34 V
14 6 V
15 -9 V
15 -31 V
15 -32 V
15 -67 V
15 -61 V
16 -80 V
15 -78 V
15 -78 V
15 -73 V
15 -70 V
15 -59 V
14 -34 V
15 -34 V
15 -11 V
15 14 V
15 21 V
15 35 V
16 49 V
15 56 V
15 58 V
15 74 V
15 70 V
15 63 V
15 52 V
14 47 V
15 35 V
15 25 V
15 3 V
15 4 V
15 -20 V
16 -42 V
15 -40 V
15 -63 V
15 -58 V
15 -68 V
15 -69 V
14 -75 V
15 -47 V
15 -64 V
15 -49 V
15 -34 V
15 -36 V
16 -21 V
15 -12 V
15 7 V
15 7 V
15 29 V
15 25 V
15 40 V
14 35 V
15 44 V
15 49 V
15 43 V
15 40 V
15 41 V
16 34 V
15 26 V
15 18 V
15 21 V
15 2 V
15 -12 V
14 -8 V
15 -18 V
15 -25 V
15 -36 V
15 -48 V
15 -45 V
16 -50 V
15 -50 V
15 -63 V
15 -56 V
15 -55 V
15 -57 V
15 -47 V
14 -54 V
15 -54 V
15 -54 V
15 -37 V
15 -43 V
15 -32 V
16 -34 V
15 -25 V
15 -19 V
15 -15 V
15 -10 V
15 -9 V
1.000 UL
LT1
457 472 M
15 3 V
16 10 V
15 19 V
15 20 V
14 28 V
15 29 V
16 37 V
15 36 V
14 54 V
15 47 V
16 53 V
14 57 V
16 47 V
14 64 V
16 55 V
15 71 V
14 46 V
16 55 V
14 55 V
15 42 V
16 44 V
14 41 V
15 18 V
16 27 V
14 10 V
15 3 V
16 -6 V
14 -8 V
15 -23 V
16 -31 V
14 -38 V
15 -37 V
16 -44 V
14 -44 V
15 -51 V
16 -44 V
14 -43 V
15 -34 V
16 -30 V
15 -20 V
14 -11 V
16 -1 V
15 15 V
14 19 V
16 33 V
15 47 V
14 49 V
16 65 V
15 68 V
14 62 V
15 70 V
15 60 V
16 69 V
15 43 V
15 31 V
14 44 V
15 10 V
15 1 V
16 -3 V
15 -38 V
15 -25 V
14 -55 V
15 -60 V
15 -76 V
15 -55 V
16 -78 V
15 -64 V
15 -47 V
14 -46 V
15 -36 V
15 -17 V
16 -6 V
15 19 V
15 35 V
14 43 V
15 56 V
15 56 V
16 75 V
15 89 V
15 75 V
14 82 V
15 57 V
15 62 V
16 30 V
15 31 V
15 7 V
14 -18 V
15 -35 V
15 -38 V
15 -62 V
16 -74 V
15 -65 V
15 -91 V
14 -75 V
15 -66 V
15 -75 V
16 -43 V
15 -35 V
15 -17 V
14 -5 V
15 25 V
15 35 V
15 46 V
15 65 V
15 71 V
16 80 V
15 76 V
15 77 V
15 69 V
15 61 V
15 47 V
15 24 V
14 13 V
15 0 V
15 -37 V
15 -29 V
15 -51 V
15 -63 V
16 -74 V
15 -77 V
15 -79 V
15 -82 V
15 -61 V
15 -61 V
14 -43 V
15 -34 V
15 -13 V
15 -5 V
15 29 V
15 28 V
16 47 V
15 56 V
15 61 V
15 71 V
15 69 V
15 67 V
15 67 V
14 40 V
15 37 V
15 24 V
15 12 V
15 -6 V
15 -8 V
16 -32 V
15 -39 V
15 -47 V
15 -51 V
15 -81 V
15 -59 V
14 -63 V
15 -75 V
15 -62 V
15 -47 V
15 -47 V
15 -26 V
16 -32 V
15 -13 V
15 8 V
15 8 V
15 19 V
15 32 V
15 33 V
14 35 V
15 56 V
15 45 V
15 44 V
15 38 V
15 47 V
16 34 V
15 30 V
15 24 V
15 12 V
15 1 V
15 -8 V
14 -8 V
15 -21 V
15 -21 V
15 -39 V
15 -43 V
15 -49 V
16 -53 V
15 -48 V
15 -60 V
15 -64 V
15 -49 V
15 -54 V
15 -70 V
14 -51 V
15 -49 V
15 -52 V
15 -48 V
15 -32 V
15 -41 V
16 -32 V
15 -27 V
15 -19 V
15 -14 V
15 -13 V
15 -7 V
1.000 UL
LT2
457 467 M
15 5 V
16 11 V
15 18 V
15 22 V
14 25 V
15 28 V
16 40 V
15 43 V
14 46 V
15 43 V
16 53 V
14 51 V
16 63 V
14 51 V
16 63 V
15 54 V
14 62 V
16 57 V
14 45 V
15 56 V
16 31 V
14 43 V
15 36 V
16 18 V
14 11 V
15 3 V
16 -4 V
14 -8 V
15 -30 V
16 -29 V
14 -31 V
15 -39 V
16 -47 V
14 -41 V
15 -45 V
16 -58 V
14 -31 V
15 -39 V
16 -26 V
15 -24 V
14 -11 V
16 3 V
15 7 V
14 21 V
16 41 V
15 39 V
14 52 V
16 61 V
15 63 V
14 66 V
15 69 V
15 58 V
16 64 V
15 55 V
15 46 V
14 29 V
15 16 V
15 -6 V
16 -3 V
15 -40 V
15 -23 V
14 -56 V
15 -71 V
15 -61 V
15 -60 V
16 -70 V
15 -70 V
15 -49 V
14 -43 V
15 -36 V
15 -27 V
16 6 V
15 14 V
15 35 V
14 36 V
15 61 V
15 63 V
16 79 V
15 77 V
15 75 V
14 80 V
15 67 V
15 53 V
16 42 V
15 24 V
15 1 V
14 -8 V
15 -42 V
15 -34 V
15 -58 V
16 -73 V
15 -85 V
15 -75 V
14 -76 V
15 -73 V
15 -63 V
16 -55 V
15 -30 V
15 -24 V
14 3 V
15 14 V
15 42 V
15 51 V
15 57 V
15 75 V
16 81 V
15 84 V
15 79 V
15 62 V
15 58 V
15 51 V
15 24 V
14 21 V
15 -3 V
15 -30 V
15 -48 V
15 -50 V
15 -69 V
16 -66 V
15 -71 V
15 -85 V
15 -73 V
15 -73 V
15 -63 V
14 -47 V
15 -24 V
15 -15 V
15 3 V
15 19 V
15 34 V
16 52 V
15 54 V
15 60 V
15 68 V
15 68 V
15 65 V
15 58 V
14 46 V
15 45 V
15 19 V
15 15 V
15 -3 V
15 -11 V
16 -25 V
15 -48 V
15 -48 V
15 -61 V
15 -71 V
15 -58 V
14 -70 V
15 -64 V
15 -53 V
15 -66 V
15 -37 V
15 -33 V
16 -23 V
15 -9 V
15 -1 V
15 7 V
15 21 V
15 36 V
15 31 V
14 41 V
15 43 V
15 51 V
15 48 V
15 47 V
15 37 V
16 26 V
15 35 V
15 14 V
15 22 V
15 3 V
15 -12 V
14 -4 V
15 -20 V
15 -36 V
15 -34 V
15 -37 V
15 -59 V
16 -43 V
15 -56 V
15 -59 V
15 -48 V
15 -64 V
15 -59 V
15 -58 V
14 -56 V
15 -52 V
15 -47 V
15 -42 V
15 -48 V
15 -33 V
16 -31 V
15 -26 V
15 -24 V
15 -15 V
15 -11 V
15 -5 V
1.000 UL
LT3
1950 1420 M
8 0 V
7 0 V
8 0 V
7 0 V
8 0 V
7 0 V
8 0 V
7 -1 V
8 0 V
7 0 V
8 0 V
7 0 V
8 -1 V
7 0 V
8 0 V
7 0 V
8 -1 V
7 0 V
8 -1 V
7 0 V
8 0 V
7 -1 V
8 0 V
7 -1 V
8 0 V
7 -1 V
8 0 V
7 -1 V
8 0 V
7 -1 V
8 -1 V
7 0 V
8 -1 V
7 -1 V
8 0 V
7 -1 V
8 -1 V
7 -1 V
8 -1 V
7 0 V
8 -1 V
7 -1 V
8 -1 V
7 -1 V
8 -1 V
7 -1 V
8 -1 V
7 -1 V
8 -1 V
7 -1 V
8 -1 V
7 -1 V
8 -1 V
7 -1 V
8 -1 V
7 -2 V
8 -1 V
7 -1 V
8 -1 V
7 -2 V
8 -1 V
7 -1 V
8 -1 V
7 -2 V
8 -1 V
7 -2 V
8 -1 V
7 -2 V
8 -1 V
7 -2 V
8 -1 V
7 -2 V
8 -1 V
7 -2 V
8 -2 V
7 -1 V
8 -2 V
7 -2 V
8 -2 V
7 -1 V
8 -2 V
7 -2 V
8 -2 V
7 -2 V
8 -2 V
7 -2 V
8 -2 V
7 -2 V
8 -2 V
7 -2 V
8 -2 V
7 -2 V
8 -2 V
7 -2 V
8 -3 V
7 -2 V
8 -2 V
7 -3 V
8 -2 V
7 -2 V
8 -3 V
7 -2 V
8 -3 V
7 -2 V
8 -3 V
7 -2 V
8 -3 V
7 -3 V
8 -2 V
7 -3 V
8 -3 V
7 -3 V
8 -3 V
7 -3 V
8 -3 V
7 -3 V
8 -3 V
7 -3 V
8 -3 V
7 -3 V
8 -3 V
7 -3 V
8 -4 V
7 -3 V
8 -3 V
7 -4 V
8 -3 V
7 -4 V
8 -3 V
7 -4 V
8 -4 V
7 -4 V
8 -3 V
7 -4 V
8 -4 V
7 -4 V
8 -4 V
7 -4 V
7 -4 V
8 -5 V
8 -4 V
7 -4 V
8 -5 V
7 -4 V
8 -5 V
7 -4 V
8 -5 V
7 -5 V
8 -5 V
7 -5 V
8 -5 V
7 -5 V
8 -5 V
7 -6 V
8 -5 V
7 -5 V
8 -6 V
7 -6 V
8 -6 V
7 -6 V
8 -6 V
7 -6 V
8 -6 V
7 -7 V
8 -6 V
7 -7 V
8 -7 V
7 -7 V
8 -7 V
7 -8 V
8 -8 V
7 -7 V
8 -8 V
7 -9 V
8 -8 V
7 -9 V
8 -9 V
7 -9 V
8 -10 V
7 -10 V
8 -10 V
7 -11 V
8 -11 V
7 -12 V
8 -12 V
7 -13 V
8 -13 V
7 -14 V
8 -15 V
7 -16 V
8 -17 V
7 -18 V
8 -20 V
7 -22 V
8 -25 V
7 -28 V
8 -34 V
7 -42 V
8 -60 V
7 -232 V
1.000 UL
LT3
1950 1420 M
-7 0 V
-8 0 V
-7 0 V
-8 0 V
-7 0 V
-8 0 V
-7 0 V
-8 -1 V
-7 0 V
-8 0 V
-7 0 V
-8 0 V
-7 -1 V
-8 0 V
-7 0 V
-8 0 V
-7 -1 V
-8 0 V
-7 -1 V
-8 0 V
-7 0 V
-8 -1 V
-7 0 V
-8 -1 V
-7 0 V
-8 -1 V
-7 0 V
-8 -1 V
-7 0 V
-8 -1 V
-7 -1 V
-8 0 V
-7 -1 V
-8 -1 V
-7 0 V
-8 -1 V
-7 -1 V
-8 -1 V
-7 -1 V
-8 0 V
-7 -1 V
-8 -1 V
-7 -1 V
-8 -1 V
-7 -1 V
-8 -1 V
-7 -1 V
-8 -1 V
-7 -1 V
-8 -1 V
-7 -1 V
-8 -1 V
-7 -1 V
-8 -1 V
-7 -1 V
-8 -2 V
-7 -1 V
-8 -1 V
-7 -1 V
-8 -2 V
-7 -1 V
-8 -1 V
-7 -1 V
-8 -2 V
-7 -1 V
-8 -2 V
-7 -1 V
-8 -2 V
-7 -1 V
-8 -2 V
-7 -1 V
-8 -2 V
-7 -1 V
-8 -2 V
-7 -2 V
-8 -1 V
-7 -2 V
-8 -2 V
-7 -2 V
-8 -1 V
-7 -2 V
-8 -2 V
-7 -2 V
-8 -2 V
-7 -2 V
-8 -2 V
-7 -2 V
-8 -2 V
-7 -2 V
-8 -2 V
-7 -2 V
-8 -2 V
-7 -2 V
-8 -2 V
-7 -3 V
-8 -2 V
-7 -2 V
-8 -3 V
-7 -2 V
-8 -2 V
-7 -3 V
-8 -2 V
-7 -3 V
-8 -2 V
-7 -3 V
-8 -2 V
-7 -3 V
-8 -3 V
-7 -2 V
-8 -3 V
-7 -3 V
-8 -3 V
-7 -3 V
-8 -3 V
-7 -3 V
-8 -3 V
-7 -3 V
-8 -3 V
-7 -3 V
-8 -3 V
-7 -3 V
-8 -3 V
-7 -4 V
-8 -3 V
-7 -3 V
-8 -4 V
-7 -3 V
-8 -4 V
-7 -3 V
-8 -4 V
-7 -4 V
-8 -4 V
-7 -3 V
-8 -4 V
-7 -4 V
-8 -4 V
-8 -4 V
-7 -4 V
-7 -4 V
-8 -5 V
-7 -4 V
-8 -4 V
-7 -5 V
-8 -4 V
-7 -5 V
-8 -4 V
-7 -5 V
-8 -5 V
-7 -5 V
-8 -5 V
-7 -5 V
-8 -5 V
-7 -5 V
-8 -6 V
-7 -5 V
-8 -5 V
-8 -6 V
-7 -6 V
-7 -6 V
-8 -6 V
-8 -6 V
-7 -6 V
-7 -6 V
-8 -7 V
-7 -6 V
-8 -7 V
-7 -7 V
-8 -7 V
-7 -7 V
-8 -8 V
-7 -8 V
-8 -7 V
-8 -8 V
-7 -9 V
-7 -8 V
-8 -9 V
-7 -9 V
-8 -9 V
-7 -10 V
-8 -10 V
-7 -10 V
-8 -11 V
-8 -11 V
-7 -12 V
-7 -12 V
-8 -13 V
-8 -13 V
-7 -14 V
-7 -15 V
-8 -16 V
-7 -17 V
-8 -18 V
-7 -20 V
-8 -22 V
-7 -25 V
-8 -28 V
-7 -34 V
-8 -42 V
-8 -60 V
450 300 L
stroke
grestore
end
showpage
}}%
\put(1950,50){\makebox(0,0){$\alpha$}}%
\put(100,1180){%
\makebox(0,0)[b]{\shortstack{$\rho(\alpha)$}}%
}%
\put(3382,200){\makebox(0,0){3}}%
\put(2905,200){\makebox(0,0){2}}%
\put(2427,200){\makebox(0,0){1}}%
\put(1950,200){\makebox(0,0){0}}%
\put(1473,200){\makebox(0,0){-1}}%
\put(995,200){\makebox(0,0){-2}}%
\put(518,200){\makebox(0,0){-3}}%
\put(400,2060){\makebox(0,0)[r]{0.3}}%
\put(400,1767){\makebox(0,0)[r]{0.25}}%
\put(400,1473){\makebox(0,0)[r]{0.2}}%
\put(400,1180){\makebox(0,0)[r]{0.15}}%
\put(400,887){\makebox(0,0)[r]{0.1}}%
\put(400,593){\makebox(0,0)[r]{0.05}}%
\put(400,300){\makebox(0,0)[r]{0}}%
\end{picture}%
\endgroup

%% file: wilsoncriticalbetagraph.tex
\begingroup%
  \makeatletter%
  \newcommand{\GNUPLOTspecial}{%
    \@sanitize\catcode`\%=14\relax\special}%
  \setlength{\unitlength}{0.1bp}%
{\GNUPLOTspecial{!
/gnudict 256 dict def
gnudict begin
/Color false def
/Solid false def
/gnulinewidth 5.000 def
/userlinewidth gnulinewidth def
/vshift -33 def
/dl {10 mul} def
/hpt_ 31.5 def
/vpt_ 31.5 def
/hpt hpt_ def
/vpt vpt_ def
/M {moveto} bind def
/L {lineto} bind def
/R {rmoveto} bind def
/V {rlineto} bind def
/vpt2 vpt 2 mul def
/hpt2 hpt 2 mul def
/Lshow { currentpoint stroke M
  0 vshift R show } def
/Rshow { currentpoint stroke M
  dup stringwidth pop neg vshift R show } def
/Cshow { currentpoint stroke M
  dup stringwidth pop -2 div vshift R show } def
/UP { dup vpt_ mul /vpt exch def hpt_ mul /hpt exch def
  /hpt2 hpt 2 mul def /vpt2 vpt 2 mul def } def
/DL { Color {setrgbcolor Solid {pop []} if 0 setdash }
 {pop pop pop Solid {pop []} if 0 setdash} ifelse } def
/BL { stroke userlinewidth 2 mul setlinewidth } def
/AL { stroke userlinewidth 2 div setlinewidth } def
/UL { dup gnulinewidth mul /userlinewidth exch def
      10 mul /udl exch def } def
/PL { stroke userlinewidth setlinewidth } def
/LTb { BL [] 0 0 0 DL } def
/LTa { AL [1 udl mul 2 udl mul] 0 setdash 0 0 0 setrgbcolor } def
/LT0 { PL [] 1 0 0 DL } def
/LT1 { PL [4 dl 2 dl] 0 1 0 DL } def
/LT2 { PL [2 dl 3 dl] 0 0 1 DL } def
/LT3 { PL [1 dl 1.5 dl] 1 0 1 DL } def
/LT4 { PL [5 dl 2 dl 1 dl 2 dl] 0 1 1 DL } def
/LT5 { PL [4 dl 3 dl 1 dl 3 dl] 1 1 0 DL } def
/LT6 { PL [2 dl 2 dl 2 dl 4 dl] 0 0 0 DL } def
/LT7 { PL [2 dl 2 dl 2 dl 2 dl 2 dl 4 dl] 1 0.3 0 DL } def
/LT8 { PL [2 dl 2 dl 2 dl 2 dl 2 dl 2 dl 2 dl 4 dl] 0.5 0.5 0.5 DL } def
/Pnt { stroke [] 0 setdash
   gsave 1 setlinecap M 0 0 V stroke grestore } def
/Dia { stroke [] 0 setdash 2 copy vpt add M
  hpt neg vpt neg V hpt vpt neg V
  hpt vpt V hpt neg vpt V closepath stroke
  Pnt } def
/Pls { stroke [] 0 setdash vpt sub M 0 vpt2 V
  currentpoint stroke M
  hpt neg vpt neg R hpt2 0 V stroke
  } def
/Box { stroke [] 0 setdash 2 copy exch hpt sub exch vpt add M
  0 vpt2 neg V hpt2 0 V 0 vpt2 V
  hpt2 neg 0 V closepath stroke
  Pnt } def
/Crs { stroke [] 0 setdash exch hpt sub exch vpt add M
  hpt2 vpt2 neg V currentpoint stroke M
  hpt2 neg 0 R hpt2 vpt2 V stroke } def
/TriU { stroke [] 0 setdash 2 copy vpt 1.12 mul add M
  hpt neg vpt -1.62 mul V
  hpt 2 mul 0 V
  hpt neg vpt 1.62 mul V closepath stroke
  Pnt  } def
/Star { 2 copy Pls Crs } def
/BoxF { stroke [] 0 setdash exch hpt sub exch vpt add M
  0 vpt2 neg V  hpt2 0 V  0 vpt2 V
  hpt2 neg 0 V  closepath fill } def
/TriUF { stroke [] 0 setdash vpt 1.12 mul add M
  hpt neg vpt -1.62 mul V
  hpt 2 mul 0 V
  hpt neg vpt 1.62 mul V closepath fill } def
/TriD { stroke [] 0 setdash 2 copy vpt 1.12 mul sub M
  hpt neg vpt 1.62 mul V
  hpt 2 mul 0 V
  hpt neg vpt -1.62 mul V closepath stroke
  Pnt  } def
/TriDF { stroke [] 0 setdash vpt 1.12 mul sub M
  hpt neg vpt 1.62 mul V
  hpt 2 mul 0 V
  hpt neg vpt -1.62 mul V closepath fill} def
/DiaF { stroke [] 0 setdash vpt add M
  hpt neg vpt neg V hpt vpt neg V
  hpt vpt V hpt neg vpt V closepath fill } def
/Pent { stroke [] 0 setdash 2 copy gsave
  translate 0 hpt M 4 {72 rotate 0 hpt L} repeat
  closepath stroke grestore Pnt } def
/PentF { stroke [] 0 setdash gsave
  translate 0 hpt M 4 {72 rotate 0 hpt L} repeat
  closepath fill grestore } def
/Circle { stroke [] 0 setdash 2 copy
  hpt 0 360 arc stroke Pnt } def
/CircleF { stroke [] 0 setdash hpt 0 360 arc fill } def
/C0 { BL [] 0 setdash 2 copy moveto vpt 90 450  arc } bind def
/C1 { BL [] 0 setdash 2 copy        moveto
       2 copy  vpt 0 90 arc closepath fill
               vpt 0 360 arc closepath } bind def
/C2 { BL [] 0 setdash 2 copy moveto
       2 copy  vpt 90 180 arc closepath fill
               vpt 0 360 arc closepath } bind def
/C3 { BL [] 0 setdash 2 copy moveto
       2 copy  vpt 0 180 arc closepath fill
               vpt 0 360 arc closepath } bind def
/C4 { BL [] 0 setdash 2 copy moveto
       2 copy  vpt 180 270 arc closepath fill
               vpt 0 360 arc closepath } bind def
/C5 { BL [] 0 setdash 2 copy moveto
       2 copy  vpt 0 90 arc
       2 copy moveto
       2 copy  vpt 180 270 arc closepath fill
               vpt 0 360 arc } bind def
/C6 { BL [] 0 setdash 2 copy moveto
      2 copy  vpt 90 270 arc closepath fill
              vpt 0 360 arc closepath } bind def
/C7 { BL [] 0 setdash 2 copy moveto
      2 copy  vpt 0 270 arc closepath fill
              vpt 0 360 arc closepath } bind def
/C8 { BL [] 0 setdash 2 copy moveto
      2 copy vpt 270 360 arc closepath fill
              vpt 0 360 arc closepath } bind def
/C9 { BL [] 0 setdash 2 copy moveto
      2 copy  vpt 270 450 arc closepath fill
              vpt 0 360 arc closepath } bind def
/C10 { BL [] 0 setdash 2 copy 2 copy moveto vpt 270 360 arc closepath fill
       2 copy moveto
       2 copy vpt 90 180 arc closepath fill
               vpt 0 360 arc closepath } bind def
/C11 { BL [] 0 setdash 2 copy moveto
       2 copy  vpt 0 180 arc closepath fill
       2 copy moveto
       2 copy  vpt 270 360 arc closepath fill
               vpt 0 360 arc closepath } bind def
/C12 { BL [] 0 setdash 2 copy moveto
       2 copy  vpt 180 360 arc closepath fill
               vpt 0 360 arc closepath } bind def
/C13 { BL [] 0 setdash  2 copy moveto
       2 copy  vpt 0 90 arc closepath fill
       2 copy moveto
       2 copy  vpt 180 360 arc closepath fill
               vpt 0 360 arc closepath } bind def
/C14 { BL [] 0 setdash 2 copy moveto
       2 copy  vpt 90 360 arc closepath fill
               vpt 0 360 arc } bind def
/C15 { BL [] 0 setdash 2 copy vpt 0 360 arc closepath fill
               vpt 0 360 arc closepath } bind def
/Rec   { newpath 4 2 roll moveto 1 index 0 rlineto 0 exch rlineto
       neg 0 rlineto closepath } bind def
/Square { dup Rec } bind def
/Bsquare { vpt sub exch vpt sub exch vpt2 Square } bind def
/S0 { BL [] 0 setdash 2 copy moveto 0 vpt rlineto BL Bsquare } bind def
/S1 { BL [] 0 setdash 2 copy vpt Square fill Bsquare } bind def
/S2 { BL [] 0 setdash 2 copy exch vpt sub exch vpt Square fill Bsquare } bind def
/S3 { BL [] 0 setdash 2 copy exch vpt sub exch vpt2 vpt Rec fill Bsquare } bind def
/S4 { BL [] 0 setdash 2 copy exch vpt sub exch vpt sub vpt Square fill Bsquare } bind def
/S5 { BL [] 0 setdash 2 copy 2 copy vpt Square fill
       exch vpt sub exch vpt sub vpt Square fill Bsquare } bind def
/S6 { BL [] 0 setdash 2 copy exch vpt sub exch vpt sub vpt vpt2 Rec fill Bsquare } bind def
/S7 { BL [] 0 setdash 2 copy exch vpt sub exch vpt sub vpt vpt2 Rec fill
       2 copy vpt Square fill
       Bsquare } bind def
/S8 { BL [] 0 setdash 2 copy vpt sub vpt Square fill Bsquare } bind def
/S9 { BL [] 0 setdash 2 copy vpt sub vpt vpt2 Rec fill Bsquare } bind def
/S10 { BL [] 0 setdash 2 copy vpt sub vpt Square fill 2 copy exch vpt sub exch vpt Square fill
       Bsquare } bind def
/S11 { BL [] 0 setdash 2 copy vpt sub vpt Square fill 2 copy exch vpt sub exch vpt2 vpt Rec fill
       Bsquare } bind def
/S12 { BL [] 0 setdash 2 copy exch vpt sub exch vpt sub vpt2 vpt Rec fill Bsquare } bind def
/S13 { BL [] 0 setdash 2 copy exch vpt sub exch vpt sub vpt2 vpt Rec fill
       2 copy vpt Square fill Bsquare } bind def
/S14 { BL [] 0 setdash 2 copy exch vpt sub exch vpt sub vpt2 vpt Rec fill
       2 copy exch vpt sub exch vpt Square fill Bsquare } bind def
/S15 { BL [] 0 setdash 2 copy Bsquare fill Bsquare } bind def
/D0 { gsave translate 45 rotate 0 0 S0 stroke grestore } bind def
/D1 { gsave translate 45 rotate 0 0 S1 stroke grestore } bind def
/D2 { gsave translate 45 rotate 0 0 S2 stroke grestore } bind def
/D3 { gsave translate 45 rotate 0 0 S3 stroke grestore } bind def
/D4 { gsave translate 45 rotate 0 0 S4 stroke grestore } bind def
/D5 { gsave translate 45 rotate 0 0 S5 stroke grestore } bind def
/D6 { gsave translate 45 rotate 0 0 S6 stroke grestore } bind def
/D7 { gsave translate 45 rotate 0 0 S7 stroke grestore } bind def
/D8 { gsave translate 45 rotate 0 0 S8 stroke grestore } bind def
/D9 { gsave translate 45 rotate 0 0 S9 stroke grestore } bind def
/D10 { gsave translate 45 rotate 0 0 S10 stroke grestore } bind def
/D11 { gsave translate 45 rotate 0 0 S11 stroke grestore } bind def
/D12 { gsave translate 45 rotate 0 0 S12 stroke grestore } bind def
/D13 { gsave translate 45 rotate 0 0 S13 stroke grestore } bind def
/D14 { gsave translate 45 rotate 0 0 S14 stroke grestore } bind def
/D15 { gsave translate 45 rotate 0 0 S15 stroke grestore } bind def
/DiaE { stroke [] 0 setdash vpt add M
  hpt neg vpt neg V hpt vpt neg V
  hpt vpt V hpt neg vpt V closepath stroke } def
/BoxE { stroke [] 0 setdash exch hpt sub exch vpt add M
  0 vpt2 neg V hpt2 0 V 0 vpt2 V
  hpt2 neg 0 V closepath stroke } def
/TriUE { stroke [] 0 setdash vpt 1.12 mul add M
  hpt neg vpt -1.62 mul V
  hpt 2 mul 0 V
  hpt neg vpt 1.62 mul V closepath stroke } def
/TriDE { stroke [] 0 setdash vpt 1.12 mul sub M
  hpt neg vpt 1.62 mul V
  hpt 2 mul 0 V
  hpt neg vpt -1.62 mul V closepath stroke } def
/PentE { stroke [] 0 setdash gsave
  translate 0 hpt M 4 {72 rotate 0 hpt L} repeat
  closepath stroke grestore } def
/CircE { stroke [] 0 setdash 
  hpt 0 360 arc stroke } def
/Opaque { gsave closepath 1 setgray fill grestore 0 setgray closepath } def
/DiaW { stroke [] 0 setdash vpt add M
  hpt neg vpt neg V hpt vpt neg V
  hpt vpt V hpt neg vpt V Opaque stroke } def
/BoxW { stroke [] 0 setdash exch hpt sub exch vpt add M
  0 vpt2 neg V hpt2 0 V 0 vpt2 V
  hpt2 neg 0 V Opaque stroke } def
/TriUW { stroke [] 0 setdash vpt 1.12 mul add M
  hpt neg vpt -1.62 mul V
  hpt 2 mul 0 V
  hpt neg vpt 1.62 mul V Opaque stroke } def
/TriDW { stroke [] 0 setdash vpt 1.12 mul sub M
  hpt neg vpt 1.62 mul V
  hpt 2 mul 0 V
  hpt neg vpt -1.62 mul V Opaque stroke } def
/PentW { stroke [] 0 setdash gsave
  translate 0 hpt M 4 {72 rotate 0 hpt L} repeat
  Opaque stroke grestore } def
/CircW { stroke [] 0 setdash 
  hpt 0 360 arc Opaque stroke } def
/BoxFill { gsave Rec 1 setgray fill grestore } def
end
}}%
\begin{picture}(3600,2160)(0,0)%
{\GNUPLOTspecial{"
gnudict begin
gsave
0 0 translate
0.100 0.100 scale
0 setgray
newpath
1.000 UL
LTb
400 300 M
63 0 V
2987 0 R
-63 0 V
400 496 M
63 0 V
2987 0 R
-63 0 V
400 691 M
63 0 V
2987 0 R
-63 0 V
400 887 M
63 0 V
2987 0 R
-63 0 V
400 1082 M
63 0 V
2987 0 R
-63 0 V
400 1278 M
63 0 V
2987 0 R
-63 0 V
400 1473 M
63 0 V
2987 0 R
-63 0 V
400 1669 M
63 0 V
2987 0 R
-63 0 V
400 1864 M
63 0 V
2987 0 R
-63 0 V
400 2060 M
63 0 V
2987 0 R
-63 0 V
400 300 M
0 63 V
0 1697 R
0 -63 V
908 300 M
0 63 V
0 1697 R
0 -63 V
1417 300 M
0 63 V
0 1697 R
0 -63 V
1925 300 M
0 63 V
0 1697 R
0 -63 V
2433 300 M
0 63 V
0 1697 R
0 -63 V
2942 300 M
0 63 V
0 1697 R
0 -63 V
3450 300 M
0 63 V
0 1697 R
0 -63 V
1.000 UL
LTb
400 300 M
3050 0 V
0 1760 V
-3050 0 V
400 300 L
1.000 UP
1.000 UL
LT0
400 357 M
0 2 V
-31 -2 R
62 0 V
-62 2 R
62 0 V
908 591 M
0 5 V
-31 -5 R
62 0 V
-62 5 R
62 0 V
478 256 R
0 8 V
-31 -8 R
62 0 V
-62 8 R
62 0 V
477 271 R
0 12 V
-31 -12 R
62 0 V
-62 12 R
62 0 V
477 288 R
0 12 V
-31 -12 R
62 0 V
-62 12 R
62 0 V
478 275 R
0 12 V
-31 -12 R
62 0 V
-62 12 R
62 0 V
477 293 R
0 25 V
-31 -25 R
62 0 V
-62 25 R
62 0 V
400 358 Pls
908 593 Pls
1417 856 Pls
1925 1137 Pls
2433 1437 Pls
2942 1724 Pls
3450 2036 Pls
1.000 UP
1.000 UL
LT1
400 379 M
0 5 V
-31 -5 R
62 0 V
-62 5 R
62 0 V
908 610 M
0 5 V
-31 -5 R
62 0 V
-62 5 R
62 0 V
478 231 R
0 14 V
-31 -14 R
62 0 V
-62 14 R
62 0 V
400 381 Crs
908 612 Crs
1417 853 Crs
1.000 UL
LT2
400 338 M
31 15 V
36 18 V
35 18 V
36 18 V
35 17 V
35 18 V
35 18 V
35 18 V
34 18 V
35 17 V
35 18 V
34 18 V
34 18 V
34 17 V
34 18 V
34 18 V
34 18 V
34 18 V
34 17 V
33 18 V
34 18 V
33 18 V
33 17 V
34 18 V
33 18 V
33 18 V
33 18 V
33 17 V
33 18 V
32 18 V
33 18 V
33 17 V
32 18 V
33 18 V
32 18 V
33 18 V
32 17 V
32 18 V
33 18 V
32 18 V
32 17 V
32 18 V
32 18 V
32 18 V
32 18 V
32 17 V
32 18 V
32 18 V
31 18 V
32 17 V
32 18 V
31 18 V
32 18 V
31 18 V
32 17 V
31 18 V
31 18 V
32 18 V
31 17 V
31 18 V
32 18 V
31 18 V
31 18 V
31 17 V
31 18 V
31 18 V
31 18 V
31 17 V
31 18 V
31 18 V
31 18 V
30 18 V
31 17 V
31 18 V
31 18 V
30 18 V
31 17 V
31 18 V
30 18 V
31 18 V
30 18 V
31 17 V
30 18 V
31 18 V
30 18 V
30 17 V
31 18 V
30 18 V
30 18 V
31 18 V
30 17 V
30 18 V
30 18 V
30 18 V
31 17 V
1 1 V
stroke
grestore
end
showpage
}}%
\put(1925,50){\makebox(0,0){$L$}}%
\put(100,1180){%
\makebox(0,0)[b]{\shortstack{$\gamma$}}%
}%
\put(3450,200){\makebox(0,0){8}}%
\put(2942,200){\makebox(0,0){7}}%
\put(2433,200){\makebox(0,0){6}}%
\put(1925,200){\makebox(0,0){5}}%
\put(1417,200){\makebox(0,0){4}}%
\put(908,200){\makebox(0,0){3}}%
\put(400,200){\makebox(0,0){2}}%
\put(350,2060){\makebox(0,0)[r]{2.2}}%
\put(350,1864){\makebox(0,0)[r]{2}}%
\put(350,1669){\makebox(0,0)[r]{1.8}}%
\put(350,1473){\makebox(0,0)[r]{1.6}}%
\put(350,1278){\makebox(0,0)[r]{1.4}}%
\put(350,1082){\makebox(0,0)[r]{1.2}}%
\put(350,887){\makebox(0,0)[r]{1}}%
\put(350,691){\makebox(0,0)[r]{0.8}}%
\put(350,496){\makebox(0,0)[r]{0.6}}%
\put(350,300){\makebox(0,0)[r]{0.4}}%
\end{picture}%
\endgroup

%% file: polyakovwilsonmatchgraph.tex
\begingroup%
  \makeatletter%
  \newcommand{\GNUPLOTspecial}{%
    \@sanitize\catcode`\%=14\relax\special}%
  \setlength{\unitlength}{0.1bp}%
{\GNUPLOTspecial{!
/gnudict 256 dict def
gnudict begin
/Color false def
/Solid false def
/gnulinewidth 5.000 def
/userlinewidth gnulinewidth def
/vshift -33 def
/dl {10 mul} def
/hpt_ 31.5 def
/vpt_ 31.5 def
/hpt hpt_ def
/vpt vpt_ def
/M {moveto} bind def
/L {lineto} bind def
/R {rmoveto} bind def
/V {rlineto} bind def
/vpt2 vpt 2 mul def
/hpt2 hpt 2 mul def
/Lshow { currentpoint stroke M
  0 vshift R show } def
/Rshow { currentpoint stroke M
  dup stringwidth pop neg vshift R show } def
/Cshow { currentpoint stroke M
  dup stringwidth pop -2 div vshift R show } def
/UP { dup vpt_ mul /vpt exch def hpt_ mul /hpt exch def
  /hpt2 hpt 2 mul def /vpt2 vpt 2 mul def } def
/DL { Color {setrgbcolor Solid {pop []} if 0 setdash }
 {pop pop pop Solid {pop []} if 0 setdash} ifelse } def
/BL { stroke userlinewidth 2 mul setlinewidth } def
/AL { stroke userlinewidth 2 div setlinewidth } def
/UL { dup gnulinewidth mul /userlinewidth exch def
      10 mul /udl exch def } def
/PL { stroke userlinewidth setlinewidth } def
/LTb { BL [] 0 0 0 DL } def
/LTa { AL [1 udl mul 2 udl mul] 0 setdash 0 0 0 setrgbcolor } def
/LT0 { PL [] 1 0 0 DL } def
/LT1 { PL [4 dl 2 dl] 0 1 0 DL } def
/LT2 { PL [2 dl 3 dl] 0 0 1 DL } def
/LT3 { PL [1 dl 1.5 dl] 1 0 1 DL } def
/LT4 { PL [5 dl 2 dl 1 dl 2 dl] 0 1 1 DL } def
/LT5 { PL [4 dl 3 dl 1 dl 3 dl] 1 1 0 DL } def
/LT6 { PL [2 dl 2 dl 2 dl 4 dl] 0 0 0 DL } def
/LT7 { PL [2 dl 2 dl 2 dl 2 dl 2 dl 4 dl] 1 0.3 0 DL } def
/LT8 { PL [2 dl 2 dl 2 dl 2 dl 2 dl 2 dl 2 dl 4 dl] 0.5 0.5 0.5 DL } def
/Pnt { stroke [] 0 setdash
   gsave 1 setlinecap M 0 0 V stroke grestore } def
/Dia { stroke [] 0 setdash 2 copy vpt add M
  hpt neg vpt neg V hpt vpt neg V
  hpt vpt V hpt neg vpt V closepath stroke
  Pnt } def
/Pls { stroke [] 0 setdash vpt sub M 0 vpt2 V
  currentpoint stroke M
  hpt neg vpt neg R hpt2 0 V stroke
  } def
/Box { stroke [] 0 setdash 2 copy exch hpt sub exch vpt add M
  0 vpt2 neg V hpt2 0 V 0 vpt2 V
  hpt2 neg 0 V closepath stroke
  Pnt } def
/Crs { stroke [] 0 setdash exch hpt sub exch vpt add M
  hpt2 vpt2 neg V currentpoint stroke M
  hpt2 neg 0 R hpt2 vpt2 V stroke } def
/TriU { stroke [] 0 setdash 2 copy vpt 1.12 mul add M
  hpt neg vpt -1.62 mul V
  hpt 2 mul 0 V
  hpt neg vpt 1.62 mul V closepath stroke
  Pnt  } def
/Star { 2 copy Pls Crs } def
/BoxF { stroke [] 0 setdash exch hpt sub exch vpt add M
  0 vpt2 neg V  hpt2 0 V  0 vpt2 V
  hpt2 neg 0 V  closepath fill } def
/TriUF { stroke [] 0 setdash vpt 1.12 mul add M
  hpt neg vpt -1.62 mul V
  hpt 2 mul 0 V
  hpt neg vpt 1.62 mul V closepath fill } def
/TriD { stroke [] 0 setdash 2 copy vpt 1.12 mul sub M
  hpt neg vpt 1.62 mul V
  hpt 2 mul 0 V
  hpt neg vpt -1.62 mul V closepath stroke
  Pnt  } def
/TriDF { stroke [] 0 setdash vpt 1.12 mul sub M
  hpt neg vpt 1.62 mul V
  hpt 2 mul 0 V
  hpt neg vpt -1.62 mul V closepath fill} def
/DiaF { stroke [] 0 setdash vpt add M
  hpt neg vpt neg V hpt vpt neg V
  hpt vpt V hpt neg vpt V closepath fill } def
/Pent { stroke [] 0 setdash 2 copy gsave
  translate 0 hpt M 4 {72 rotate 0 hpt L} repeat
  closepath stroke grestore Pnt } def
/PentF { stroke [] 0 setdash gsave
  translate 0 hpt M 4 {72 rotate 0 hpt L} repeat
  closepath fill grestore } def
/Circle { stroke [] 0 setdash 2 copy
  hpt 0 360 arc stroke Pnt } def
/CircleF { stroke [] 0 setdash hpt 0 360 arc fill } def
/C0 { BL [] 0 setdash 2 copy moveto vpt 90 450  arc } bind def
/C1 { BL [] 0 setdash 2 copy        moveto
       2 copy  vpt 0 90 arc closepath fill
               vpt 0 360 arc closepath } bind def
/C2 { BL [] 0 setdash 2 copy moveto
       2 copy  vpt 90 180 arc closepath fill
               vpt 0 360 arc closepath } bind def
/C3 { BL [] 0 setdash 2 copy moveto
       2 copy  vpt 0 180 arc closepath fill
               vpt 0 360 arc closepath } bind def
/C4 { BL [] 0 setdash 2 copy moveto
       2 copy  vpt 180 270 arc closepath fill
               vpt 0 360 arc closepath } bind def
/C5 { BL [] 0 setdash 2 copy moveto
       2 copy  vpt 0 90 arc
       2 copy moveto
       2 copy  vpt 180 270 arc closepath fill
               vpt 0 360 arc } bind def
/C6 { BL [] 0 setdash 2 copy moveto
      2 copy  vpt 90 270 arc closepath fill
              vpt 0 360 arc closepath } bind def
/C7 { BL [] 0 setdash 2 copy moveto
      2 copy  vpt 0 270 arc closepath fill
              vpt 0 360 arc closepath } bind def
/C8 { BL [] 0 setdash 2 copy moveto
      2 copy vpt 270 360 arc closepath fill
              vpt 0 360 arc closepath } bind def
/C9 { BL [] 0 setdash 2 copy moveto
      2 copy  vpt 270 450 arc closepath fill
              vpt 0 360 arc closepath } bind def
/C10 { BL [] 0 setdash 2 copy 2 copy moveto vpt 270 360 arc closepath fill
       2 copy moveto
       2 copy vpt 90 180 arc closepath fill
               vpt 0 360 arc closepath } bind def
/C11 { BL [] 0 setdash 2 copy moveto
       2 copy  vpt 0 180 arc closepath fill
       2 copy moveto
       2 copy  vpt 270 360 arc closepath fill
               vpt 0 360 arc closepath } bind def
/C12 { BL [] 0 setdash 2 copy moveto
       2 copy  vpt 180 360 arc closepath fill
               vpt 0 360 arc closepath } bind def
/C13 { BL [] 0 setdash  2 copy moveto
       2 copy  vpt 0 90 arc closepath fill
       2 copy moveto
       2 copy  vpt 180 360 arc closepath fill
               vpt 0 360 arc closepath } bind def
/C14 { BL [] 0 setdash 2 copy moveto
       2 copy  vpt 90 360 arc closepath fill
               vpt 0 360 arc } bind def
/C15 { BL [] 0 setdash 2 copy vpt 0 360 arc closepath fill
               vpt 0 360 arc closepath } bind def
/Rec   { newpath 4 2 roll moveto 1 index 0 rlineto 0 exch rlineto
       neg 0 rlineto closepath } bind def
/Square { dup Rec } bind def
/Bsquare { vpt sub exch vpt sub exch vpt2 Square } bind def
/S0 { BL [] 0 setdash 2 copy moveto 0 vpt rlineto BL Bsquare } bind def
/S1 { BL [] 0 setdash 2 copy vpt Square fill Bsquare } bind def
/S2 { BL [] 0 setdash 2 copy exch vpt sub exch vpt Square fill Bsquare } bind def
/S3 { BL [] 0 setdash 2 copy exch vpt sub exch vpt2 vpt Rec fill Bsquare } bind def
/S4 { BL [] 0 setdash 2 copy exch vpt sub exch vpt sub vpt Square fill Bsquare } bind def
/S5 { BL [] 0 setdash 2 copy 2 copy vpt Square fill
       exch vpt sub exch vpt sub vpt Square fill Bsquare } bind def
/S6 { BL [] 0 setdash 2 copy exch vpt sub exch vpt sub vpt vpt2 Rec fill Bsquare } bind def
/S7 { BL [] 0 setdash 2 copy exch vpt sub exch vpt sub vpt vpt2 Rec fill
       2 copy vpt Square fill
       Bsquare } bind def
/S8 { BL [] 0 setdash 2 copy vpt sub vpt Square fill Bsquare } bind def
/S9 { BL [] 0 setdash 2 copy vpt sub vpt vpt2 Rec fill Bsquare } bind def
/S10 { BL [] 0 setdash 2 copy vpt sub vpt Square fill 2 copy exch vpt sub exch vpt Square fill
       Bsquare } bind def
/S11 { BL [] 0 setdash 2 copy vpt sub vpt Square fill 2 copy exch vpt sub exch vpt2 vpt Rec fill
       Bsquare } bind def
/S12 { BL [] 0 setdash 2 copy exch vpt sub exch vpt sub vpt2 vpt Rec fill Bsquare } bind def
/S13 { BL [] 0 setdash 2 copy exch vpt sub exch vpt sub vpt2 vpt Rec fill
       2 copy vpt Square fill Bsquare } bind def
/S14 { BL [] 0 setdash 2 copy exch vpt sub exch vpt sub vpt2 vpt Rec fill
       2 copy exch vpt sub exch vpt Square fill Bsquare } bind def
/S15 { BL [] 0 setdash 2 copy Bsquare fill Bsquare } bind def
/D0 { gsave translate 45 rotate 0 0 S0 stroke grestore } bind def
/D1 { gsave translate 45 rotate 0 0 S1 stroke grestore } bind def
/D2 { gsave translate 45 rotate 0 0 S2 stroke grestore } bind def
/D3 { gsave translate 45 rotate 0 0 S3 stroke grestore } bind def
/D4 { gsave translate 45 rotate 0 0 S4 stroke grestore } bind def
/D5 { gsave translate 45 rotate 0 0 S5 stroke grestore } bind def
/D6 { gsave translate 45 rotate 0 0 S6 stroke grestore } bind def
/D7 { gsave translate 45 rotate 0 0 S7 stroke grestore } bind def
/D8 { gsave translate 45 rotate 0 0 S8 stroke grestore } bind def
/D9 { gsave translate 45 rotate 0 0 S9 stroke grestore } bind def
/D10 { gsave translate 45 rotate 0 0 S10 stroke grestore } bind def
/D11 { gsave translate 45 rotate 0 0 S11 stroke grestore } bind def
/D12 { gsave translate 45 rotate 0 0 S12 stroke grestore } bind def
/D13 { gsave translate 45 rotate 0 0 S13 stroke grestore } bind def
/D14 { gsave translate 45 rotate 0 0 S14 stroke grestore } bind def
/D15 { gsave translate 45 rotate 0 0 S15 stroke grestore } bind def
/DiaE { stroke [] 0 setdash vpt add M
  hpt neg vpt neg V hpt vpt neg V
  hpt vpt V hpt neg vpt V closepath stroke } def
/BoxE { stroke [] 0 setdash exch hpt sub exch vpt add M
  0 vpt2 neg V hpt2 0 V 0 vpt2 V
  hpt2 neg 0 V closepath stroke } def
/TriUE { stroke [] 0 setdash vpt 1.12 mul add M
  hpt neg vpt -1.62 mul V
  hpt 2 mul 0 V
  hpt neg vpt 1.62 mul V closepath stroke } def
/TriDE { stroke [] 0 setdash vpt 1.12 mul sub M
  hpt neg vpt 1.62 mul V
  hpt 2 mul 0 V
  hpt neg vpt -1.62 mul V closepath stroke } def
/PentE { stroke [] 0 setdash gsave
  translate 0 hpt M 4 {72 rotate 0 hpt L} repeat
  closepath stroke grestore } def
/CircE { stroke [] 0 setdash 
  hpt 0 360 arc stroke } def
/Opaque { gsave closepath 1 setgray fill grestore 0 setgray closepath } def
/DiaW { stroke [] 0 setdash vpt add M
  hpt neg vpt neg V hpt vpt neg V
  hpt vpt V hpt neg vpt V Opaque stroke } def
/BoxW { stroke [] 0 setdash exch hpt sub exch vpt add M
  0 vpt2 neg V hpt2 0 V 0 vpt2 V
  hpt2 neg 0 V Opaque stroke } def
/TriUW { stroke [] 0 setdash vpt 1.12 mul add M
  hpt neg vpt -1.62 mul V
  hpt 2 mul 0 V
  hpt neg vpt 1.62 mul V Opaque stroke } def
/TriDW { stroke [] 0 setdash vpt 1.12 mul sub M
  hpt neg vpt 1.62 mul V
  hpt 2 mul 0 V
  hpt neg vpt -1.62 mul V Opaque stroke } def
/PentW { stroke [] 0 setdash gsave
  translate 0 hpt M 4 {72 rotate 0 hpt L} repeat
  Opaque stroke grestore } def
/CircW { stroke [] 0 setdash 
  hpt 0 360 arc Opaque stroke } def
/BoxFill { gsave Rec 1 setgray fill grestore } def
end
}}%
\begin{picture}(3600,2160)(0,0)%
{\GNUPLOTspecial{"
gnudict begin
gsave
0 0 translate
0.100 0.100 scale
0 setgray
newpath
1.000 UL
LTb
450 300 M
63 0 V
2937 0 R
-63 0 V
450 652 M
63 0 V
2937 0 R
-63 0 V
450 1004 M
63 0 V
2937 0 R
-63 0 V
450 1356 M
63 0 V
2937 0 R
-63 0 V
450 1708 M
63 0 V
2937 0 R
-63 0 V
450 2060 M
63 0 V
2937 0 R
-63 0 V
518 300 M
0 63 V
0 1697 R
0 -63 V
995 300 M
0 63 V
0 1697 R
0 -63 V
1473 300 M
0 63 V
0 1697 R
0 -63 V
1950 300 M
0 63 V
0 1697 R
0 -63 V
2427 300 M
0 63 V
0 1697 R
0 -63 V
2905 300 M
0 63 V
0 1697 R
0 -63 V
3382 300 M
0 63 V
0 1697 R
0 -63 V
1.000 UL
LTb
450 300 M
3000 0 V
0 1760 V
-3000 0 V
450 300 L
1.000 UL
LT0
457 1213 M
15 23 V
16 40 V
15 58 V
15 68 V
14 61 V
15 60 V
16 40 V
15 18 V
14 -18 V
15 -26 V
16 -48 V
14 -69 V
16 -60 V
14 -53 V
16 -47 V
15 -29 V
14 7 V
16 16 V
14 46 V
15 57 V
16 65 V
14 60 V
15 58 V
16 28 V
14 16 V
15 -11 V
16 -35 V
14 -47 V
15 -74 V
16 -48 V
14 -65 V
15 -45 V
16 -24 V
14 6 V
15 38 V
16 43 V
14 51 V
15 58 V
16 73 V
15 64 V
14 35 V
16 -14 V
15 -19 V
14 -29 V
16 -62 V
15 -57 V
14 -73 V
16 -58 V
15 -37 V
14 -17 V
15 13 V
15 43 V
16 41 V
15 68 V
15 65 V
14 61 V
15 48 V
15 39 V
16 -5 V
15 -11 V
15 -82 V
14 -55 V
15 -48 V
15 -77 V
15 -53 V
16 -22 V
15 -8 V
15 10 V
14 51 V
15 53 V
15 69 V
16 78 V
15 51 V
15 29 V
14 15 V
15 -8 V
15 -41 V
16 -48 V
15 -63 V
15 -69 V
14 -58 V
15 -41 V
15 -15 V
16 5 V
15 34 V
15 55 V
14 70 V
15 69 V
15 59 V
15 44 V
16 21 V
15 -16 V
15 -28 V
14 -44 V
15 -71 V
15 -49 V
16 -85 V
15 -44 V
15 -18 V
14 18 V
15 19 V
15 32 V
15 79 V
15 77 V
15 44 V
16 59 V
15 23 V
15 14 V
15 -33 V
15 -54 V
15 -60 V
15 -64 V
14 -78 V
15 -53 V
15 -21 V
15 -16 V
15 28 V
15 35 V
16 54 V
15 80 V
15 56 V
15 71 V
15 19 V
15 3 V
14 -4 V
15 -36 V
15 -45 V
15 -83 V
15 -53 V
15 -76 V
16 -34 V
15 -15 V
15 -5 V
15 40 V
15 57 V
15 68 V
15 61 V
14 68 V
15 27 V
15 27 V
15 -7 V
15 -22 V
15 -44 V
16 -57 V
15 -73 V
15 -59 V
15 -55 V
15 -35 V
15 -5 V
14 13 V
15 40 V
15 53 V
15 54 V
15 78 V
15 56 V
16 38 V
15 14 V
15 -6 V
15 -29 V
15 -44 V
15 -68 V
15 -84 V
14 -48 V
15 -48 V
15 -12 V
15 -22 V
15 21 V
15 49 V
16 58 V
15 62 V
15 56 V
15 53 V
15 42 V
15 9 V
14 2 V
15 -33 V
15 -71 V
15 -58 V
15 -61 V
15 -59 V
16 -37 V
15 -33 V
15 -8 V
15 35 V
15 39 V
15 51 V
15 57 V
14 73 V
15 49 V
15 43 V
15 18 V
15 -12 V
15 -40 V
16 -50 V
15 -52 V
15 -78 V
15 -57 V
15 -46 V
15 -26 V
1.000 UL
LT1
457 331 M
15 1 V
16 5 V
15 16 V
15 11 V
14 19 V
15 22 V
16 34 V
15 35 V
14 45 V
15 46 V
16 62 V
14 70 V
16 69 V
14 81 V
16 79 V
15 89 V
14 75 V
16 85 V
14 56 V
15 58 V
16 50 V
14 21 V
15 17 V
16 -20 V
14 -17 V
15 -33 V
16 -34 V
14 -36 V
15 -36 V
16 -6 V
14 -2 V
15 20 V
16 26 V
14 71 V
15 55 V
16 87 V
14 77 V
15 51 V
16 50 V
15 31 V
14 -13 V
16 -33 V
15 -50 V
14 -53 V
16 -68 V
15 -54 V
14 -5 V
16 17 V
15 40 V
14 85 V
15 79 V
15 74 V
16 90 V
15 25 V
15 24 V
14 -9 V
15 -62 V
15 -81 V
16 -93 V
15 -59 V
15 -25 V
14 4 V
15 49 V
15 105 V
15 85 V
16 98 V
15 52 V
15 54 V
14 -22 V
15 -34 V
15 -107 V
16 -95 V
15 -86 V
15 -46 V
14 3 V
15 55 V
15 76 V
16 129 V
15 98 V
15 72 V
14 30 V
15 -41 V
15 -70 V
16 -103 V
15 -98 V
15 -90 V
14 -37 V
15 14 V
15 79 V
15 107 V
16 104 V
15 122 V
15 35 V
14 8 V
15 -53 V
15 -122 V
16 -121 V
15 -101 V
15 -52 V
14 -2 V
15 53 V
15 116 V
15 89 V
15 122 V
15 61 V
16 -3 V
15 -51 V
15 -76 V
15 -118 V
15 -119 V
15 -68 V
15 -7 V
14 24 V
15 84 V
15 113 V
15 99 V
15 83 V
15 19 V
16 -24 V
15 -97 V
15 -97 V
15 -105 V
15 -94 V
15 -33 V
14 -8 V
15 45 V
15 94 V
15 71 V
15 105 V
15 55 V
16 26 V
15 -58 V
15 -66 V
15 -96 V
15 -118 V
15 -68 V
15 -43 V
14 1 V
15 12 V
15 69 V
15 91 V
15 62 V
15 52 V
16 38 V
15 -11 V
15 -53 V
15 -70 V
15 -93 V
15 -84 V
14 -78 V
15 -49 V
15 16 V
15 4 V
15 36 V
15 58 V
16 65 V
15 35 V
15 43 V
15 1 V
15 -6 V
15 -52 V
15 -58 V
14 -71 V
15 -78 V
15 -80 V
15 -52 V
15 -40 V
15 -31 V
16 19 V
15 2 V
15 25 V
15 43 V
15 47 V
15 30 V
14 23 V
15 20 V
15 -45 V
15 -6 V
15 -39 V
15 -58 V
16 -80 V
15 -63 V
15 -80 V
15 -82 V
15 -81 V
15 -76 V
15 -76 V
14 -60 V
15 -68 V
15 -49 V
15 -42 V
15 -45 V
15 -25 V
16 -24 V
15 -23 V
15 -12 V
15 -9 V
15 -8 V
15 -2 V
1.000 UL
LT2
457 323 M
15 4 V
16 5 V
15 10 V
15 12 V
14 18 V
15 24 V
16 34 V
15 30 V
14 43 V
15 52 V
16 66 V
14 64 V
16 74 V
14 83 V
16 79 V
15 93 V
14 83 V
16 81 V
14 80 V
15 50 V
16 38 V
14 40 V
15 13 V
16 -12 V
14 -24 V
15 -33 V
16 -41 V
14 -41 V
15 -40 V
16 -14 V
14 -4 V
15 20 V
16 30 V
14 71 V
15 69 V
16 90 V
14 68 V
15 70 V
16 46 V
15 18 V
14 -11 V
16 -36 V
15 -58 V
14 -58 V
16 -77 V
15 -30 V
14 -24 V
16 19 V
15 45 V
14 96 V
15 80 V
15 78 V
16 81 V
15 49 V
15 11 V
14 -33 V
15 -79 V
15 -74 V
16 -81 V
15 -66 V
15 -30 V
14 24 V
15 51 V
15 96 V
15 109 V
16 87 V
15 72 V
15 22 V
14 -27 V
15 -55 V
15 -102 V
16 -91 V
15 -77 V
15 -43 V
14 -2 V
15 68 V
15 99 V
16 104 V
15 117 V
15 54 V
14 14 V
15 -21 V
15 -95 V
16 -121 V
15 -93 V
15 -67 V
14 -27 V
15 11 V
15 88 V
15 122 V
16 105 V
15 79 V
15 46 V
14 -6 V
15 -63 V
15 -102 V
16 -112 V
15 -96 V
15 -68 V
14 -3 V
15 65 V
15 113 V
15 89 V
15 107 V
15 73 V
16 14 V
15 -51 V
15 -91 V
15 -125 V
15 -100 V
15 -81 V
15 -13 V
14 28 V
15 73 V
15 107 V
15 99 V
15 79 V
15 44 V
16 -18 V
15 -66 V
15 -111 V
15 -101 V
15 -110 V
15 -48 V
14 -10 V
15 41 V
15 78 V
15 96 V
15 95 V
15 65 V
16 20 V
15 -22 V
15 -73 V
15 -89 V
15 -105 V
15 -88 V
15 -55 V
14 -17 V
15 26 V
15 61 V
15 80 V
15 93 V
15 43 V
16 33 V
15 2 V
15 -42 V
15 -79 V
15 -69 V
15 -111 V
14 -67 V
15 -52 V
15 -24 V
15 21 V
15 41 V
15 59 V
16 63 V
15 50 V
15 49 V
15 16 V
15 -17 V
15 -52 V
15 -64 V
14 -75 V
15 -70 V
15 -90 V
15 -55 V
15 -59 V
15 -9 V
16 5 V
15 22 V
15 32 V
15 44 V
15 39 V
15 39 V
14 24 V
15 9 V
15 -14 V
15 -22 V
15 -54 V
15 -51 V
16 -86 V
15 -72 V
15 -83 V
15 -92 V
15 -87 V
15 -80 V
15 -78 V
14 -66 V
15 -57 V
15 -50 V
15 -47 V
15 -34 V
15 -28 V
16 -25 V
15 -19 V
15 -13 V
15 -8 V
15 -5 V
15 -3 V
stroke
grestore
end
showpage
}}%
\put(1950,50){\makebox(0,0){$\alpha$}}%
\put(100,1180){%
\makebox(0,0)[b]{\shortstack{$\rho(\alpha)$}}%
}%
\put(3382,200){\makebox(0,0){3}}%
\put(2905,200){\makebox(0,0){2}}%
\put(2427,200){\makebox(0,0){1}}%
\put(1950,200){\makebox(0,0){0}}%
\put(1473,200){\makebox(0,0){-1}}%
\put(995,200){\makebox(0,0){-2}}%
\put(518,200){\makebox(0,0){-3}}%
\put(400,2060){\makebox(0,0)[r]{0.25}}%
\put(400,1708){\makebox(0,0)[r]{0.2}}%
\put(400,1356){\makebox(0,0)[r]{0.15}}%
\put(400,1004){\makebox(0,0)[r]{0.1}}%
\put(400,652){\makebox(0,0)[r]{0.05}}%
\put(400,300){\makebox(0,0)[r]{0}}%
\end{picture}%
\endgroup

%% file: plot_l1_40.tex
\setlength{\unitlength}{0.240900pt}
\ifx\plotpoint\undefined\newsavebox{\plotpoint}\fi
\sbox{\plotpoint}{\rule[-0.200pt]{0.400pt}{0.400pt}}%
\begin{picture}(1650,944)(0,0)
\font\gnuplot=cmr10 at 12pt
\gnuplot
\sbox{\plotpoint}{\rule[-0.200pt]{0.400pt}{0.400pt}}%
\put(350.0,150.0){\rule[-0.200pt]{4.818pt}{0.400pt}}
\put(325,150){\makebox(0,0)[r]{\ \ {$0$}}}
\put(1555.0,150.0){\rule[-0.200pt]{4.818pt}{0.400pt}}
\put(350.0,282.0){\rule[-0.200pt]{4.818pt}{0.400pt}}
\put(325,282){\makebox(0,0)[r]{\ \ {$0.1$}}}
\put(1555.0,282.0){\rule[-0.200pt]{4.818pt}{0.400pt}}
\put(350.0,415.0){\rule[-0.200pt]{4.818pt}{0.400pt}}
\put(325,415){\makebox(0,0)[r]{\ \ {$0.2$}}}
\put(1555.0,415.0){\rule[-0.200pt]{4.818pt}{0.400pt}}
\put(350.0,547.0){\rule[-0.200pt]{4.818pt}{0.400pt}}
\put(325,547){\makebox(0,0)[r]{\ \ {$0.3$}}}
\put(1555.0,547.0){\rule[-0.200pt]{4.818pt}{0.400pt}}
\put(350.0,679.0){\rule[-0.200pt]{4.818pt}{0.400pt}}
\put(325,679){\makebox(0,0)[r]{\ \ {$0.4$}}}
\put(1555.0,679.0){\rule[-0.200pt]{4.818pt}{0.400pt}}
\put(350.0,812.0){\rule[-0.200pt]{4.818pt}{0.400pt}}
\put(325,812){\makebox(0,0)[r]{\ \ {$0.5$}}}
\put(1555.0,812.0){\rule[-0.200pt]{4.818pt}{0.400pt}}
\put(350.0,944.0){\rule[-0.200pt]{4.818pt}{0.400pt}}
\put(325,944){\makebox(0,0)[r]{\ \ {$0.6$}}}
\put(1555.0,944.0){\rule[-0.200pt]{4.818pt}{0.400pt}}
\put(350.0,150.0){\rule[-0.200pt]{0.400pt}{4.818pt}}
\put(350,100){\makebox(0,0){\ {$0$}}}
\put(350.0,924.0){\rule[-0.200pt]{0.400pt}{4.818pt}}
\put(503.0,150.0){\rule[-0.200pt]{0.400pt}{4.818pt}}
\put(503,100){\makebox(0,0){\ {$0.5$}}}
\put(503.0,924.0){\rule[-0.200pt]{0.400pt}{4.818pt}}
\put(656.0,150.0){\rule[-0.200pt]{0.400pt}{4.818pt}}
\put(656,100){\makebox(0,0){\ {$1$}}}
\put(656.0,924.0){\rule[-0.200pt]{0.400pt}{4.818pt}}
\put(809.0,150.0){\rule[-0.200pt]{0.400pt}{4.818pt}}
\put(809,100){\makebox(0,0){\ {$1.5$}}}
\put(809.0,924.0){\rule[-0.200pt]{0.400pt}{4.818pt}}
\put(963.0,150.0){\rule[-0.200pt]{0.400pt}{4.818pt}}
\put(963,100){\makebox(0,0){\ {$2$}}}
\put(963.0,924.0){\rule[-0.200pt]{0.400pt}{4.818pt}}
\put(1116.0,150.0){\rule[-0.200pt]{0.400pt}{4.818pt}}
\put(1116,100){\makebox(0,0){\ {$2.5$}}}
\put(1116.0,924.0){\rule[-0.200pt]{0.400pt}{4.818pt}}
\put(1269.0,150.0){\rule[-0.200pt]{0.400pt}{4.818pt}}
\put(1269,100){\makebox(0,0){\ {$3$}}}
\put(1269.0,924.0){\rule[-0.200pt]{0.400pt}{4.818pt}}
\put(1422.0,150.0){\rule[-0.200pt]{0.400pt}{4.818pt}}
\put(1422,100){\makebox(0,0){\ {$3.5$}}}
\put(1422.0,924.0){\rule[-0.200pt]{0.400pt}{4.818pt}}
\put(1575.0,150.0){\rule[-0.200pt]{0.400pt}{4.818pt}}
\put(1575,100){\makebox(0,0){\ {$4$}}}
\put(1575.0,924.0){\rule[-0.200pt]{0.400pt}{4.818pt}}
\put(350.0,150.0){\rule[-0.200pt]{295.102pt}{0.400pt}}
\put(1575.0,150.0){\rule[-0.200pt]{0.400pt}{191.275pt}}
\put(350.0,944.0){\rule[-0.200pt]{295.102pt}{0.400pt}}
\put(150,647){\makebox(0,0){\Large{$\langle v \rangle$}}}
\put(912,25){\makebox(0,0){\large{$\gamma$}}}
\put(350.0,150.0){\rule[-0.200pt]{0.400pt}{191.275pt}}
\put(456,161){\usebox{\plotpoint}}
\put(446.0,161.0){\rule[-0.200pt]{4.818pt}{0.400pt}}
\put(446.0,161.0){\rule[-0.200pt]{4.818pt}{0.400pt}}
\put(563,173){\usebox{\plotpoint}}
\put(553.0,173.0){\rule[-0.200pt]{4.818pt}{0.400pt}}
\put(553.0,173.0){\rule[-0.200pt]{4.818pt}{0.400pt}}
\put(669.0,182.0){\usebox{\plotpoint}}
\put(659.0,182.0){\rule[-0.200pt]{4.818pt}{0.400pt}}
\put(659.0,183.0){\rule[-0.200pt]{4.818pt}{0.400pt}}
\put(775.0,193.0){\rule[-0.200pt]{0.400pt}{0.482pt}}
\put(765.0,193.0){\rule[-0.200pt]{4.818pt}{0.400pt}}
\put(765.0,195.0){\rule[-0.200pt]{4.818pt}{0.400pt}}
\put(882.0,199.0){\rule[-0.200pt]{0.400pt}{1.204pt}}
\put(872.0,199.0){\rule[-0.200pt]{4.818pt}{0.400pt}}
\put(872.0,204.0){\rule[-0.200pt]{4.818pt}{0.400pt}}
\put(988.0,210.0){\rule[-0.200pt]{0.400pt}{1.204pt}}
\put(978.0,210.0){\rule[-0.200pt]{4.818pt}{0.400pt}}
\put(978.0,215.0){\rule[-0.200pt]{4.818pt}{0.400pt}}
\put(1094.0,226.0){\rule[-0.200pt]{0.400pt}{2.650pt}}
\put(1084.0,226.0){\rule[-0.200pt]{4.818pt}{0.400pt}}
\put(1084.0,237.0){\rule[-0.200pt]{4.818pt}{0.400pt}}
\put(1201.0,248.0){\rule[-0.200pt]{0.400pt}{5.059pt}}
\put(1191.0,248.0){\rule[-0.200pt]{4.818pt}{0.400pt}}
\put(1191.0,269.0){\rule[-0.200pt]{4.818pt}{0.400pt}}
\put(1307.0,278.0){\rule[-0.200pt]{0.400pt}{7.709pt}}
\put(1297.0,278.0){\rule[-0.200pt]{4.818pt}{0.400pt}}
\put(1297.0,310.0){\rule[-0.200pt]{4.818pt}{0.400pt}}
\put(1318.0,279.0){\rule[-0.200pt]{0.400pt}{5.059pt}}
\put(1308.0,279.0){\rule[-0.200pt]{4.818pt}{0.400pt}}
\put(1308.0,300.0){\rule[-0.200pt]{4.818pt}{0.400pt}}
\put(1328.0,312.0){\rule[-0.200pt]{0.400pt}{8.191pt}}
\put(1318.0,312.0){\rule[-0.200pt]{4.818pt}{0.400pt}}
\put(1318.0,346.0){\rule[-0.200pt]{4.818pt}{0.400pt}}
\put(1339.0,551.0){\rule[-0.200pt]{0.400pt}{12.045pt}}
\put(1329.0,551.0){\rule[-0.200pt]{4.818pt}{0.400pt}}
\put(1329.0,601.0){\rule[-0.200pt]{4.818pt}{0.400pt}}
\put(1350.0,392.0){\rule[-0.200pt]{0.400pt}{9.395pt}}
\put(1340.0,392.0){\rule[-0.200pt]{4.818pt}{0.400pt}}
\put(1340.0,431.0){\rule[-0.200pt]{4.818pt}{0.400pt}}
\put(1360.0,340.0){\rule[-0.200pt]{0.400pt}{10.359pt}}
\put(1350.0,340.0){\rule[-0.200pt]{4.818pt}{0.400pt}}
\put(1350.0,383.0){\rule[-0.200pt]{4.818pt}{0.400pt}}
\put(1371.0,541.0){\rule[-0.200pt]{0.400pt}{16.381pt}}
\put(1361.0,541.0){\rule[-0.200pt]{4.818pt}{0.400pt}}
\put(1361.0,609.0){\rule[-0.200pt]{4.818pt}{0.400pt}}
\put(1381.0,715.0){\rule[-0.200pt]{0.400pt}{4.336pt}}
\put(1371.0,715.0){\rule[-0.200pt]{4.818pt}{0.400pt}}
\put(1371.0,733.0){\rule[-0.200pt]{4.818pt}{0.400pt}}
\put(1392.0,548.0){\rule[-0.200pt]{0.400pt}{19.754pt}}
\put(1382.0,548.0){\rule[-0.200pt]{4.818pt}{0.400pt}}
\put(1382.0,630.0){\rule[-0.200pt]{4.818pt}{0.400pt}}
\put(1403.0,731.0){\rule[-0.200pt]{0.400pt}{3.613pt}}
\put(1393.0,731.0){\rule[-0.200pt]{4.818pt}{0.400pt}}
\put(1393.0,746.0){\rule[-0.200pt]{4.818pt}{0.400pt}}
\put(1413.0,745.0){\rule[-0.200pt]{0.400pt}{3.854pt}}
\put(1403.0,745.0){\rule[-0.200pt]{4.818pt}{0.400pt}}
\put(456,161){\circle*{18}}
\put(563,173){\circle*{18}}
\put(669,182){\circle*{18}}
\put(775,194){\circle*{18}}
\put(882,201){\circle*{18}}
\put(988,213){\circle*{18}}
\put(1094,231){\circle*{18}}
\put(1201,258){\circle*{18}}
\put(1307,294){\circle*{18}}
\put(1318,289){\circle*{18}}
\put(1328,329){\circle*{18}}
\put(1339,576){\circle*{18}}
\put(1350,411){\circle*{18}}
\put(1360,361){\circle*{18}}
\put(1371,575){\circle*{18}}
\put(1381,724){\circle*{18}}
\put(1392,589){\circle*{18}}
\put(1403,739){\circle*{18}}
\put(1413,753){\circle*{18}}
\put(1403.0,761.0){\rule[-0.200pt]{4.818pt}{0.400pt}}
\put(456.0,150.0){\rule[-0.200pt]{0.400pt}{8.191pt}}
\put(446.0,150.0){\rule[-0.200pt]{4.818pt}{0.400pt}}
\put(446.0,184.0){\rule[-0.200pt]{4.818pt}{0.400pt}}
\put(563.0,162.0){\rule[-0.200pt]{0.400pt}{7.468pt}}
\put(553.0,162.0){\rule[-0.200pt]{4.818pt}{0.400pt}}
\put(553.0,193.0){\rule[-0.200pt]{4.818pt}{0.400pt}}
\put(669.0,155.0){\rule[-0.200pt]{0.400pt}{6.022pt}}
\put(659.0,155.0){\rule[-0.200pt]{4.818pt}{0.400pt}}
\put(659.0,180.0){\rule[-0.200pt]{4.818pt}{0.400pt}}
\put(775.0,151.0){\rule[-0.200pt]{0.400pt}{3.613pt}}
\put(765.0,151.0){\rule[-0.200pt]{4.818pt}{0.400pt}}
\put(765.0,166.0){\rule[-0.200pt]{4.818pt}{0.400pt}}
\put(882.0,152.0){\rule[-0.200pt]{0.400pt}{2.409pt}}
\put(872.0,152.0){\rule[-0.200pt]{4.818pt}{0.400pt}}
\put(872.0,162.0){\rule[-0.200pt]{4.818pt}{0.400pt}}
\put(1094.0,155.0){\rule[-0.200pt]{0.400pt}{1.927pt}}
\put(1084.0,155.0){\rule[-0.200pt]{4.818pt}{0.400pt}}
\put(1084.0,163.0){\rule[-0.200pt]{4.818pt}{0.400pt}}
\put(1201.0,162.0){\rule[-0.200pt]{0.400pt}{2.168pt}}
\put(1191.0,162.0){\rule[-0.200pt]{4.818pt}{0.400pt}}
\put(1191.0,171.0){\rule[-0.200pt]{4.818pt}{0.400pt}}
\put(1307.0,176.0){\rule[-0.200pt]{0.400pt}{1.927pt}}
\put(1297.0,176.0){\rule[-0.200pt]{4.818pt}{0.400pt}}
\put(1297.0,184.0){\rule[-0.200pt]{4.818pt}{0.400pt}}
\put(1318.0,164.0){\rule[-0.200pt]{0.400pt}{2.650pt}}
\put(1308.0,164.0){\rule[-0.200pt]{4.818pt}{0.400pt}}
\put(1308.0,175.0){\rule[-0.200pt]{4.818pt}{0.400pt}}
\put(1328.0,170.0){\rule[-0.200pt]{0.400pt}{3.613pt}}
\put(1318.0,170.0){\rule[-0.200pt]{4.818pt}{0.400pt}}
\put(1318.0,185.0){\rule[-0.200pt]{4.818pt}{0.400pt}}
\put(1339.0,278.0){\rule[-0.200pt]{0.400pt}{7.227pt}}
\put(1329.0,278.0){\rule[-0.200pt]{4.818pt}{0.400pt}}
\put(1329.0,308.0){\rule[-0.200pt]{4.818pt}{0.400pt}}
\put(1350.0,199.0){\rule[-0.200pt]{0.400pt}{3.613pt}}
\put(1340.0,199.0){\rule[-0.200pt]{4.818pt}{0.400pt}}
\put(1340.0,214.0){\rule[-0.200pt]{4.818pt}{0.400pt}}
\put(1360.0,185.0){\rule[-0.200pt]{0.400pt}{3.613pt}}
\put(1350.0,185.0){\rule[-0.200pt]{4.818pt}{0.400pt}}
\put(1350.0,200.0){\rule[-0.200pt]{4.818pt}{0.400pt}}
\put(1371.0,261.0){\rule[-0.200pt]{0.400pt}{8.672pt}}
\put(1361.0,261.0){\rule[-0.200pt]{4.818pt}{0.400pt}}
\put(1361.0,297.0){\rule[-0.200pt]{4.818pt}{0.400pt}}
\put(1381.0,359.0){\rule[-0.200pt]{0.400pt}{2.409pt}}
\put(1371.0,359.0){\rule[-0.200pt]{4.818pt}{0.400pt}}
\put(1371.0,369.0){\rule[-0.200pt]{4.818pt}{0.400pt}}
\put(1392.0,267.0){\rule[-0.200pt]{0.400pt}{9.154pt}}
\put(1382.0,267.0){\rule[-0.200pt]{4.818pt}{0.400pt}}
\put(1382.0,305.0){\rule[-0.200pt]{4.818pt}{0.400pt}}
\put(1403.0,366.0){\rule[-0.200pt]{0.400pt}{3.132pt}}
\put(1393.0,366.0){\rule[-0.200pt]{4.818pt}{0.400pt}}
\put(1393.0,379.0){\rule[-0.200pt]{4.818pt}{0.400pt}}
\put(1413.0,369.0){\rule[-0.200pt]{0.400pt}{2.650pt}}
\put(1403.0,369.0){\rule[-0.200pt]{4.818pt}{0.400pt}}
\put(456,152){\circle{18}}
\put(563,178){\circle{18}}
\put(669,168){\circle{18}}
\put(775,158){\circle{18}}
\put(882,157){\circle{18}}
\put(1094,159){\circle{18}}
\put(1201,167){\circle{18}}
\put(1307,180){\circle{18}}
\put(1318,170){\circle{18}}
\put(1328,178){\circle{18}}
\put(1339,293){\circle{18}}
\put(1350,207){\circle{18}}
\put(1360,192){\circle{18}}
\put(1371,279){\circle{18}}
\put(1381,364){\circle{18}}
\put(1392,286){\circle{18}}
\put(1403,373){\circle{18}}
\put(1413,374){\circle{18}}
\put(1403.0,380.0){\rule[-0.200pt]{4.818pt}{0.400pt}}
\end{picture}

%% file: plot_histl1b.tex
\setlength{\unitlength}{0.240900pt}
\ifx\plotpoint\undefined\newsavebox{\plotpoint}\fi
\sbox{\plotpoint}{\rule[-0.200pt]{0.400pt}{0.400pt}}%
\begin{picture}(1650,944)(0,0)
\font\gnuplot=cmr10 at 12pt
\gnuplot
\sbox{\plotpoint}{\rule[-0.200pt]{0.400pt}{0.400pt}}%
\put(350.0,150.0){\rule[-0.200pt]{4.818pt}{0.400pt}}
\put(325,150){\makebox(0,0)[r]{\ \ {$0$}}}
\put(1555.0,150.0){\rule[-0.200pt]{4.818pt}{0.400pt}}
\put(350.0,349.0){\rule[-0.200pt]{4.818pt}{0.400pt}}
\put(325,349){\makebox(0,0)[r]{\ \ {$100$}}}
\put(1555.0,349.0){\rule[-0.200pt]{4.818pt}{0.400pt}}
\put(350.0,547.0){\rule[-0.200pt]{4.818pt}{0.400pt}}
\put(325,547){\makebox(0,0)[r]{\ \ {$200$}}}
\put(1555.0,547.0){\rule[-0.200pt]{4.818pt}{0.400pt}}
\put(350.0,746.0){\rule[-0.200pt]{4.818pt}{0.400pt}}
\put(325,746){\makebox(0,0)[r]{\ \ {$300$}}}
\put(1555.0,746.0){\rule[-0.200pt]{4.818pt}{0.400pt}}
\put(350.0,944.0){\rule[-0.200pt]{4.818pt}{0.400pt}}
\put(325,944){\makebox(0,0)[r]{\ \ {$400$}}}
\put(1555.0,944.0){\rule[-0.200pt]{4.818pt}{0.400pt}}
\put(350.0,150.0){\rule[-0.200pt]{0.400pt}{4.818pt}}
\put(350,100){\makebox(0,0){\ {$0$}}}
\put(350.0,924.0){\rule[-0.200pt]{0.400pt}{4.818pt}}
\put(573.0,150.0){\rule[-0.200pt]{0.400pt}{4.818pt}}
\put(573,100){\makebox(0,0){\ {$0.1$}}}
\put(573.0,924.0){\rule[-0.200pt]{0.400pt}{4.818pt}}
\put(795.0,150.0){\rule[-0.200pt]{0.400pt}{4.818pt}}
\put(795,100){\makebox(0,0){\ {$0.2$}}}
\put(795.0,924.0){\rule[-0.200pt]{0.400pt}{4.818pt}}
\put(1018.0,150.0){\rule[-0.200pt]{0.400pt}{4.818pt}}
\put(1018,100){\makebox(0,0){\ {$0.3$}}}
\put(1018.0,924.0){\rule[-0.200pt]{0.400pt}{4.818pt}}
\put(1241.0,150.0){\rule[-0.200pt]{0.400pt}{4.818pt}}
\put(1241,100){\makebox(0,0){\ {$0.4$}}}
\put(1241.0,924.0){\rule[-0.200pt]{0.400pt}{4.818pt}}
\put(1464.0,150.0){\rule[-0.200pt]{0.400pt}{4.818pt}}
\put(1464,100){\makebox(0,0){\ {$0.5$}}}
\put(1464.0,924.0){\rule[-0.200pt]{0.400pt}{4.818pt}}
\put(350.0,150.0){\rule[-0.200pt]{295.102pt}{0.400pt}}
\put(1575.0,150.0){\rule[-0.200pt]{0.400pt}{191.275pt}}
\put(350.0,944.0){\rule[-0.200pt]{295.102pt}{0.400pt}}
\put(150,647){\makebox(0,0){\Large{N}}}
\put(912,25){\makebox(0,0){\large{$|{\bar l}_1|$}}}
\put(350.0,150.0){\rule[-0.200pt]{0.400pt}{191.275pt}}
\put(362,237){\circle*{18}}
\put(386,327){\circle*{18}}
\put(409,501){\circle*{18}}
\put(432,597){\circle*{18}}
\put(455,668){\circle*{18}}
\put(478,537){\circle*{18}}
\put(501,666){\circle*{18}}
\put(524,652){\circle*{18}}
\put(548,335){\circle*{18}}
\put(571,426){\circle*{18}}
\put(594,335){\circle*{18}}
\put(617,273){\circle*{18}}
\put(640,200){\circle*{18}}
\put(663,178){\circle*{18}}
\put(687,267){\circle*{18}}
\put(710,216){\circle*{18}}
\put(733,214){\circle*{18}}
\put(756,249){\circle*{18}}
\put(779,305){\circle*{18}}
\put(802,186){\circle*{18}}
\put(826,206){\circle*{18}}
\put(849,267){\circle*{18}}
\put(872,277){\circle*{18}}
\put(895,337){\circle*{18}}
\put(918,313){\circle*{18}}
\put(941,275){\circle*{18}}
\put(964,200){\circle*{18}}
\put(988,307){\circle*{18}}
\put(1011,329){\circle*{18}}
\put(1034,319){\circle*{18}}
\put(1057,269){\circle*{18}}
\put(1080,315){\circle*{18}}
\put(1103,227){\circle*{18}}
\put(1127,233){\circle*{18}}
\put(1150,184){\circle*{18}}
\put(1173,239){\circle*{18}}
\put(1196,303){\circle*{18}}
\put(1219,456){\circle*{18}}
\put(1242,589){\circle*{18}}
\put(1265,599){\circle*{18}}
\put(1289,801){\circle*{18}}
\put(1312,694){\circle*{18}}
\put(1335,487){\circle*{18}}
\put(1358,472){\circle*{18}}
\put(1381,223){\circle*{18}}
\put(1404,233){\circle*{18}}
\put(1428,243){\circle*{18}}
\put(1451,190){\circle*{18}}
\put(1474,225){\circle*{18}}
\put(1497,206){\circle*{18}}
\put(1520,152){\circle*{18}}
\end{picture}

%% file: plot_a1V.tex
\setlength{\unitlength}{0.240900pt}
\ifx\plotpoint\undefined\newsavebox{\plotpoint}\fi
\sbox{\plotpoint}{\rule[-0.200pt]{0.400pt}{0.400pt}}%
\begin{picture}(1650,944)(0,0)
\font\gnuplot=cmr10 at 12pt
\gnuplot
\sbox{\plotpoint}{\rule[-0.200pt]{0.400pt}{0.400pt}}%
\put(350.0,150.0){\rule[-0.200pt]{4.818pt}{0.400pt}}
\put(325,150){\makebox(0,0)[r]{\ \ {$0$}}}
\put(1555.0,150.0){\rule[-0.200pt]{4.818pt}{0.400pt}}
\put(350.0,282.0){\rule[-0.200pt]{4.818pt}{0.400pt}}
\put(325,282){\makebox(0,0)[r]{\ \ {$0.1$}}}
\put(1555.0,282.0){\rule[-0.200pt]{4.818pt}{0.400pt}}
\put(350.0,415.0){\rule[-0.200pt]{4.818pt}{0.400pt}}
\put(325,415){\makebox(0,0)[r]{\ \ {$0.2$}}}
\put(1555.0,415.0){\rule[-0.200pt]{4.818pt}{0.400pt}}
\put(350.0,547.0){\rule[-0.200pt]{4.818pt}{0.400pt}}
\put(325,547){\makebox(0,0)[r]{\ \ {$0.3$}}}
\put(1555.0,547.0){\rule[-0.200pt]{4.818pt}{0.400pt}}
\put(350.0,679.0){\rule[-0.200pt]{4.818pt}{0.400pt}}
\put(325,679){\makebox(0,0)[r]{\ \ {$0.4$}}}
\put(1555.0,679.0){\rule[-0.200pt]{4.818pt}{0.400pt}}
\put(350.0,812.0){\rule[-0.200pt]{4.818pt}{0.400pt}}
\put(325,812){\makebox(0,0)[r]{\ \ {$0.5$}}}
\put(1555.0,812.0){\rule[-0.200pt]{4.818pt}{0.400pt}}
\put(350.0,944.0){\rule[-0.200pt]{4.818pt}{0.400pt}}
\put(325,944){\makebox(0,0)[r]{\ \ {$0.6$}}}
\put(1555.0,944.0){\rule[-0.200pt]{4.818pt}{0.400pt}}
\put(510.0,150.0){\rule[-0.200pt]{0.400pt}{4.818pt}}
\put(510,100){\makebox(0,0){\ {$2.5$}}}
\put(510.0,924.0){\rule[-0.200pt]{0.400pt}{4.818pt}}
\put(776.0,150.0){\rule[-0.200pt]{0.400pt}{4.818pt}}
\put(776,100){\makebox(0,0){\ {$3$}}}
\put(776.0,924.0){\rule[-0.200pt]{0.400pt}{4.818pt}}
\put(1042.0,150.0){\rule[-0.200pt]{0.400pt}{4.818pt}}
\put(1042,100){\makebox(0,0){\ {$3.5$}}}
\put(1042.0,924.0){\rule[-0.200pt]{0.400pt}{4.818pt}}
\put(1309.0,150.0){\rule[-0.200pt]{0.400pt}{4.818pt}}
\put(1309,100){\makebox(0,0){\ {$4$}}}
\put(1309.0,924.0){\rule[-0.200pt]{0.400pt}{4.818pt}}
\put(1575.0,150.0){\rule[-0.200pt]{0.400pt}{4.818pt}}
\put(1575,100){\makebox(0,0){\ {$4.5$}}}
\put(1575.0,924.0){\rule[-0.200pt]{0.400pt}{4.818pt}}
\put(350.0,150.0){\rule[-0.200pt]{295.102pt}{0.400pt}}
\put(1575.0,150.0){\rule[-0.200pt]{0.400pt}{191.275pt}}
\put(350.0,944.0){\rule[-0.200pt]{295.102pt}{0.400pt}}
\put(150,647){\makebox(0,0){\Large{$\langle \delta u \rangle$}}}
\put(912,25){\makebox(0,0){\large{$\gamma$}}}
\put(350.0,150.0){\rule[-0.200pt]{0.400pt}{191.275pt}}
\put(750.0,179.0){\rule[-0.200pt]{0.400pt}{3.854pt}}
\put(740.0,179.0){\rule[-0.200pt]{4.818pt}{0.400pt}}
\put(740.0,195.0){\rule[-0.200pt]{4.818pt}{0.400pt}}
\put(843.0,188.0){\rule[-0.200pt]{0.400pt}{3.854pt}}
\put(833.0,188.0){\rule[-0.200pt]{4.818pt}{0.400pt}}
\put(833.0,204.0){\rule[-0.200pt]{4.818pt}{0.400pt}}
\put(861.0,195.0){\rule[-0.200pt]{0.400pt}{4.577pt}}
\put(851.0,195.0){\rule[-0.200pt]{4.818pt}{0.400pt}}
\put(851.0,214.0){\rule[-0.200pt]{4.818pt}{0.400pt}}
\put(879.0,194.0){\rule[-0.200pt]{0.400pt}{5.059pt}}
\put(869.0,194.0){\rule[-0.200pt]{4.818pt}{0.400pt}}
\put(869.0,215.0){\rule[-0.200pt]{4.818pt}{0.400pt}}
\put(898.0,196.0){\rule[-0.200pt]{0.400pt}{4.577pt}}
\put(888.0,196.0){\rule[-0.200pt]{4.818pt}{0.400pt}}
\put(888.0,215.0){\rule[-0.200pt]{4.818pt}{0.400pt}}
\put(917.0,190.0){\rule[-0.200pt]{0.400pt}{5.059pt}}
\put(907.0,190.0){\rule[-0.200pt]{4.818pt}{0.400pt}}
\put(907.0,211.0){\rule[-0.200pt]{4.818pt}{0.400pt}}
\put(935.0,511.0){\rule[-0.200pt]{0.400pt}{7.709pt}}
\put(925.0,511.0){\rule[-0.200pt]{4.818pt}{0.400pt}}
\put(925.0,543.0){\rule[-0.200pt]{4.818pt}{0.400pt}}
\put(953.0,278.0){\rule[-0.200pt]{0.400pt}{15.418pt}}
\put(943.0,278.0){\rule[-0.200pt]{4.818pt}{0.400pt}}
\put(943.0,342.0){\rule[-0.200pt]{4.818pt}{0.400pt}}
\put(972.0,314.0){\rule[-0.200pt]{0.400pt}{18.549pt}}
\put(962.0,314.0){\rule[-0.200pt]{4.818pt}{0.400pt}}
\put(962.0,391.0){\rule[-0.200pt]{4.818pt}{0.400pt}}
\put(991.0,576.0){\rule[-0.200pt]{0.400pt}{4.577pt}}
\put(981.0,576.0){\rule[-0.200pt]{4.818pt}{0.400pt}}
\put(981.0,595.0){\rule[-0.200pt]{4.818pt}{0.400pt}}
\put(1009.0,387.0){\rule[-0.200pt]{0.400pt}{17.104pt}}
\put(999.0,387.0){\rule[-0.200pt]{4.818pt}{0.400pt}}
\put(999.0,458.0){\rule[-0.200pt]{4.818pt}{0.400pt}}
\put(1027.0,597.0){\rule[-0.200pt]{0.400pt}{4.577pt}}
\put(1017.0,597.0){\rule[-0.200pt]{4.818pt}{0.400pt}}
\put(750,187){\circle*{18}}
\put(843,196){\circle*{18}}
\put(861,204){\circle*{18}}
\put(879,204){\circle*{18}}
\put(898,206){\circle*{18}}
\put(917,200){\circle*{18}}
\put(935,527){\circle*{18}}
\put(953,310){\circle*{18}}
\put(972,352){\circle*{18}}
\put(991,585){\circle*{18}}
\put(1009,423){\circle*{18}}
\put(1027,607){\circle*{18}}
\put(1017.0,616.0){\rule[-0.200pt]{4.818pt}{0.400pt}}
\put(473.0,159.0){\rule[-0.200pt]{0.400pt}{4.577pt}}
\put(463.0,159.0){\rule[-0.200pt]{4.818pt}{0.400pt}}
\put(463.0,178.0){\rule[-0.200pt]{4.818pt}{0.400pt}}
\put(658.0,174.0){\rule[-0.200pt]{0.400pt}{4.336pt}}
\put(648.0,174.0){\rule[-0.200pt]{4.818pt}{0.400pt}}
\put(648.0,192.0){\rule[-0.200pt]{4.818pt}{0.400pt}}
\put(843.0,202.0){\rule[-0.200pt]{0.400pt}{4.336pt}}
\put(833.0,202.0){\rule[-0.200pt]{4.818pt}{0.400pt}}
\put(833.0,220.0){\rule[-0.200pt]{4.818pt}{0.400pt}}
\put(861.0,176.0){\rule[-0.200pt]{0.400pt}{5.782pt}}
\put(851.0,176.0){\rule[-0.200pt]{4.818pt}{0.400pt}}
\put(851.0,200.0){\rule[-0.200pt]{4.818pt}{0.400pt}}
\put(879.0,190.0){\rule[-0.200pt]{0.400pt}{7.468pt}}
\put(869.0,190.0){\rule[-0.200pt]{4.818pt}{0.400pt}}
\put(869.0,221.0){\rule[-0.200pt]{4.818pt}{0.400pt}}
\put(898.0,407.0){\rule[-0.200pt]{0.400pt}{13.972pt}}
\put(888.0,407.0){\rule[-0.200pt]{4.818pt}{0.400pt}}
\put(888.0,465.0){\rule[-0.200pt]{4.818pt}{0.400pt}}
\put(917.0,249.0){\rule[-0.200pt]{0.400pt}{6.986pt}}
\put(907.0,249.0){\rule[-0.200pt]{4.818pt}{0.400pt}}
\put(907.0,278.0){\rule[-0.200pt]{4.818pt}{0.400pt}}
\put(935.0,220.0){\rule[-0.200pt]{0.400pt}{6.986pt}}
\put(925.0,220.0){\rule[-0.200pt]{4.818pt}{0.400pt}}
\put(925.0,249.0){\rule[-0.200pt]{4.818pt}{0.400pt}}
\put(953.0,372.0){\rule[-0.200pt]{0.400pt}{17.345pt}}
\put(943.0,372.0){\rule[-0.200pt]{4.818pt}{0.400pt}}
\put(943.0,444.0){\rule[-0.200pt]{4.818pt}{0.400pt}}
\put(972.0,569.0){\rule[-0.200pt]{0.400pt}{4.577pt}}
\put(962.0,569.0){\rule[-0.200pt]{4.818pt}{0.400pt}}
\put(962.0,588.0){\rule[-0.200pt]{4.818pt}{0.400pt}}
\put(991.0,383.0){\rule[-0.200pt]{0.400pt}{18.549pt}}
\put(981.0,383.0){\rule[-0.200pt]{4.818pt}{0.400pt}}
\put(981.0,460.0){\rule[-0.200pt]{4.818pt}{0.400pt}}
\put(1009.0,581.0){\rule[-0.200pt]{0.400pt}{6.504pt}}
\put(999.0,581.0){\rule[-0.200pt]{4.818pt}{0.400pt}}
\put(999.0,608.0){\rule[-0.200pt]{4.818pt}{0.400pt}}
\put(1027.0,588.0){\rule[-0.200pt]{0.400pt}{5.782pt}}
\put(1017.0,588.0){\rule[-0.200pt]{4.818pt}{0.400pt}}
\put(473,169){\circle{18}}
\put(658,183){\circle{18}}
\put(843,211){\circle{18}}
\put(861,188){\circle{18}}
\put(879,206){\circle{18}}
\put(898,436){\circle{18}}
\put(917,264){\circle{18}}
\put(935,235){\circle{18}}
\put(953,408){\circle{18}}
\put(972,579){\circle{18}}
\put(991,421){\circle{18}}
\put(1009,595){\circle{18}}
\put(1027,600){\circle{18}}
\put(1017.0,612.0){\rule[-0.200pt]{4.818pt}{0.400pt}}
\put(750.0,244.0){\rule[-0.200pt]{0.400pt}{12.768pt}}
\put(740.0,244.0){\rule[-0.200pt]{4.818pt}{0.400pt}}
\put(740.0,297.0){\rule[-0.200pt]{4.818pt}{0.400pt}}
\put(843.0,322.0){\rule[-0.200pt]{0.400pt}{15.418pt}}
\put(833.0,322.0){\rule[-0.200pt]{4.818pt}{0.400pt}}
\put(833.0,386.0){\rule[-0.200pt]{4.818pt}{0.400pt}}
\put(861.0,313.0){\rule[-0.200pt]{0.400pt}{13.249pt}}
\put(851.0,313.0){\rule[-0.200pt]{4.818pt}{0.400pt}}
\put(851.0,368.0){\rule[-0.200pt]{4.818pt}{0.400pt}}
\put(879.0,386.0){\rule[-0.200pt]{0.400pt}{14.454pt}}
\put(869.0,386.0){\rule[-0.200pt]{4.818pt}{0.400pt}}
\put(869.0,446.0){\rule[-0.200pt]{4.818pt}{0.400pt}}
\put(898.0,358.0){\rule[-0.200pt]{0.400pt}{15.899pt}}
\put(888.0,358.0){\rule[-0.200pt]{4.818pt}{0.400pt}}
\put(888.0,424.0){\rule[-0.200pt]{4.818pt}{0.400pt}}
\put(917.0,429.0){\rule[-0.200pt]{0.400pt}{12.768pt}}
\put(907.0,429.0){\rule[-0.200pt]{4.818pt}{0.400pt}}
\put(907.0,482.0){\rule[-0.200pt]{4.818pt}{0.400pt}}
\put(935.0,437.0){\rule[-0.200pt]{0.400pt}{12.768pt}}
\put(925.0,437.0){\rule[-0.200pt]{4.818pt}{0.400pt}}
\put(925.0,490.0){\rule[-0.200pt]{4.818pt}{0.400pt}}
\put(953.0,391.0){\rule[-0.200pt]{0.400pt}{13.249pt}}
\put(943.0,391.0){\rule[-0.200pt]{4.818pt}{0.400pt}}
\put(943.0,446.0){\rule[-0.200pt]{4.818pt}{0.400pt}}
\put(972.0,415.0){\rule[-0.200pt]{0.400pt}{15.177pt}}
\put(962.0,415.0){\rule[-0.200pt]{4.818pt}{0.400pt}}
\put(962.0,478.0){\rule[-0.200pt]{4.818pt}{0.400pt}}
\put(991.0,461.0){\rule[-0.200pt]{0.400pt}{13.972pt}}
\put(981.0,461.0){\rule[-0.200pt]{4.818pt}{0.400pt}}
\put(981.0,519.0){\rule[-0.200pt]{4.818pt}{0.400pt}}
\put(1027.0,448.0){\rule[-0.200pt]{0.400pt}{10.840pt}}
\put(1017.0,448.0){\rule[-0.200pt]{4.818pt}{0.400pt}}
\put(1017.0,493.0){\rule[-0.200pt]{4.818pt}{0.400pt}}
\put(1120.0,532.0){\rule[-0.200pt]{0.400pt}{10.359pt}}
\put(1110.0,532.0){\rule[-0.200pt]{4.818pt}{0.400pt}}
\put(1110.0,575.0){\rule[-0.200pt]{4.818pt}{0.400pt}}
\put(1212.0,641.0){\rule[-0.200pt]{0.400pt}{5.059pt}}
\put(1202.0,641.0){\rule[-0.200pt]{4.818pt}{0.400pt}}
\put(1202.0,662.0){\rule[-0.200pt]{4.818pt}{0.400pt}}
\put(1305.0,644.0){\rule[-0.200pt]{0.400pt}{10.118pt}}
\put(1295.0,644.0){\rule[-0.200pt]{4.818pt}{0.400pt}}
\put(1295.0,686.0){\rule[-0.200pt]{4.818pt}{0.400pt}}
\put(1398.0,670.0){\rule[-0.200pt]{0.400pt}{7.709pt}}
\put(1388.0,670.0){\rule[-0.200pt]{4.818pt}{0.400pt}}
\put(750,270){\makebox(0,0){$+$}}
\put(843,354){\makebox(0,0){$+$}}
\put(861,341){\makebox(0,0){$+$}}
\put(879,416){\makebox(0,0){$+$}}
\put(898,391){\makebox(0,0){$+$}}
\put(917,456){\makebox(0,0){$+$}}
\put(935,464){\makebox(0,0){$+$}}
\put(953,419){\makebox(0,0){$+$}}
\put(972,446){\makebox(0,0){$+$}}
\put(991,490){\makebox(0,0){$+$}}
\put(1027,470){\makebox(0,0){$+$}}
\put(1120,554){\makebox(0,0){$+$}}
\put(1212,652){\makebox(0,0){$+$}}
\put(1305,665){\makebox(0,0){$+$}}
\put(1398,686){\makebox(0,0){$+$}}
\put(1388.0,702.0){\rule[-0.200pt]{4.818pt}{0.400pt}}
\end{picture}

%% file: plot_l1.tex
\setlength{\unitlength}{0.240900pt}
\ifx\plotpoint\undefined\newsavebox{\plotpoint}\fi
\sbox{\plotpoint}{\rule[-0.200pt]{0.400pt}{0.400pt}}%
\begin{picture}(1650,944)(0,0)
\font\gnuplot=cmr10 at 12pt
\gnuplot
\sbox{\plotpoint}{\rule[-0.200pt]{0.400pt}{0.400pt}}%
\put(350.0,150.0){\rule[-0.200pt]{4.818pt}{0.400pt}}
\put(325,150){\makebox(0,0)[r]{\ \ {$0$}}}
\put(1555.0,150.0){\rule[-0.200pt]{4.818pt}{0.400pt}}
\put(350.0,272.0){\rule[-0.200pt]{4.818pt}{0.400pt}}
\put(325,272){\makebox(0,0)[r]{\ \ {$0.1$}}}
\put(1555.0,272.0){\rule[-0.200pt]{4.818pt}{0.400pt}}
\put(350.0,394.0){\rule[-0.200pt]{4.818pt}{0.400pt}}
\put(325,394){\makebox(0,0)[r]{\ \ {$0.2$}}}
\put(1555.0,394.0){\rule[-0.200pt]{4.818pt}{0.400pt}}
\put(350.0,516.0){\rule[-0.200pt]{4.818pt}{0.400pt}}
\put(325,516){\makebox(0,0)[r]{\ \ {$0.3$}}}
\put(1555.0,516.0){\rule[-0.200pt]{4.818pt}{0.400pt}}
\put(350.0,639.0){\rule[-0.200pt]{4.818pt}{0.400pt}}
\put(325,639){\makebox(0,0)[r]{\ \ {$0.4$}}}
\put(1555.0,639.0){\rule[-0.200pt]{4.818pt}{0.400pt}}
\put(350.0,761.0){\rule[-0.200pt]{4.818pt}{0.400pt}}
\put(325,761){\makebox(0,0)[r]{\ \ {$0.5$}}}
\put(1555.0,761.0){\rule[-0.200pt]{4.818pt}{0.400pt}}
\put(350.0,883.0){\rule[-0.200pt]{4.818pt}{0.400pt}}
\put(325,883){\makebox(0,0)[r]{\ \ {$0.6$}}}
\put(1555.0,883.0){\rule[-0.200pt]{4.818pt}{0.400pt}}
\put(491.0,150.0){\rule[-0.200pt]{0.400pt}{4.818pt}}
\put(491,100){\makebox(0,0){\ {$2.5$}}}
\put(491.0,924.0){\rule[-0.200pt]{0.400pt}{4.818pt}}
\put(727.0,150.0){\rule[-0.200pt]{0.400pt}{4.818pt}}
\put(727,100){\makebox(0,0){\ {$3$}}}
\put(727.0,924.0){\rule[-0.200pt]{0.400pt}{4.818pt}}
\put(963.0,150.0){\rule[-0.200pt]{0.400pt}{4.818pt}}
\put(963,100){\makebox(0,0){\ {$3.5$}}}
\put(963.0,924.0){\rule[-0.200pt]{0.400pt}{4.818pt}}
\put(1198.0,150.0){\rule[-0.200pt]{0.400pt}{4.818pt}}
\put(1198,100){\makebox(0,0){\ {$4$}}}
\put(1198.0,924.0){\rule[-0.200pt]{0.400pt}{4.818pt}}
\put(1434.0,150.0){\rule[-0.200pt]{0.400pt}{4.818pt}}
\put(1434,100){\makebox(0,0){\ {$4.5$}}}
\put(1434.0,924.0){\rule[-0.200pt]{0.400pt}{4.818pt}}
\put(350.0,150.0){\rule[-0.200pt]{295.102pt}{0.400pt}}
\put(1575.0,150.0){\rule[-0.200pt]{0.400pt}{191.275pt}}
\put(350.0,944.0){\rule[-0.200pt]{295.102pt}{0.400pt}}
\put(150,647){\makebox(0,0){\Large{$\langle|{\bar l}_1|\rangle$}}}
\put(912,25){\makebox(0,0){\large{$\gamma$}}}
\put(350.0,150.0){\rule[-0.200pt]{0.400pt}{191.275pt}}
\put(524.0,475.0){\rule[-0.200pt]{0.400pt}{6.504pt}}
\put(514.0,475.0){\rule[-0.200pt]{4.818pt}{0.400pt}}
\put(514.0,502.0){\rule[-0.200pt]{4.818pt}{0.400pt}}
\put(622.0,483.0){\rule[-0.200pt]{0.400pt}{6.745pt}}
\put(612.0,483.0){\rule[-0.200pt]{4.818pt}{0.400pt}}
\put(612.0,511.0){\rule[-0.200pt]{4.818pt}{0.400pt}}
\put(720.0,553.0){\rule[-0.200pt]{0.400pt}{8.913pt}}
\put(710.0,553.0){\rule[-0.200pt]{4.818pt}{0.400pt}}
\put(710.0,590.0){\rule[-0.200pt]{4.818pt}{0.400pt}}
\put(753.0,576.0){\rule[-0.200pt]{0.400pt}{7.950pt}}
\put(743.0,576.0){\rule[-0.200pt]{4.818pt}{0.400pt}}
\put(743.0,609.0){\rule[-0.200pt]{4.818pt}{0.400pt}}
\put(786.0,573.0){\rule[-0.200pt]{0.400pt}{7.950pt}}
\put(776.0,573.0){\rule[-0.200pt]{4.818pt}{0.400pt}}
\put(776.0,606.0){\rule[-0.200pt]{4.818pt}{0.400pt}}
\put(818.0,593.0){\rule[-0.200pt]{0.400pt}{8.672pt}}
\put(808.0,593.0){\rule[-0.200pt]{4.818pt}{0.400pt}}
\put(808.0,629.0){\rule[-0.200pt]{4.818pt}{0.400pt}}
\put(851.0,611.0){\rule[-0.200pt]{0.400pt}{8.913pt}}
\put(841.0,611.0){\rule[-0.200pt]{4.818pt}{0.400pt}}
\put(841.0,648.0){\rule[-0.200pt]{4.818pt}{0.400pt}}
\put(884.0,615.0){\rule[-0.200pt]{0.400pt}{8.672pt}}
\put(874.0,615.0){\rule[-0.200pt]{4.818pt}{0.400pt}}
\put(874.0,651.0){\rule[-0.200pt]{4.818pt}{0.400pt}}
\put(917.0,628.0){\rule[-0.200pt]{0.400pt}{9.395pt}}
\put(907.0,628.0){\rule[-0.200pt]{4.818pt}{0.400pt}}
\put(907.0,667.0){\rule[-0.200pt]{4.818pt}{0.400pt}}
\put(949.0,615.0){\rule[-0.200pt]{0.400pt}{9.636pt}}
\put(939.0,615.0){\rule[-0.200pt]{4.818pt}{0.400pt}}
\put(939.0,655.0){\rule[-0.200pt]{4.818pt}{0.400pt}}
\put(1113.0,713.0){\rule[-0.200pt]{0.400pt}{10.600pt}}
\put(1103.0,713.0){\rule[-0.200pt]{4.818pt}{0.400pt}}
\put(1103.0,757.0){\rule[-0.200pt]{4.818pt}{0.400pt}}
\put(1277.0,780.0){\rule[-0.200pt]{0.400pt}{6.263pt}}
\put(1267.0,780.0){\rule[-0.200pt]{4.818pt}{0.400pt}}
\put(1267.0,806.0){\rule[-0.200pt]{4.818pt}{0.400pt}}
\put(1440.0,775.0){\rule[-0.200pt]{0.400pt}{12.286pt}}
\put(1430.0,775.0){\rule[-0.200pt]{4.818pt}{0.400pt}}
\put(524,489){\makebox(0,0){$+$}}
\put(622,497){\makebox(0,0){$+$}}
\put(720,572){\makebox(0,0){$+$}}
\put(753,592){\makebox(0,0){$+$}}
\put(786,590){\makebox(0,0){$+$}}
\put(818,611){\makebox(0,0){$+$}}
\put(851,630){\makebox(0,0){$+$}}
\put(884,633){\makebox(0,0){$+$}}
\put(917,647){\makebox(0,0){$+$}}
\put(949,635){\makebox(0,0){$+$}}
\put(1113,735){\makebox(0,0){$+$}}
\put(1277,793){\makebox(0,0){$+$}}
\put(1440,800){\makebox(0,0){$+$}}
\put(1430.0,826.0){\rule[-0.200pt]{4.818pt}{0.400pt}}
\put(459.0,313.0){\rule[-0.200pt]{0.400pt}{4.336pt}}
\put(449.0,313.0){\rule[-0.200pt]{4.818pt}{0.400pt}}
\put(449.0,331.0){\rule[-0.200pt]{4.818pt}{0.400pt}}
\put(704.0,397.0){\rule[-0.200pt]{0.400pt}{9.877pt}}
\put(694.0,397.0){\rule[-0.200pt]{4.818pt}{0.400pt}}
\put(694.0,438.0){\rule[-0.200pt]{4.818pt}{0.400pt}}
\put(786.0,457.0){\rule[-0.200pt]{0.400pt}{12.045pt}}
\put(776.0,457.0){\rule[-0.200pt]{4.818pt}{0.400pt}}
\put(776.0,507.0){\rule[-0.200pt]{4.818pt}{0.400pt}}
\put(802.0,466.0){\rule[-0.200pt]{0.400pt}{10.118pt}}
\put(792.0,466.0){\rule[-0.200pt]{4.818pt}{0.400pt}}
\put(792.0,508.0){\rule[-0.200pt]{4.818pt}{0.400pt}}
\put(818.0,525.0){\rule[-0.200pt]{0.400pt}{11.322pt}}
\put(808.0,525.0){\rule[-0.200pt]{4.818pt}{0.400pt}}
\put(808.0,572.0){\rule[-0.200pt]{4.818pt}{0.400pt}}
\put(835.0,504.0){\rule[-0.200pt]{0.400pt}{12.768pt}}
\put(825.0,504.0){\rule[-0.200pt]{4.818pt}{0.400pt}}
\put(825.0,557.0){\rule[-0.200pt]{4.818pt}{0.400pt}}
\put(851.0,556.0){\rule[-0.200pt]{0.400pt}{10.600pt}}
\put(841.0,556.0){\rule[-0.200pt]{4.818pt}{0.400pt}}
\put(841.0,600.0){\rule[-0.200pt]{4.818pt}{0.400pt}}
\put(868.0,569.0){\rule[-0.200pt]{0.400pt}{8.431pt}}
\put(858.0,569.0){\rule[-0.200pt]{4.818pt}{0.400pt}}
\put(858.0,604.0){\rule[-0.200pt]{4.818pt}{0.400pt}}
\put(884.0,508.0){\rule[-0.200pt]{0.400pt}{12.286pt}}
\put(874.0,508.0){\rule[-0.200pt]{4.818pt}{0.400pt}}
\put(874.0,559.0){\rule[-0.200pt]{4.818pt}{0.400pt}}
\put(900.0,541.0){\rule[-0.200pt]{0.400pt}{12.045pt}}
\put(890.0,541.0){\rule[-0.200pt]{4.818pt}{0.400pt}}
\put(890.0,591.0){\rule[-0.200pt]{4.818pt}{0.400pt}}
\put(917.0,593.0){\rule[-0.200pt]{0.400pt}{9.877pt}}
\put(907.0,593.0){\rule[-0.200pt]{4.818pt}{0.400pt}}
\put(907.0,634.0){\rule[-0.200pt]{4.818pt}{0.400pt}}
\put(949.0,569.0){\rule[-0.200pt]{0.400pt}{12.286pt}}
\put(939.0,569.0){\rule[-0.200pt]{4.818pt}{0.400pt}}
\put(939.0,620.0){\rule[-0.200pt]{4.818pt}{0.400pt}}
\put(1031.0,645.0){\rule[-0.200pt]{0.400pt}{10.840pt}}
\put(1021.0,645.0){\rule[-0.200pt]{4.818pt}{0.400pt}}
\put(1021.0,690.0){\rule[-0.200pt]{4.818pt}{0.400pt}}
\put(1113.0,761.0){\rule[-0.200pt]{0.400pt}{4.818pt}}
\put(1103.0,761.0){\rule[-0.200pt]{4.818pt}{0.400pt}}
\put(1103.0,781.0){\rule[-0.200pt]{4.818pt}{0.400pt}}
\put(1195.0,763.0){\rule[-0.200pt]{0.400pt}{7.950pt}}
\put(1185.0,763.0){\rule[-0.200pt]{4.818pt}{0.400pt}}
\put(1185.0,796.0){\rule[-0.200pt]{4.818pt}{0.400pt}}
\put(1277.0,801.0){\rule[-0.200pt]{0.400pt}{6.504pt}}
\put(1267.0,801.0){\rule[-0.200pt]{4.818pt}{0.400pt}}
\put(459,322){\circle{18}}
\put(704,418){\circle{18}}
\put(786,482){\circle{18}}
\put(802,487){\circle{18}}
\put(818,549){\circle{18}}
\put(835,531){\circle{18}}
\put(851,578){\circle{18}}
\put(868,586){\circle{18}}
\put(884,534){\circle{18}}
\put(900,566){\circle{18}}
\put(917,614){\circle{18}}
\put(949,595){\circle{18}}
\put(1031,667){\circle{18}}
\put(1113,771){\circle{18}}
\put(1195,779){\circle{18}}
\put(1277,815){\circle{18}}
\put(1267.0,828.0){\rule[-0.200pt]{4.818pt}{0.400pt}}
\put(704.0,256.0){\rule[-0.200pt]{0.400pt}{6.263pt}}
\put(694.0,256.0){\rule[-0.200pt]{4.818pt}{0.400pt}}
\put(694.0,282.0){\rule[-0.200pt]{4.818pt}{0.400pt}}
\put(786.0,300.0){\rule[-0.200pt]{0.400pt}{7.709pt}}
\put(776.0,300.0){\rule[-0.200pt]{4.818pt}{0.400pt}}
\put(776.0,332.0){\rule[-0.200pt]{4.818pt}{0.400pt}}
\put(827.0,322.0){\rule[-0.200pt]{0.400pt}{15.418pt}}
\put(817.0,322.0){\rule[-0.200pt]{4.818pt}{0.400pt}}
\put(817.0,386.0){\rule[-0.200pt]{4.818pt}{0.400pt}}
\put(868.0,544.0){\rule[-0.200pt]{0.400pt}{11.804pt}}
\put(858.0,544.0){\rule[-0.200pt]{4.818pt}{0.400pt}}
\put(858.0,593.0){\rule[-0.200pt]{4.818pt}{0.400pt}}
\put(908.0,661.0){\rule[-0.200pt]{0.400pt}{4.818pt}}
\put(898.0,661.0){\rule[-0.200pt]{4.818pt}{0.400pt}}
\put(704,269){\circle*{18}}
\put(786,316){\circle*{18}}
\put(827,354){\circle*{18}}
\put(868,569){\circle*{18}}
\put(908,671){\circle*{18}}
\put(898.0,681.0){\rule[-0.200pt]{4.818pt}{0.400pt}}
\end{picture}

%% file: plot_l2.tex
\setlength{\unitlength}{0.240900pt}
\ifx\plotpoint\undefined\newsavebox{\plotpoint}\fi
\sbox{\plotpoint}{\rule[-0.200pt]{0.400pt}{0.400pt}}%
\begin{picture}(1650,944)(0,0)
\font\gnuplot=cmr10 at 12pt
\gnuplot
\sbox{\plotpoint}{\rule[-0.200pt]{0.400pt}{0.400pt}}%
\put(350.0,150.0){\rule[-0.200pt]{4.818pt}{0.400pt}}
\put(325,150){\makebox(0,0)[r]{\ \ {$0$}}}
\put(1555.0,150.0){\rule[-0.200pt]{4.818pt}{0.400pt}}
\put(350.0,309.0){\rule[-0.200pt]{4.818pt}{0.400pt}}
\put(325,309){\makebox(0,0)[r]{\ \ {$0.2$}}}
\put(1555.0,309.0){\rule[-0.200pt]{4.818pt}{0.400pt}}
\put(350.0,468.0){\rule[-0.200pt]{4.818pt}{0.400pt}}
\put(325,468){\makebox(0,0)[r]{\ \ {$0.4$}}}
\put(1555.0,468.0){\rule[-0.200pt]{4.818pt}{0.400pt}}
\put(350.0,626.0){\rule[-0.200pt]{4.818pt}{0.400pt}}
\put(325,626){\makebox(0,0)[r]{\ \ {$0.6$}}}
\put(1555.0,626.0){\rule[-0.200pt]{4.818pt}{0.400pt}}
\put(350.0,785.0){\rule[-0.200pt]{4.818pt}{0.400pt}}
\put(325,785){\makebox(0,0)[r]{\ \ {$0.8$}}}
\put(1555.0,785.0){\rule[-0.200pt]{4.818pt}{0.400pt}}
\put(350.0,944.0){\rule[-0.200pt]{4.818pt}{0.400pt}}
\put(325,944){\makebox(0,0)[r]{\ \ {$1$}}}
\put(1555.0,944.0){\rule[-0.200pt]{4.818pt}{0.400pt}}
\put(350.0,150.0){\rule[-0.200pt]{0.400pt}{4.818pt}}
\put(350,100){\makebox(0,0){\ {$0$}}}
\put(350.0,924.0){\rule[-0.200pt]{0.400pt}{4.818pt}}
\put(503.0,150.0){\rule[-0.200pt]{0.400pt}{4.818pt}}
\put(503,100){\makebox(0,0){\ {$25$}}}
\put(503.0,924.0){\rule[-0.200pt]{0.400pt}{4.818pt}}
\put(656.0,150.0){\rule[-0.200pt]{0.400pt}{4.818pt}}
\put(656,100){\makebox(0,0){\ {$50$}}}
\put(656.0,924.0){\rule[-0.200pt]{0.400pt}{4.818pt}}
\put(809.0,150.0){\rule[-0.200pt]{0.400pt}{4.818pt}}
\put(809,100){\makebox(0,0){\ {$75$}}}
\put(809.0,924.0){\rule[-0.200pt]{0.400pt}{4.818pt}}
\put(963.0,150.0){\rule[-0.200pt]{0.400pt}{4.818pt}}
\put(963,100){\makebox(0,0){\ {$100$}}}
\put(963.0,924.0){\rule[-0.200pt]{0.400pt}{4.818pt}}
\put(1116.0,150.0){\rule[-0.200pt]{0.400pt}{4.818pt}}
\put(1116,100){\makebox(0,0){\ {$125$}}}
\put(1116.0,924.0){\rule[-0.200pt]{0.400pt}{4.818pt}}
\put(1269.0,150.0){\rule[-0.200pt]{0.400pt}{4.818pt}}
\put(1269,100){\makebox(0,0){\ {$150$}}}
\put(1269.0,924.0){\rule[-0.200pt]{0.400pt}{4.818pt}}
\put(1422.0,150.0){\rule[-0.200pt]{0.400pt}{4.818pt}}
\put(1422,100){\makebox(0,0){\ {$175$}}}
\put(1422.0,924.0){\rule[-0.200pt]{0.400pt}{4.818pt}}
\put(1575.0,150.0){\rule[-0.200pt]{0.400pt}{4.818pt}}
\put(1575,100){\makebox(0,0){\ {$200$}}}
\put(1575.0,924.0){\rule[-0.200pt]{0.400pt}{4.818pt}}
\put(350.0,150.0){\rule[-0.200pt]{295.102pt}{0.400pt}}
\put(1575.0,150.0){\rule[-0.200pt]{0.400pt}{191.275pt}}
\put(350.0,944.0){\rule[-0.200pt]{295.102pt}{0.400pt}}
\put(150,647){\makebox(0,0){\Large{$\langle | {\bar l}_2 |\rangle$}}}
\put(912,25){\makebox(0,0){\large{$\gamma$}}}
\put(350.0,150.0){\rule[-0.200pt]{0.400pt}{191.275pt}}
\put(361,194){\usebox{\plotpoint}}
\put(351.0,194.0){\rule[-0.200pt]{4.818pt}{0.400pt}}
\put(351.0,194.0){\rule[-0.200pt]{4.818pt}{0.400pt}}
\put(456.0,234.0){\rule[-0.200pt]{0.400pt}{1.927pt}}
\put(446.0,234.0){\rule[-0.200pt]{4.818pt}{0.400pt}}
\put(446.0,242.0){\rule[-0.200pt]{4.818pt}{0.400pt}}
\put(563.0,251.0){\rule[-0.200pt]{0.400pt}{3.854pt}}
\put(553.0,251.0){\rule[-0.200pt]{4.818pt}{0.400pt}}
\put(553.0,267.0){\rule[-0.200pt]{4.818pt}{0.400pt}}
\put(669.0,263.0){\rule[-0.200pt]{0.400pt}{5.782pt}}
\put(659.0,263.0){\rule[-0.200pt]{4.818pt}{0.400pt}}
\put(659.0,287.0){\rule[-0.200pt]{4.818pt}{0.400pt}}
\put(775.0,307.0){\rule[-0.200pt]{0.400pt}{6.504pt}}
\put(765.0,307.0){\rule[-0.200pt]{4.818pt}{0.400pt}}
\put(765.0,334.0){\rule[-0.200pt]{4.818pt}{0.400pt}}
\put(882.0,289.0){\rule[-0.200pt]{0.400pt}{4.577pt}}
\put(872.0,289.0){\rule[-0.200pt]{4.818pt}{0.400pt}}
\put(872.0,308.0){\rule[-0.200pt]{4.818pt}{0.400pt}}
\put(988.0,385.0){\rule[-0.200pt]{0.400pt}{7.468pt}}
\put(978.0,385.0){\rule[-0.200pt]{4.818pt}{0.400pt}}
\put(978.0,416.0){\rule[-0.200pt]{4.818pt}{0.400pt}}
\put(1094.0,507.0){\rule[-0.200pt]{0.400pt}{9.154pt}}
\put(1084.0,507.0){\rule[-0.200pt]{4.818pt}{0.400pt}}
\put(1084.0,545.0){\rule[-0.200pt]{4.818pt}{0.400pt}}
\put(1201.0,390.0){\rule[-0.200pt]{0.400pt}{8.672pt}}
\put(1191.0,390.0){\rule[-0.200pt]{4.818pt}{0.400pt}}
\put(1191.0,426.0){\rule[-0.200pt]{4.818pt}{0.400pt}}
\put(1307.0,479.0){\rule[-0.200pt]{0.400pt}{15.899pt}}
\put(1297.0,479.0){\rule[-0.200pt]{4.818pt}{0.400pt}}
\put(1297.0,545.0){\rule[-0.200pt]{4.818pt}{0.400pt}}
\put(1413.0,649.0){\rule[-0.200pt]{0.400pt}{5.300pt}}
\put(1403.0,649.0){\rule[-0.200pt]{4.818pt}{0.400pt}}
\put(1403.0,671.0){\rule[-0.200pt]{4.818pt}{0.400pt}}
\put(1520.0,682.0){\rule[-0.200pt]{0.400pt}{7.227pt}}
\put(1510.0,682.0){\rule[-0.200pt]{4.818pt}{0.400pt}}
\put(361,194){\circle{18}}
\put(456,238){\circle{18}}
\put(563,259){\circle{18}}
\put(669,275){\circle{18}}
\put(775,320){\circle{18}}
\put(882,299){\circle{18}}
\put(988,401){\circle{18}}
\put(1094,526){\circle{18}}
\put(1201,408){\circle{18}}
\put(1307,512){\circle{18}}
\put(1413,660){\circle{18}}
\put(1520,697){\circle{18}}
\put(1510.0,712.0){\rule[-0.200pt]{4.818pt}{0.400pt}}
\put(371.0,180.0){\usebox{\plotpoint}}
\put(361.0,180.0){\rule[-0.200pt]{4.818pt}{0.400pt}}
\put(361.0,181.0){\rule[-0.200pt]{4.818pt}{0.400pt}}
\put(456.0,188.0){\rule[-0.200pt]{0.400pt}{0.964pt}}
\put(446.0,188.0){\rule[-0.200pt]{4.818pt}{0.400pt}}
\put(446.0,192.0){\rule[-0.200pt]{4.818pt}{0.400pt}}
\put(563.0,206.0){\rule[-0.200pt]{0.400pt}{2.409pt}}
\put(553.0,206.0){\rule[-0.200pt]{4.818pt}{0.400pt}}
\put(553.0,216.0){\rule[-0.200pt]{4.818pt}{0.400pt}}
\put(775.0,202.0){\rule[-0.200pt]{0.400pt}{2.650pt}}
\put(765.0,202.0){\rule[-0.200pt]{4.818pt}{0.400pt}}
\put(765.0,213.0){\rule[-0.200pt]{4.818pt}{0.400pt}}
\put(988.0,200.0){\rule[-0.200pt]{0.400pt}{1.686pt}}
\put(978.0,200.0){\rule[-0.200pt]{4.818pt}{0.400pt}}
\put(978.0,207.0){\rule[-0.200pt]{4.818pt}{0.400pt}}
\put(1068.0,248.0){\rule[-0.200pt]{0.400pt}{3.854pt}}
\put(1058.0,248.0){\rule[-0.200pt]{4.818pt}{0.400pt}}
\put(1058.0,264.0){\rule[-0.200pt]{4.818pt}{0.400pt}}
\put(1148.0,256.0){\rule[-0.200pt]{0.400pt}{4.818pt}}
\put(1138.0,256.0){\rule[-0.200pt]{4.818pt}{0.400pt}}
\put(1138.0,276.0){\rule[-0.200pt]{4.818pt}{0.400pt}}
\put(1201.0,319.0){\rule[-0.200pt]{0.400pt}{6.504pt}}
\put(1191.0,319.0){\rule[-0.200pt]{4.818pt}{0.400pt}}
\put(1191.0,346.0){\rule[-0.200pt]{4.818pt}{0.400pt}}
\put(1307.0,259.0){\rule[-0.200pt]{0.400pt}{7.950pt}}
\put(1297.0,259.0){\rule[-0.200pt]{4.818pt}{0.400pt}}
\put(1297.0,292.0){\rule[-0.200pt]{4.818pt}{0.400pt}}
\put(1413.0,557.0){\rule[-0.200pt]{0.400pt}{6.504pt}}
\put(1403.0,557.0){\rule[-0.200pt]{4.818pt}{0.400pt}}
\put(1403.0,584.0){\rule[-0.200pt]{4.818pt}{0.400pt}}
\put(1520.0,609.0){\rule[-0.200pt]{0.400pt}{3.373pt}}
\put(1510.0,609.0){\rule[-0.200pt]{4.818pt}{0.400pt}}
\put(371,181){\circle*{18}}
\put(456,190){\circle*{18}}
\put(563,211){\circle*{18}}
\put(775,207){\circle*{18}}
\put(988,204){\circle*{18}}
\put(1068,256){\circle*{18}}
\put(1148,266){\circle*{18}}
\put(1201,333){\circle*{18}}
\put(1307,275){\circle*{18}}
\put(1413,571){\circle*{18}}
\put(1520,616){\circle*{18}}
\put(1510.0,623.0){\rule[-0.200pt]{4.818pt}{0.400pt}}
\end{picture}

%% file: plot_histl2b.tex
\setlength{\unitlength}{0.240900pt}
\ifx\plotpoint\undefined\newsavebox{\plotpoint}\fi
\sbox{\plotpoint}{\rule[-0.200pt]{0.400pt}{0.400pt}}%
\begin{picture}(1650,944)(0,0)
\font\gnuplot=cmr10 at 12pt
\gnuplot
\sbox{\plotpoint}{\rule[-0.200pt]{0.400pt}{0.400pt}}%
\put(350.0,150.0){\rule[-0.200pt]{4.818pt}{0.400pt}}
\put(325,150){\makebox(0,0)[r]{\ \ {$0$}}}
\put(1555.0,150.0){\rule[-0.200pt]{4.818pt}{0.400pt}}
\put(350.0,263.0){\rule[-0.200pt]{4.818pt}{0.400pt}}
\put(325,263){\makebox(0,0)[r]{\ \ {$50$}}}
\put(1555.0,263.0){\rule[-0.200pt]{4.818pt}{0.400pt}}
\put(350.0,377.0){\rule[-0.200pt]{4.818pt}{0.400pt}}
\put(325,377){\makebox(0,0)[r]{\ \ {$100$}}}
\put(1555.0,377.0){\rule[-0.200pt]{4.818pt}{0.400pt}}
\put(350.0,490.0){\rule[-0.200pt]{4.818pt}{0.400pt}}
\put(325,490){\makebox(0,0)[r]{\ \ {$150$}}}
\put(1555.0,490.0){\rule[-0.200pt]{4.818pt}{0.400pt}}
\put(350.0,604.0){\rule[-0.200pt]{4.818pt}{0.400pt}}
\put(325,604){\makebox(0,0)[r]{\ \ {$200$}}}
\put(1555.0,604.0){\rule[-0.200pt]{4.818pt}{0.400pt}}
\put(350.0,717.0){\rule[-0.200pt]{4.818pt}{0.400pt}}
\put(325,717){\makebox(0,0)[r]{\ \ {$250$}}}
\put(1555.0,717.0){\rule[-0.200pt]{4.818pt}{0.400pt}}
\put(350.0,831.0){\rule[-0.200pt]{4.818pt}{0.400pt}}
\put(325,831){\makebox(0,0)[r]{\ \ {$300$}}}
\put(1555.0,831.0){\rule[-0.200pt]{4.818pt}{0.400pt}}
\put(350.0,944.0){\rule[-0.200pt]{4.818pt}{0.400pt}}
\put(325,944){\makebox(0,0)[r]{\ \ {$350$}}}
\put(1555.0,944.0){\rule[-0.200pt]{4.818pt}{0.400pt}}
\put(350.0,150.0){\rule[-0.200pt]{0.400pt}{4.818pt}}
\put(350,100){\makebox(0,0){\ {$0$}}}
\put(350.0,924.0){\rule[-0.200pt]{0.400pt}{4.818pt}}
\put(656.0,150.0){\rule[-0.200pt]{0.400pt}{4.818pt}}
\put(656,100){\makebox(0,0){\ {$0.1$}}}
\put(656.0,924.0){\rule[-0.200pt]{0.400pt}{4.818pt}}
\put(963.0,150.0){\rule[-0.200pt]{0.400pt}{4.818pt}}
\put(963,100){\makebox(0,0){\ {$0.2$}}}
\put(963.0,924.0){\rule[-0.200pt]{0.400pt}{4.818pt}}
\put(1269.0,150.0){\rule[-0.200pt]{0.400pt}{4.818pt}}
\put(1269,100){\makebox(0,0){\ {$0.3$}}}
\put(1269.0,924.0){\rule[-0.200pt]{0.400pt}{4.818pt}}
\put(1575.0,150.0){\rule[-0.200pt]{0.400pt}{4.818pt}}
\put(1575,100){\makebox(0,0){\ {$0.4$}}}
\put(1575.0,924.0){\rule[-0.200pt]{0.400pt}{4.818pt}}
\put(350.0,150.0){\rule[-0.200pt]{295.102pt}{0.400pt}}
\put(1575.0,150.0){\rule[-0.200pt]{0.400pt}{191.275pt}}
\put(350.0,944.0){\rule[-0.200pt]{295.102pt}{0.400pt}}
\put(150,647){\makebox(0,0){\Large{N}}}
\put(912,25){\makebox(0,0){\large{$|{\bar l}_2|$}}}
\put(350.0,150.0){\rule[-0.200pt]{0.400pt}{191.275pt}}
\put(363,320){\circle*{18}}
\put(383,436){\circle*{18}}
\put(403,279){\circle*{18}}
\put(423,422){\circle*{18}}
\put(443,601){\circle*{18}}
\put(463,633){\circle*{18}}
\put(483,631){\circle*{18}}
\put(503,556){\circle*{18}}
\put(523,524){\circle*{18}}
\put(543,413){\circle*{18}}
\put(563,347){\circle*{18}}
\put(583,336){\circle*{18}}
\put(603,443){\circle*{18}}
\put(623,411){\circle*{18}}
\put(643,293){\circle*{18}}
\put(663,252){\circle*{18}}
\put(683,232){\circle*{18}}
\put(703,179){\circle*{18}}
\put(723,164){\circle*{18}}
\put(743,159){\circle*{18}}
\put(763,218){\circle*{18}}
\put(783,295){\circle*{18}}
\put(803,202){\circle*{18}}
\put(823,195){\circle*{18}}
\put(843,234){\circle*{18}}
\put(863,189){\circle*{18}}
\put(882,266){\circle*{18}}
\put(902,454){\circle*{18}}
\put(922,379){\circle*{18}}
\put(942,449){\circle*{18}}
\put(962,363){\circle*{18}}
\put(982,663){\circle*{18}}
\put(1002,744){\circle*{18}}
\put(1022,552){\circle*{18}}
\put(1042,576){\circle*{18}}
\put(1062,656){\circle*{18}}
\put(1082,536){\circle*{18}}
\put(1102,456){\circle*{18}}
\put(1122,354){\circle*{18}}
\put(1142,191){\circle*{18}}
\put(1162,207){\circle*{18}}
\put(1182,336){\circle*{18}}
\put(1202,381){\circle*{18}}
\put(1222,474){\circle*{18}}
\put(1242,388){\circle*{18}}
\put(1262,354){\circle*{18}}
\put(1282,313){\circle*{18}}
\put(1302,297){\circle*{18}}
\put(1322,241){\circle*{18}}
\put(1342,243){\circle*{18}}
\put(1362,152){\circle*{18}}
\end{picture}

%% file: version4.bbl
\begin{thebibliography}{99}






\bibitem{GW}
D. Gross and E. Witten,
Phys. Rev. D 21 (1980) 446.

\bibitem{NNlat05}
R. Narayanan and H. Neuberger,
{\it Plenary Talk at Lattice 2005},
hep-lat/0509014.

\bibitem{Campo98}
M. Campostrini,
Nucl. Phys. Proc. Suppl. 73 (1999) 724
[hep-lat/9809072].

\bibitem{OxG01}
B. Lucini and M. Teper,
JHEP  0106 (2001) 050 [hep-lat/0103027].

\bibitem{OxT05}
B. Lucini, M. Teper and U. Wenger,
JHEP 0502 (2005) 033 [hep-lat/0502003].

\bibitem{OxT03}
B. Lucini, M. Teper and U. Wenger,
JHEP 0401 (2004) 061 [hep-lat/0307017].

\bibitem{NN03}
J. Kiskis, R. Narayanan and H. Neuberger,
Phys.Lett. B574 (2003) 65 [hep-lat/0308033]. \\
R. Narayanan and H. Neuberger,
Phys. Rev. Lett. 91 (2003) 081601 [hep-lat/0303023].

\bibitem{NNect04}
R. Narayanan and H. Neuberger,
in {\it Large $N_c$ QCD 2004} (World Scientific, Ed. Goity et. al.)
hep-lat/0501031.

\bibitem{DurOle}
B. Durhuus and P. Olesen,
Nucl. Phys. B184 (1981) 461.

\bibitem{BasGriVian}
A. Bassetto, L. Griguolo and F. Vian,
Nucl. Phys. B559 (1999) 563 [hep-th/9906125].

\bibitem{MTect04}
M. Teper,
in {\it Large $N_c$ QCD 2004} (World Scientific, Ed. Goity et. al.)
[hep-th/0412005].

\bibitem{thooft}
G. 't Hooft, Nucl. Phys. B72 (1974) 461, B75 (1974) 461.

\bibitem{Nd3}
M.Teper,
Phys. Rev. D59 (1999) 014512 [hep-lat/9804008]. \\
B. Lucini and M.Teper,
Phys. Rev. D66 (2002) 097502 [hep-lat/0206027].

\bibitem{OxG04}
B. Lucini, M. Teper and U. Wenger,
JHEP 0406 (2004) 012 [hep-lat/0404008].

\bibitem{BurTepVai}
F. Bursa, M.Teper and H. Vairinhos,
PoS (LAT2005) 282 [hep-lat/0509092] and work in progress.

\bibitem{Creutz}
M. Creutz, {\it Quarks, gluons and lattices} (CUP, 1983).

\bibitem{SmixedD4N2}
G. Bhanot and M. Creutz,
Phys. Rev. D 24 (1981) 3212. \\
R.W.B. Ardill, M. Creutz and K.J.M. Moriarty
J. Phys. G 10 (1984) 867. \\
P. Stephenson,
hep-lat/9509070,  hep-lat/9604008.

\bibitem{SmixedD4N3}
T. Blum et al,
Nucl. Phys. B442 (1995) 301 [hep-lat/9412038].

\bibitem{SmixedMono}
R. C. Brower, D. A. Kessler and H. Levine,
Phys. Rev. Lett. 47 (1981) 621. \\
R. C. Brower, D. A. Kessler, H. Levine, M. Nauenberg and T. Schalk,
Phys. Rev. D26 (1982) 959. \\
I. G. Halliday and A. Schwimmer,
Phys. Lett. B101 (1981) 327; B102 (1981) 337. \\
L. Caneschi, I. G. Halliday and A. Schwimmer,
Nucl. Phys. B200 (1982) 409;  Phys. Lett. B117 (1982) 427.


\bibitem{d2beta_a}
Tien-lun Chen, Chung-I Tan and Xi-te Zheng,
Phys. Lett. B109 (1982) 383.


\bibitem{Z_N2+1d}
G. Bhanot and M. Creutz,
Phys. Rev. D 21 (1980) 2892.\\
P. Provero and S. Vinti,
Mod. Phys. Lett. A6 (1991) 157.

\bibitem{BaigCuervo}
M. Baig and A. Cuervo
Nucl. Phys. Proc. Suppl. 4 (1988) 21.

\bibitem{dimred4}
K. Kajantie, M. Laine, K. Rummukainen and M. Shaposhnikov,
Nucl. Phys. B503 (1997) 357
[hep-ph/9704416]. \\
M. Laine and Y. Schroder,
JHEP 0503 (2005) 067
[hep-ph/0503061].

\bibitem{TcD3}
Ph. de Forcrand and O. Jahn,
hep-lat/0309153. \\
K. Holland,
hep-lat/0509041.\\
J. Liddle and M. Teper,
hep-lat/0509082.

\bibitem{dimred3}
P.Bialas, A. Morel, B.Petersson, K. Petrov and T.Reisz,
Nucl. Phys. B581 (2000) 477
[hep-lat/0003004]; Nucl. Phys. B603 (2001) 369
[hep-lat/0012019]. \\
P.Bialas, A. Morel and B.Petersson,
Prog. Theor. Phys. Suppl. 153 (2004) 220
[hep-lat/0403020].

\bibitem{aharony}
O. Aharony, J. Marsano, S. Minwalla, K. Papadodimas, M. van Raamsdook
and T. Wiseman, hep-th/0508077

\bibitem{largeN}
S. Coleman, 1979 Erice Lectures. \\
E. Witten, {\it Nucl. Phys.} {\bf B160} (1979) 57. \\
A. Manohar, 1997 Les Houches Lectures, hep-ph/9802419. \\ 
Y. Makeenko, hep-th/0001047. \\
G. 't Hooft, in {\it Large N QCD}, Ed. R. F. Lebed 
(World Scientific, 2002)  (hep-th/0204069).

\bibitem{RMT}
M. Stephanov, J. Verbaarschot and T. Wettig,
hep-ph/0509286.

\bibitem{sigkref}
T. Bhattacharya, A. Gocksch, C. Korthals Altes and R. Pisarski,
Phys. Rev. Lett. 66(1991) 998; Nucl. Phys. B383 (1992) 497
(hep-ph/9205231). \\
P. Giovannangeli and C. P. Korthals Altes,
Nucl. Phys. B608 (2001) 203 
(hep-ph/0102022).\\
P. Giovannangeli and C. P. Korthals Altes,
Nucl. Phys. B608 (2001) 203 
(hep-ph/0102022); hep-ph/0412322. \\
C. Korthals Altes, A. Michels, M. Stephanov and M. Teper,
Phys. Rev. D55 (1997) 1047 
(hep-lat/9606021). 

\bibitem{NN06}
R. Narayanan and H. Neuberger,
hep-th/0601210.

\end{thebibliography}
